\documentclass{mn2e}
\usepackage{times}
\usepackage{amssymb}
\usepackage{epsfig}
\usepackage{lscape}

\title[BLAST 250${\bf \rm \mu m}$ galaxies in GOODS-South]{The BLAST 250${\rm {\bf \mu m}}$-selected galaxy population in GOODS-South}

\author[J.S.~Dunlop, et al.]
{J.\,S. Dunlop$^{1,2}$\thanks{Email: jsd@roe.ac.uk}, 
P.\,A.\,R.~Ade$^{3}$, J.\,J.~Bock$^{4}$, E.\,L.~Chapin$^{2}$, M.~Cirasuolo$^1$, 
K.\,E.\,K.~Coppin$^5$, \and M.\,J.~Devlin$^{6}$, M.~Griffin$^{3}$, T.\,R.~Greve$^7$, 
J.\,O.~Gundersen$^{8}$, M.~Halpern$^{2}$, P.\,C.~Hargrave$^{3}$, \and D.\,H.~Hughes$^{9}$, 
R.\,J.~Ivison$^{10}$, J.~Klein$^{6}$, A.~Kovacs$^{11}$, G.~Marsden$^{2}$, P.~Mauskopf$^{3}$, \and 
C.\,B.~Netterfield$^{12,13}$, L.~Olmi$^{14,15}$, E.~Pascale$^{3}$, G.~Patanchon$^{16}$, M.~Rex$^{6}$, 
D.~Scott$^{2}$, \and C.~Semisch$^{6}$, I.~Smail$^5$, T.\,A.~Targett$^{2}$, 
N.~Thomas$^{12}$, M.\,D.\,P.~Truch$^{6}$, C.~Tucker$^{4}$, \and G.\,S.~Tucker$^{17}$, 
M.\,P.~Viero$^{12}$, F.~Walter$^{7}$, J.\,L.~Wardlow$^{5}$, A.~Weiss$^{11}$, D.\,V.~Wiebe$^{2}$\\
\footnotesize\\
$^{1}$ SUPA\thanks{Scottish Universities Physics Alliance}, 
Institute for Astronomy, University of Edinburgh, 
Royal Observatory, Edinburgh, EH9 3HJ, UK\\
$^{2}$Department of Physics \& Astronomy, University of British Columbia, 6224 Agricultural Road, Vancouver, BC V6T 1Z1, Canada\\
$^{3}$School of Physics \& Astronomy, Cardiff University, 5 The Parade, Cardiff, CF24 3AA, UK.\\
$^{4}$Jet Propulsion Laboratory, Pasadena, CA 91109-8099, USA.\\
$^{5}$Institute for Computational Cosmology, Durham University, South Road, Durham, DH1 3LE, UK.\\
$^{6}$Department of Physics \& Astronomy, University of Pennsylvania, 209 South 33rd Street, Philadelphia, PA, 19104, USA.\\
$^{7}$Max-Planck Institut fur Astronomie, Konigstuhl 17, 69117 Heidelberg, Germany.\\
$^{8}$Department of Physics, University of Miami, 1320 Campo Sano Drive, Coral Gables, FL 33146, USA.\\
$^{9}$Instituto Nacional de Astrof\'isica \'Optica y Electr\'onica (INAOE), Aptdo. Postal 51 y 72000 Puebla, Mexico.\\
$^{10}$UK ATC, Royal Observatory, Edinburgh, EH9 3HJ, UK\\
$^{11}$Max-Planck Institut fur Radioastronomy, Auf dem Hugel 69, 53121 Bonn, Germany.\\
$^{12}$Department of Astronomy \& Astrophysics, University of Toronto, 50 St. George Street Toronto, ON M5S~3H4, Canada.\\
$^{13}$Department of Physics, University of Toronto, 60 St. George Street, Toronto, ON M5S~1A7, Canada.\\
$^{14}$University of Puerto Rico, Rio Piedras Campus, Physics Dept., Box 23343, UPR station, Puerto Rico.\\
$^{15}$IRA-INAF, Largo E. Fermi 5, I-50125, Firenze, Italy.\\
$^{16}$Laboratoire APC, 10, rue Alice Domon et L{\'e}onie Duquet 75205 Paris, France.\\
$^{17}$Department of Physics, Brown University, 182 Hope Street, Providence, RI 02912, USA.}

\date{Accepted ... ; Received ... ; in original form ...}

\pagerange{000--000}

\begin{document}
\hsize=6truein

\maketitle

\vspace*{-0.1cm}

\begin{abstract}
We identify and investigate the nature of the 20 brightest 250\,${\rm \mu m}$ sources detected by
the Balloon-borne Large Aperture Submillimetre Telescope (BLAST) 
within the central 150\,arcmin$^2$ of the GOODS-South field. Aided by the available deep VLA 1.4\,GHz radio imaging, 
reaching $S_{\rm 1.4} \simeq 40\,{\rm \mu Jy}$ (4-$\sigma$), we have identified 
radio counterparts for 17/20 of the 250\,${\rm \mu m}$ sources. The resulting enhanced 
positional accuracy of $\simeq 1$\,arcsec has then allowed us to exploit the deep optical ({\it HST}), 
near-infrared (VLT) and mid-infrared ({\it Spitzer}) imaging of GOODS-South to establish secure galaxy 
counterparts for the 17 radio-identified sources, and plausible galaxy candidates for the 
3 radio-unidentified sources. 
Confusion is a serious issue for this deep BLAST 250\,${\rm \mu m}$ survey, due to the large size of the beam.
Nevertheless, we argue that our chosen counterparts are significant, and often dominant contributors to the measured BLAST flux densities.
For all of these 20 galaxies we 
have been able to determine spectroscopic (8) or photometric (12) redshifts. The result 
is the first near-complete redshift distribution for a deep 250\,${\rm \mu m}$-selected 
galaxy sample. This reveals that 250\,${\rm \mu m}$ surveys reaching detection limits 
of $\simeq 40$\,mJy have a median redshift $z \simeq 1$, and contain not only low-redshift spirals/LIRGs, but also the extreme $z \simeq 2$ dust-enshrouded starburst galaxies 
previously discovered at sub-millimetre wavelengths. Inspection of the LABOCA 870\,${\rm \mu m}$ imaging 
of GOODS-South yields detections of $\simeq 1/3$ of the proposed BLAST sources (all at $z > 1.5$), and reveals 250/870\,${\rm \mu m}$ 
flux-density ratios consistent  
with a standard 40\,K modified black-body fit with a dust emissivity index $\beta = 1.5$.
Based on their IRAC colours, we find that virtually all of the BLAST galaxy identifications 
appear better described as analogues of the M82 starburst galaxy, or Sc star-forming discs rather than 
highly obscured ULIRGs. This is perhaps as expected at low redshift, where the 250\,$\mu m$ BLAST 
selection function is biased towards spectral energy distributions which peak longward of 
$\lambda_{rest} = 100\, {\rm \mu m}$. However, it also appears largely true at $z \simeq 2$. 
 
\end{abstract}

\begin{keywords}
galaxies: active -- galaxies: starburst -- galaxies: photometry -- galaxies: fundamental parameters -- infrared: galaxies.
\end{keywords}

\vspace*{6cm}

%%%%%%%%%%%%%%%%%%%%%%%%%%%%%%%%%%%%%%%%%%%%%%%%%%%%%%%%%%%%%%%%%%%%%%%%%%%%
\section{INTRODUCTION}
\label{INTRO}
%%%%%%%%%%%%%%%%%%%%%%%%%%%%%%%%%%%%%%%%%%%%%%%%%%%%%%%%%%%%%%%%%%%%%%%%%%%%

The observed far-infrared background peaks around a wavelength  $\lambda \simeq 200-250\,{\rm \mu m}$ (Puget et al.~1996; Fixsen et al.~1998). 
Over the past 10 years, deep extragalactic surveys have 
closed in on this wavelength regime from both shorter and longer wavelengths, and there has been dramatic progress in uncovering populations of sources which undoubtedly 
{\it contribute} to this background. Specifically, at wavelengths $\simeq 4$ times longer than the peak ($\lambda \simeq 850 - 1200\,{\rm \mu m}$), 
surveys with the Sub-millimetre Common-User Bolometer Array (SCUBA), Mambo, AzTEC and now LABOCA (e.g. Coppin et al., 2006; Bertoldi et al. 2007; 
Weiss et al. 2009; Austermann et al. 2010) have been used to successfully assemble substantial samples of submm-selected extragalactic sources, 
and the spectral energy distributions (SEDs) of these sources are certainly rising steeply towards shorter wavelengths. 
Conversely, at 
wavelengths $\simeq
4$ times shorter than the peak,
the {\it Spitzer Space Telescope} has proved effective at extragalactic source selection up to wavelengths $\lambda \simeq 70\,{\rm
\mu m}$ (Magnelli et al., 2009). But until now, the effective production of deep extragalactic samples in the central wavelength range $\lambda \simeq 100 - 500\,{\rm
\mu m}$ has been prohibited by the atmosphere and/or limitations on instrument sensitivity and telescope aperture.

This situation should shortly be transformed by the SPIRE instrument on the 3.5-m diameter {\it Herschel Space Observatory} 
(Griffin et al.~2007). However, a powerful first insight into the nature of the
$250\,{\rm \mu m}$-selected galaxy population is already being provided by the
the Balloon-borne Large Aperture Submillimetre Telescope (BLAST). 

BLAST is a 1.8\,m diameter stratospheric balloon telescope that operates at an altitude of approximately 35\,km, above most of the atmospheric water 
vapour which essentially prohibits effective far-infrared observations from the ground.
BLAST is thus half the size of {\it Herschel}, and it is equipped
with a camera which is a prototype of the SPIRE camera, enabling simultaneous broad-band imaging with central wavelengths 
of 250, 350 and 500\,${\rm \mu m}$. In 2006, BLAST undertook an 11-day flight from Antarctica, during which it executed an 
observing programme which included the first deep far-infrared imaging survey ever undertaken within the Extended {\it Chandra} Deep Field South (ECDFS).

This unique survey (described in more detail in Section 2) has already been the subject of several investigations. First results on the far-infrared 
number counts and the resolution of the background were presented by Devlin et al. (2009), with more detailed investigations of these 
two key topics being presented by Patanchon et al. (2009) and Marsden et al. (2009). Pascale et al. (2009) used the survey to constrain
cosmic star-formation history, while Viero et al. (2009) studied correlations in the background. Dye et al. (2009) and Ivison et al. (2010) 
used the supporting radio (VLA) and mid-infrared {\it Spitzer} imaging of the field to attempt to identify countparts to the brighter BLAST 
sources, and to explore the far-infrared/radio correlation in distant galaxies.

The study by Dye et al. (2009) showed that the BLAST beam, at least at $250\, {\rm \mu m}$, is still small enough 
to allow the successful identification of a reasonable fraction of BLAST sources with radio and mid-infrared counterparts. However, the supporting data 
over the full 9\,deg$^2$ BLAST survey area are of variable quality, and Dye et al. attempted to secure identifications for sources selected at 
all 3 BLAST wavelengths. As a result, they succeeded in identifying counterparts for only $\simeq 55$\% of the sources in the BLAST catalogues 
detected at $> 5\,\sigma$ in at least one waveband. This inevitably limits the conclusions that can be drawn about the redshift distribution
of the BLAST source population, although Dye et al. concluded that $\simeq 75$\% of the sources in the shallow+deep catalogue lie at $z < 1$.

The aim of the study presented here is to establish a near-complete redshift distribution 
for a subset of the BLAST $250\, {\rm \mu m}$ sources selected down to a fainter flux-density limit of $S_{250} 
\simeq 40$\,mJy, and subsequently to establish the basic physical properties
(stellar mass, size, morphology, star-formation history, AGN content) of the galaxies which host the observed $250\, {\rm \mu m}$ emission.
To try to achieve this, we have deliberately confined our attention to the small sub-area of the BLAST map 
which contains the very best optical ({\it HST} ACS), near-infrared (VLT ISAAC) and mid-infrared ({\it Spitzer}) imaging data, and 
the highest density of optical spectroscopic redshifts. This area, the $\simeq 150$\,arcmin$^2$ GOODS-South field (Dickinson et al. 2004), is also where 
the radio (VLA) imaging is most sensitive, and it has been mapped in its entirety at $870\, {\rm \mu m}$ by LABOCA (Weiss et al. 2009), and at 
1.1\,mm by AzTEC (Scott et al. 2010). GOODS-South is also the chosen location for some of the first deep high-resolution near-infrared imaging currently being undertaken
with WFC3 on the refurbished {\it HST}, and is the future target of planned deep sub-mm and far-infrared imaging with SCUBA2 (Holland et al. 2006) 
and SPIRE+PACS on {\it Herschel}.

Thus, by confining our attention to GOODS-South we can explore just what can be achieved given the best possible 
supporting data. Conversely, we can also establish what level of supporting data is actually {\it required} for an effective, complete 
study of the future, larger, 
$250\, {\rm \mu m}$-selected galaxy samples which will be produced 
by {\it Herschel}. In addition, the wealth of existing information 
on other galaxy populations in this field (see Section 2) facilitates the comparison of BLAST galaxies with reference samples of 
field galaxies selected at similar redshifts and/or stellar masses, and to explore the frequency 
of mergers. Finally, the recent acquisition of deep sub-millimetre and millimetre wavelength imaging in the same field allows a 
first exploration of how the $250\, {\rm \mu m}$ and $870/1100\, {\rm \mu m}$ populations are related. 

To assemble a useful sample of $250\, {\rm \mu m}$ sources within this field, we have pushed 
the BLAST source selection significance threshold to $\sim 3.5\sigma$. However, the resulting sample is still sufficiently
small (20 sources) that we can afford to explore the properties of individual sources in some detail. As a result this 
work should be regarded as a pilot study of the mix of source populations which can be uncovered in a deep $250\, {\rm \mu m}$
survey, and an exploration of how best to overcome the problems that are encountered in trying to identify and study the faint 
far-infrared sources in maps which are inevitably highly confused (note that the {\it Herschel} beam at $500\, {\rm \mu m}$ is essentially 
the same size as the BLAST beam at $250\, {\rm \mu m}$). The broader statistical robustness of the results presented
here is clearly limited both by small number statistics, and by cosmic variance.

The layout of the paper is as follows. In Section 2 we describe the selection and robustness of the BLAST $250\, {\rm \mu m}$ sub-sample within 
the central region of GOODS-South, and then summarize the wealth of available multi-frequency imaging and spectroscopy in the field. In Section 3 we 
describe the process of obtaining radio and optical counterparts 
for the BLAST sources, and highlight the care required to overcome the problems associated with source blending
in the confused BLAST maps. In Section 4 we provide or derive redshifts for all of the sources, including secondary/alternative counterparts.
Then, in Section 5, we present multi-colour images of the BLAST galaxies and provide notes on each of the individual sources.
Finally, in Section 6 we review our results and discuss the robustness of our conclusions. In particular 
we present and discuss the implications of the striking associations found between the BLAST $250\, {\rm \mu m}$ sources
and the LABOCA  $870\, {\rm \mu m}$ sources recently uncovered by Weiss et al. (2009). We also discuss the robustness of our derived redshift distribution
for $250\, {\rm \mu m}$-selected sources, and briefly compare the mid-infrared colours of BLAST galaxies with 
those of other known galaxies at both high and low redshifts.
Our conclusions are summarized in Section 7.

%%%%%%%%%%%%%%%%%%%%%%%%%%%%%%%%%%%%%%%%%%%%%%%%%%%%%%%%%%%%%%%%%%%%%%%%%%%%
\section{DATA} 
\label{DATA}
%%%%%%%%%%%%%%%%%%%%%%%%%%%%%%%%%%%%%%%%%%%%%%%%%%%%%%%%%%%%%%%%%%%%%%%%%%%%
\begin{figure*}
\begin{tabular}{ll}
\includegraphics[width=0.48\textwidth]{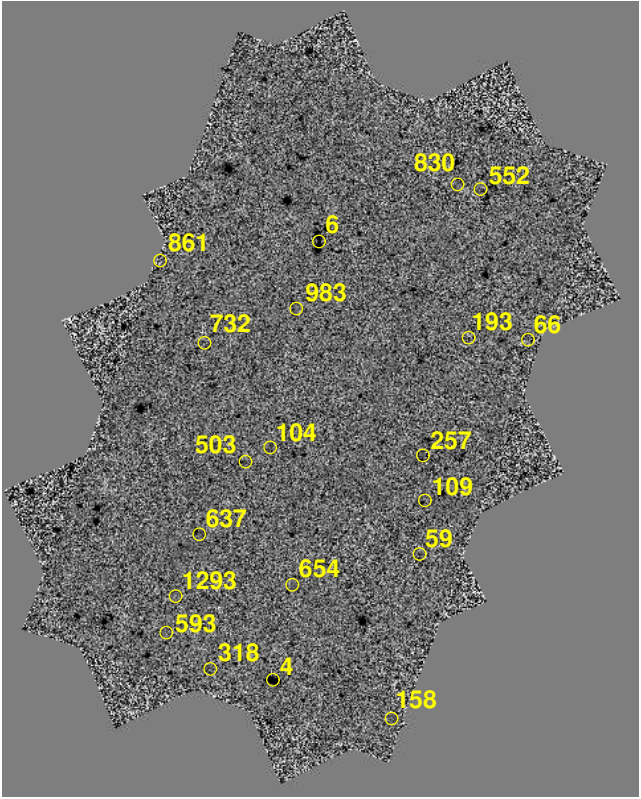} &
\includegraphics[width=0.48\textwidth]{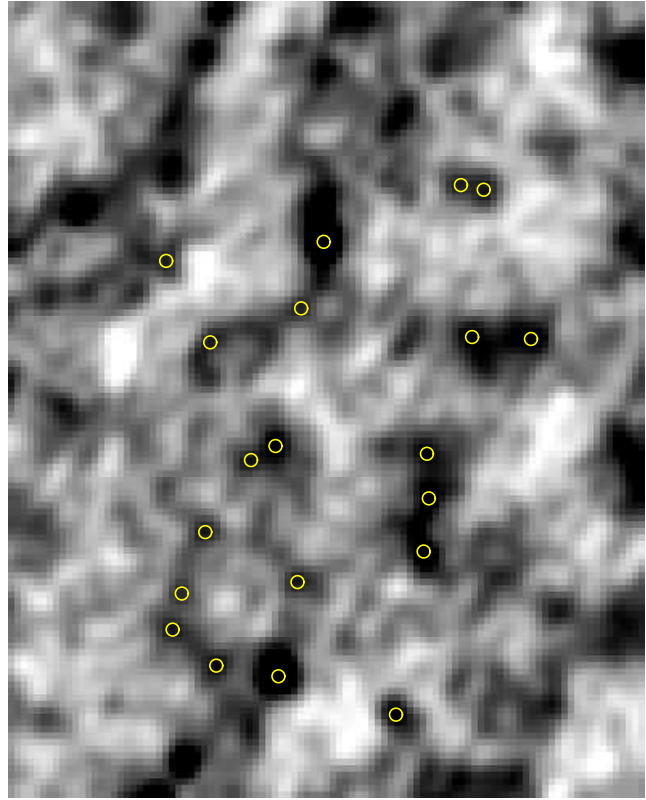}\\
\end{tabular}
\caption{The location of the 20 BLAST 250\,${\rm \mu m}$ sources in GOODS-South 
superimposed on the {\it HST} ACS $z$-band imaging mosaic of the field (LH plot) and on 
a greyscale of the BLAST 250\,${\rm \mu m}$ map itself (RH plot). 
Sources are numbered by BLAST source ID (from the full BLAST map), as given in Table 1, with higher numbers indicating lower 250\,${\rm \mu m}$ 
flux densities. The circles have a radius of 
15\,arcsec, which is the radius adopted for the search for radio and 
optical counterparts to the BLAST sources. The actual BLAST beam 
has a FWHM of 36\,arcsec at 250\,${\rm \mu m}$}.
\end{figure*}

\subsection{The BLAST 250\,${\rm {\bf \mu m}}$ source sub-sample}

The 250\,${\rm  \mu m}$ source sample selected for study here represents a small subset 
of the far-infrared (250\,${\rm \mu m}$, 350\,${\rm \mu m}$, 500\,${\rm \mu m}$) sample uncovered in 
the ECDFS by the BLAST Antarctic flight of December 2006
(Devlin et al. 2009). An area of 8.7\,${\rm deg^2}$ centred on ${\rm RA\, 03^h \, 32^m}$, ${\rm Dec\, -28^{\circ}\, 12^{\prime}}$ (J2000)
was mapped by BLAST to mean sensitivities of $\sigma_{250}~=~36\,$mJy, $\sigma_{350}~=~30\,$mJy, and $\sigma_{500}~=~20\,$mJy per beam.
Within this `BLAST GOODS South Wide' (BGS-Wide) map, a deeper image, reaching $\sigma_{250} = 11\,$mJy, $\sigma_{350} = 9\,$mJy, and $\sigma_{500} = 6\,$mJy per beam
was made of the central 0.8$\,{\rm deg^2}$ centred on the southern field of the Great Observatories Origins Deep Survey (GOODS; Dickinson 
et al. 2003) at ${\rm RA\, 03^h \, 32^m \, 30^s}$, ${\rm Dec\, -28^{\circ}\, 48^{\prime}}$ (J2000). It is from the core of
this `BLAST GOODS South Deep' (BGS-Deep) map that we have extracted a `complete' sub-sample of the BLAST 250\,${\rm \mu m}$ sources
which lie within the sub-region with the very best supporting multi-frequency data. As shown in Fig. 1, we have 
chosen to define this area as the $\simeq 150$ sq. arcmin region of sky covered by the GOODS-South {\it HST} ACS 4-band optical imaging (Giavalisco et al. 2004).

Because our primary aim here is to study the optical/infrared galaxy counterparts of the BLAST sources, we have confined our attention
to  250\,${\rm \mu m}$ sources, because the larger BLAST beam sizes at 350 and 500\,${\rm \mu m}$ make it even more difficult 
to reliably identify all but the brightest longer-wavelength sources (the 
BLAST06 beams are best fit by Gaussians with FWHM of 36, 42, and 60\,arcsec at 
250\,${\rm \mu m}$, 350\,${\rm \mu m}$ and 500\,${\rm \mu m}$ respectively; Pascale
et al. 2008). Our study thus differs from, and complements that of Dye et al. (2009), who sought identifications
for all 351 BLAST sources detected at $>5\sigma$ in any one of the three BLAST wavebands across the whole of BLAST GOODS South survey. 
We have also extracted a significantly deeper sub-sample of sources for study, selecting 
all 250\,${\rm \mu m}$ sources which lie within the region shown in Fig. 1 and have a signal:noise ratio greater than 3.25.
Pushing the BLAST data in this way allows us to construct a sample of useful size from within the relatively small area
covered by the highest quality supporting datasets, and also allows a first investigation of the nature of the fainter 250\,${\rm \mu m}$
source population which will be studied in detail by {\it Herschel}.

Imposing this signal:noise cut produced a sample of 24 sources. We chose this significance threshold because, 
below this level, there are only 9 sources with signal:noise $>$ 2, and only three of these were reproduced at $> 2\,\sigma$ in an alternative map reduction (and all three
at $< 2.8\sigma$). In other words, this threshold 
appears to mark a clear (not unexpected) point below which source reliability degenerates markedly, at least in this 
particular region of the BLAST map. 

Of these 24 sources, we then rejected three
(176, 228, 777) as inappropriate for this study because their nominal positions moved by greater than 20 arcsec between different map reductions. This does not necessarily
mean these sources are not real peaks in the map (they range in S/N from 5$\sigma$ to 3.5$\sigma$), but simply that their positions
appear to be too sensitive to the precise choice of map-production/source-extraction parameters to allow reliable determination of the most likely radio or optical counterpart
(possibly because they are extreme examples of confusion from a number of fainter sources).
There certainly appears to be 
something different about these sources because all other sources in our sample moved by less than 5 arcsec between alternative map reductions.
Finally, we also rejected the second least significant source (1158), because it was not found at $> 2\, \sigma$ in the alternative map reduction.

This leaves a `clean' sample of 20 sources for further study. The 250\,${\rm \mu m}$ positions of the BLAST 
sources are indicated in Fig.~1, and tabulated in Table 1, along with their  250\,${\rm \mu m}$ flux densities and signal:noise ratios. 
Given that the 4 rejected sources displayed a range of 250${\rm \mu m}$ signal:noise, the remaining 
sample can be regarded as representative down to a flux-density detection limit of $S_{250} \simeq 35\, {\rm mJy}$ in the zero-mean map ($S_{250} \simeq 45\, {\rm mJy}$
relative to the local background -- see below).
The mean signal:noise ratio for these 20 sources is 5.6, but this is biased by the brightest 2 or 3 sources. 
The median signal:noise is 4.2, and this number is adopted as representative where appropriate later in this paper. 

Before searching for identifications, it is worth pausing to consider the robustness of the far-infrared sources listed in Table 1.

The 250\,${\rm \mu m}$ flux densities given in Table 1 are the `raw' values as extracted from the BLAST map via convolution 
with the relevant noise-weighted point-spread function (Truch et al 2009), effectively a weighted fit of the beam to
the raw unsmoothed data to derive the best (maximum likelihood) estimate of the flux density of a point source in the map (e.g. Serjeant et al. 2003).
Due to the steepness of the 250\,${\rm \mu m}$ source counts, and the large BLAST beam, 
the flux densities of the individual sources (which would be seen in a higher resolution map) will undoubtedly have been {\it boosted} by the effects 
of confusion (i.e. flux contributions from fainter sources) and by Eddington
bias (a combination of noise and steep source counts), especially for the lower signal:noise detections 
(Scott et al. 2002; Coppin et al. 2006; Pantachon et al. 2009). Reading fluxes from the zero-mean map 
can, however, partially offset the flux contributions from fainter confused sources, which in effect constitute the local background for any 
brighter source in the map. However, detailed simulations are clearly 
required to establish robust statistical corrections for sources at a given flux density (Patanchon et al. 2009).

The other issue to consider is the noise. The flux-density noise levels quoted in Table 1 include only the instrumental noise which, within this deep
section of the map, is typically $\sigma_{250} \simeq 11$\,mJy per beam. However, analysis of the pixel flux-density distribution within the relevant 
150\,arcmin$^2$ region under study here, shows that the spatial noise in the map is $\sigma_{250} \simeq 15$\,mJy per beam (as measured from a Gaussian 
fit to the negative tail of the flux-density distribution). This implies that the confusion noise is comparable to the instrumental noise,
at $\sigma_{250} \simeq 10$\,mJy per beam. The situation is therefore similar to that encountered with the deepest 850\,${\rm \mu m}$ maps made of the Hubble Deep Field with
SCUBA on the JCMT (Hughes et al. 1998; Peacock et al. 2000).

It might therefore seem appropriate to increase the noise level to $\sigma_{250} \simeq 15$\,mJy per beam when assessing the significance of our sources in the actual
map. However, if we choose to do this, for consistency we must also adopt the local background in the map which examination shows to be typically $-11$\,mJy per beam. This is 
not surprising, as the mode of the distribution of flux-density values in the zero-mean map lies at $-12$\,mJy per beam. Thus, performing aperture photometry 
on the map typically increases flux-densities by $11$\,mJy per beam, and boosts the assumed noise level to  $\sigma_{250} \simeq 15$\,mJy per beam. 
Adoption of these two corrections reduces only slightly the formal significance level of the sources listed in Table 1, and still leaves our faintest source 
above the 3$\sigma$ threshold.

Given these complications, in this paper we make little direct use of the 250\,${\rm  \mu m}$ flux densities, other than 
to discuss the use of 250/870\,${\rm \mu m}$ flux-density ratio as a consistency check on derived source redshift (see Section 6.1).
We also do not attempt to exploit the 350 or 500\,${\rm \mu m}$ measurements of our 250\,${\rm \mu m}$-selected sources, as the larger beams, and relatively
low sensitivity (at least to 250\,${\rm \mu m}$-selected sources) of BLAST at these longer wavelengths prohibits the extraction of usefully accurate flux densities.
We thus do not quote 350 or 500\,${\rm \mu m}$ flux densities for the sources in Table 1 (although values for the brighter sources can be found
in Dye et al. 2009). We do, however, check, on a source-by-source basis, for detections or non-detections at these longer wavelengths, as described in Section 5.
Finally, we note that as there are $\simeq 500$ BLAST 250\,${\rm \mu m}$ beams in the 150 arcmin$^2$ area under study here, we would expect $\simeq 0.5$ non-existent 
sources to clear our adopted 3.3$\sigma$ threshold purely by chance. Relative to the 250\,${\rm \mu m}$ map mode of $-12$\,mJy, the MUSIC region under study here
contains one 3$\sigma$ negative peak. We thus conclude that, while flux boosting may be substantial for individual sources, our adopted signal:noise threshold
is reasonable.

\begin{table*}
\begin{tabular}{crccrr}
\hline
BLAST SOURCE NAME& BLAST & RA$_{250}$ & DEC$_{250}$ & S/N & $S_{250}$\\
(IAU)&ID\phantom{ST} & (J2000) & (J2000) & (250) & /mJy \\
\hline
BLAST J033235$-$275530 (${\rm 250\,\mu m}$) &4\phantom{-1}     &  53.146220  &    $-$27.925272  &  16.2 & 177 \\ 
BLAST J033229$-$274414 (${\rm 250\,\mu m}$) &6\phantom{-1}       &  53.123900  &  $-$27.737326  &  14.4 & 157 \\ 
BLAST J033218$-$275216 (${\rm 250\,\mu m}$) &59\phantom{-1}      &  53.075139  &  $-$27.871343  &  6.7  & 74 \\ 
BLAST J033205$-$274645 (${\rm 250\,\mu m}$) &66\phantom{-1}      &  53.022647  &  $-$27.779397  &  6.6  & 72 \\ 
BLAST J033235$-$274932 (${\rm 250\,\mu m}$) &104\phantom{-1}     &  53.147540  &  $-$27.825642  &  6.0  & 65 \\ 
BLAST J033217$-$275054 (${\rm 250\,\mu m}$) &109\phantom{-1}     &  53.072586  &  $-$27.848349  &  5.9  & 64 \\ 
BLAST J033221$-$275630 (${\rm 250\,\mu m}$) &158\phantom{-1}     &  53.088715  &  $-$27.941911  &  5.2  & 58 \\ 
BLAST J033212$-$274642 (${\rm 250\,\mu m}$) &193\phantom{-1}     &  53.051434  &  $-$27.778541  &  4.9  & 54 \\ 
BLAST J033217$-$274944 (${\rm 250\,\mu m}$) &257\phantom{-1}     &  53.073511  &  $-$27.829046  &  4.6  & 51 \\ 
BLAST J033242$-$275514 (${\rm 250\,\mu m}$) &318\phantom{-1}     &  53.176619  &  $-$27.920633  &  4.4  & 48 \\ 
BLAST J033238$-$274954 (${\rm 250\,\mu m}$) &503\phantom{-1}     &  53.159514  &  $-$27.831733  &  4.0  & 44 \\ 
BLAST J033213$-$274302 (${\rm 250\,\mu m}$) &552\phantom{-1}     &  53.054825  &  $-$27.717303  &  3.9  & 43 \\ 
BLAST J033247$-$275418 (${\rm 250\,\mu m}$) &593\phantom{-1}     &  53.197956  &  $-$27.905028  &  3.9  & 43 \\ 
BLAST J033243$-$275146 (${\rm 250\,\mu m}$) &637\phantom{-1}     &  53.181931  &  $-$27.862863  &  3.8  & 42 \\ 
BLAST J033232$-$275304 (${\rm 250\,\mu m}$) &654\phantom{-1}     &  53.136852  &  $-$27.884531  &  3.8  & 42 \\ 
BLAST J033243$-$274650 (${\rm 250\,\mu m}$) &732\phantom{-1}     &  53.179377  &  $-$27.780768  &  3.7  & 40 \\ 
BLAST J033213$-$274246 (${\rm 250\,\mu m}$) &830\phantom{-1}     &  53.056853  &  $-$27.712811  &  3.6  & 40 \\ 
BLAST J033248$-$274443 (${\rm 250\,\mu m}$) &861\phantom{-1}     &  53.200852  &  $-$27.745501  &  3.6  & 39 \\ 
BLAST J033232$-$274558 (${\rm 250\,\mu m}$) &983\phantom{-1}     &  53.134915  &  $-$27.766112  &  3.4  & 38 \\ 
BLAST J033246$-$275321 (${\rm 250\,\mu m}$) &1293\phantom{-1}    &  53.193438  &  $-$27.889410  &  3.3  & 36 \\ 
\hline
\end{tabular}
\caption{The BLAST 250\,${\rm \mu m}$-selected sample in GOODS-South. The positions are derived from optimal beam 
fitting to the BLAST map. $S_{250}$ is the flux density derived from the zero-mean BLAST map.}
\end{table*}

\subsection{Supporting multi-frequency data}

\subsubsection{Radio: VLA 1.4-GHz imaging}
As demonstrated by the follow-up of sub-millimetre surveys with SCUBA, very deep VLA imaging is necessary (and frequently sufficient) for 
the successful identification of the galaxy counterparts of sources detected with the large beams 
of current far-infrared/sub-millimetre facilities (e.g. Ivison et al. 2002, Ivison et al. 2007). 
This works for three reasons. First, star-forming galaxies produce copious quantities of synchrotron emission. 
Second, even in the deepest available radio maps, 1.4\,GHz sources still have sufficiently low surface-density that positional 
coincidences within a `reasonable' search radius are generally statistically rare. Third, if a robust radio counterpart is found,
the $\simeq 1\,$arcsec positional accuracy provided by the VLA at $1.4\,$GHz essentially always yields an unambiguous optical/infrared
galaxy counterpart for further study.

Very deep ($\sigma_{1.4} \simeq 7.5\,{\rm \mu Jy}$),
high-resolution 1.4\,GHz imaging of GOODS-South is now available at the centre of the ECDFS  
as described by Kellermann et al. (2008) and Miller et al. (2008). As discussed by Dye et al. (2009), the published 
source catalogue adopts a very conservative detection threshold of 
7$\sigma$. We therefore re-analysed the image of Miller et al. (2008) using the techniques described by Ibar et al. (2009)
to create a radio catalogue down to a $4\,\sigma$ limit of 
30\,${\rm \mu Jy}$ (at which depth the cumulative source density on the sky is $\simeq 0.8{\, \rm arcmin^{-2}}$).
This catalogue was then searched for potential radio counterparts to the BLAST sources  using the method described 
below in Section 3. 

In cases where the radio emission appears heavily resolved (e.g. for the brightest
source, BLAST 4, where at first sight there appear to be 6 individual VLA candidate counterparts) we measured the total radio flux density using
{\sc TVSTAT} within {\it AIPS}. We checked these values against the flux densities measured with the 16-arcsec beam of the Australia Telescope Compact Array (ACTA) (Norris et al. 2006),
and found them to be in good agreement for the 4 sources which proved sufficiently bright to have also been detected at 1.4\,GHz by ACTA (Afonso et al. 2006).

\subsubsection{Optical: HST ACS imaging}
Deep optical imaging over the 150\,arcmin$^2$ area shown in Fig.~1 has been obtained with the
Advanced Camera for Surveys (ACS) on board {\it HST} in 4 different filters: $F435W\,(B_{435})$, $F606W\,(V_{606})$, $F775W\,(i_{775})$ and 
$F850LP\,(z_{850})$. These data were taken as part of GOODS (Giavalisco et al. 2004) and reach 5\,$\sigma$ limiting 
(AB) magnitudes (within a 1\,arcsec diameter aperture) of 26.9, 26.9, 26.1, and 25.8, respectively. 
By definition, all 20 of the BLAST sources lie within the area covered by the GOODS-South {\it HST} ACS optical imaging.
One of the BLAST sources (BLAST 732) lies within the even deeper ACS imaging provided by the Hubble Ultra Deep Field
(HUDF: Beckwith et al. 2006), over an area of 11\,arcmin$^2$ centred on
 ${\rm RA\, 03^h \, 32^m \, 39s}$, ${\rm Dec\, -27^{\circ}\, 47^{\prime}\, 29.1^{\prime \prime}}$ (J2000).

\subsubsection{Near-infrared data: VLT and HST imaging}
Deep near-infrared ($J,H,K$) imaging of almost all of the {\it HST} ACS field illustrated
in Fig. 1 has now been completed with the ISAAC camera on the VLT (Retzlaff et al. 2010).
This imaging covers $\simeq 143\,{\rm arcmin^2}$, of which $\simeq 136\,{\rm arcmin^2}$  
overlaps with the {\it HST} ACS imaging. All but 2 of the BLAST sources under study here lie within
the area covered by the ISAAC images.

This near-infrared imaging is of excellent quality, with the FWHM of the PSF having a median value of 
$\simeq 0.5\,$arcsec (it varies from $\simeq 0.37$ to $\simeq 0.7\,$ arcsec across the field; Bouwens et al. 2008).
Partly as a result of this good image quality, the point source sensitivity of this imaging is very deep,
reaching 5$\,\sigma$ detection levels of $J \simeq 25$, and $H,K_s \simeq 24.2$ (AB magnitudes).

As described by Bouwens et al. (2008), various {\it HST} NICMOS $J_{110}$ and $H_{160}$ imaging programs have been 
undertaken in GOODS-South. However, the coverage offered by this imaging is too patchy to be 
of much use in the present study, and frustratingly (but not unexpectedly) no BLAST source lies within the 5.8\,arcmin$^2$ ultra-deep
NICMOS sub-image of the HUDF undertaken by Thompson et al. (2005). However, BLAST 732 does lie within the new Wide Field Camera 3
(WFC3) $Y_{105}$, $J_{125}$, $H_{160}$ imaging of the HUDF, which reaches a 5-$\sigma$ limiting magnitude (AB) of 29 in all three
near-infrared bands (McLure et al. 2010; Oesch et al. 2010; Bouwens et al. 2010).   

\subsubsection{Mid-infrared: Spitzer imaging}
Again as part of GOODS, ultra-deep {\it Spitzer} imaging with the Infrared Array Camera (IRAC: Fazio et al. 2004) has been obtained 
over the whole of the ACS GOODS-South field illustrated in Fig.~1 (proposal ID 194, Dickinson et al., in preparation), in 
all 4 IRAC channels (3.6, 4.5, 5.6, and 8.0\,${\rm \mu m}$).
The IRAC 5$\,\sigma$ detection limits are $S_{3.6} \simeq 25.9$, $S_{4.5} \simeq 25.5$, $S_{5.6} \simeq 23.3$, $S_{8.0} \simeq 22.9$
(AB magnitudes). 

The 24\,${\rm \mu m}$ {\it Spitzer} MIPS data originally obtained as part of GOODS has 
been augmented and incorporated within the {\it Spitzer} Far-Infrared Deep Extragalactic 
Legacy (FIDEL)\footnote{PI M. Dickinson, see {\tt http://www.noao.edu/noao/fidel/}} survey (Magnelli et al. 2009), and reach  5-$\sigma$ detection limits of
$S_{24} \simeq 30{\, \rm \mu Jy}$.

\subsubsection{Sub-mm: LABOCA 870\,${\rm \mu m}$ survey}

The full $30 \times 30$\,arcmin ECDFS has recently been mapped at a wavelength of 
870\,${\rm \mu m}$ by the LABOCA ECDFS Sub-millimetre Survey (LESS) (Weiss et al. 2009).
The 12-m diameter 
of the APEX telescope (Gusten et al. 2006) delivers a 19.2-arcsec FWHM beam.
The LESS image has a uniform depth of $\sigma_{870} \simeq 1.2 {\rm \, mJy\,beam^{-1}}$ (as measured in the map, and hence including 
confusion noise) and the positions and flux densities
of the LABOCA 
sources have been determined in a similar manner to that described above for the BLAST source extraction (Weiss et al. 2009).
 
The combined analysis of the BLAST and LABOCA LESS source lists and maps in GOODS-South is the subject of a major study which will be presented elsewhere
(Chapin et al. 2010).
Here we have confined our attention to checking which of the 20 BLAST 250\,${\rm \mu m}$ sources in Fig. 1 are also present
in the LESS 870\,${\rm \mu m}$ catalogue presented by Weiss et al. (2009), and briefly exploring the implications
of the detections and non-detections (see Section 6.1).
 
\subsubsection{Spectroscopic redshifts}
In recent years the GOODS-South field has been the target of a series of spectroscopic campaigns
(Dickinson et al. 2004, Stanway et al. 2004, Strolger et al. 2004, van der Wel et al. 2004 
Szokoly et al. 2004, Le Fevre et al. 2004, 2005, Mignoli et al. 2005, Doherty et al. 2005,
Roche et al. 2006, Ravikumar et al. 2007, Vanzella et al. 2009, Tayler et al. 2009, Eales et al. 2009). Of particular potential importance  
for the present study has been the completion of the ESO/GOODS VLT/FORS2 programme in the GOODS-South field
(Vanzella et al. 2005, 2006, 2008). 

The combined impact of these programs is that over 1000 galaxies in the GOODS-South field shown in Fig. 1 now possess reliable spectroscopic redshifts.
As a result, and unsurprisingly, it transpires that all except one of the BLAST galaxy counterparts which we identify below with a low-redshift
(i.e. $z < 1$) galaxy already possesses a known spectroscopic redshift.

However, what is perhaps surprising, especially given the work of Vanzella et al., is that only one of the potential higher-redshift BLAST
galaxy counterparts currently has a spectroscopic redshift (BLAST 1293, $z = 1.382$; Vanzella et al. 2008). As explored further below,
this is primarily a consequence of the fact that the high-redshift BLAST galaxy counterparts are too faint/red for optical redshifts to be determined, even 
in the VLT/FORS2 campaign.
The exploitation of the aforementioned multi-waveband photometry for the production of photometric redshift estimates thus
remains important for this and future work on the study of sources selected at far-infrared/sub-millimetre wavelengths.

\subsubsection{Photometric redshifts}
Several photometric redshift catalogues have now been published for the ECDFS in general and for the central GOODS-South field
in particular.  

The COMBO 17 project (Wolf et al. 2004, 2008) covers the whole $30 \times 30$\,arcmin area of the ECDFS, and exploits 
17-band (5 wide and 12 narrow) optical imaging obtained with the Wide Field Imager (WFI) at the ESO 2.2-m telescope. 
As explained by Wolf et al. (2004, 2008) these `very low resolution spectra' 
allow solid ($\sigma_z \simeq 0.02-0.05$) photometric redshifts to be obtained down to a magnitude limit 
of $R \simeq 23$. While the COMBO 17 catalogue contains estimated redshifts reaching $z \simeq 2$, 
in practice the depth of the optical imaging, and lack of near-infrared information limits its usefulness 
to $z < 1$ where the optical spectroscopy in GOODS-South is very complete. Consequently, a 
COMBO 17 redshift was only adopted for one BLAST source (BLAST 257-1, $z = 0.689$).

Over the $30 \times 30$\,arcmin ECDFS, photometric redshift catalogues based on the available 
broad-band optical+near-infrared data have been produced by 
the MUSYC (Multiwavelength Survey by Yale-Chile) collaboration (e.g. Taylor et al. 2009), and (for 24~${\rm \mu m}$
sources) by Rowan-Robinson et al. (2008). However, of more importance for the current study are 
the photometric redshift catalogues assembled for the deep central $\simeq 150\, {\rm arcmin^2}$ area by the 
GOODS-MUSIC project (Grazian et al. 2006), by the alternative analyses of Caputi et al.
(2006) and Dunlop, Cirasuolo \& McLure (2007), and most recently by Brammer et al. (2008), applying the {\sc EAZY} algorithm to the 
FIREWORKS data (Wuyts et al. 2008). 

We utilised results from these studies but, as described in Section 4, we also decided to derive new redshift 
estimates for each of the BLAST source candidate identifications. This was partly necessary simply because not all of the sources  
feature in existing photo-$z$ catalogues. However, it also allowed us to check the impact 
of deriving photometry through different apertures, and to explore the effect on $z_{phot}$ and $\delta z_{phot}$ of allowing extreme
values of extinction $A_V$ (see Dunlop, Cirasuolo \& McLure 2007).

\begin{table*}
\begin{tabular}{rrrccrrll}
\hline
BLAST & $d_{B-ID}$    & VLA   & RA$_{1.4}$ &  DEC$_{1.4}$  & $S_{1.4}$     & S/N & \phantom{00}$P$ & Notes\\
ID\phantom{ST}   & /arcsec  & ID    & (J2000)    &  (J2000)      &/${\rm \mu Jy}$&(1.4)& &\\
\hline
4\phantom{-2}     &  1.5     &  499  &  53.146221  &  $-$27.925692  &  680   &  20.0  & 0.0002\\ 
6-1   & 10.8      &  72   &  53.124515  &  $-$27.740279  &  1100  &  40.9 & 0.004\\
(6-2   &  9.5     &  100  &  53.124986  &  $-$27.734862  &  458   &  26.4 & 0.008)\\ 
59-1  & 12.2      &  423  &  53.079433  &  $-$27.870691  &  91   &  7.9  & 0.063\\
(59-2  & 13.5     &  240  &  53.071577  &  $-$27.872494  & 102   &  12.3 & 0.065)\\
66\phantom{-2}    &  7.2     &  290  &  53.020439  &  $-$27.779829  &  126   &  10.2 & 0.021\\
104\phantom{-2}   &  10.2    &  370  &  53.150748  &  $-$27.825553  &  89   &  8.5  & 0.050\\
109\phantom{-2}   &  7.6     &  531  &  53.074414  &  $-$27.849717  &  86   &  6.6  & 0.035\\ 
158\phantom{-2}   &  8.6     &  987  &  53.090100  &  $-$27.939860  &  38   &  4.6  & 0.087\\
193\phantom{-2}   &  7.2     &       &  53.053608  &  $-$27.778025  & $<40$ &       &       & Opt/IR ID\phantom{)}\\
257\phantom{-2}   &  10.4    &  585  &  53.075206  &  $-$27.831523  &  46   &  6.2  & 0.092\\
318\phantom{-2}   &  10.8    &  361  &  53.180015  &  $-$27.920681  &  92   &  8.8  & 0.053\\
503-1 &   9.7    &  132  &  53.157190  &  $-$27.833468  &  170  &  21.0 & 0.025\\
(503-2 &   7.4    &  315  &  53.161745  &  $-$27.832355  &  65   &  9.7  & 0.044)\\
552\phantom{-2}   &   8.9    &  932  &  53.052277  &  $-$27.718325  &   33  & 4.7   & 0.099\\
593\phantom{-2}   &  6.6     &  677  &  53.199965  &  $-$27.904560  &  44   &  5.7  & 0.054\\
637-1 &  5.8     &  110  &  53.183594  &  $-$27.862207  &  290  &  23.9 & 0.006\\ 
(637-2 &  9.6     &  74   &  53.184536  &  $-$27.861512  &  257  &  39.9 & 0.015)\\
654\phantom{-2}   &  4.1     &       &  53.136577  &  $-$27.885657  & $<40$ &       &       & Opt/IR ID\phantom{)}    \\
732-1 &  13.5    &       &  53.181458  &  $-$27.777472  & $<40$ &       &       & Opt/IR ID\phantom{)}     \\
(732-2 &  5.4     &       &  53.180542  &  $-$27.779686  & $<40$ &       &       & Opt/IR ID)     \\
(732-3 &  9.5     &       &  53.182018  &  $-$27.779537  & $<40$ &       &       & Opt/IR ID)     \\
830-1 &  7.1     &  239  &  53.055171  &  $-$27.711515  &  89   &  12.4 & 0.030\\
(830-2 &  4.2     &  1179 &  53.057837  &  $-$27.713600  &  30   &   4.3 & 0.043)\\
861\phantom{-2}   &  11.9    &  570  &  53.198276  &  $-$27.747894  &  59   &  6.4  & 0.088\\
983\phantom{-2}   &  12.4    &  405  &  53.136761  &  $-$27.769166  &  100   &  8.1  & 0.059\\
1293\phantom{-2}  &  5.2     &  211  &  53.193040  &  $-$27.890814  &  91   &  13.2 & 0.018\\
\hline
      &          &       &             &              &       & $\Sigma P = $      & 0.784 \\
\hline
\end{tabular}
\caption{Galaxy identifications derived from associations with 1.4\,GHz VLA sources (17 objects) or analysis of the 
optical-infrared photometry of possible candidates (3 sources) within a search radius of 15\,arcsec. 6 sources have 
more than one statistically acceptable galaxy counterpart (given in parenthesis). Note that $\Sigma P = $ refers to the sum of the individual
values of $P$ for only the 17 primary (i.e. lower $P$ value) radio identifications.}
\end{table*}

\begin{figure*}
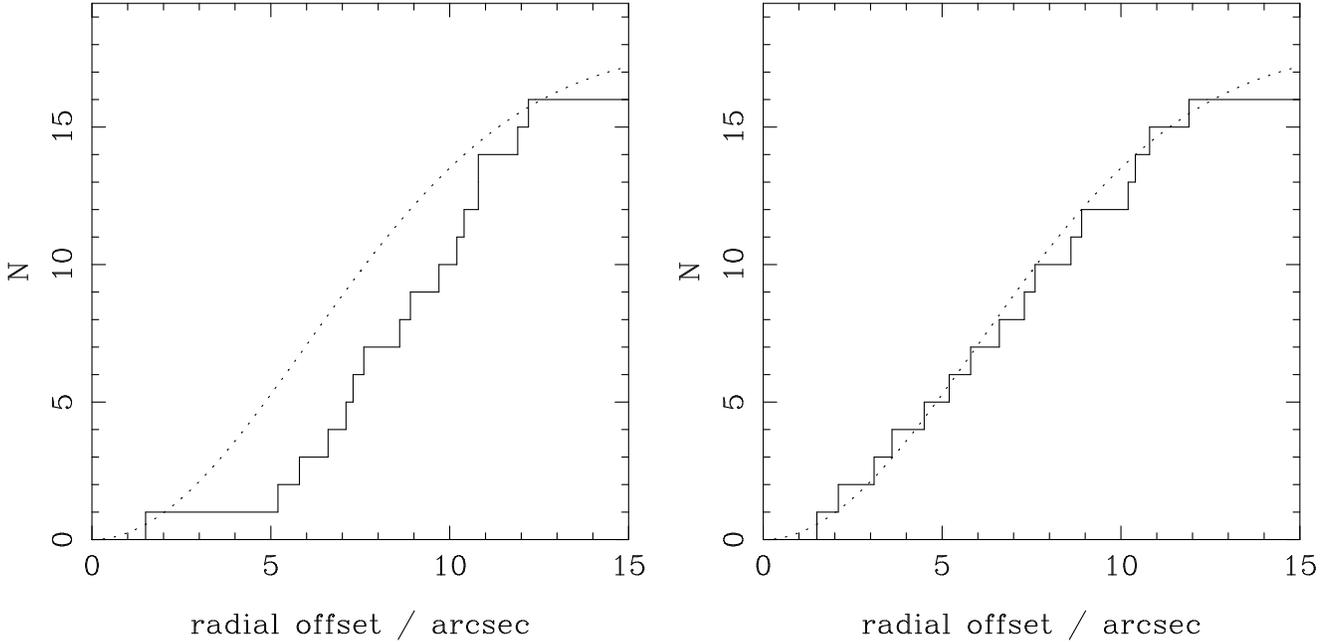

\begin{tabular}{llll}

\includegraphics[width=0.48\textwidth]{offset1.ps}&
\includegraphics[width=0.48\textwidth]{offset2.ps}\\
\end{tabular}
\caption{Left: Cumulative distribution of angular offsets between BLAST 250 and radio positions for the 17 radio identifications, compared with the expected
distribution for 18 sources given the typical deboosted S/N of the BLAST detections and the size of the 250\,${\rm \mu m}$ Gaussian beam (i.e. $\sigma = 6$\,arcsec). 
Right: The equivalent plot after moving the radio position to the 
mean of the radio position of the two alternative IDs for the 4 BLAST sources 6, 59, 503 and 830, which all appear to be potential 2-source
blends within the BLAST beam. Moving 
these 4 sources restores the expected behaviour, showing that the distribution of offsets is as expected.}
\end{figure*}

%%%%%%%%%%%%%%%%%%%%%%%%%%%%%%%%%%%%%%%%%%%%%%%%%%%%%%%%%%%%%%%%%%%%%%%%%%%%
\section{SOURCE IDENTIFICATION}
\label{ID}
%%%%%%%%%%%%%%%%%%%%%%%%%%%%%%%%%%%%%%%%%%%%%%%%%%%%%%%%%%%%%%%%%%%%%%%%%%%%

\subsection{Radio identifications}

There is no doubt that, even with the smallest BLAST beam, (FWHM=36\,arcsec at 250\,${\rm \mu m}$) identifying secure unambiguous 
optical/IR galaxy counterparts to the far-infrared sources is a challenge. It is also clearly the case that some (possibly large) subset of 
the BLAST sources may be the product of the blended far-infrared flux density from two or more galaxies at the same, or differing redshifts.

For some of the brighter BLAST sources identified by Dye et al. (2009) and Ivison et al. (2010), sufficiently bright and rare 24\,${\rm \mu m}$ counterparts 
exist to allow the unambiguous identification of the BLAST sources with {\it Spitzer} mid-infrared sources. However, at the fainter flux densities 
which are the subject of the present study, the sheer number-density of MIPS sources in the deep {\it Spitzer} imaging makes the isolation 
of statistically significant unambiguous mid-infrared counterparts virtually impossible.

We have therefore focussed here on the search for associations with VLA 1.4\,GHz sources which, even at the aforementioned deep flux-density limit of 
$S_{1.4} \simeq 30\, {\rm \mu Jy}$, remain sufficiently rare on the sky ($N(S_{1.4} > 30\, {\rm \mu Jy}) \simeq 0.8\, {\rm arcmin^{-2}}$) for 
random associations to be statistically unlikely.

\subsubsection{Selection of candidate radio counterparts}

We searched for radio-source counterparts within a radius of 15\,arcsec of each BLAST 250\,${\rm \mu m}$ source.
This choice of search radius is not arbitrary, and can be justified in a number of ways. 

First, we note that 
the anticipated positional uncertainty in a source detected with an (undeboosted)  signal:noise ratio, S/N, with a Gaussian beam of FWHM
$\theta$, in the presence of background power-law cumulative number counts of the form $N(> f) \propto f^{-\gamma}$ is given by 
(Ivison et al. 2007):

$$\sigma_{pos} (= \Delta \alpha = \Delta \delta) = \frac{0.6 \theta}{({\rm S/N}^2 - (2\gamma+4))^{1/2}}.$$

\noindent
Into this expression we insert the FWHM of the BLAST 250\,${\rm \mu m}$ beam $\theta = 36\,{\rm arcsec}$, our modal value of S/N = 4.2, 
and an assumed faint count power-law index $\gamma = 1$ (slightly sub-Euclidean
number counts appear appropriate at fainter flux densities, as a double power-law fit transits from $\gamma = 2.5$ to $\gamma = 0.8$ at 
$S_{250} \simeq 30$\,mJy; Patanchon et al. 2009).
The result is $\sigma_{pos} = 6\,{\rm arcsec}$. 15\,arcsec is then the 2.5${\rm \sigma}$ search radius 
which is expected to contain $\simeq 95$\% of all genuine radio counterparts (see Ivison et al. 2007). Note here that the number count 
correction term is applied to effectively deboost the source flux density for Eddington bias, and has the effect of reducing a raw S/N = 4.2 to an effective 
deboosted S/N = 3.4.

Second, while expanding the search radius further might result in one or two additional radio counterparts, the increased number of 
ways in which a counterpart can be selected inevitably weakens the statistical security of associations which could still have been found within
a smaller search radius (see discussion of probability of mis-identification below). From a series of simulations we have established 
that, given the depth of the radio data available in this field, the number of statistically secure radio identifications is optimised 
by adoption of a search radius between 12 and 15 arcsec. Fortunately this number is (just) consistent with the 
2 to 2.5\,$\sigma$ positional uncertainty in the BLAST 250\,${\rm \mu m}$ sources under consideration here. This provides some retrospective
justification for our decision to confine our study of the faint sources to the BLAST 250\,${\rm \mu m}$ catalogue. It 
would not be possible to successfully undertake this type of analysis for $\simeq 4\sigma$ sources if the beam was much bigger than 36 arcsec 
(as is the case, for example, at 500\,${\rm \mu m}$).

Third, after determining the most likely galaxy identifications, one can check that the distribution of angular offsets is consistent  
with that expected given the anticipated positional uncertainty in the BLAST sources. We perform this check below, after careful 
consideration of the confusing effects of source blending.

Finally, we experimented with scaling the search radius with BLAST source S/N, 
but did not find that this significantly affected the 
results. In addition, such refinement of the search radius may place too 
much confidence in the precise accuracy of the positions
of the brighter BLAST sources. This is not to say there is any evidence 
for serious systematic errors in the BLAST positions. Indeed
the positional offset of only 1.5\,arcsec to the galaxy counterpart of the 
brightest BLAST source in the field provides confidence that any additional 
pointing errors in the 
BLAST 250\,${\rm \mu m}$ positions are minimal. 
Rather, this is dangerous 
because, as discussed 
below (see also Ivison et al. 2010), many of the brightest
sources may consist of blends. In this situation, blind 
application of equation (1) can result in statistically robust 
primary and secondary radio identifications being missed because of an 
inadequate search radius (e.g. for BLAST 6, rigorous application of equation
(1) would yield a 2.5$\sigma$ search radius of 3.8\,arcsec, and a failure to
find either of the two contributing galaxies -- see Section 3.1.3).

\subsubsection{Calculation of probability of mis-identification}

Having chosen a candidate radio identification within our 15\,arcsec search radius, we then 
calculate the probability that such a coincidence could have occurred by chance. Following Downes et al. (1986; see also Dunlop
et al. 1989), we calculated the raw Poisson probability that a 
radio source of the observed 1.4\,GHz flux density would be discovered by chance at the measured distance from
the nominal BLAST source position. We then correct this {\it a priori} probability for the number of ways such a statistical 
coincidence could have been discovered given the available search parameter space defined by the maximum search radius
(15\,arcsec), the radio-source number density at the limiting search flux-density ($N(S_{1.4} > 30\, {\rm \mu Jy}) \simeq 0.8\, {\rm arcmin^{-2}}$), and 
the form of the radio source counts over the flux-density 
range of interest (here we adopt a power-law index of $1.4$). For the deep data under study here, this correction
is often substantial, typically increasing $P$ by a factor of a few. This technique has been applied previously
to estimate the robustness of radio identifications for SCUBA sources (e.g. Ivison et al. 2002, 2007), and should yield
similar results to the method adopted by Dye et al. (2009) based on Monte Carlo simulations. 

The value of $P$ thus calculated 
is the probability that the observed association is the result of chance. We stress that even a very low value of $P$ does 
not prove that the radio source {\it is} the BLAST source. Nevertheless, a low value of $P$ clearly does imply that the radio source
is likely related to the BLAST source in some way. This could be true for several reasons. First, the radio source could be {\it the single} 
true counterpart of the BLAST source. Alternatively it could be one of group of 2 or more galaxies 
which contribute to the BLAST flux-density peak (of particular relevance here, given the large beam-size). Third, the statistical 
result could be a consequence of some secondary association (e.g. clustering with the true BLAST sources, or the result of gravitational lensing). 
Despite these worries over statistical interpretation, we re-emphasize that our search for radio counterparts is not motivated purely by their statistical rarity and 
good positional accuracy, but also by the physical evidence that dust-enshrouded star-forming galaxies also 
produce copious quantities of synchrotron emission (resulting in the well-known far-infrared:radio luminosity correlation -- e.g. Ivison
et al. 2010). A sensible hypothesis, therefore, is that an apparently statistically secure radio counterpart to 
a BLAST source is at least a significant contributor to the observed 250\,${\rm \mu m}$ emission.
 
\subsubsection{Results}

The results of this process are tabulated in Table 2. Radio counterparts were found for 17 of the 20 BLAST sources, with multiple counterparts
found for 5 of these 17. Half of the BLAST sources have radio counterparts with $P < 0.05$, and 
all 17 radio counterparts have $P < 0.1$. While it would be nice to insist on a significance threshold of 
$P < 0.05$, we here adopt $P < 0.1$ as the best that can be realistically achieved given the large search radius required by the BLAST beam.
Reassuringly, a sum of the $P$ values for the 17 primary identifications yields only $\Sigma P \simeq 0.8$,
suggesting that $\simeq 1$ source has been misidentified. At the same time our adopted search radius would lead us to expect 
to have missed $\simeq 1$ true radio identification at larger radii, so our failure to find a radio counterpart for 3 BLAST sources is not 
altogether surprising. The technique used to isolate the possible optical counterparts for these 3 remaining BLAST sources (listed in Table 2)
is explained below in subsection 3.2 (although we note here the possibility that 1 or 2 of these 
250\,${\rm \mu m}$ of these sources could be erroneous -- see Section 6).

As a consistency check, in the left-hand panel of Fig. 2, we plot the cumulative distribution of angular offsets between BLAST 250\,${\rm \mu m}$ and radio/optical positions for the 
17 primary identifications listed in Table 2, and compare this with the `expected'
distribution for the afore-mentioned circular Gaussian with $\sigma = 6$\,arcsec (assuming 18 sources in total, to allow for the fact 
our search radius should only include 95\% of identifications). While application of the Kolmogorov-Smirnov (KS) test reveals these distributions
to only differ at slightly less than the 2$\sigma$ level, this comparison has clearly not worked as well as for SCUBA sources (Pope et al. 2006; Ivison et al. 2007). Closer 
inspection of Table 2 reveals that, despite low values of $P$, the positional offsets for some of the brighter BLAST+radio associations are 
surprisingly large. This point is perhaps best illustrated by comparison of the identifications for the brightest two BLAST sources in our
sub-sample. For BLAST 4, 
the radio identification is completely secure ($P=0.0002$) and, as expected, lies only 1.5 arcsec from the BLAST 250\,${\rm \mu m}$ position (predicted 
positional uncertainty $\sigma_{pos} =
1.35\,{\rm arcsec}$ for this 16-$\sigma$ BLAST source). By contrast, the most likely identification 
for BLAST 6, while still statistically compelling ($P=0.004$), lies 11 arcsec from the BLAST 250\,${\rm \mu m}$ position, a 7\,$\sigma$ deviation 
in positional offset for a 14\,$\sigma$ BLAST source.

The key to resolving this apparent discrepancy lies in the fact that BLAST 6 has a second (also statistically compelling) radio identification
lying almost equi-distant from the 250\,${\rm \mu m}$ position, in the diametrically opposite direction. It turns out that BLAST 6 consists of 
an interacting pair of galaxies, whose positions (as marked by their VLA centroids) are separated by 20\,arcsec, and which both lie at 
$z =0.076$ (see Section 4). Clearly the radio identification procedure adopted here has yielded the correct redshift, but the position of the 250\,${\rm \mu m}$
source has either been distorted by contributions from both galaxies, or arises from dust which lies in the region between 
the two optical galaxies. Either way, the apparently excessive distance to either alternative radio identification
should not be naively interpreted as a failure of the radio identification technique, or as casting doubt on the BLAST positional accuracy. 
Thus, in the case of multiple identifications, each of which appears 
statistically secure, a fairer assessment of the identification procedure is provided by comparing 
the BLAST position with the {\it mean} position of the two alternative radio identifications, especially when such identifications straddle the 
nominal 250\,${\rm \mu m}$ position.
This is in fact the case for 4 of the 5 BLAST sources listed in Table 2 which have two alternative radio identifications -- BLAST 6, 59, 503 and 830.
If, for these 4 sources, we replace the primary identification radio source position with the mean radio position of both 
alternative radio counterparts, the cumulative distribution of positional offsets changes to that presented in the right-hand panel of Fig. 2. This shows that
allowance for the possibility of 250\,${\rm \mu m}$ source blending in just these 4 apparently multiply-identified sources, is sufficient to bring
the distribution of angular offsets completely into line with expectation.

This analysis provides increased confidence in the statistical robustness of our identification procedure. However, it does serve to 
highlight the potential problems of confusion arising from the large size of the BLAST beam. In particular it shows that, even when our identification
procedure can lead to an unambiguous redshift, it is unclear what fraction of the 250\,${\rm \mu m}$ emission can be reliably attributed to a given 
radio identification. To explore this further we attempted to distinguish whether, even for a clear-cut high S/N case such as BLAST 6, we could 
distinguish whether or not the BLAST source was a single source, or a blend of both galaxies. The result, summarized and discussed in Fig. 3, is 
that we cannot, even when we are confident that the positions of the potentially blended sources are well known, and separated by more than half  
the FWHM of the BLAST beam. Thus it is clear that we need to be 
very cautious in attempting to combine 250\,${\rm \mu m}$ and 1.4\,GHz data to determine,
 for example, the far-infrared:radio flux ratio for individual BLAST sources, despite the fact that for only two sources in the current 
sample (BLAST 59 and BLAST 503)
does the choice between two alternative radio identifications significantly influence the inferred redshift (see Section 4). There are clear lessons here 
that care will need to be taken in deriving the far-infrared spectral energy distribution of sources uncovered by {\it Herschel}, especially
at 500\,${\rm \mu m}$ where the beam-size is comparable to that which applies here at 250\,${\rm \mu m}$.

Finally, in Fig. 4 we plot the radio flux-density distribution of the primary identifications listed in Table 2. 
This figure demonstrates that most of the radio identifications have $S_{1.4} < 100\, {\rm \mu Jy}$ and that several 
have flux densities close to the 30\,${\rm \mu}$Jy limit. This is another reason that the existence of 3 radio-unidentified sources in our 20-source sample 
should not be regarded as surprising. Indeed, Fig. 4 makes the generic point that near-complete radio identification 
of a $250{\rm \mu m}$-selected galaxy sample requires 1.4\,GHz radio data reaching at least the same sensitivity in ${\rm \mu Jy}$ as is achieved at 
$250{\rm \mu m}$ in mJy.

\begin{figure}
\includegraphics[width=0.46\textwidth]{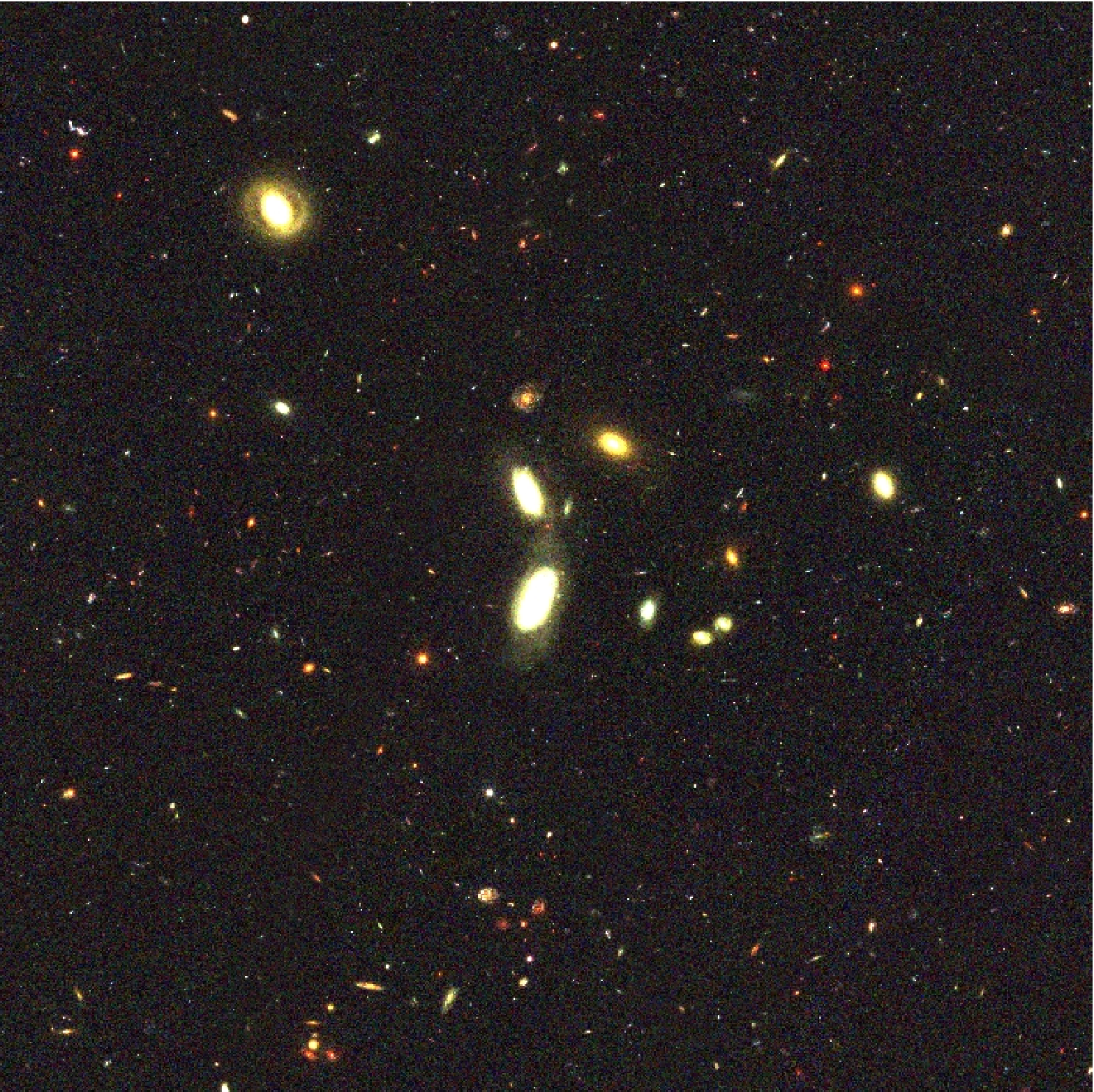}

\includegraphics[width=0.46\textwidth]{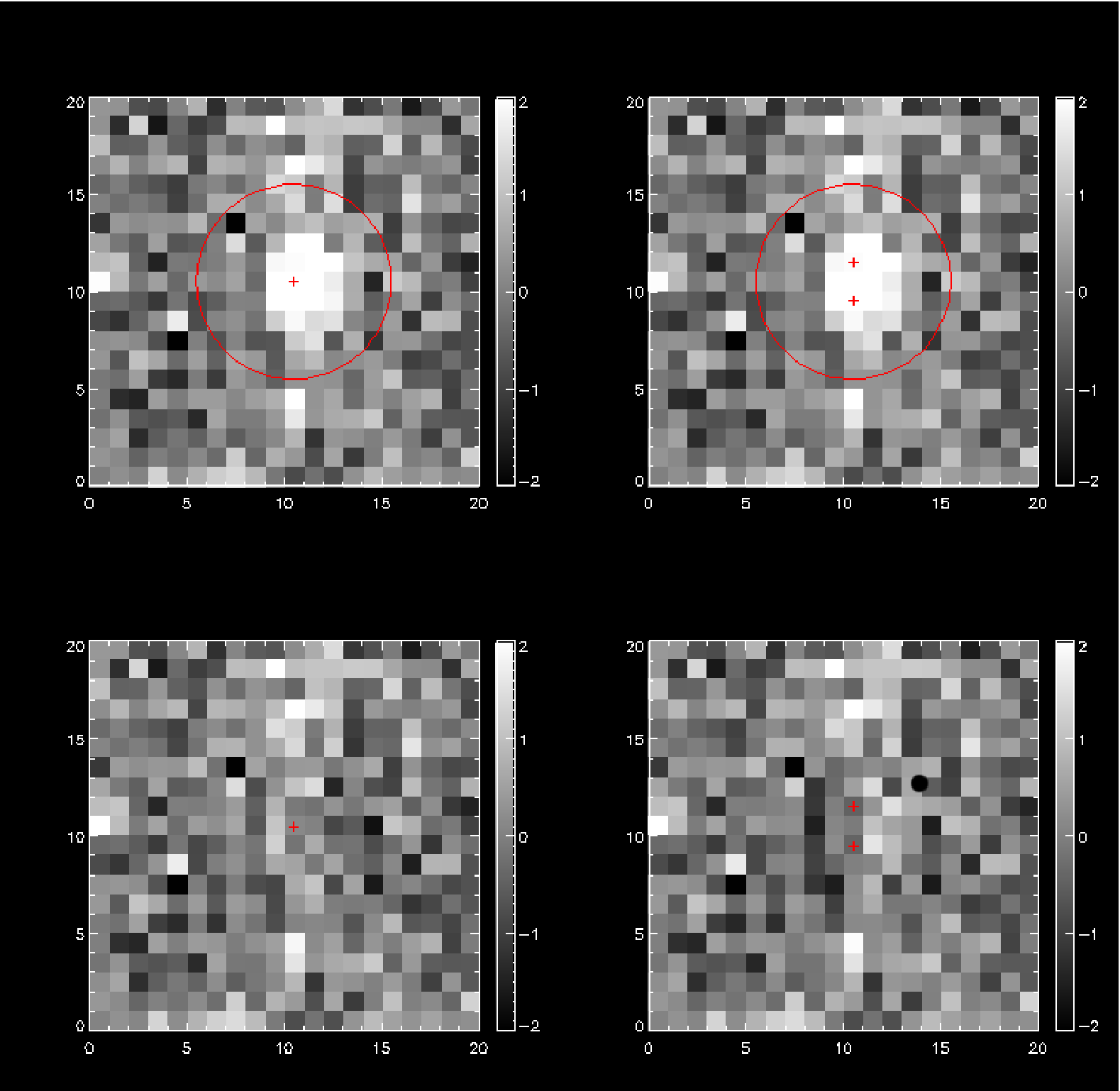}
\caption{Results of an attempt to deblend the 250\,${\rm \mu m}$ image of BLAST 6. At the top we show a $200 \times 200$ arcsec $B+V+z$ {\it HST} ACS colour
image of the field centred on BLAST 6. 
The upper 2 grey-scale panels then both show the same region as imaged by BLAST at 
250\,${\rm \mu m}$ (with 10\, arcsec pixels). The single cross in the upper-left image shows the position of a proposed single source which 
when convolved with the BLAST beam and fitted to the data yields a best-fit flux density of $S_{250} = 157$\,mJy. When 
subtracted from the data this leaves the residual image shown in the lower-left panel. The two crosses in the upper-right image show the proposed positions of 
two distinct sources (i.e. the positions of the radio sources 6-1 and 6-2) which, 
when convolved with the BLAST beam and fitted to the data, yield best-fit flux densities of $S_{250} = 91$\,mJy (lower source, 6-1 in Table 2) and 
$S_{250} = 75$mJy (upper source, 6-2 in Table 2). When subtracted from the data this combined-source model leaves the residual image shown in the lower-right panel.
Unfortunately, due to the BLAST beam, even in this rather well-defined test case it is not possible to distinguish whether the single
or double source model is a better description of the BLAST data, as the single-source fit yields $\chi^2 = 520$, and the double-source fit yields $\chi^2 = 530$ (for 400
pixels).
Thus, while the redshift of this source is not in doubt, and it is clear that both galaxies are related to the 250\,${\rm \mu m}$ emission, it is
not possible to decide whether this emission arises primarly between the two optical galaxies, or is a blend of emission from both nuclei.}
\end{figure}

\begin{figure}
\includegraphics[width=0.48\textwidth]{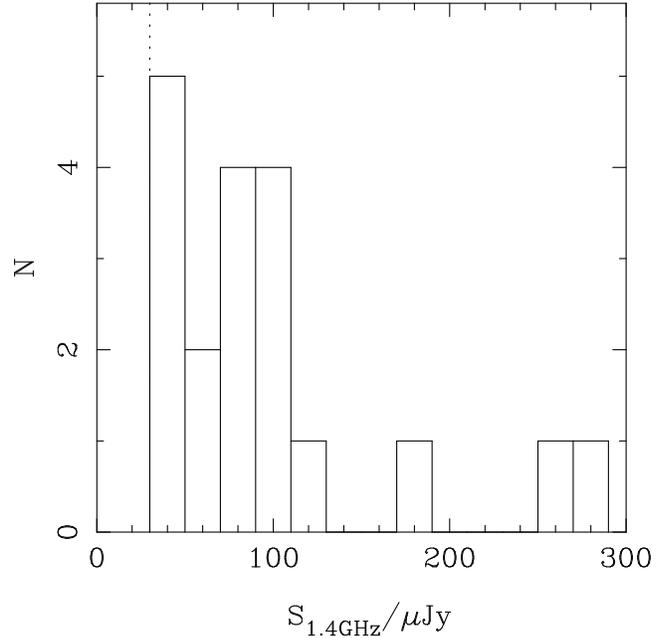}
\caption{The distribution of 1.4\,GHz radio flux density, in 20\,${\rm \mu}$Jy-wide bins, for the
radio-identified sources in the 20-source BLAST 250\,${\rm \mu m}$ 
GOODS-South sample (excluding the two very bright sources at $z < 0.1$).
All radio flux densities are as given in Table 2.
The dotted line marks the 4$\sigma$ 30\,${\rm \mu}$Jy limit of our radio
candidate search in the VLA data. Given this distribution of flux densities, and expecially the fact that several identifications have flux densities just above 
the 30\,${\rm \mu}$Jy limit, the existence of 3 radio-unidentified sources in our 20-source sample is not surprising. This figure demonstrates that near-complete radio identification 
of a $250\,{\rm \mu m}$-selected galaxy sample requires 1.4\,GHz radio data reaching at least the same sensitivity in ${\rm \mu Jy}$ as is achieved at 
$250\,{\rm \mu m}$ in mJy.}
\end{figure}

\subsection{Optical/infrared galaxy counterparts}

For the radio-identified sources, determining the correct optical/near-infrared galaxy counterpart is then relatively straightforward, due to the high
accuracy of the VLA positions. Most, but not all, of these galaxies are listed in the GOODS-MUSIC catalogue, and we give GOODS-MUSIC ID numbers 
for each object in Table 3. For those sources which do not feature in GOODS MUSIC we were able to successfully identify the galaxy counterpart
in the $K_s$-band and IRAC imaging (one source, 593, is completely undetected in the ACS optical imaging, even at $z_{850}$), and then performed our 
own photometry using the available multi-frequency imaging. 

This leaves the 3 radio-unidentified BLAST sources, 193, 654, and 732. We must bear in mind the real possibility that, given the S/N ratio of the source
catalogue under study, at least one of these 250\,${\rm \mu m}$ sources may not be real. However, as discussed in more detail in the source notes (see Section 5), 193 
appears to be confirmed at 350\,${\rm \mu m}$, 654 appears to be confirmed by a detection at $500\,{\rm \mu m}$, and 732 has apparently been detected at 
870\,${\rm \mu m}$ by LABOCA. We also checked that modest extension of the radio search radius (to 20\,arcsec) would still not have yielded a possible
radio counterpart for these sources. Thus it seems more likely that these sources may have radio detections lying just below the current radio image threshold,
possibly because they lie within the high-redshift tail of the source population.

We therefore attempted to establish `best-bet' optical/infrared counterparts for these sources by deriving a photometric redshift for every 
possible galaxy counterpart within the same 15\,arcsec search radius used for the radio identifications, and then confining our attention to candidates
with $z > 1.5$. The reason for restricting the potential 
redshift range to $z > 1.5$ is that all three of these BLAST sources, 
if real, have 250\,${\rm \mu m}$/1.4\,GHz flux ratios $S_{250}/S_{1.4} > 1000$,
a value which is not exceeded by any known galaxy spectral energy distribution
until $z > 1.5$.
As described in more detail in section 5, this yields unique candidates for 193 and 654, and 3 candidates for 732. The positions and 
GOODS-MUSIC ID numbers (where available) for these objects are also listed in Tables 2 and 3.

The identifications for these sources must inevitably be regarded as more speculative than most of the radio identifications 
(indeed we later reject 654 on the basis of no detection at 870${\rm \mu m}$ or 24${\rm \mu m}$). However, the selected
objects provide a plausible galaxy counterpart for each of these sources which is at least consistent with the other available photometric and 
redshift infomation.

The optical, near-infrared and Spitzer photometric data for all 27 
galaxies listed in Tables 2 and 3 are presented 
in Appendix A.

%%%%%%%%%%%%%%%%%%%%%%%%%%%%%%%%%%%%%%%%%%%%%%%%%%%%%%%%%%%%%%%%%%%%%%%%%%%%
\section{REDSHIFTS}
\label{REDSHIFTS}
%%%%%%%%%%%%%%%%%%%%%%%%%%%%%%%%%%%%%%%%%%%%%%%%%%%%%%%%%%%%%%%%%%%%%%%%%%%%

\subsection{Spectroscopic redshifts}
A search of the latest redshift catalogues resulting from the numerous spectroscopic surveys within GOODS-South (see Section 2.2.6)
produced redshifts for 13 of the radio-identified galaxy counterparts listed in Table 2. These redshifts are presented in Table 3, where it can be seen
that 8 are for primary counterparts, and 5 are for secondary counterparts. The highest spectroscopic redshift is $z~=~1.382$. 

\subsection{Photometric redshifts}
For 637-2 we adopted the COMBO 17 redshift estimate of $z~=~0.689$, but no other galaxy counterpart lacking a spectroscopic 
redshift has a robust COMBO 17 redshift. Reference to the 
GOODS-MUSIC catalogue revealed that this is because the remaining candidates all appear to lie at $z > 1$.

Rather than simply adopting the redshift estimates for the higher redshift galaxies from the existing published catalogues, we chose 
to re-analyse the photometry of each proposed identification to derive our own values of $z_{phot}$. We chose to do this for 5 reasons.
First, not all of the candidates are listed in the existing catalogues, because they are too faint, and for these we had to extract 
our own photometry from the imaging data and perform the first estimate of their redshifts. Second, given we are potentially dealing with very 
dusty galaxies, we wished to explore the effect on $z_{phot}$ of allowing extinction ($A_V$) to range to large values (as previously explored 
in a different context by Dunlop et al. 2007). Third, we wished to to marginalise over the full parameter space to derive meaningful 
confidence intervals on the estimated redshifts (errors on individual redshift determinations are frequently not 
provided in the published catalogues). Fourth, we simply wanted to check the robustness of the published redshift estimates, and explore
the effect of extracting the photometry through, for example, apertures of different sizes. Fifth, we wish to derive physical parameters such as galaxy age 
and stellar mass, to facilitate the further study of the physical properties
of the BLAST galaxies.

The SED fitting procedure we applied to derive the photometric redshifts and physical properties of the BLAST galaxy counterparts
is based largely on the public package {\sc Hyperz} (Bolzonella et al. 2000).
The observed photometry was fitted with synthetic galaxy templates generated with the stellar population models of Charlot \& Bruzual (e.g. Bruzual 2007). We used a variety
of star-formation histories: instantaneous burst and exponentially declining star-formation with e-folding times $0.1 \le \tau(\rm Gyr) \le 10$, assuming solar
metallicity and a Salpeter initial mass function (IMF). For dust reddening we adopted the prescription from Calzetti et al. (2000) within the range $0 \le A_V \le
4$. We also included absorption due to $\rm HI$ clouds along the line of sight in the intergalactic medium, according to Madau (1995).

Our best estimate of $z_{phot}$ for every source (including those with spectroscopic redshifts)
is tabulated in Table 3. For every galaxy which lacks a spectroscopic redshift 
we show in Fig. 5 the best-fitting spectral energy distribution 
and a plot of $\chi^2$ versus redshift marginalised over all other 
fitted parameters (age, star-formation history, dust extinction, and normalization). The $1\sigma$ confidence interval in redshift 
is given below each plot.

As a test of the robustness of our photometric redshifts we compare our values for $z_{phot}$ with the 
available spectroscopic redshifts in Fig. 6. This only provides a direct test out to $z \simeq 1.5$, but confirms good agreement and 
reveals no catastropic outliers. In this same figure we also compare our results for the higher redshift sources with the results in the GOODS-MUSIC
catalogue, where such values are available. While the scatter in the comparison is inevitably larger, the results still provide 
confidence that the values of $z _{phot}$ for the higher redshift sources are also reasonably robust.

In Fig. 7 we provide 
multi-wavelength postage stamp images of the primary BLAST identifications, ranked by redshift. Similar visual information 
is provided for the secondary/alternative counterparts in Fig. 8.

The final redshift distribution for the GOODS-South 250\,${\rm \mu m}$ `brightest-counterpart' sample is presented in Fig. 10, and discussed in Section 6.2

\begin{table}
\begin{tabular}{rrrrl}
\hline
BLAST & GOODS &$z_{spec}$ & $z_{phot}$ & $S_{870\mu m}$ \\
ID\phantom{ST}    & MUSIC             &  &          & /${\rm mJy}$ \\
\hline
4\phantom{-2}&555     &  0.038   & 0.05    &  $<$4.0\\
6-1   & 13855 & 0.076    & 0.08    &  $<$4.0\\
(6-2   & 13853 & 0.076   & 0.08    &  $<$4.0)\\
59-1  & 4107&          & 2.29    &\phantom{$<$}$9.3 \pm 1.2$\\ 
(59-2  &3920 &  1.097   & 1.11    & \phantom{$<$}$6.8 \pm 1.3$)\\
66\phantom{-2}&10764    &          & 1.94    &\phantom{$<$}$7.7 \pm 1.2$\\ 
104\phantom{-2}&7347   &  0.547   & 0.56    &  $<$4.0\\
109\phantom{-2}&5261   &  0.124   & 0.11    &  $<$4.0\\
158\phantom{-2}&136   &          & 1.85    &\phantom{$<$}$5.5 \pm 1.3$\\ 
193\phantom{-2}&30093   &          & 1.81    & $<$4.0\\
257\phantom{-2}&6771   &          & 0.69    &  $<$4.0\\
318\phantom{-2}&899   &          & 2.09    &\phantom{$<$}$5.9 \pm 1.3$\\ 
503-1 & 6790 &           & 1.96    &  $<$4.0\\
(503-2 & 6756 &0.241   & 0.31    &  $<$4.0)\\
552\phantom{-2}& 30025   &          & 1.68    &$<$4.0\\ 
593\phantom{-2}&   &          & $>2.50$ & \phantom{$<$}$8.9 \pm 1.2$\\
637-1 &  4484& 0.279    & 0.26    &  $<$4.0\\
(637-2 &  4578 & 0.279   & 0.27    &  $<$4.0)\\
654\phantom{-2}& 2977   &          & 2.62    &  $<$4.0\\
732-1 &   &         &2.97    &  \phantom{$<$}$6.8 \pm 1.2$\\ 
(732-2 & 30080 &         & 2.63    &) \\
(732-3 & 10787     &   &  2.40   &) \\
830-1 &     15626 &0.605    & 0.54    & $<$4.0\\
(830-2 & 15382 & 0.735   & 0.52    & $<$4.0)\\
861\phantom{-2}&13175   &          & 1.95    & $<$4.0\\
983\phantom{-2}&11348   &  0.366   & 0.41    & $<$4.0\\
1293\phantom{-2}&2645  &  1.382   & 1.37    & $<$4.0\\
\hline
\end{tabular}
\caption{Spectroscopic and photometric redshifts for the BLAST galaxy identifications (including 
alternative identifications listed in parenthesis). 
Only for 2 of the 6 sources with alternative identifications does the choice 
significantly affect the inferred redshift (59 and 503).
Also given is the sub-millimetre flux density ($S_{\rm 870}$) or 3-$\sigma$ limit
taken from the LABOCA LESS survey of GOODS-South by Weiss et al. (2009). Of the 11 proposed 
BLAST sources with $z > 1.5$, LABOCA has detected 6. In contrast, 
no 870\,${\rm \mu m}$ detections were achieved for any of the 
BLAST sources identified with galaxies at $z < 1.5$. As discussed in 
Section 6, this is largely as expected, and 
provides an independent sanity check on the basic form of the redshift distribution presented here.}
\end{table}

\section{Notes on individual sources}

\noindent
{\bf BLAST 4:} $z_{\rm spec} = 0.038$\\  
The brightest 250\,${\rm \mu m}$ source in the field is unambiguously identified 
with the brightest optical/near-infrared galaxy in GOODS-South. 
This is a very low-redshift, edge-on disc galaxy in which the effect of patchy 
dust obscuration can clearly be seen in the {\it HST} ACS colour composite image shown 
in the first panel of Fig. 7.
Comparison with the VLA imaging apparently reveals 5 statistically secure alternative 
radio counterparts (VLA sources 499, 524, 572, 701, 986, all of which have $P < 0.05$), 
but in fact all of these lie
near to the centroid of the optical galaxy, and it is clear that the VLA imaging 
has fragmented a more extended radio source. We have adopted VLA source 499 as 
the formal radio ID (position listed in Table 2); this is both the closest candidate and has 
the smallest value of $P$ (and is the same radio ID adopted by Dye et al. 2009). 
A comparison of the sum of the flux densities of the 5 
VLA counterparts ($\Sigma S_{\rm 1.4} = 202$\,mJy) with the flux density derived from 
the lower resolution (16\,arsec FWHM) ACTA imaging of the field (ACTA $S_{\rm 1.4} = 310 \pm 48\, {\rm \mu Jy}$; Norris et al. 2006)
confirms that much of the radio emission from the galaxy may have been resolved out by the
VLA imaging. Our own sum of all the flux-density in the VLA map yields $S_{\rm 1.4} = 680 \pm 34\, {\rm \mu Jy}$,
and we adopt this as the best available estimate of
1.4\,GHz flux density. This lowers the inferred $S_{\rm 250}:S_{\rm 1.4}$ flux 
density ratio to $\simeq 260$ from the erroneously high value of $\simeq 3000$ implied by the numbers
quoted in Dye et al. (2009). 
Given the angular size of the galaxy, and the lack of deep low-resolution radio imaging with a beam size comparable 
to BLAST, this ratio is still probably an over-estimate of the true value.
The fact that the BLAST position lies only 1.5\,arcsec from the centroid
of the optical/radio galaxy provides confidence in the BLAST 250\,${\rm \mu m}$ positional
accuracy because, at S/N = 16, the rms positional error for a BLAST 250\,${\rm \mu m}$ source 
is expected to be 1.35\,arcsec (see Section 3.1). 
Thus, at least in this region of the 250\,${\rm \mu m}$ map, any systematic pointing error 
appears impressively small. The source is also clearly detected at 350\,${\rm \mu m}$
and 500\,${\rm \mu m}$, although the 500\,${\rm \mu m}$ position lies 15\,arcsec distant 
from the true galaxy centroid. This is not hugely unexpected given the rms positional error
of 7.5\,arcsec anticipated from a 5.5$\sigma$ detection with the 60-arcsec FWHM 500\,${\rm \mu m}$
BLAST beam, but serves as a reminder of why it is virtually impossible 
to securely identify BLAST 500\,${\rm \mu m}$ sources which lack 250\,${\rm \mu m}$
counterparts.\\

\noindent
{\bf BLAST 6:} $z_{\rm spec} = 0.076$\\
The second brightest 250\,${\rm \mu m}$ source in the field is clearly associated 
with an interacting galaxy pair. Both the 250\,${\rm \mu m}$ 
and 350\,${\rm \mu m}$ positions appear to lie almost equi-distant between 
two comparably bright, apparently interacting galaxies at the same redshift (see Fig. 1 and Fig. 3). The result
is two statistically significant alternative radio IDs (VLA sources 72 and 100), 
and we have adopted the southern source (VLA 72) as the chosen ID because 
it has a marginally smaller value of $P$  
(the 500\,${\rm \mu m}$ position also favours
the southern galaxy). It is this galaxy which is shown in Fig.~7, with the alternative galaxy identification
shown in Fig.~8. Because 
both objects lie at the same redshift, the precise choice obviously does not actually affect our final redshift 
distribution at all. However, in truth the inferred positional offset of 10.4\,arcsec
between the radio and  250\,${\rm \mu m}$ positions is unexpectedly large for a 14\,$\sigma$ 
BLAST source and a 40$\sigma$ VLA source. Thus as discussed in detail in Section 3.1.3, 
it is essentially certain that the 250\,${\rm \mu m}$ emission either receives comparable 
contributions from both galaxies, or in fact does actually arise from a region between 
the two optical galaxies. Unfortunately, as shown in Fig. 3, the BLAST data do not allow us
to distinguish between these two scenarios. As with BLAST 4, the integrated 1.4\,GHz flux density of
this source (quoted in Table 2 for both galaxies) is much larger than the nominal value derived from the
VLA radio catalogue by Dye et al. (2009). \\

\begin{figure*}
\begin{tabular}{llll}

\includegraphics[width=0.37\textwidth]{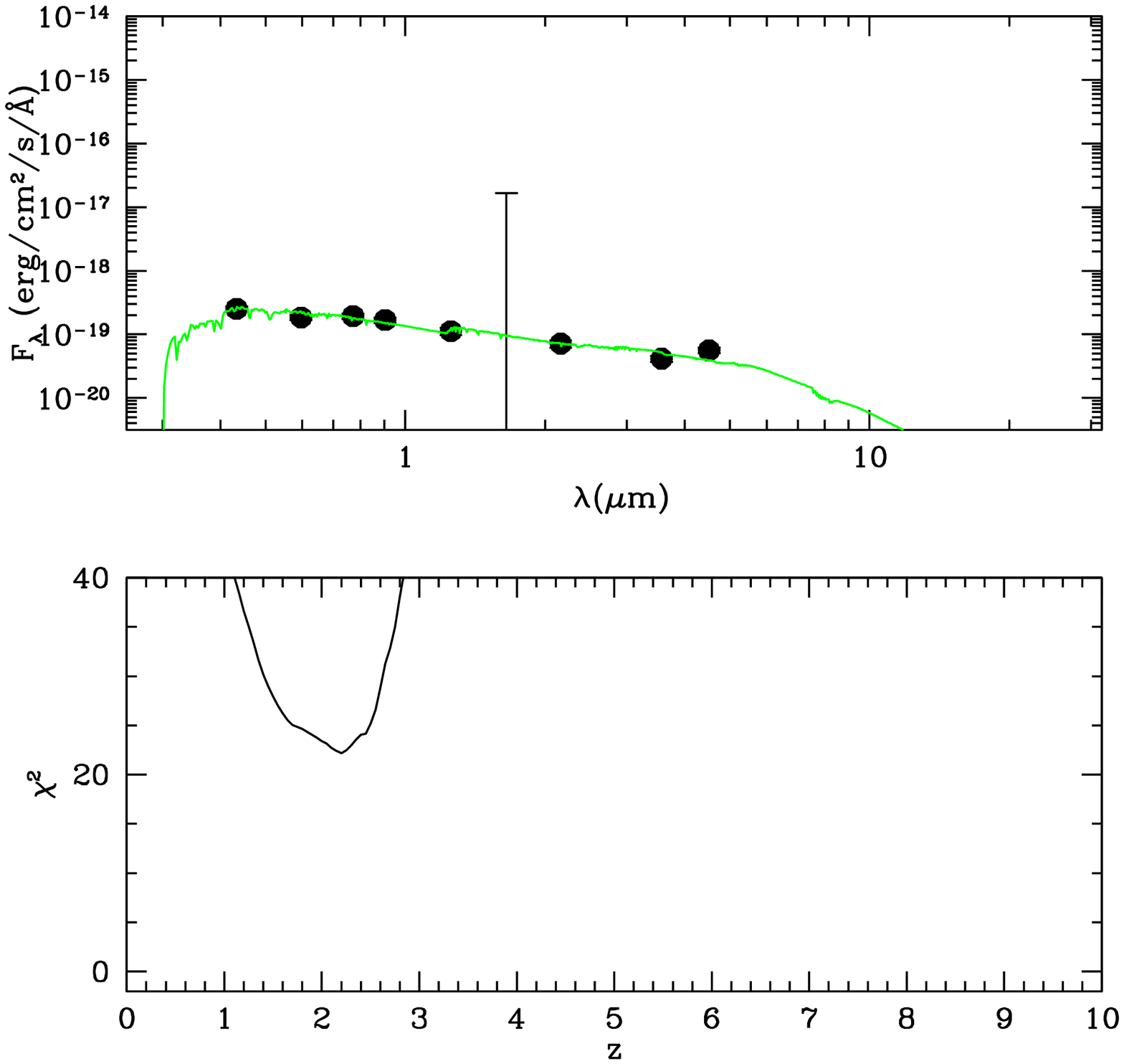}&
\includegraphics[width=0.37\textwidth]{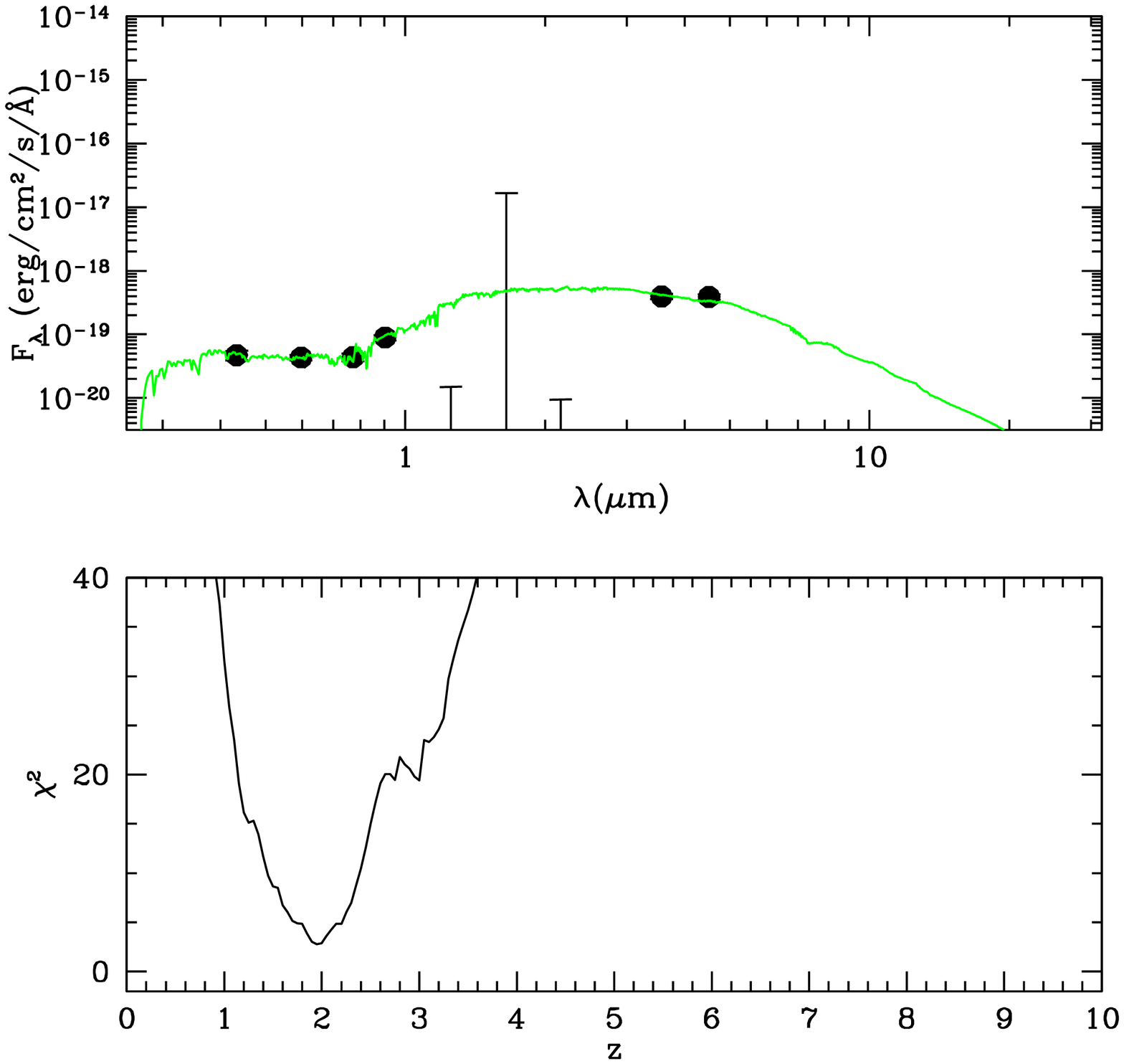}\\
\hspace*{1cm}{\bf BLAST 59-1:} $z_{\rm est} = 2.29\, (2.05-2.35)$ &\hspace*{1cm} {\bf BLAST 66:} $z_{\rm est} = 1.94\, (1.85-2.10)$
\\
\\
\\                          
\includegraphics[width=0.37\textwidth]{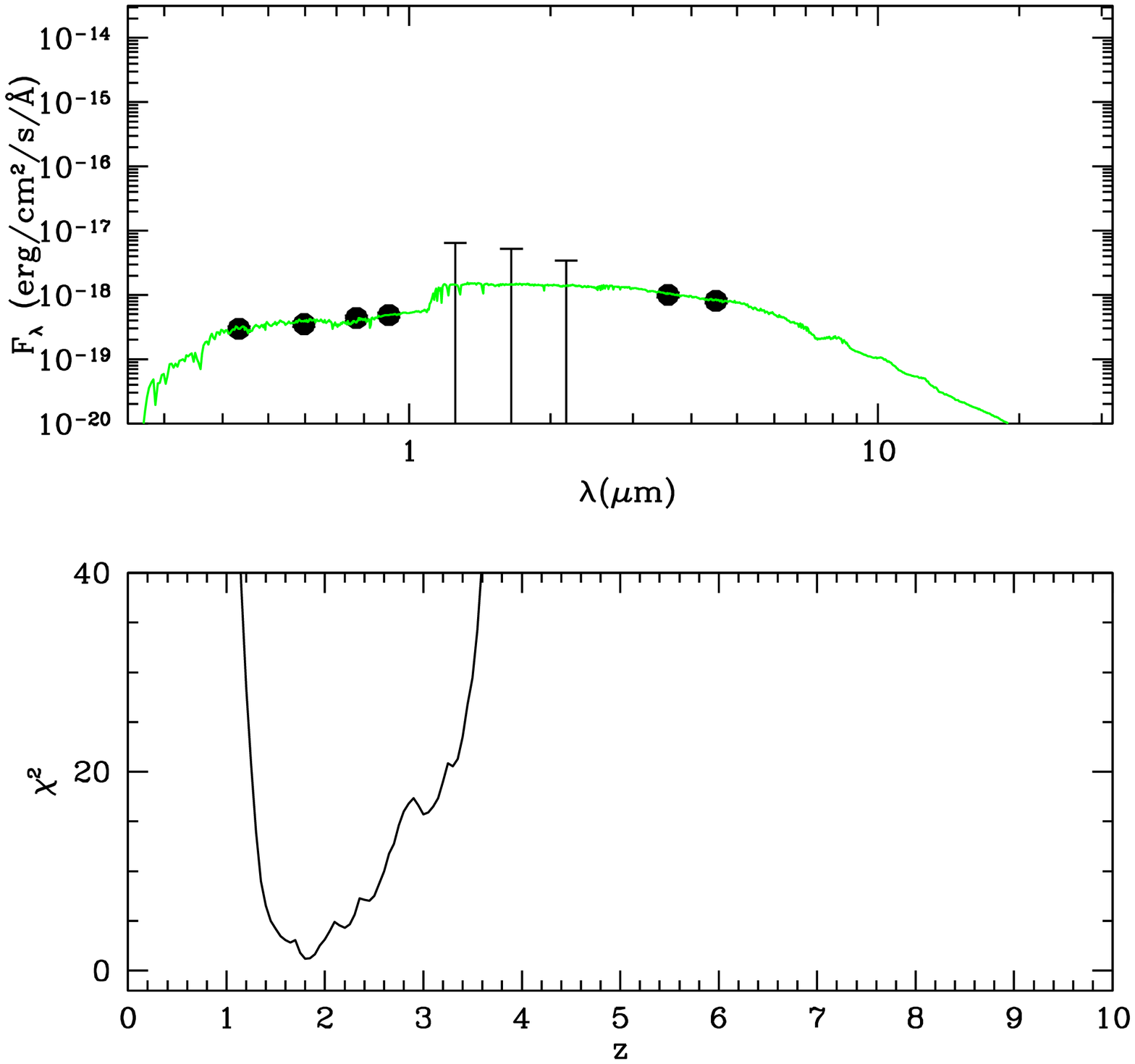}&
\includegraphics[width=0.37\textwidth]{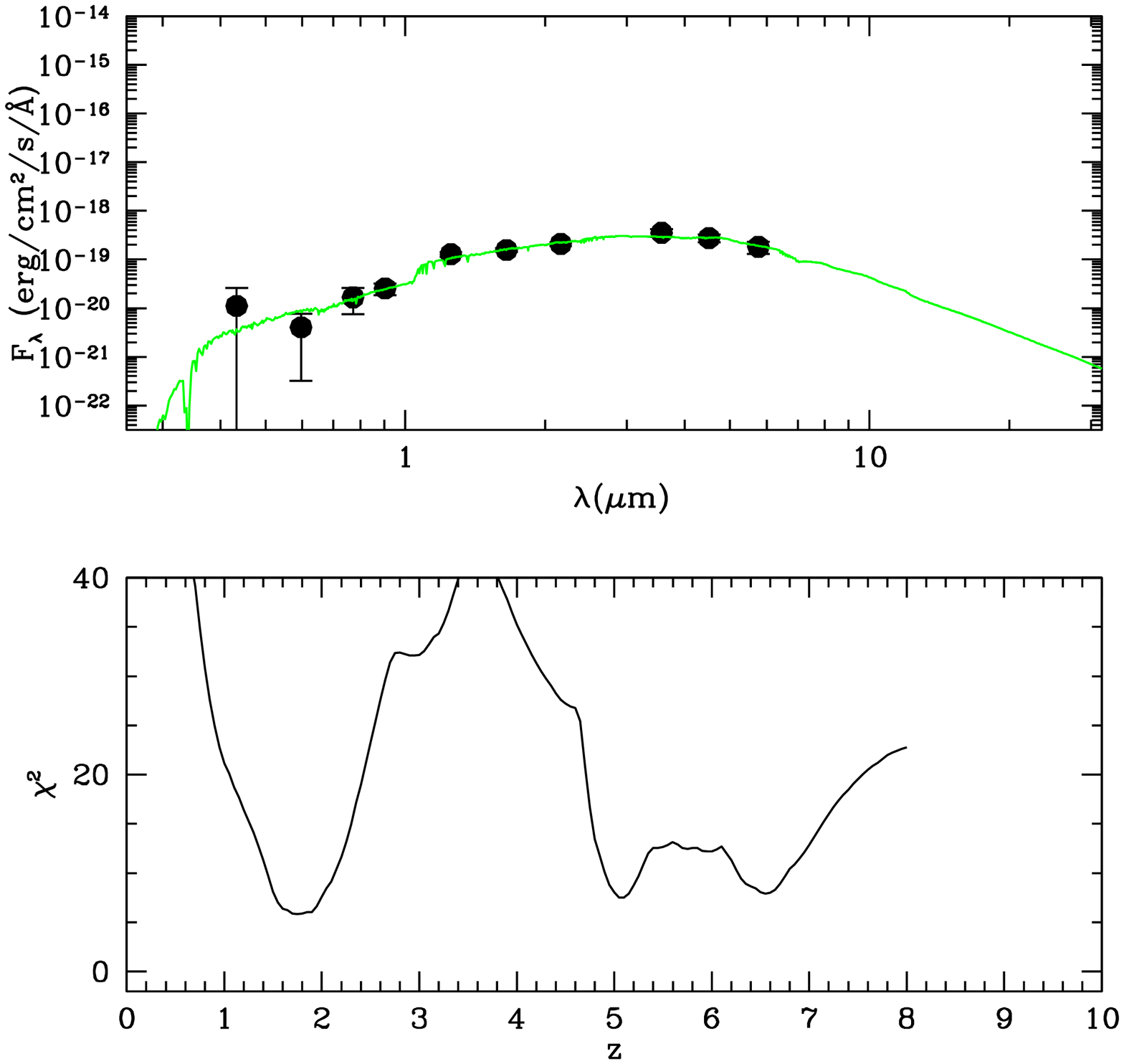}\\
\hspace*{1cm}{\bf BLAST 158:} $z_{\rm est} = 1.85\, (1.70-1.95)$ &\hspace*{1cm} {\bf BLAST 193:} $z_{\rm est} = 1.81\, (1.55-2.00)$
\\
\\
\end{tabular}
%\addtocounter{figure}{-1}
\caption{Photometric redshift determination for the 12 galaxy identifications in the 20-source BLAST 
250\,${\rm \mu m}$ GOODS-South sub-sample which 
lack spectroscopic redshifts. For each source the upper plot shows the best galaxy spectral
energy distribution fit to the {\it HST} optical, ISAAC near-infrared and {\it Spitzer} 3.6\,${\rm \mu m}$ and 4.5\,${\rm \mu m}$
photometry. The lower plot shows how $\chi^2$ varies with redshift, marginalised over galaxy age,
star-formation history, and dust reddening (allowing the extinction to range up to $A_V = 4$). The BLAST ID number 
of each source, and its estimated redshift (with $1\sigma$ error range) are given under each two-panel
plot.}
\end{figure*}

\begin{figure*}
\begin{tabular}{llll}

\includegraphics[width=0.37\textwidth]{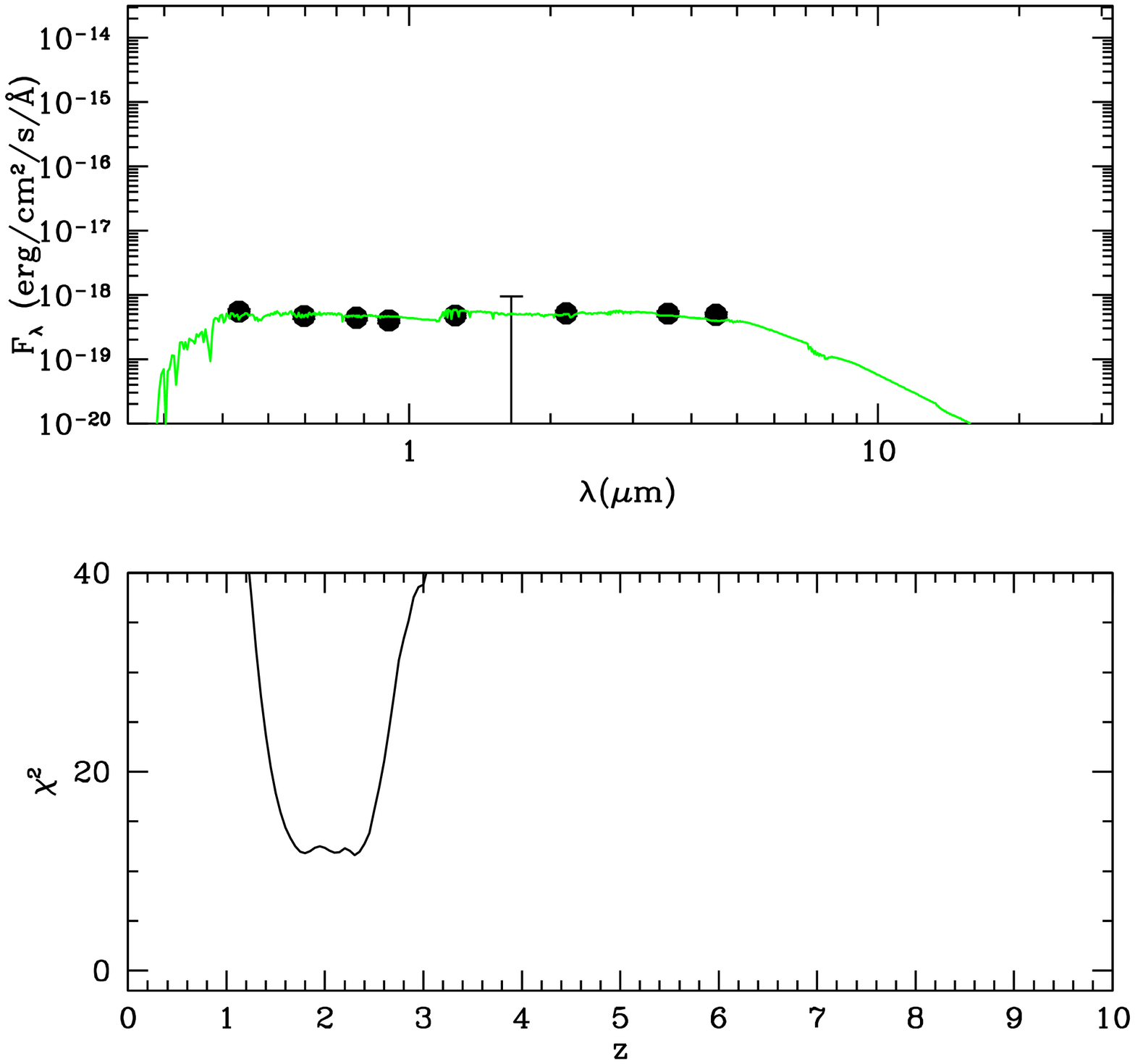}&
\includegraphics[width=0.37\textwidth]{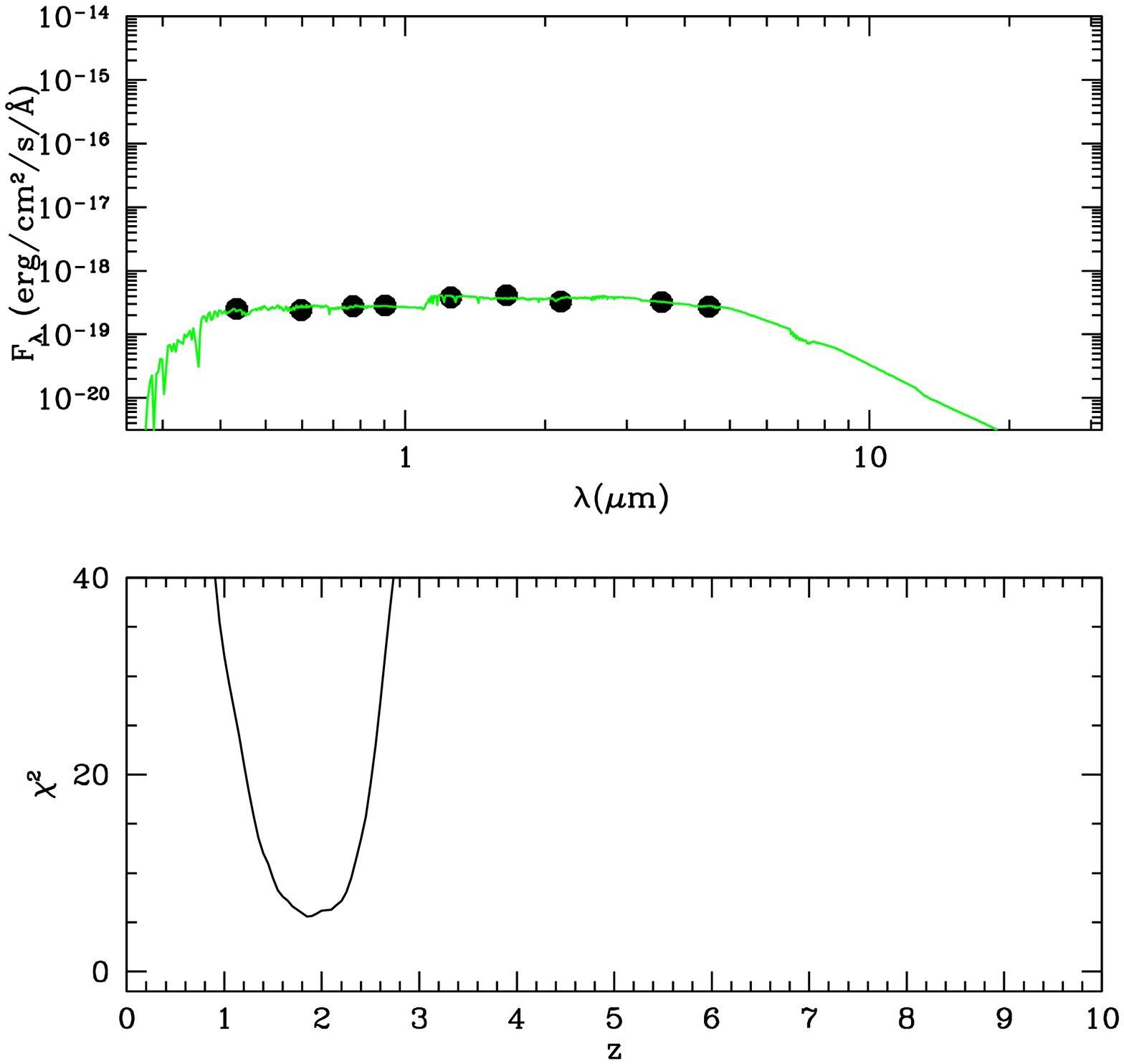}\\
\hspace*{1cm}{\bf BLAST 318:} $z_{\rm est} = 2.09\, (1.65-2.40)$ & \hspace*{1cm}{\bf BLAST 503-1:} $z_{\rm est} = 1.96\, (1.70-2.15)$
\\
\\
\\
\includegraphics[width=0.37\textwidth]{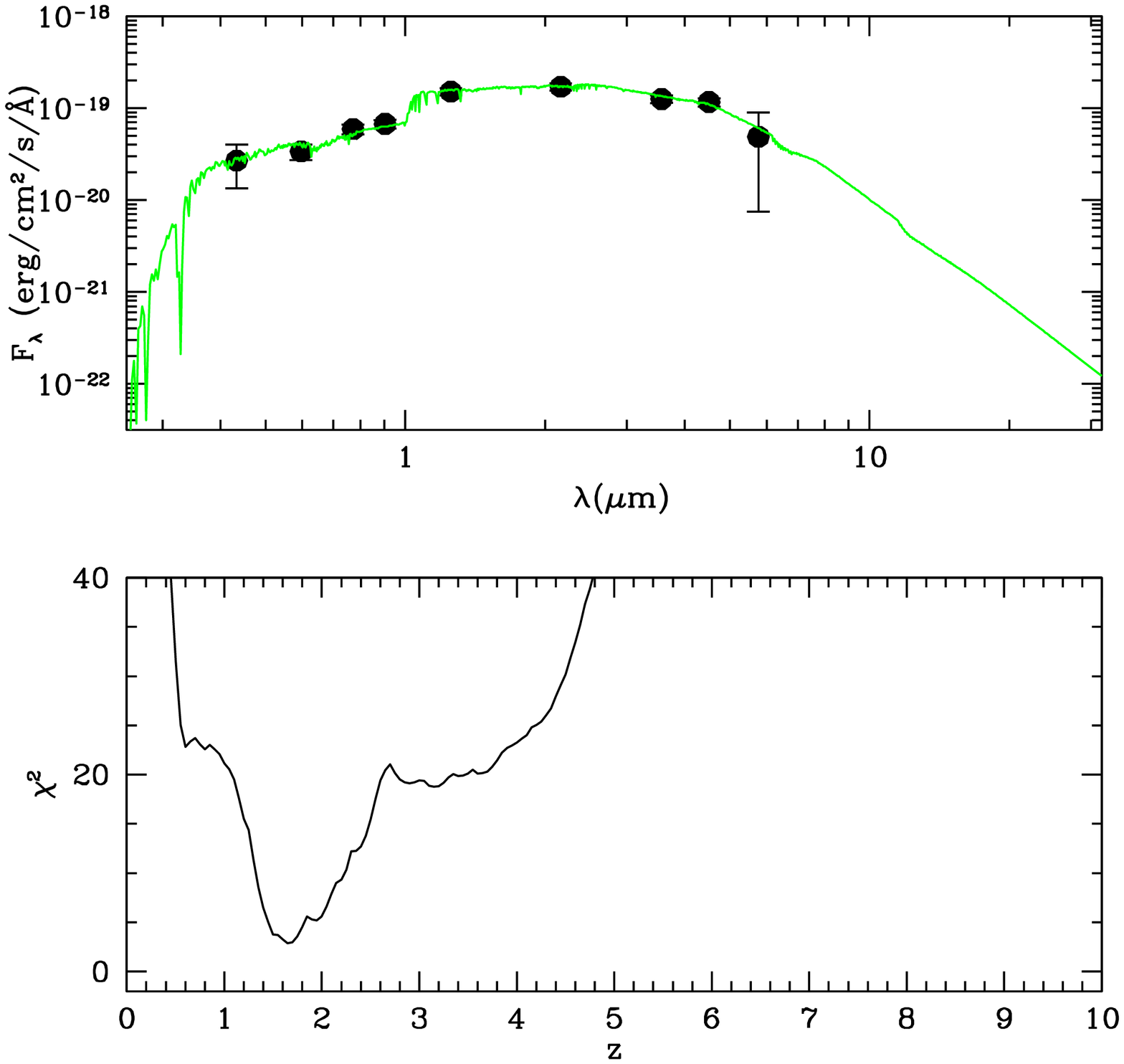}&
\includegraphics[width=0.37\textwidth]{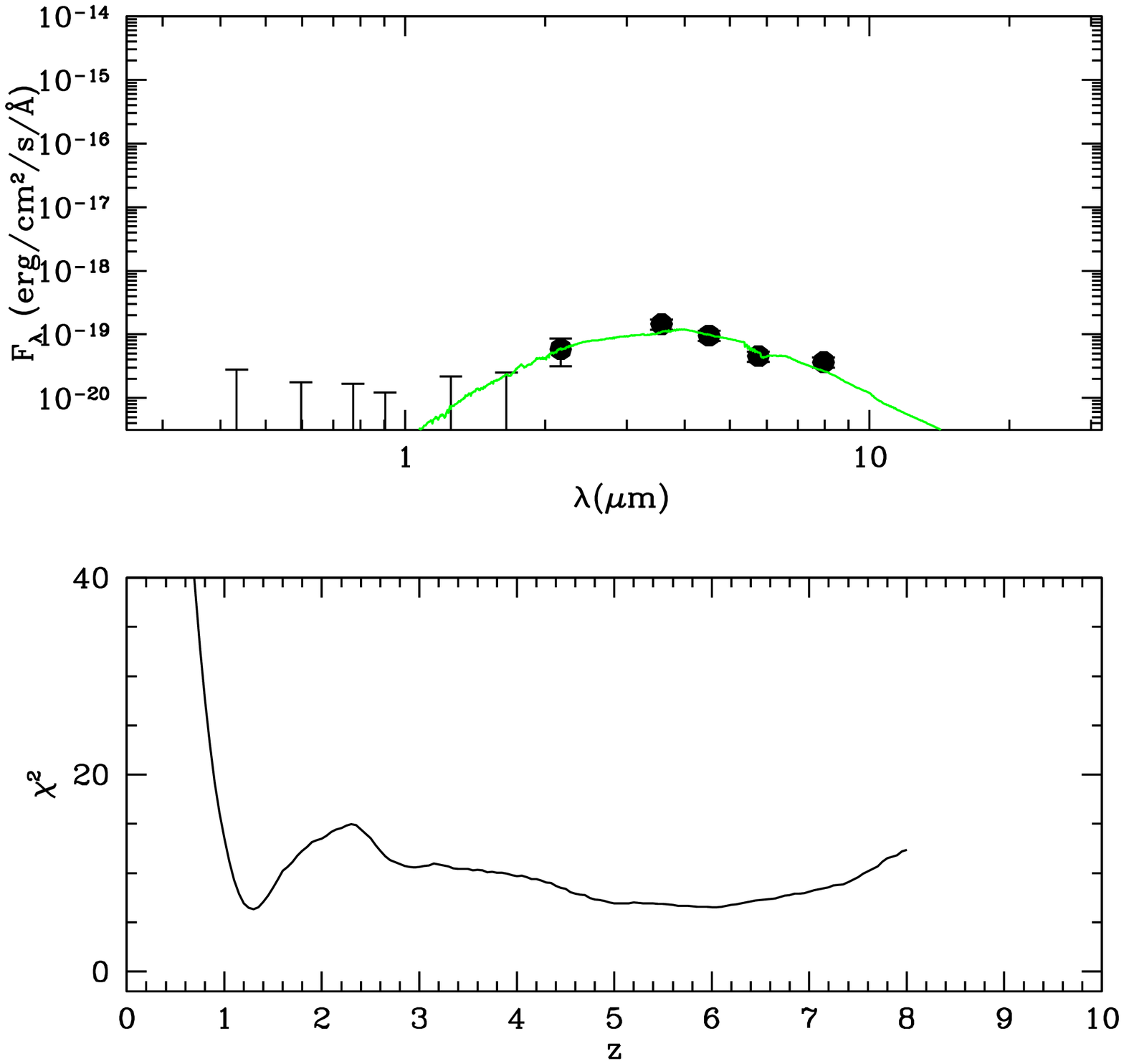}\\
\hspace*{1cm}{\bf BLAST 552:} $z_{\rm est} = 1.68\, (1.45-1.80)$ &\hspace*{1cm} {\bf BLAST 593:} $z_{\rm est} > 2.5\, (2\sigma )$
\\
\\
\end{tabular}
\addtocounter{figure}{-1}
\caption{continued.}
\end{figure*}

\begin{figure*}
\begin{tabular}{llll}

\includegraphics[width=0.37\textwidth]{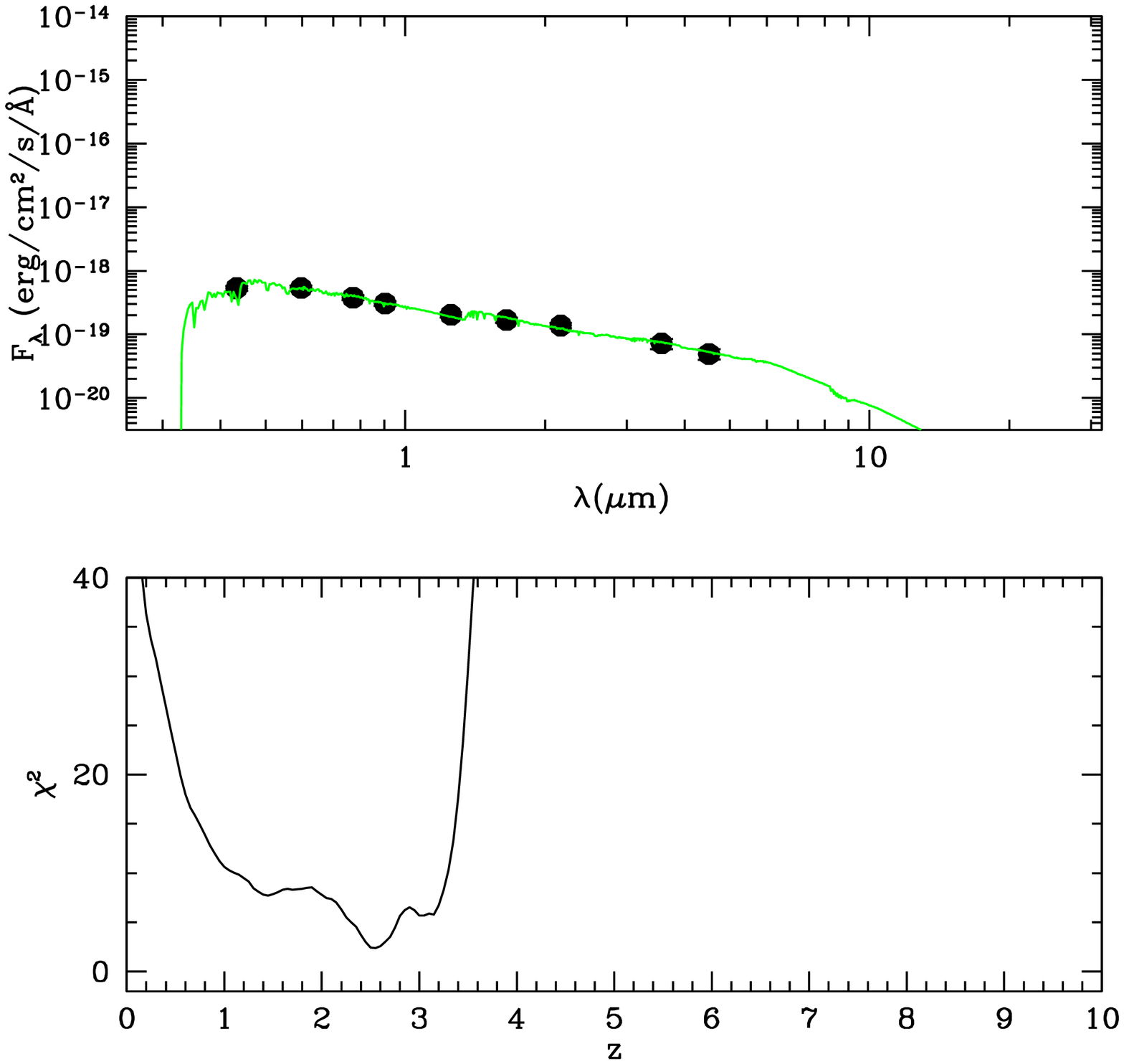}&
\includegraphics[width=0.37\textwidth]{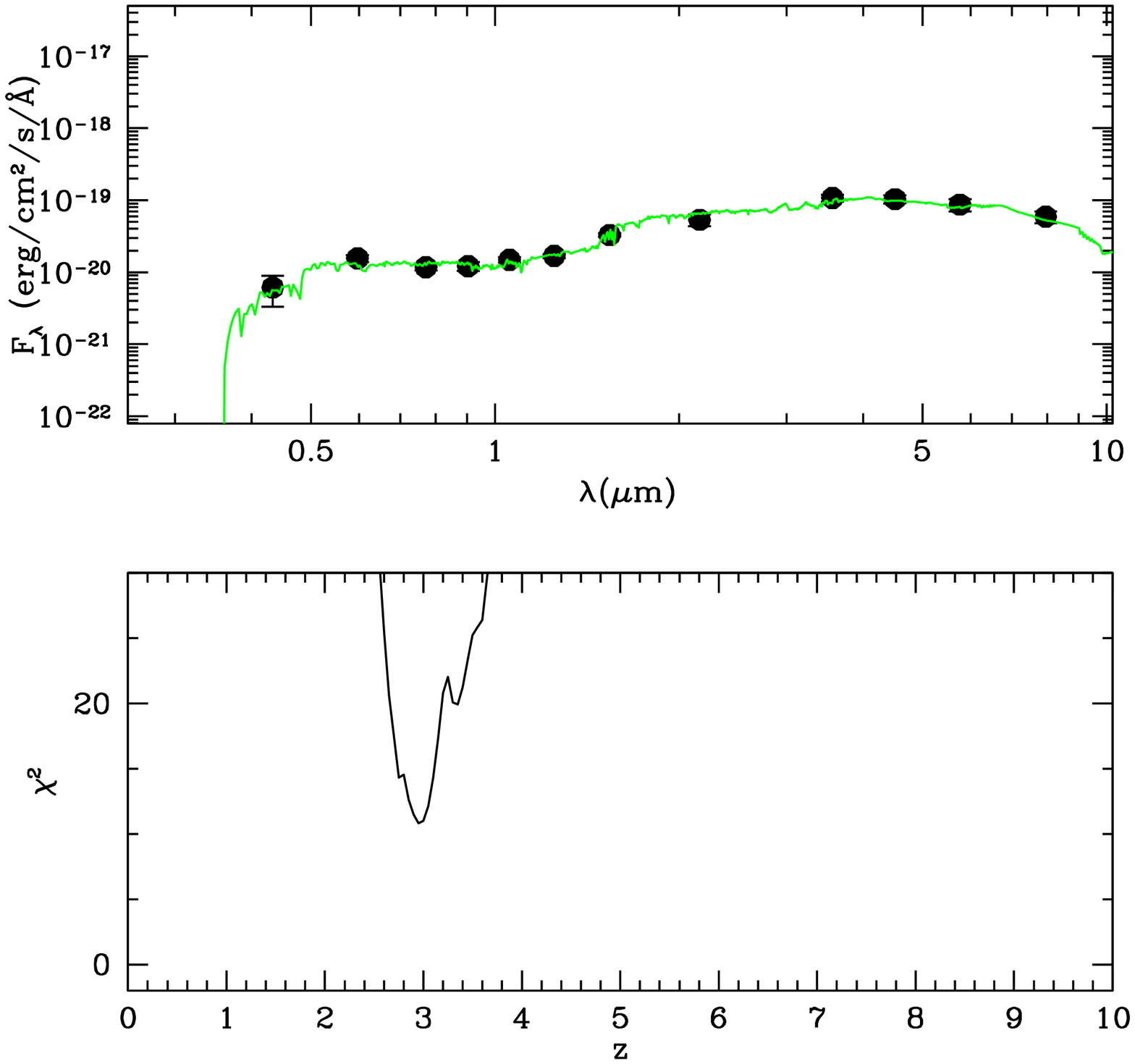}\\
\hspace*{1cm}{\bf BLAST 654:} $z_{\rm est} = 2.62\, (2.40-2.70)$ & \hspace*{1cm}{\bf BLAST 732-1:} $z_{\rm est} = 2.97\, (2.85-3.05)$
\\
\\
\\
\includegraphics[width=0.37\textwidth]{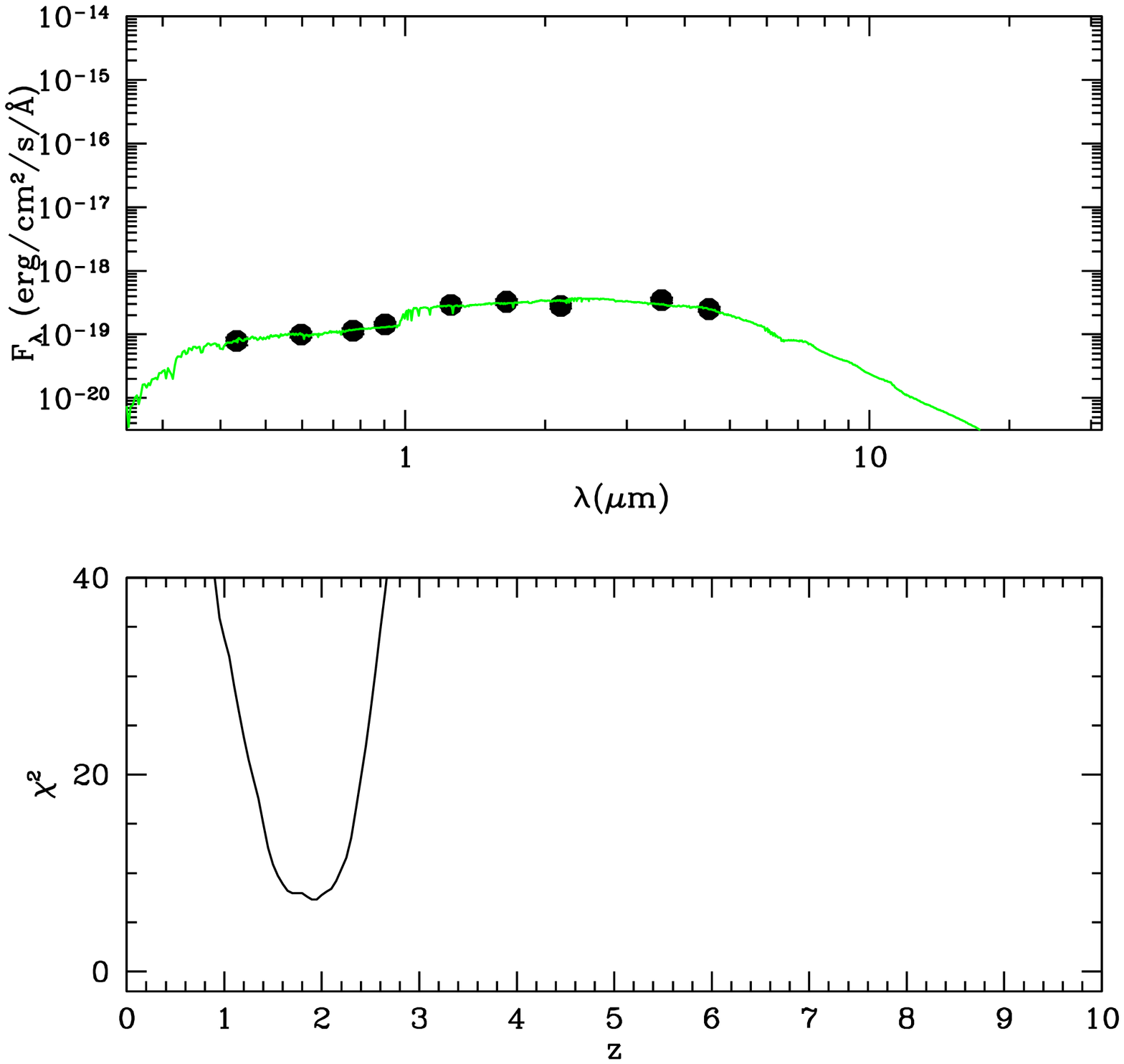}&
\includegraphics[width=0.37\textwidth]{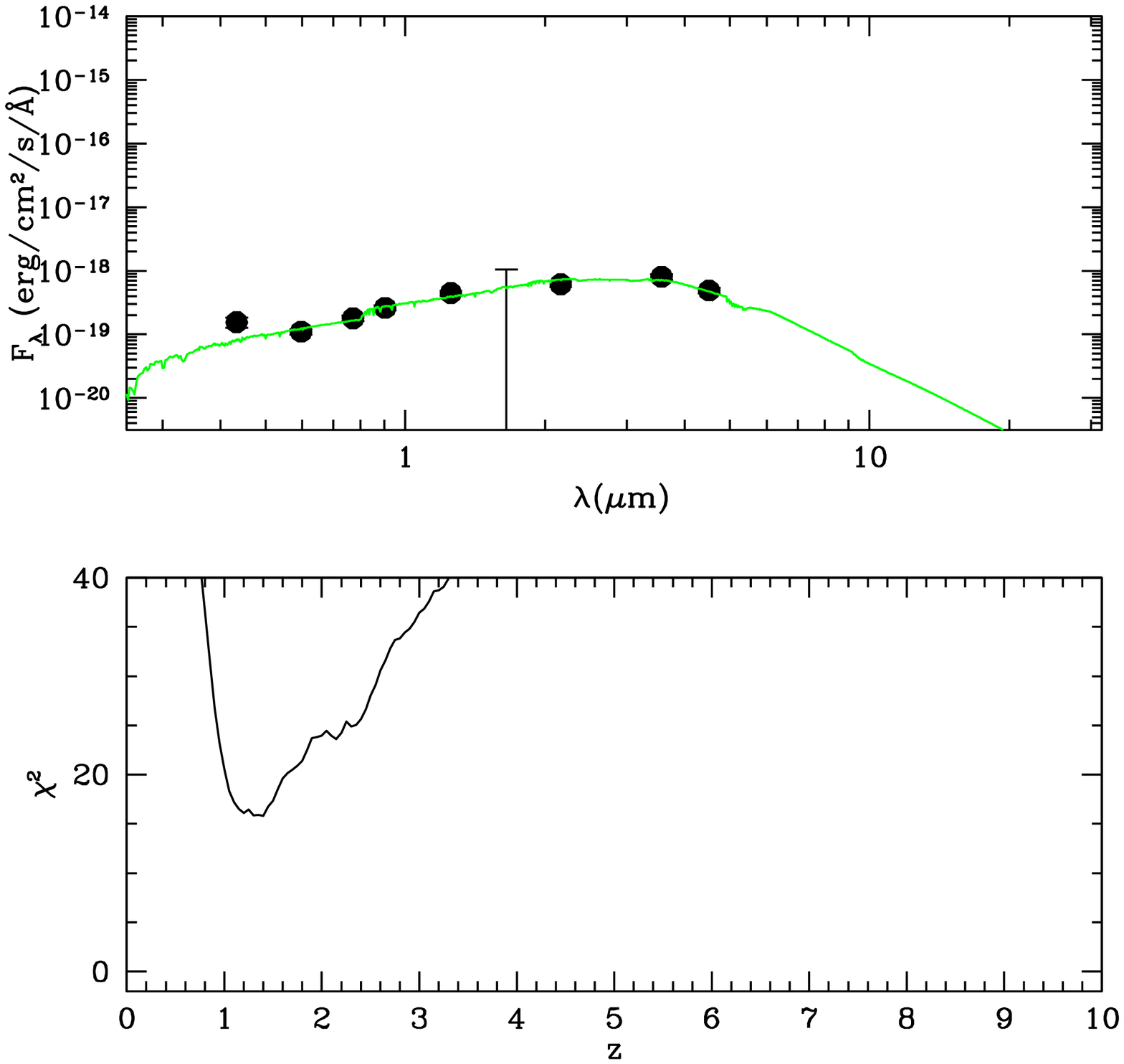}\\
\hspace*{1cm}{\bf BLAST 861:} $z_{\rm est} = 1.95\, (1.62-2.18)$ &\hspace*{1cm} {\bf BLAST 1293:} $z_{\rm est} = 1.37\, (1.35 - 1.55)$
\\
\\
\end{tabular}
\addtocounter{figure}{-1}
\caption{continued.}
\end{figure*}

\noindent
{\bf BLAST 59:} $z_{\rm est} = 2.29$ (or $z_{\rm spec} = 1.097$)\\
Stepping down by a factor of two in 250\,${\rm \mu m}$ brightness, the third 
brightest BLAST source has a flux density of $S_{250} \simeq 75$mJy, more 
typical of what is expected in a field of this size given the source counts (Devlin et al. 2009).
However, for this object, selection of the correct identification is extremely difficult. 
The BLAST centroid lies roughly equidistant between 3 alternative radio counterparts.
The statistically preferred radio ID for this source lies to the east of the 250\,${\rm \mu m}$ position,
but gains additional support from the 500\,${\rm \mu m}$ position. Statistically, the association
seems reasonably secure. The radio source 
has an optical/infrared counterpart in the GOODS MUSIC catalogue (shown in Fig.~7), but 
this has no spectroscopic redshift. The GOODS MUSIC estimated redshift is $z = 2.13$, while
our own analysis of the photometry yields $z = 2.29$ $(2.05-2.35)$ as shown in the 
first panel of Fig. 5. However, given the position of the 250\,${\rm \mu m}$ source 
it is extremely likely that at least some of the far-infrared flux is  
contributed by the other radio-identified objects, which are galaxies lying within
a known large-scale structure at $z \simeq 1.09$. We thus also retain (as 59-2) the second most 
likely radio identification, VLA 240, which is a galaxy at $z = 1.097$ which displays AGN emission, and lies 
to the southwest of the 250\,${\rm \mu m}$ centroid (Fig.~8). This is the one case in the sample where both 
alternative radio counterparts also appear to be associated with individual LABOCA sources, strengthening 
the case that the 250\,${\rm \mu m}$ emission results from the blending of at least these two sources.\\

\noindent
{\bf BLAST 66:} $z_{\rm est} = 1.94$\\
This 250\,${\rm \mu m}$ source has clear detections at 350\,${\rm \mu m}$ and 500\,${\rm \mu m}$,
both reassuringly close to the 250\,${\rm \mu m}$ position. The radio identification is 
robust and unambiguous. The source lies just outside the ISAAC near-infrared imaging, but the ACS 
and {\it Spitzer} IRAC photometry is adequate to yield a unique and well-constrained 
redshift solution at $z \simeq 2$, as shown in Fig. 5. The high-redshift nature of this source receives support from its 
detection by LABOCA at 870\,${\rm \mu m}$. The multi-colour postage stamps in
Fig. 7 reveal a faint complex source at optical wavelengths, which is bright and unconfused
at 3.6\,${\rm \mu m}$. This is the third of the three sources in this deep GOODS-South sample which
were also identified by Dye et al. (2009), and not surprisingly they deduced the same VLA 
source as the correct ID. Confusingly however, they list a COMBO17 redshift of $z = 1.16$ which is
clearly inconsistent with the photometric redshift constraints derived here. However, upon 
inspection of the latest COMBO17 catalogue it can be seen that the estimated redshift of 
this source (COMBO17 35066) is highly uncertain (unsurprisingly so, given that 
the object is very faint, with
$R \simeq 25.8$) and that the peak of the redshift probability distribution
is listed as $z = 1.84$, consistent with our value. 
The redshift listed by Dye et al.
(2009) can thus be safely rejected.\\

\noindent
{\bf BLAST 104:} $z_{\rm spec} = 0.547$\\
This 250\,${\rm \mu m}$ source has detections at 350\,${\rm \mu m}$ and 500\,${\rm \mu m}$
which reinforce the BLAST position. The VLA identification is centred on a red galaxy 
with a spectroscopic redshift of  $z = 0.547$. The closest companion seen in Fig. 7 lies at $z_{est} = 1.44$, but the object ENE of the BLAST position has $z_{est} = 0.54$ and lies
within 28\,kpc of the BLAST source at this redshift (although it has no radio detection, so presumably
is contributing less to the far-infrared flux density). The next nearest radio counterpart lies 15.7\,arcsec
north-east of the BLAST position, and thus just outside our adopted search radius. This object has a 
spectroscopic redshift $z = 2.578$.\\

\noindent
{\bf BLAST 109:} $z_{\rm spec} = 0.124$\\
This 250\,${\rm \mu m}$ source is not found in the BLAST catalogues at longer wavelengths,
suggesting it lies at only moderate redshift. There are two formally-significant radio 
counterparts, but one lies close to our search boundary at $d = 15$ arcsec. We adopt 
the closer radio counterpart, with the lower value of $P$ as the most likely
identification. This ties the BLAST source to a low-redshift, edge-on disc galaxy.\\

\noindent
{\bf BLAST 158:} $z_{\rm est} = 1.85$\\
Secure 350\,${\rm \mu m}$ and 500\,${\rm \mu m}$ counterparts to the 250\,${\rm \mu m}$
source confirm its position as just off the southern edge of the ISAAC $K$-band 
mosaic. The radio ID is faint, but unique and reasonably secure. As with source 66, 
the available ACS optical and {\it Spitzer} IRAC photometry are sufficient to 
provide a reasonably solid estimated redshift for the associated galaxy, 
at $z_{est} = 1.85$, and the high-redshift nature of this source receives support from its 
detection by LABOCA at 870\,${\rm \mu m}$. In Fig. 7 the galaxy looks like an extremely 
complex faint system at optical wavelengths, but it is bright at IRAC wavelengths.\\

\noindent
{\bf BLAST 193:} $z_{\rm est} = 1.81$\\
This 250\,${\rm \mu m}$ source has a solid detection at 350\,${\rm \mu m}$, 
but has no candidate VLA counterpart within our adopted search radius of 15\,arcsec
(in fact none within 20\,arcsec). 
We therefore
fitted galaxy models to the optical-infrared photometry of all objects in the ISAAC $K$-band image within 15\,arcsec
of the 250\,${\rm \mu m}$ position. This search produced five potential counterparts. Only one of these lies
at $z > 1.5$, and we found this object to also be extremely red. This transpires to be one of only 6 ultra-obscured $K$-selected galaxies
in GOODS-South studied in detail by Dunlop et al. (2007; object 1865), and lies 
only 7\,arcsec from the far-infrared position.\\

\noindent
{\bf BLAST 257:} $z_{\rm est} = 0.689$\\
This 250\,${\rm \mu m}$ source is not found in the BLAST catalogues at longer wavelengths,
suggesting it lies at only moderate redshift. This is the second source in the sample 
which has two alternative radio identifications associated with galaxies at the 
same redshift. 
The preferred VLA counterpart (585) has a COMBO17 redshift $z_{phot} = 0.689$, while 
the alternative radio identification (VLA source 852, 19\,arcsec distant (so outside our formal search radius) 
lies in a galaxy with $z_{spec} = 0.664$. It seems likely 
that both galaxies contribute to the 250\,${\rm \mu m}$ flux density, but the precise choice of ID
does not affect the final redshift distribution. We adopt the galaxy associated with VLA source
585 as the statistically most likely association.\\

\noindent
{\bf BLAST 318:} $z_{\rm est} = 2.09$\\
This 250\,${\rm \mu m}$ source has clear detections at 350\,${\rm \mu m}$ and 500\,${\rm \mu m}$,
both reassuringly close to the 250\,${\rm \mu m}$ position. The radio identification is 
robust and unambiguous, and associated with an interacting galaxy pair (Fig. 7) which 
the photometry constrains to lie at $z \simeq 2$ (Fig.~5). The high-redshift nature of this source 
gains support from its detection at 870\,${\rm \mu m}$ by LABOCA.
VLT FORS2 spectroscopy has been 
attempted for this object, but yielded no redshift.\\

 \begin{figure}
\includegraphics[width=0.48\textwidth]{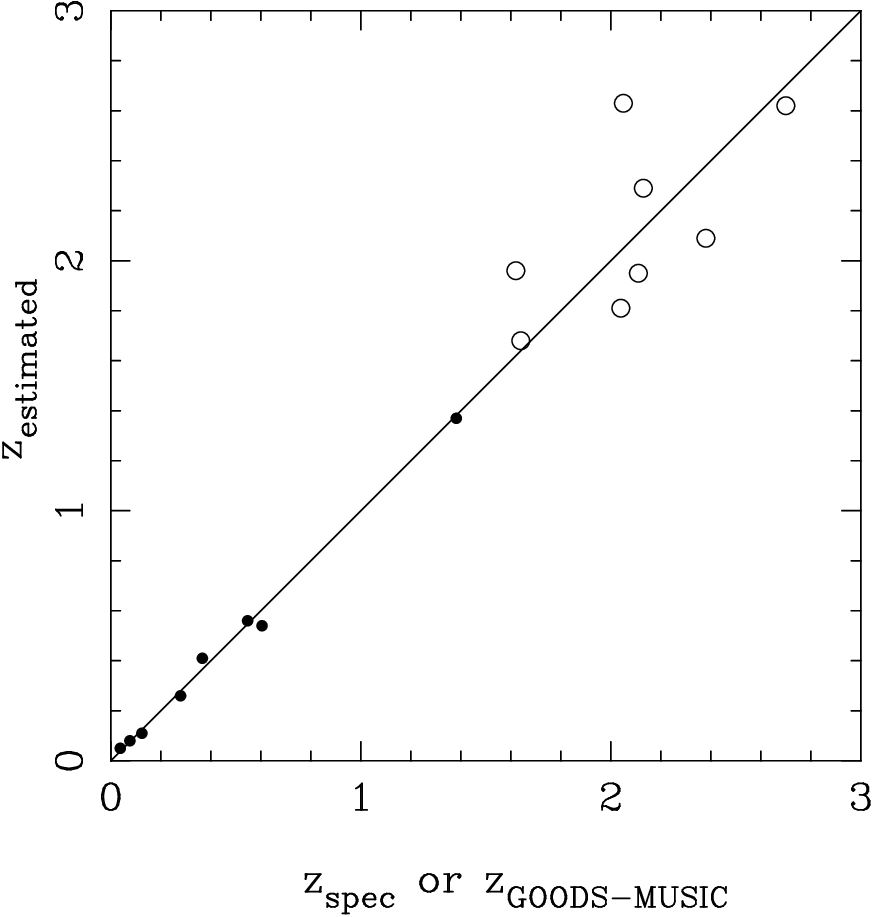}
\caption{A test of the robustness of the new optical-infrared photometric redshifts derived here for the galaxy identifications
of the 250\,${\rm \mu m}$ GOODS-South sample. Our new photometric redshifts are compared with the precise spectroscopic values for
8 out of the 9 sources at $z < 1.5$ (solid symbols), and against the published GOODS-MUSIC photometric redshifts for the 8 other sources in the sample 
for which these are available (open symbols). The former comparison is a true test of the accuracy of our redshift estimates (at least at low redshift), 
while the latter comparison demonstrates the extent to which independent photometric redshifts agree for the fainter higher-redshift 
sources in the sample. The agreement is very good, which is perhaps not surprising given the generally well-defined and unique minima in $\chi^2$ 
for the individual fits shown in Fig.5.}
\end{figure}

\noindent
{\bf BLAST 503:} $z_{\rm est} = 1.96$ (or $z_{\rm spec} = 0.241$)\\
This source is not found in the BLAST catalogues at longer wavelengths, which would 
suggests it lies at only moderate redshift. However, while there is a possible radio
counterpart with a spectroscopic redshift $z= 0.241$ (VLA 315), 
the statistically preferred option is VLA 132. The optical to infrared
images shown in Fig. 7 appear to reveal a complex multiple interacting system, which again 
is well constrained by the photometry to lie at $z \simeq 2$ (Fig.7). The 250\,${\rm \mu m}$ 
flux density may receive contributions from both sources, but in this case our choice
of ID obviously does (strongly) influence the adopted redshift. There is no LABOCA 
detection to help support a high redshift, but neither does the non-detection
strongly exclude it (see Section 6.1).\\

\noindent
{\bf BLAST 552:} $z_{\rm est} = 1.68$\\
This source may be detected at 350\,${\rm \mu m}$ and 500\,${\rm \mu m}$, but the positional agreement
is poor. We have adopted VLA 932 as the most likely radio identification, but have concerns 
here about the extent to which the 250\,${\rm \mu m}$ emission in this region is confused 
by the close proximity of BLAST 830. This is one of the more dubious 
250\,${\rm \mu m}$ sources in the sample. For this reason, and because this source does not appear to have 
been detected at 870\,${\rm \mu m}$ (see Section 6) it has, in the end, been excluded from the final 
proposed redshift distribution for the sample presented in Fig. 10.\\

\noindent
{\bf BLAST 593:} $z_{\rm est} > 2.5$\\
This source has no catalogued 350\,${\rm \mu m}$ counterpart, but is a clear detection at 
500\,${\rm \mu m}$, and is nearly coincident with the second brightest LABOCA 
source in the field. In addition it has a solid and unambiguous radio counterpart in VLA 405.
The host galaxy of this radio emission is completely invisible in the optical imaging,
perhaps just visible in $K$, but clearly seen at 3.6\,${\rm \mu m}$ (Fig. 7). All evidence points towards 
a high-redshift dusty galaxy (perhaps the most distant in our sample), 
but the lack of detections over a wide range in wavelength means that the photometric
redshift is poorly constrained (see Fig. 5), and thus we adopt a lower limit for $z_{est}$.\\

\noindent
{\bf BLAST 637:} $z_{\rm spec} = 0.279$\\
This 250\,${\rm \mu m}$ source is not found in the BLAST catalogues at longer wavelengths,
suggesting it lies at only moderate redshift. There are two alternative formally-significant radio 
counterparts, and we select the closer and marginally more significant 
VLA 110 in favour of VLA 74. Both radio sources are
associated with galaxies with a spectroscopic redshift $z = 0.279$, and Fig. 7 
indicates that this is another interacting galaxy pair. However, unlike the situation 
for BLAST 6, the far-infrared position does not lie between the two objects, and favours 
association with the more luminous galaxy. Clearly the choice of galaxy counterpart  does not influence
the adopted redshift.\\

\noindent
{\bf BLAST 654:} $z_{\rm est} = 2.62$\\
This source has no catalogued 350\,${\rm \mu m}$ source, but is detected at 
500\,${\rm \mu m}$ only $\simeq 4$\,arcsec distant
from the 250\,${\rm \mu m}$ position. This  
500\,${\rm \mu m}$ detection, combined with the lack of any radio detection (no counterparts within 30\,arcsec), suggest
the source lies at high redshift. We therefore
fitted galaxy models to the optical-infrared photometry of all objects in the ISAAC $K$-band image within 15\,arcsec
of the 250\,${\rm \mu m}$ position. This search produced seven potential counterparts. Only one of these lies
at $z > 1.5$, and it is the second closest to the 250\,${\rm \mu m}$ position (4.1\,arcsec). It is also favoured
by the 500\,${\rm \mu m}$ position.
We have therefore adopted this galaxy (GOODS MUSIC 2977, at $z_{est} = 2.62$) 
as the best candidate identification in the available data. The GOODS-MUSIC redshift 
for this galaxy is $z = 2.7$, in excellent agreement with our own results. The COMBO 17 redshift is 
$z = 0.115$, revealing the severe limitations of the COMBO 17 catalogue at faint magnitudes ($R > 24$).
As described in Section 6, however, this source does not have a catalogued LABOCA 870${\rm \mu m}$ 
counterpart (as might be expect at high redshift), and it is the only primary BLAST ID with no flux at 
24${\rm \mu m}$. For these reasons, while we present the information we have gathered on this potential
galaxy counterpart, we exclude it from the final redshift distribution for the sample.\\

\noindent
{\bf BLAST 732:} $z_{\rm est} = 2.97$ (or $z_{\rm est} = 2.63$)\\
This source does not have a counterpart at 350\,${\rm \mu m}$ or 500\,${\rm \mu m}$,
but the 250\,${\rm \mu m}$ flux-density is too faint for this to offer a useful redshift constraint.
However, the lack of any radio detection suggests
the source, if real, lies at high redshift, and indeed the 250\,${\rm \mu m}$ position lies only $\simeq 10$\,arcsec from
the fourth brightest 870\,${\rm \mu m}$ LABOCA source in GOODS-South (Weiss et al. 2009).
We therefore fitted galaxy models to the optical-infrared photometry of all objects in the ISAAC $K$-band within 15\,arcsec
of the 250\,${\rm \mu m}$ position. This search produced six potential counterparts, three of
which have estimated redshifts $z > 1.5$. Of these 3, only one has a 24\,${\rm \mu m}$ counterpart, so we adopt this
as the primary identification. This object, which we designate 732-1, is a very red galaxy which was too faint 
to be included in the GOODS-MUSIC catalogue, but which fortuitously lies with the Hubble Ultra Deep Field 
recently imaged with WFC3 on HST (McLure et al. 2010). The resulting high-accuracy photometry for this source, tabulated in 
Table A1, yields a very robust photometric redshift $z_{est} = 2.97)$, as ilustrated in Fig. 5. The alternative 
IDs are 732-2, which is GOODS MUSIC 30080 ($z_{est} = 2.63$), and 
732-3= GOODS MUSIC 10787 ($z_{est} = 2.40$). Choosing between these alternatives clearly does not significantly
affect the final redshift distribution, and it is certainly possible that 
both the 250 and 870\,${\rm \mu m}$ emission may arise from a blend of a number of sources at $z \simeq 3.0$.\\

\noindent
{\bf BLAST 830:} $z_{\rm spec} = 0.605$ (or $z_{\rm spec} = 0.735$)\\
This source does not have a counterpart at 350\,${\rm \mu m}$ or 500\,${\rm \mu m}$,
but the 250\,${\rm \mu m}$ flux density is too faint for this to offer a useful redshift constraint.
There are two alternative, statistically-significant radio counterparts, and we select 
VLA 239 in favour of VLA 1179. The chosen identification
appears to be an interacting galaxy with spectroscopic redshift $z = 0.605$. The 
edge-on disc galaxy a few arcsec NE lies at $z = 0.735$ and so is not physically
associated with the adopted identification. However, the 250\,${\rm \mu m}$ flux density may receive contributions 
from both these objects, and from VLA 1179. Whatever the exact division of blended emission, it is
clear that this 250\,${\rm \mu m}$ arises primarily from sources at $z \simeq 0.6 - 0.75$.\\

\noindent
{\bf BLAST 861:} $z_{\rm est} = 1.95$\\
This source does not have a counterpart at 350\,${\rm \mu m}$ or 500\,${\rm \mu m}$,
but the 250\,${\rm \mu m}$ flux-density is too faint for this to offer a useful redshift constraint.
The radio indentification, at a radius of 11.9\,arcsec, only just passes the adopted
significance threshold $P < 0.1$. This identifies the BLAST source with a 
faint, red,
possibly interacting galaxy, for which the available photometry delivers a 
unique and well-constrained estimated redshift at $z \simeq 2$. As with BLAST 66, the published COMBO17 
photometric redshift $z_{phot} = 0.881$ is very poorly constrained, and can be safely rejected.\\

\noindent
{\bf BLAST 983:} $z_{\rm spec} = 0.366$\\
This 250\,${\rm \mu m}$ source is not found in the BLAST catalogues at longer wavelengths,
suggesting it lies at only moderate redshift. The radio identication is formally
secure with $P < 0.1$. It lies rather far from the 250\,${\rm \mu m}$ position, almost
at our limiting search radius, but this is not unreasonable given that it is the 
second least significant source in our sample. The resulting galaxy identification 
at $z = 0.366$ can clearly be seen to be dusty in Fig. 7. Only 3\,arcsec distant is
another less massive red galaxy at essentially the same redshift ($z = 0.368$).\\
 
\noindent
{\bf BLAST 1293:} $z_{\rm spec} = 1.382$\\
This source does not have a catalogued counterpart at 350\,${\rm \mu m}$ or 500\,${\rm \mu m}$,
but the 250\,${\rm \mu m}$ flux-density is too faint for this to offer a useful redshift constraint.
There are two alternative, formally-significant, radio counterparts, and we select 
VLA 211 in preference to VLA 148. While the 250\,${\rm \mu m}$ source could be a blend of both 
objects, our statistically chosen identification transpires to be a bright source at 24\,${\rm \mu m}$, while the 
alternative radio counterpart is undetected in the mid-infrared.
As can be seen from Fig.~7, our selected identification
is a red, high-redshift, apparently interacting galaxy. The GOODS-MUSIC redshift for this object is $z = 1.99$,
but our own photometric redshift for this source 
is $z_{phot} \simeq 1.37$ (see last panel of Fig.~5). The latest release 
of VLT FORS2 GOODS-South redshifts yields $z_{spec} = 1.382$ (Vanzella et al. 2008).\\

\clearpage

\begin{figure*}
\begin{tabular}{llll}

\includegraphics[width=0.24\textwidth]{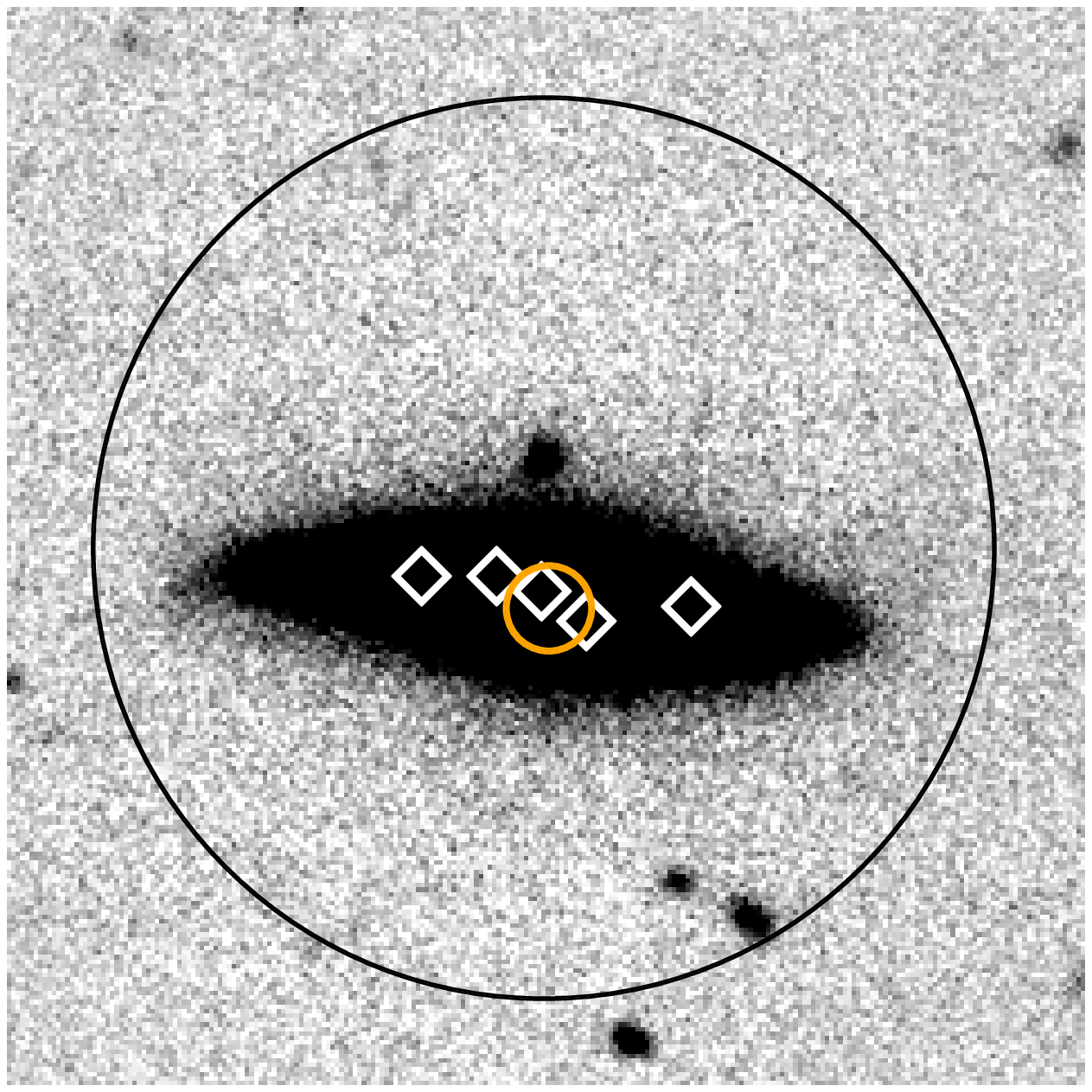}&
\includegraphics[width=0.24\textwidth]{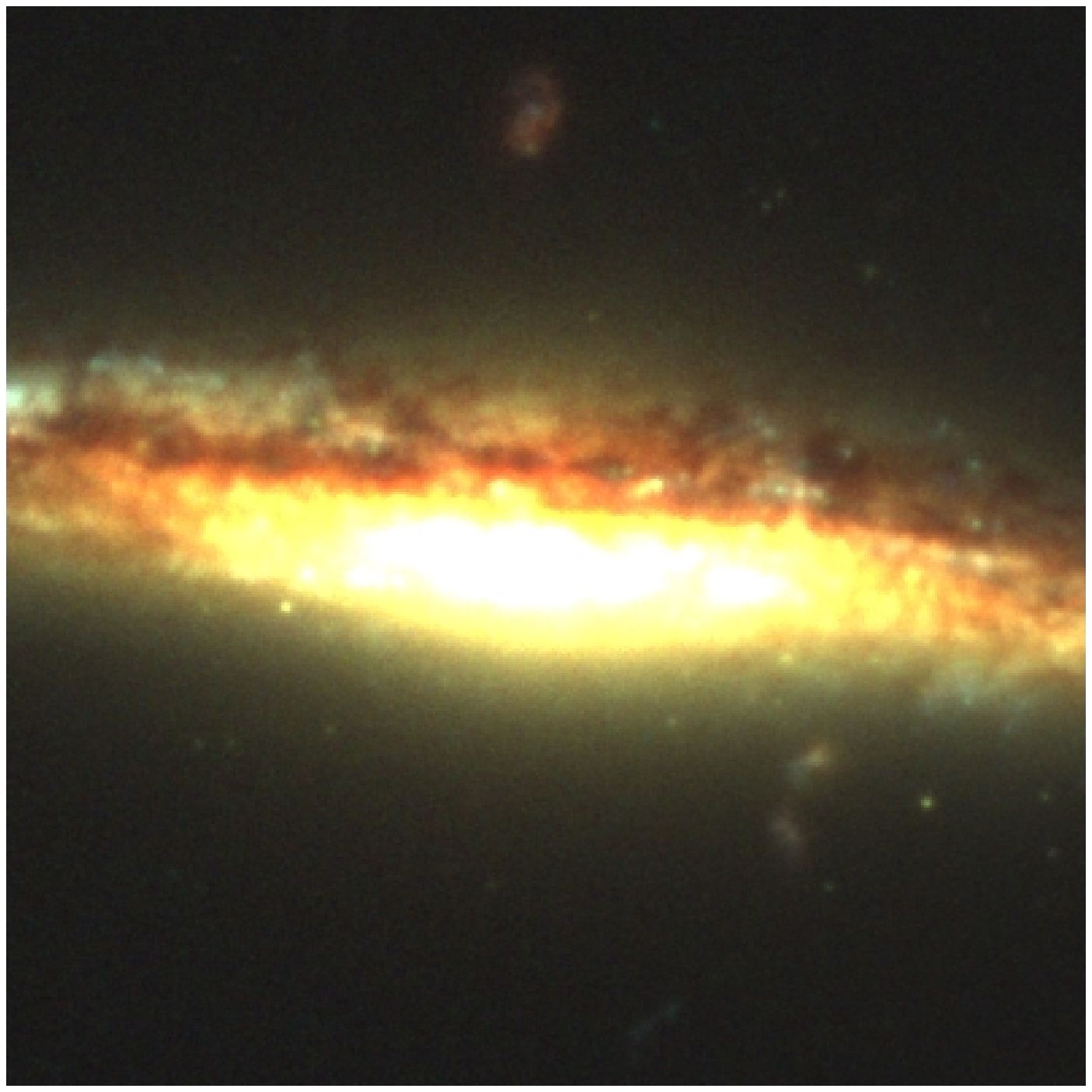}&
\includegraphics[width=0.24\textwidth]{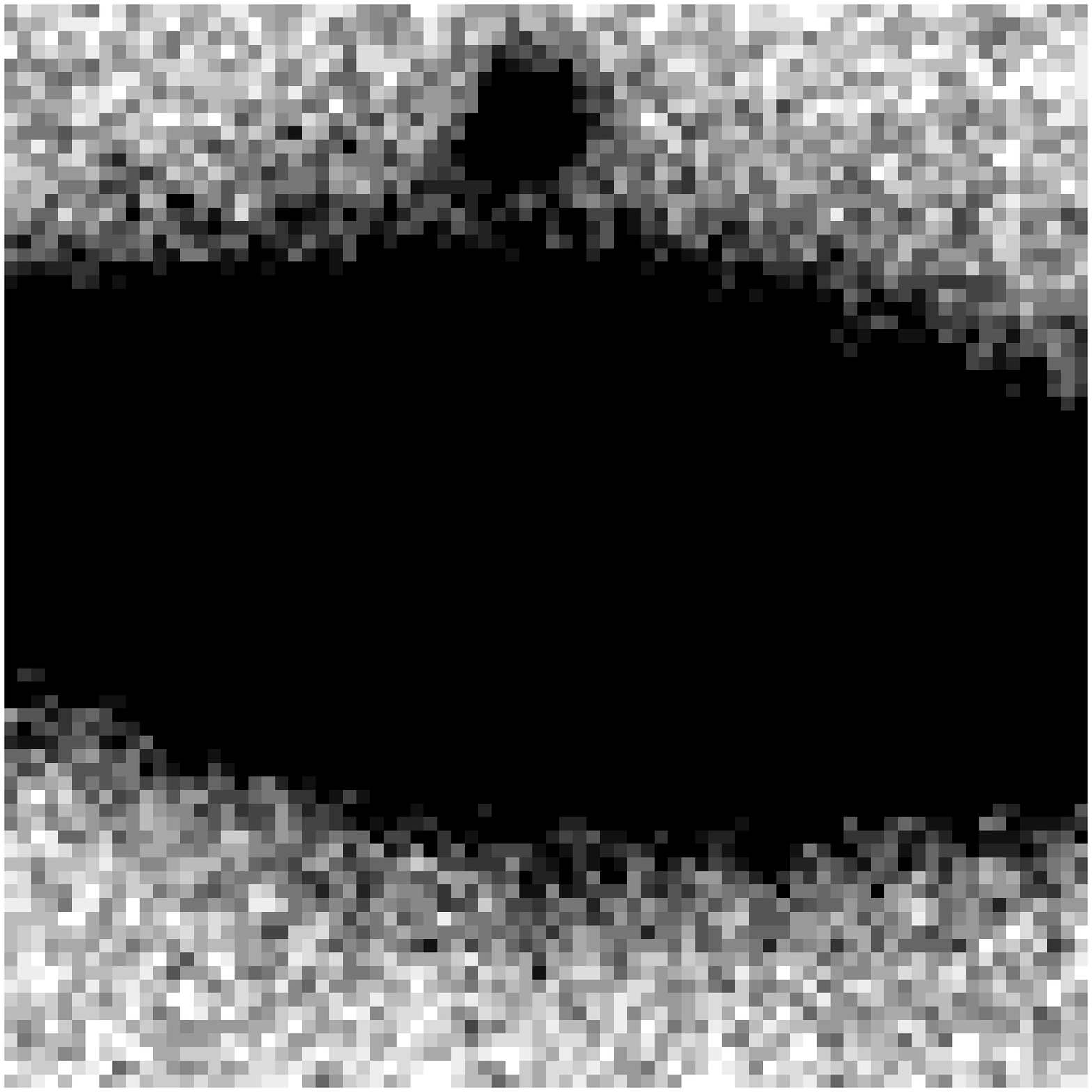}&
\includegraphics[width=0.24\textwidth]{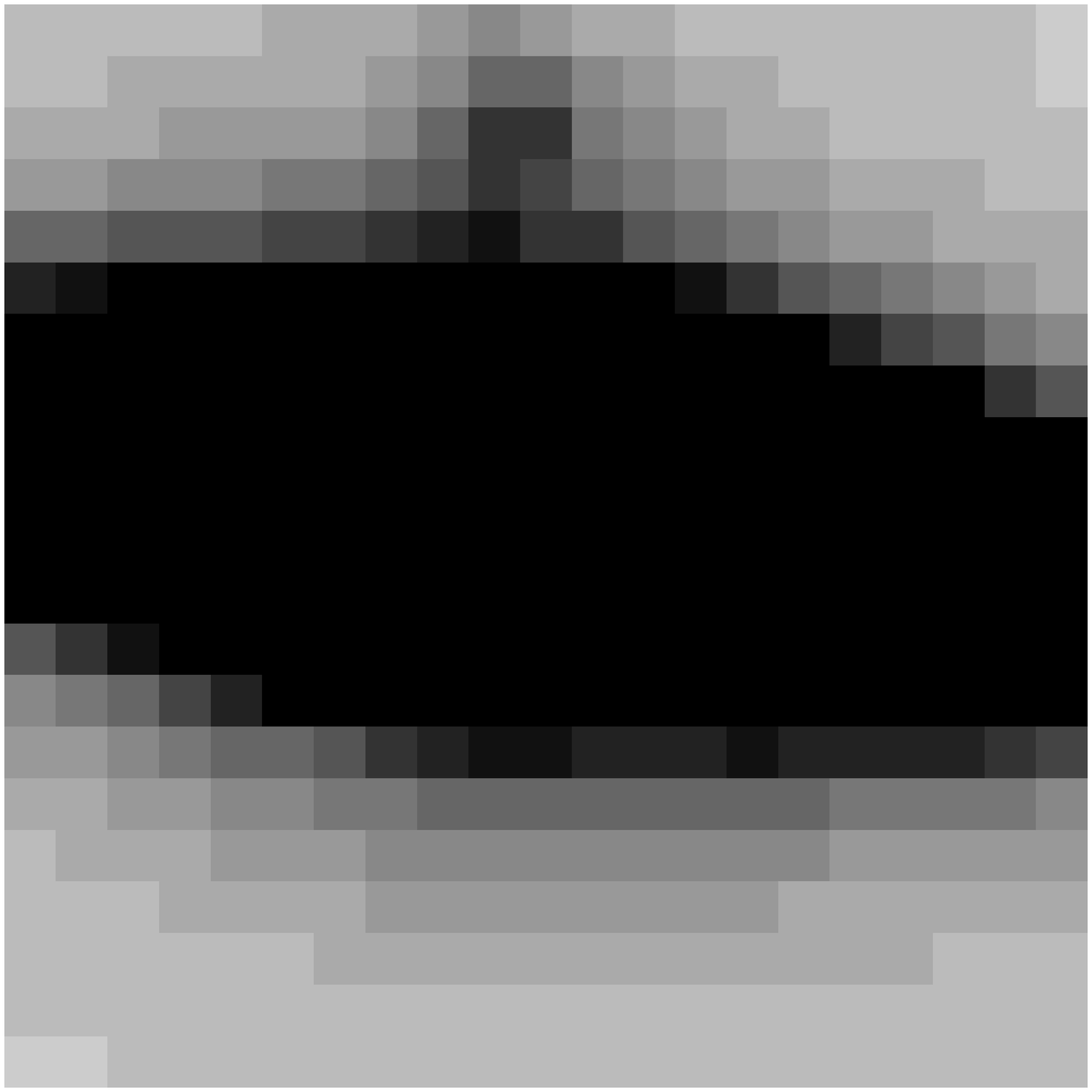}\\
{\bf BLAST 4:} $z_{\rm spec} = 0.038$\\
\\
\\                                
\includegraphics[width=0.24\textwidth]{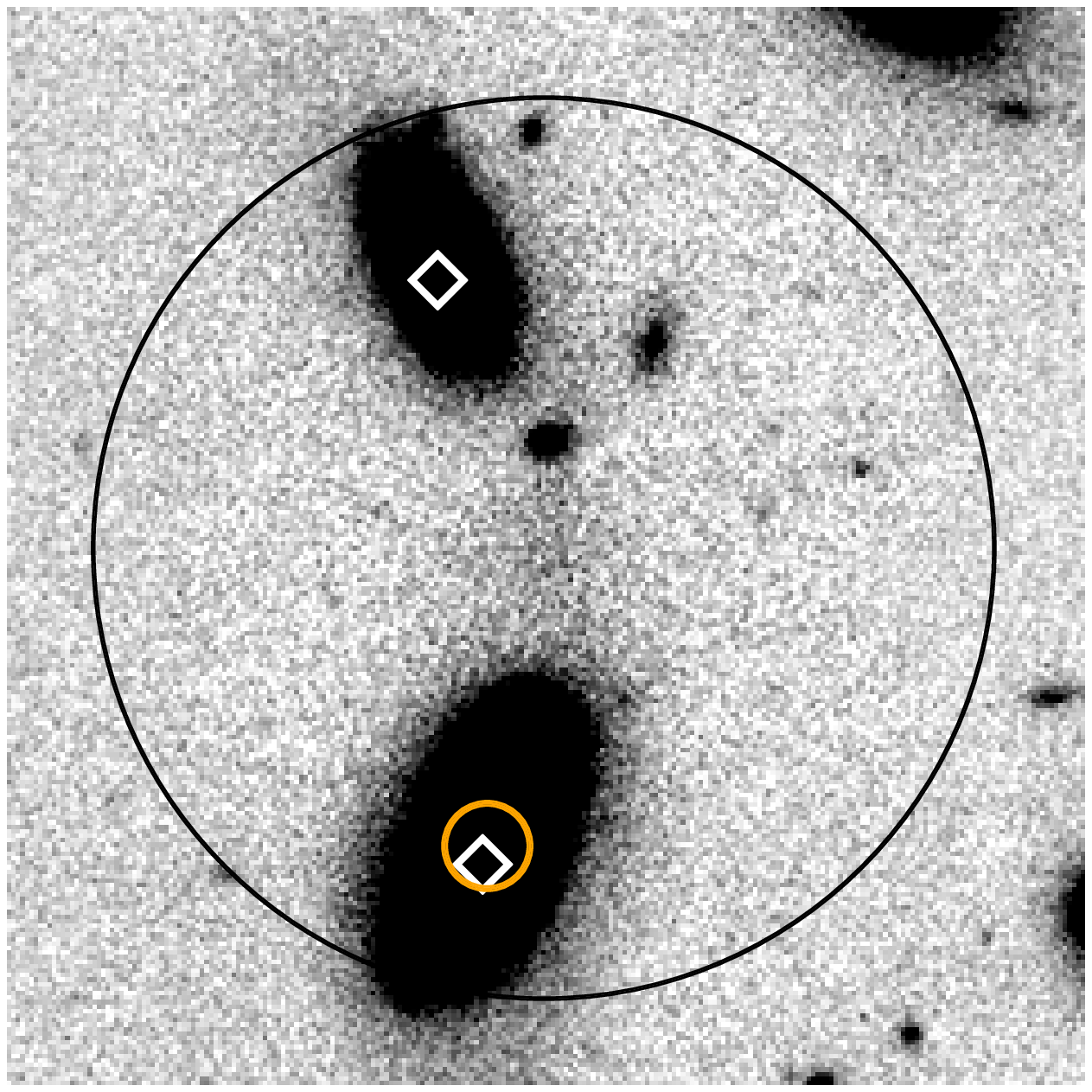}&
\includegraphics[width=0.24\textwidth]{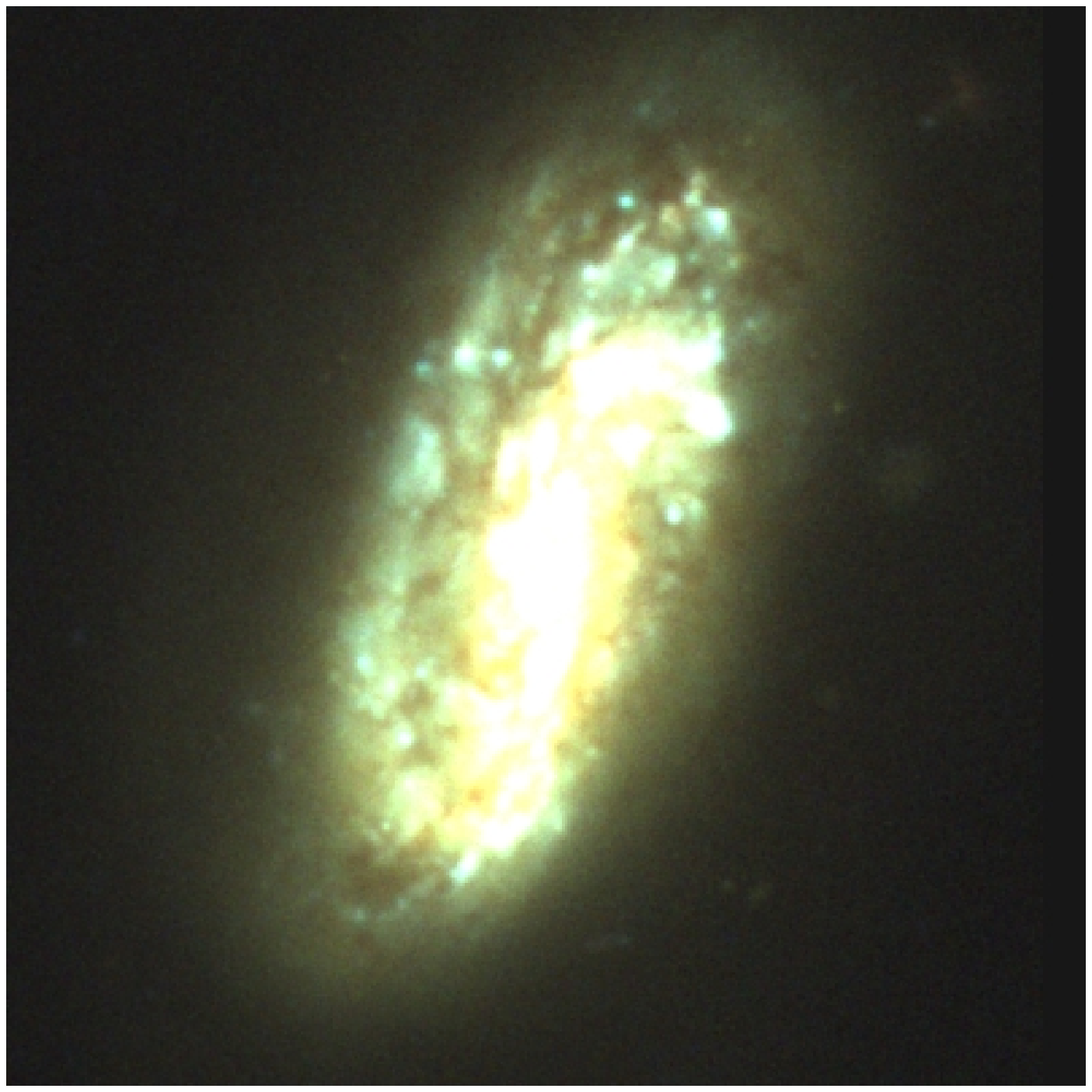}&
\includegraphics[width=0.24\textwidth]{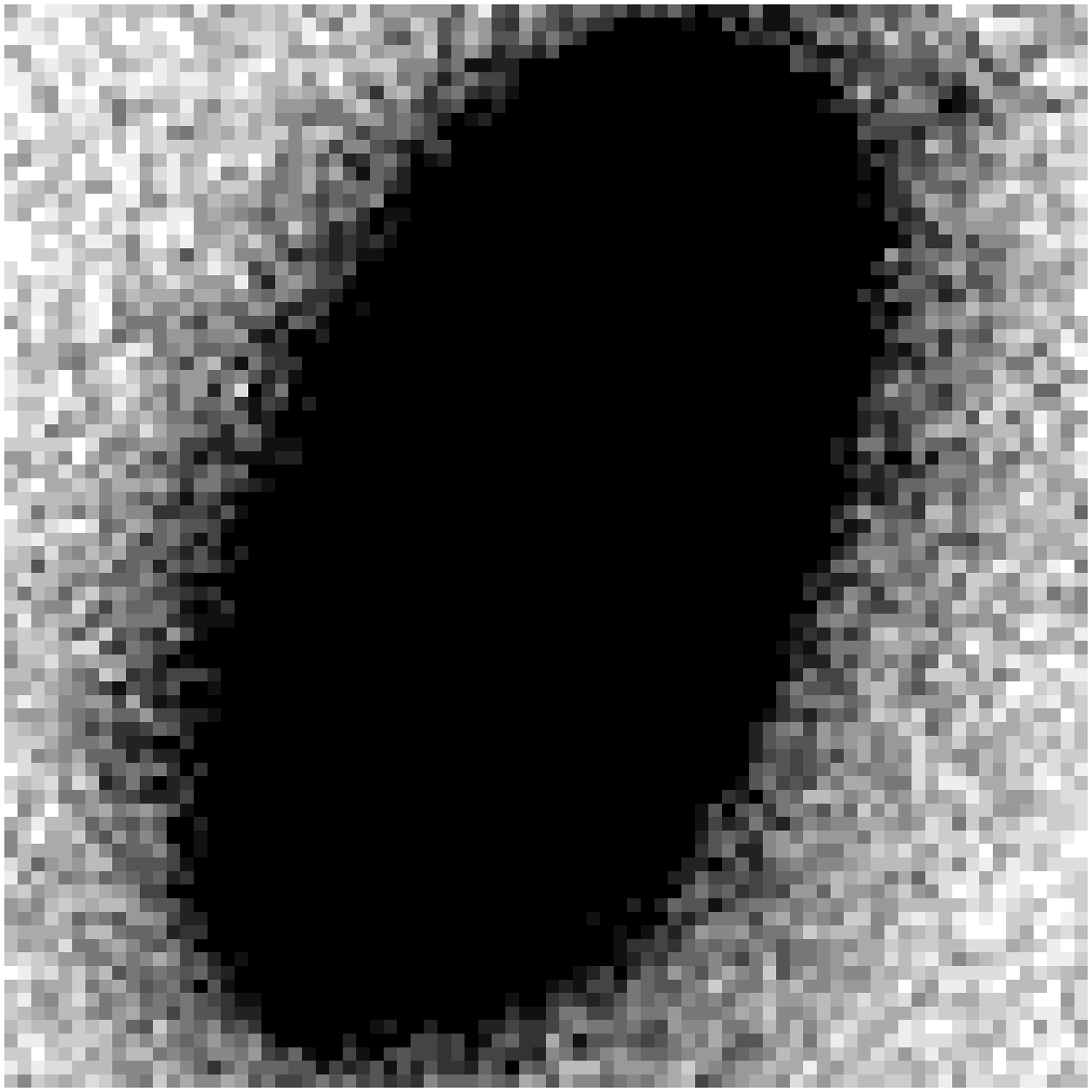}&
\includegraphics[width=0.24\textwidth]{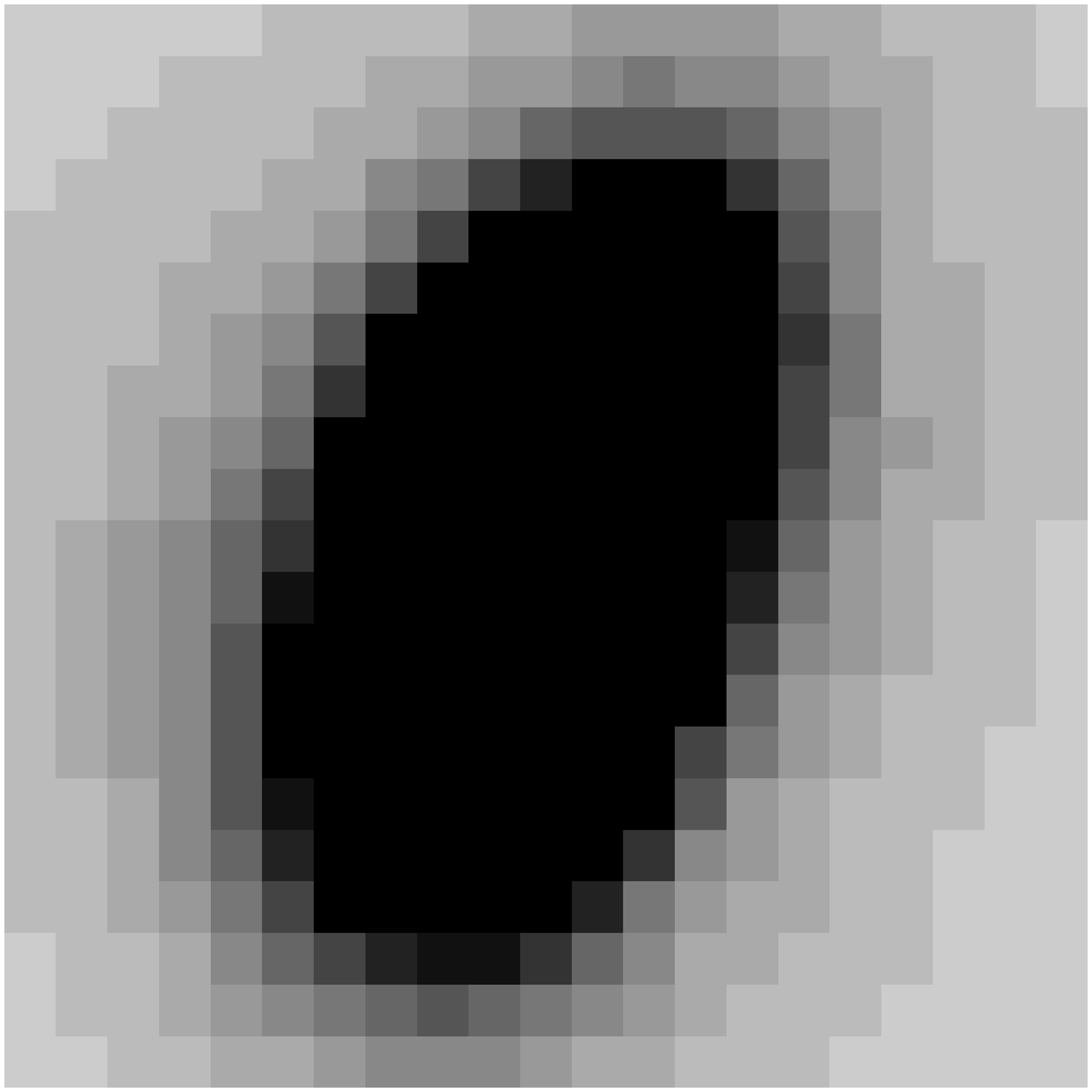}\\
{\bf BLAST 6-1:} $z_{\rm spec} = 0.076$\\
\\
\\      
\includegraphics[width=0.24\textwidth]{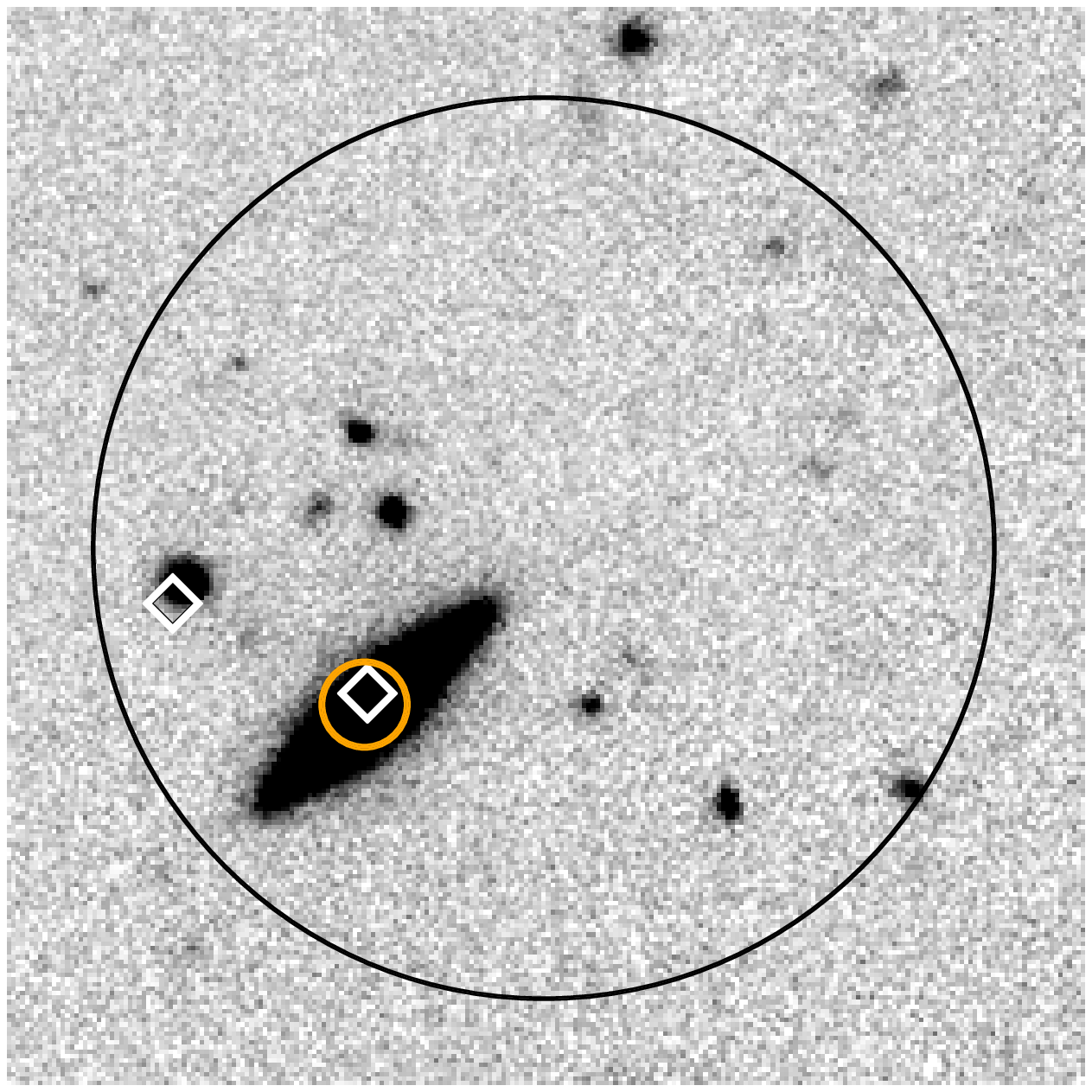}&
\includegraphics[width=0.24\textwidth]{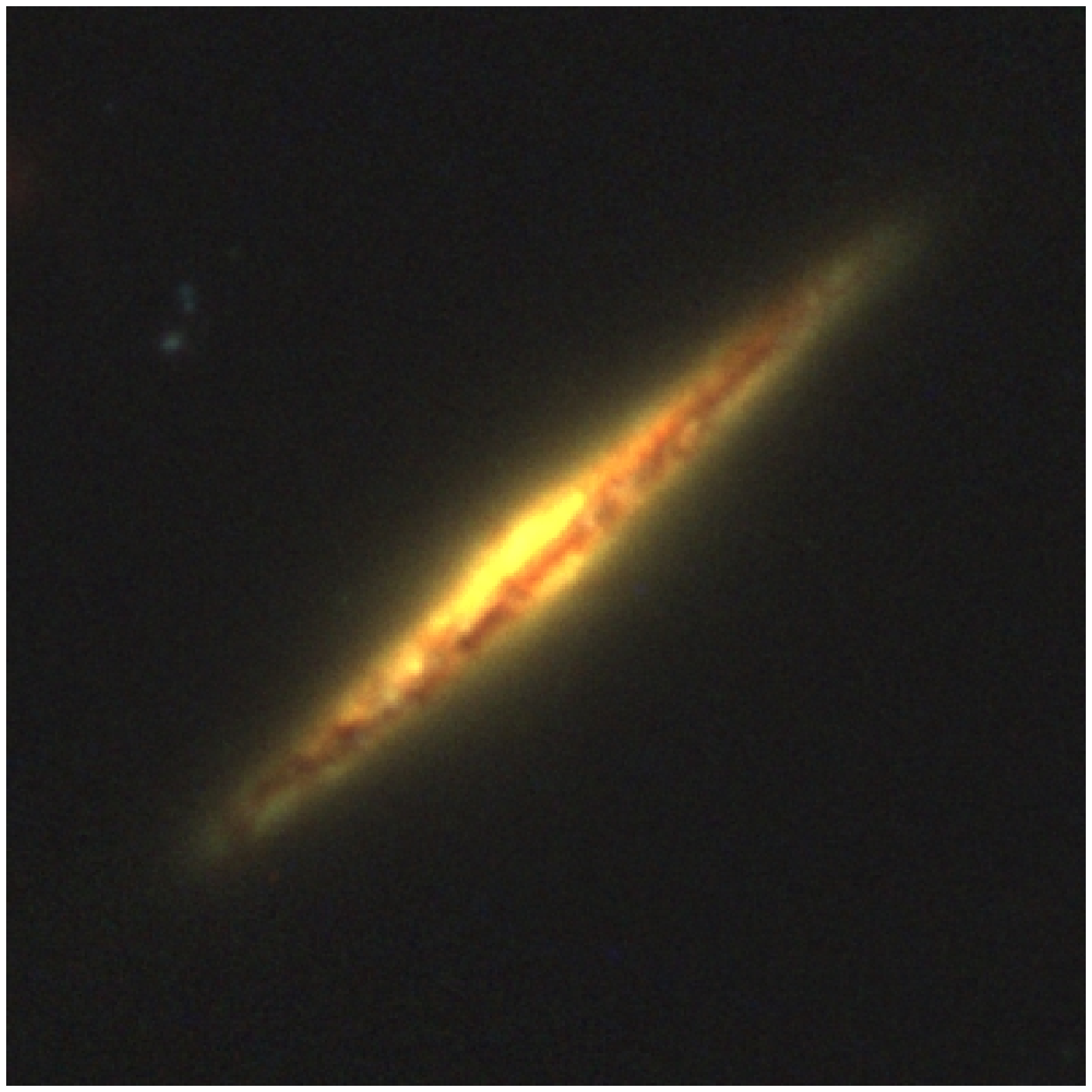}&
\includegraphics[width=0.24\textwidth]{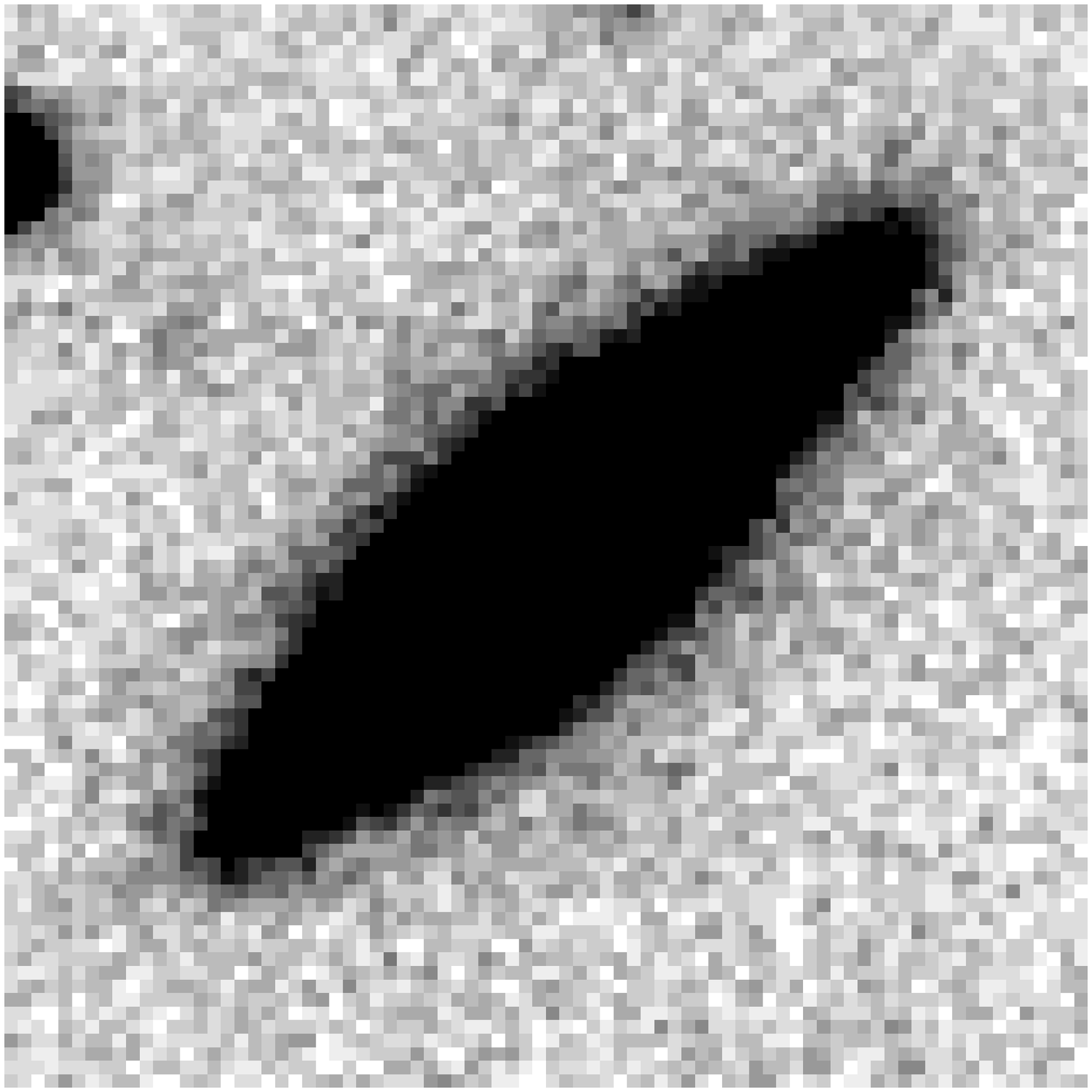}&
\includegraphics[width=0.24\textwidth]{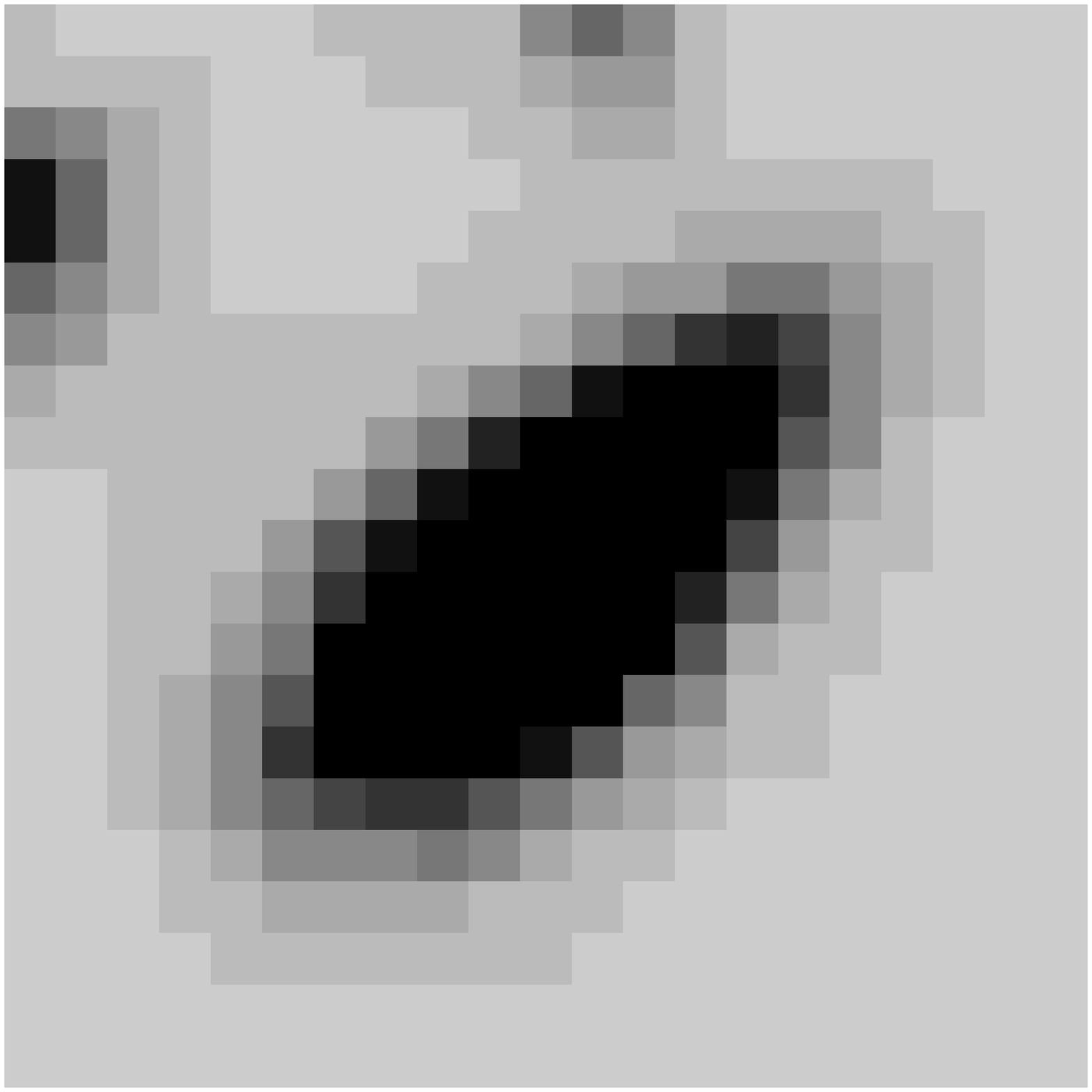}\\
{\bf BLAST 109:} $z_{\rm spec} = 0.124$\\
\\
\\
\includegraphics[width=0.24\textwidth]{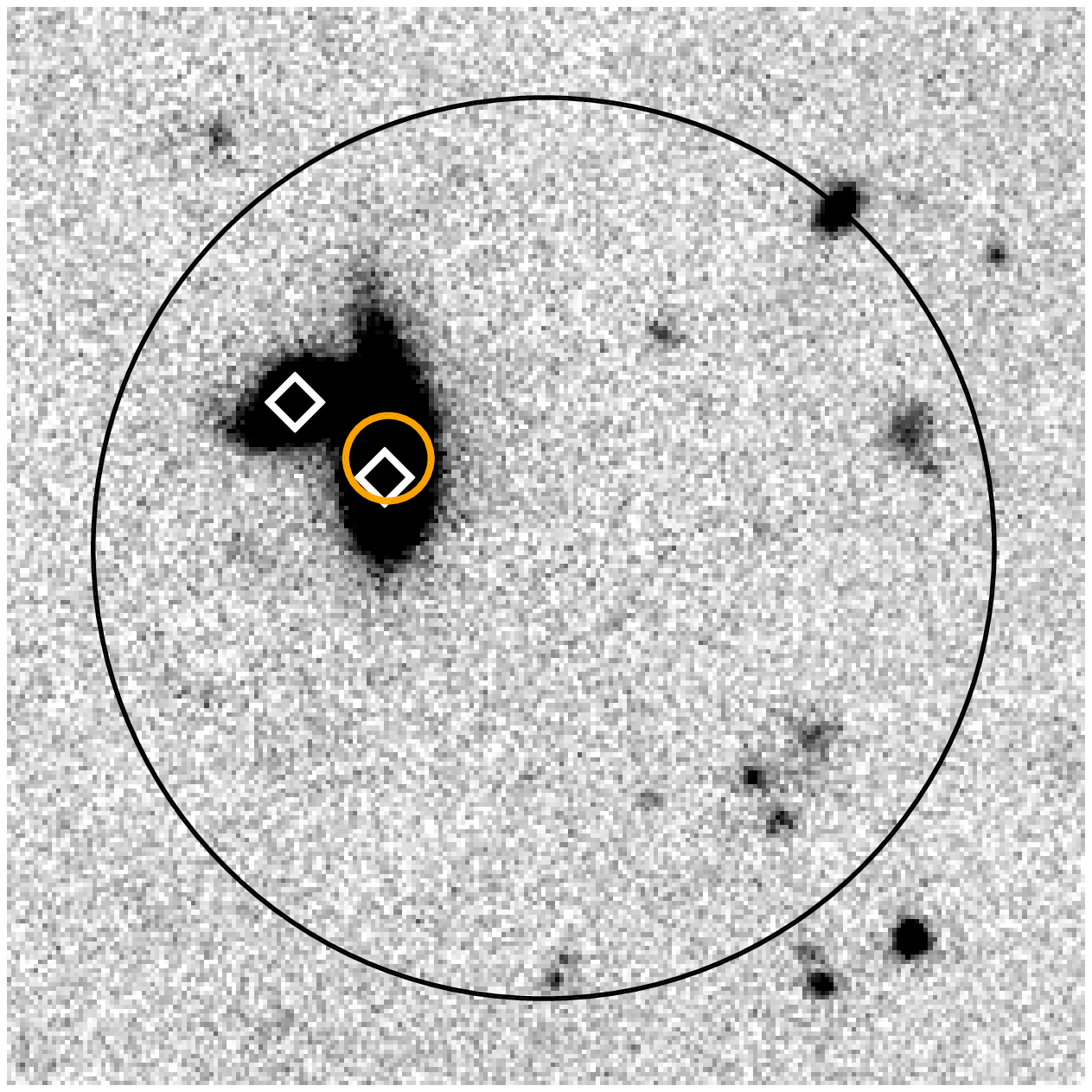}&
\includegraphics[width=0.24\textwidth]{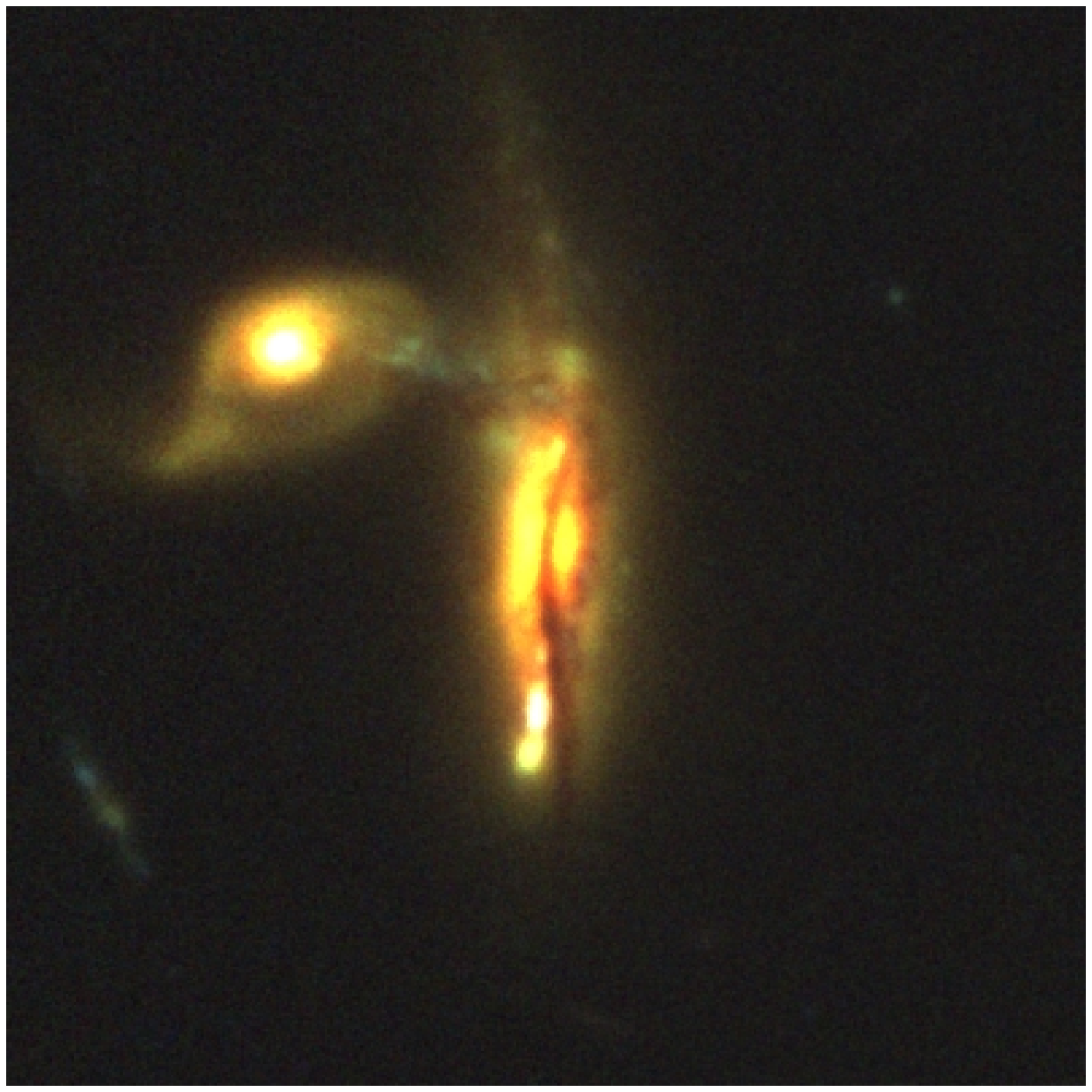}&
\includegraphics[width=0.24\textwidth]{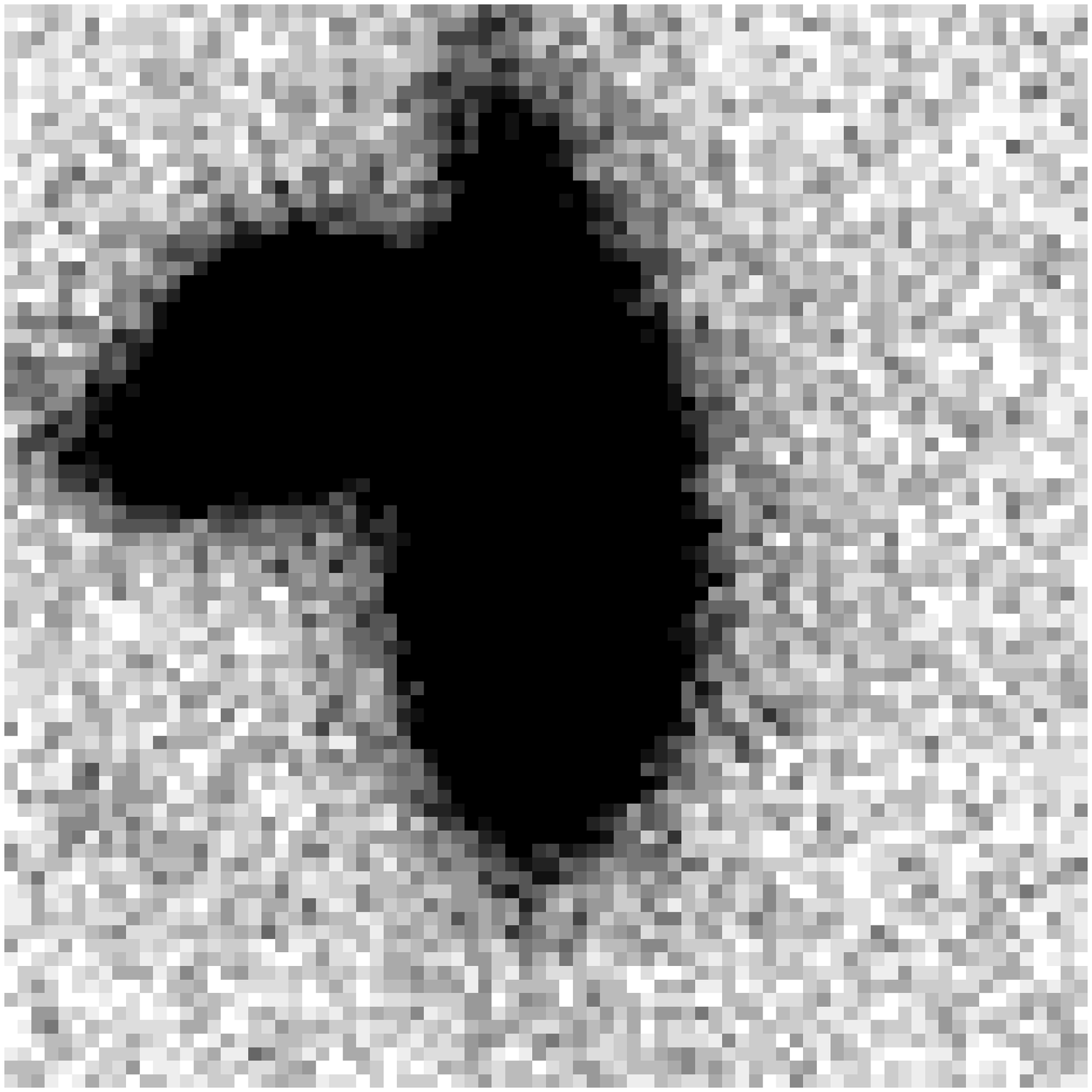}&
\includegraphics[width=0.24\textwidth]{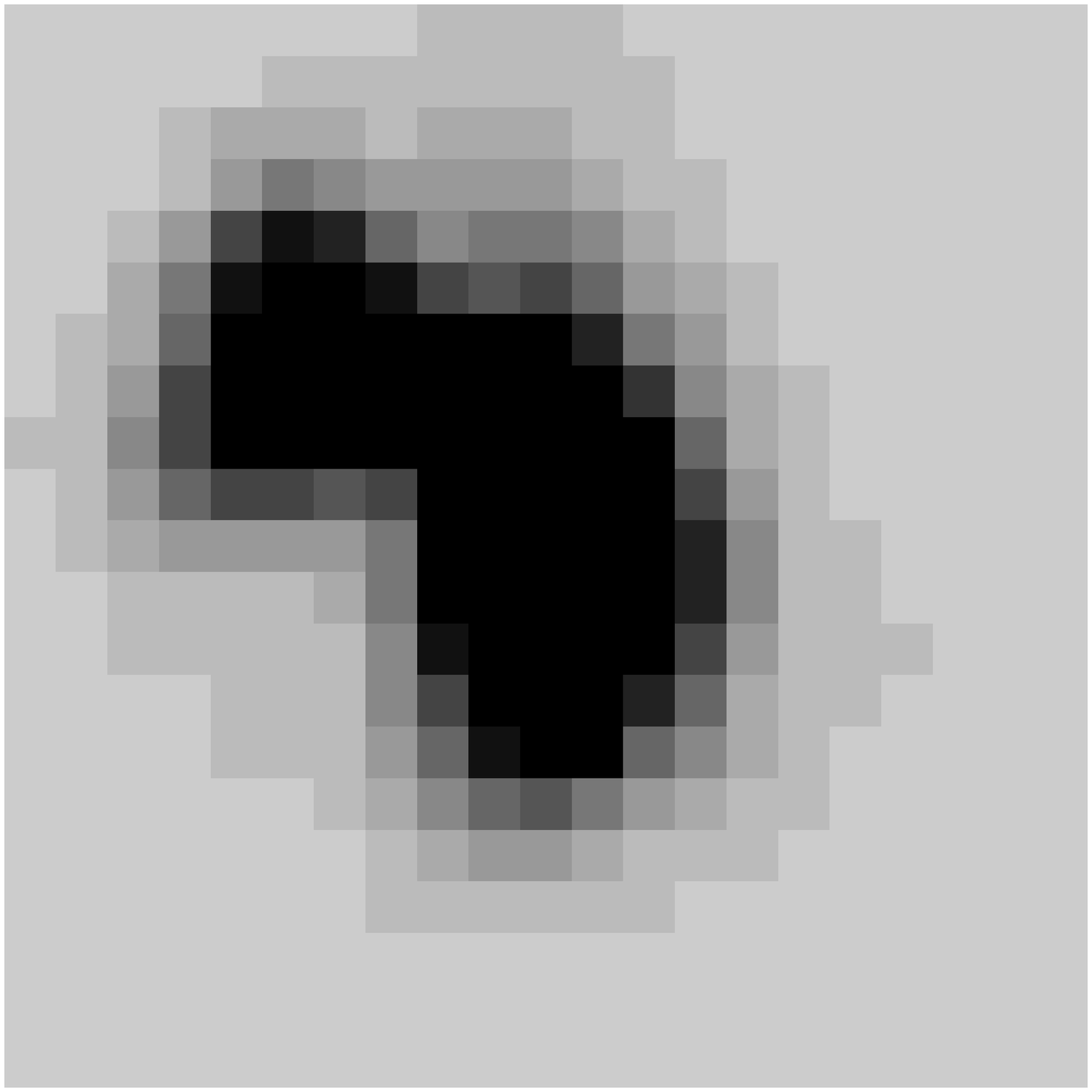}\\
{\bf BLAST 637-1:} $z_{\rm spec} = 0.279$\\

\end{tabular}
\caption{Optical/infared postage stamp images of the primary galaxy counterparts we have identified for the 20 
BLAST 250\,${\rm \mu m}$ sources in GOODS-South, ranked by redshift (with BLAST source 
number and redshift given below each left-hand stamp).
Each row of stamps commences with a 36 x 36 arcsec {\it HST} ACS $z$-band image, centred on the BLAST source position, with the identification search 
area indicated by the 15 arcsec radius circle. White diamonds mark the available VLA candidate counterparts, with the selected counterpart 
marked by the orange circle. The remaining three stamps in each row are centred on the position of the 
selected counterpart, and are 12 x 12 arcsec in size. From left-to-right these show a $B+V+z$ {\it HST} ACS colour image, 
the $K_s$-band VLT ISAAC image and the 3.6\,${\rm \mu m}$ {\it Spitzer} IRAC image. Display levels on the colour images have been set to 
$+100\sigma, -1\sigma$ for a $z = 0$ source, and then these values are reduced by the factor $(1+z)^3$ at progressively higher redshifts (to offset 
surface brightness dimming). In all grey-scale plots black is set to $+8\sigma$, and white to $-8\sigma$ where $\sigma$ is the pixel rms.
As explained in Section 2, two sources lie just outside the available near-infrared imaging, and hence lack a $K_s$-band postage stamp.}

\label{RESULTS}
\end{figure*}

\begin{figure*}
\begin{tabular}{llll}

\includegraphics[width=0.24\textwidth]{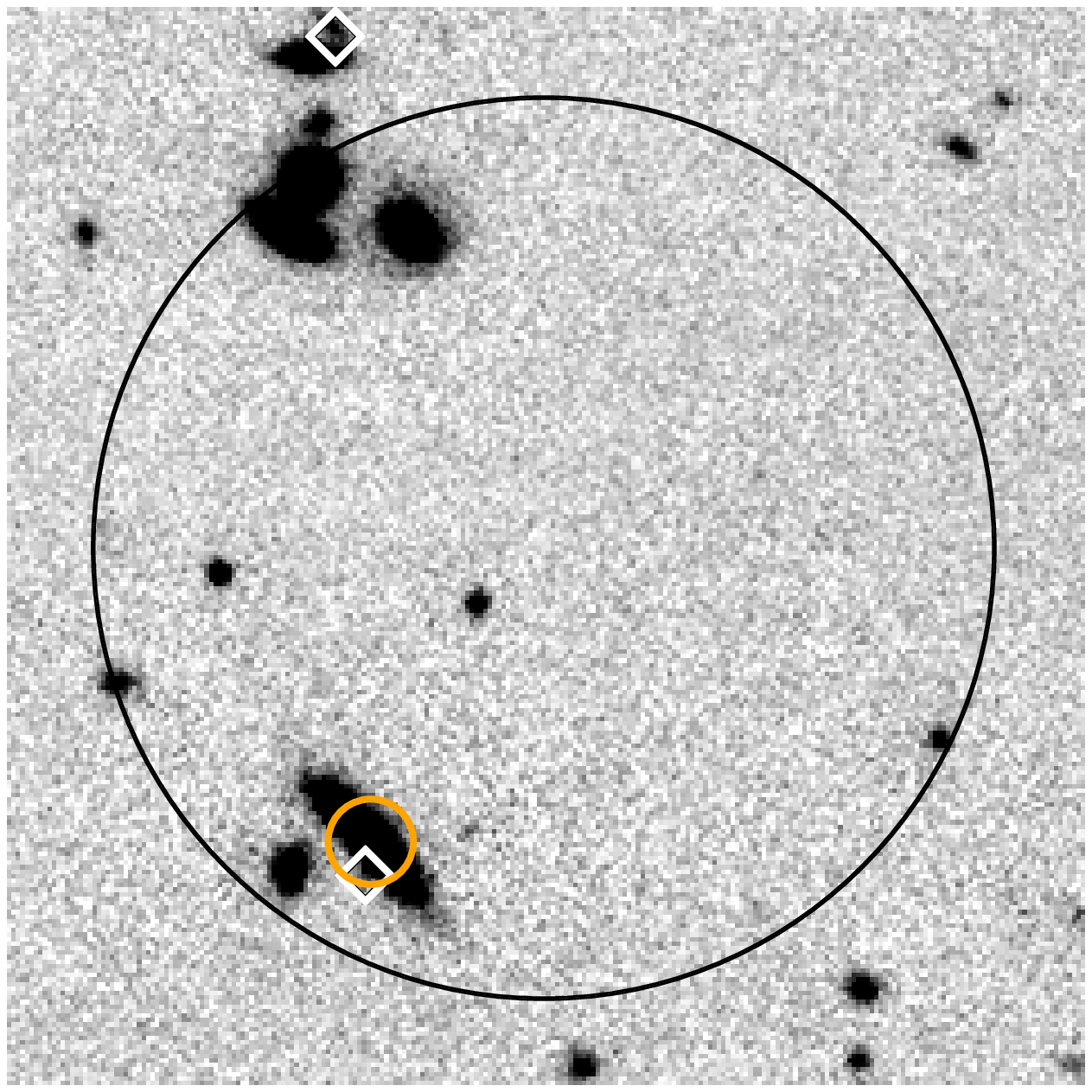}&
\includegraphics[width=0.24\textwidth]{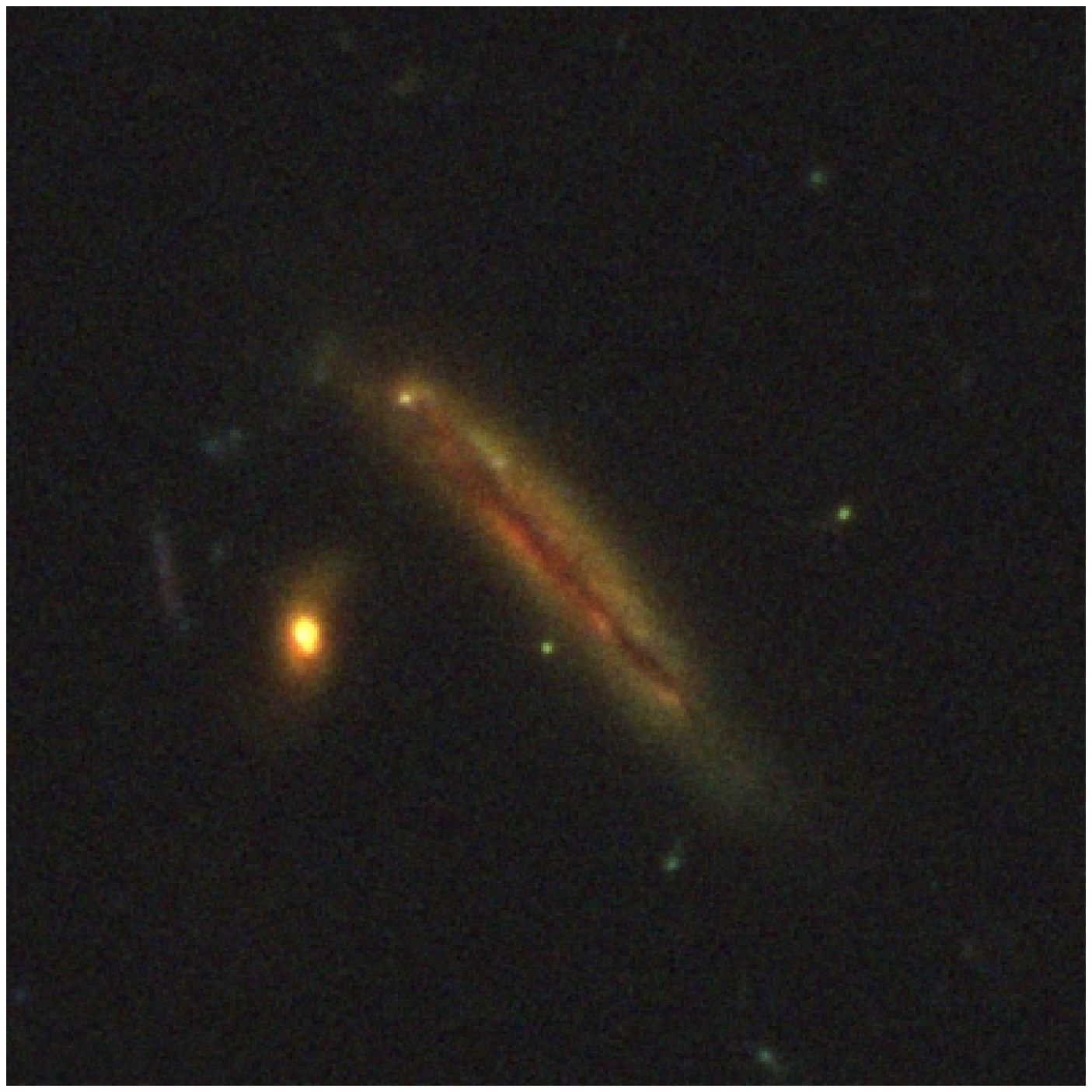}&
\includegraphics[width=0.24\textwidth]{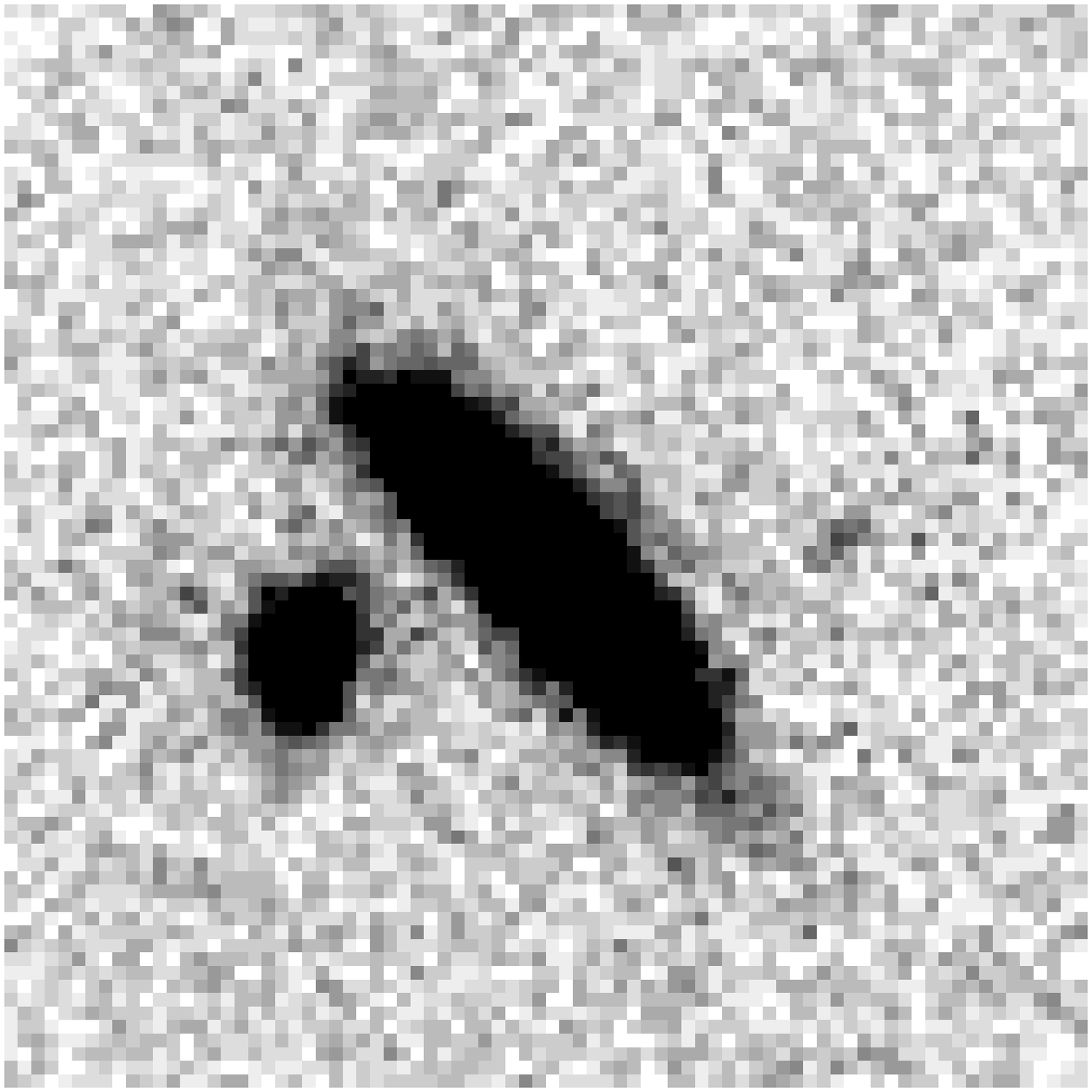}&
\includegraphics[width=0.24\textwidth]{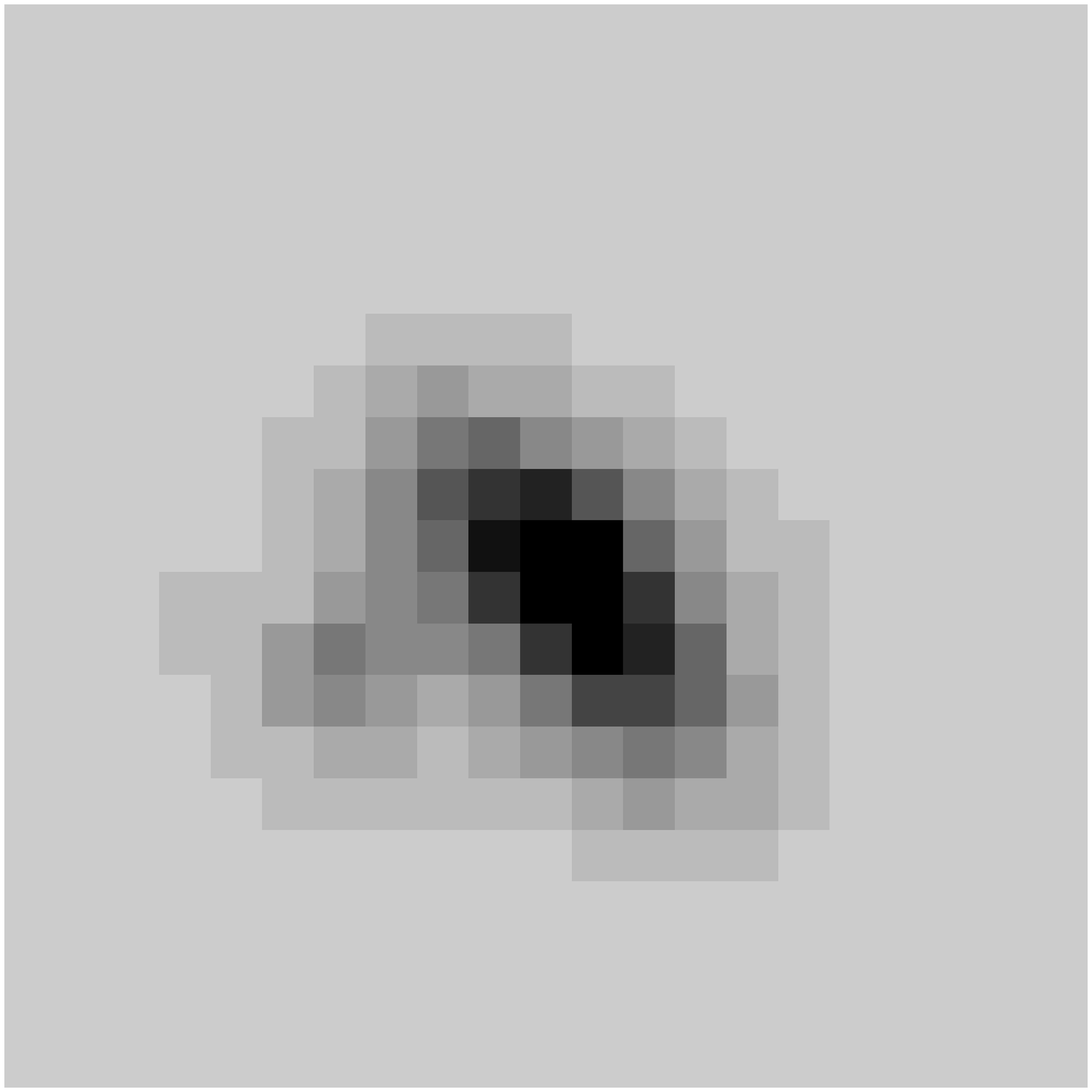}\\
{\bf BLAST 983:} $z_{\rm spec} = 0.366$\\
\\
\\
\includegraphics[width=0.24\textwidth]{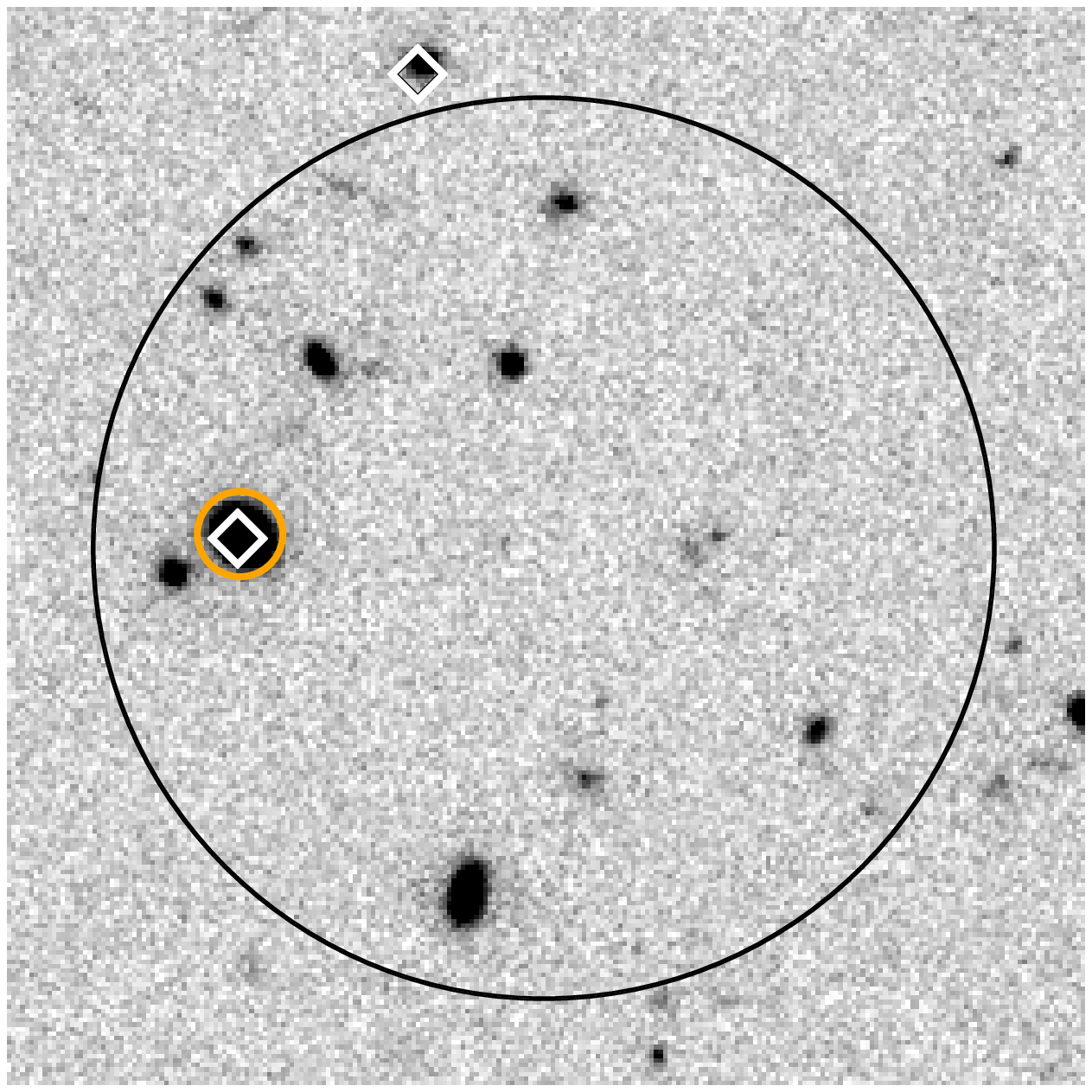}&
\includegraphics[width=0.24\textwidth]{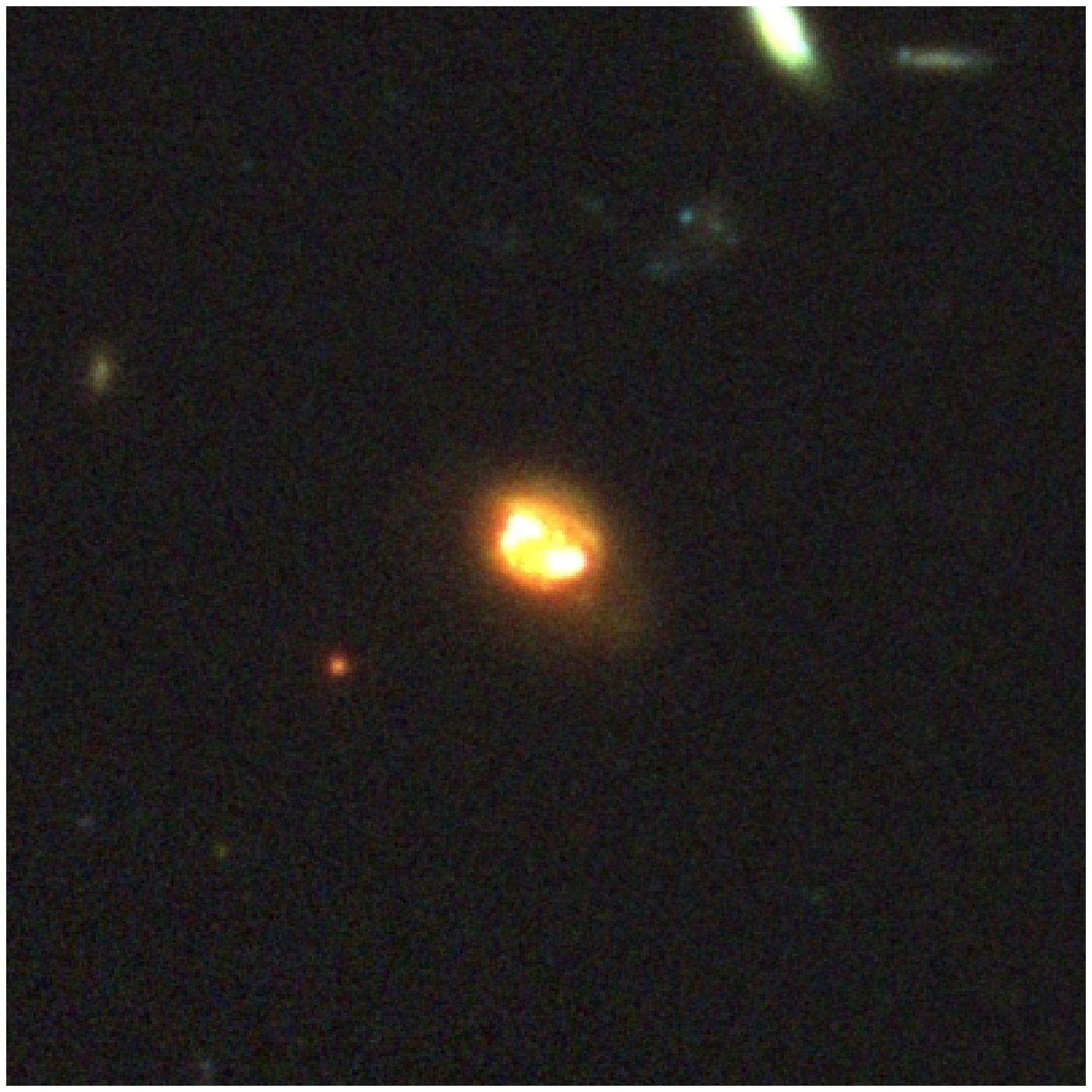}&
\includegraphics[width=0.24\textwidth]{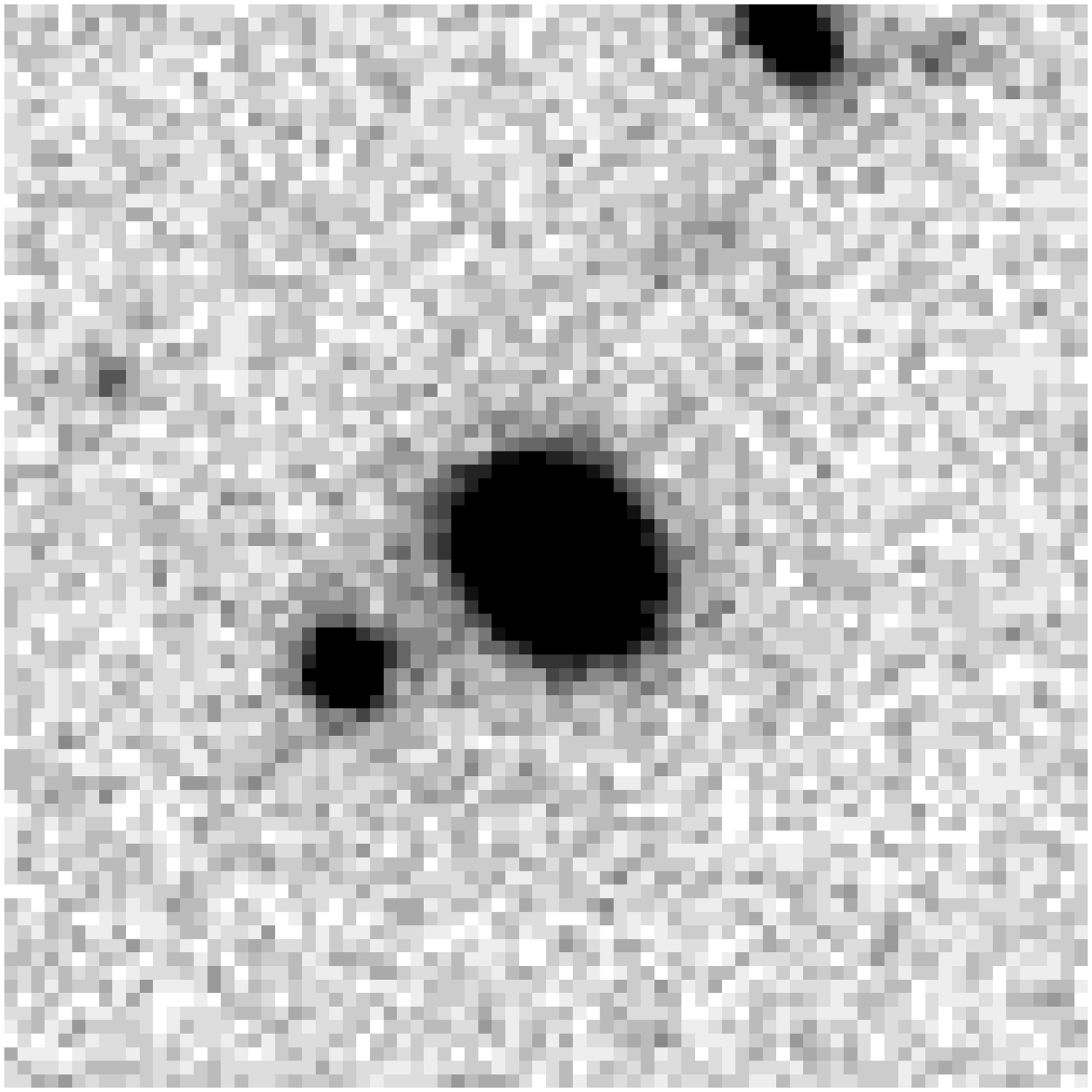}&
\includegraphics[width=0.24\textwidth]{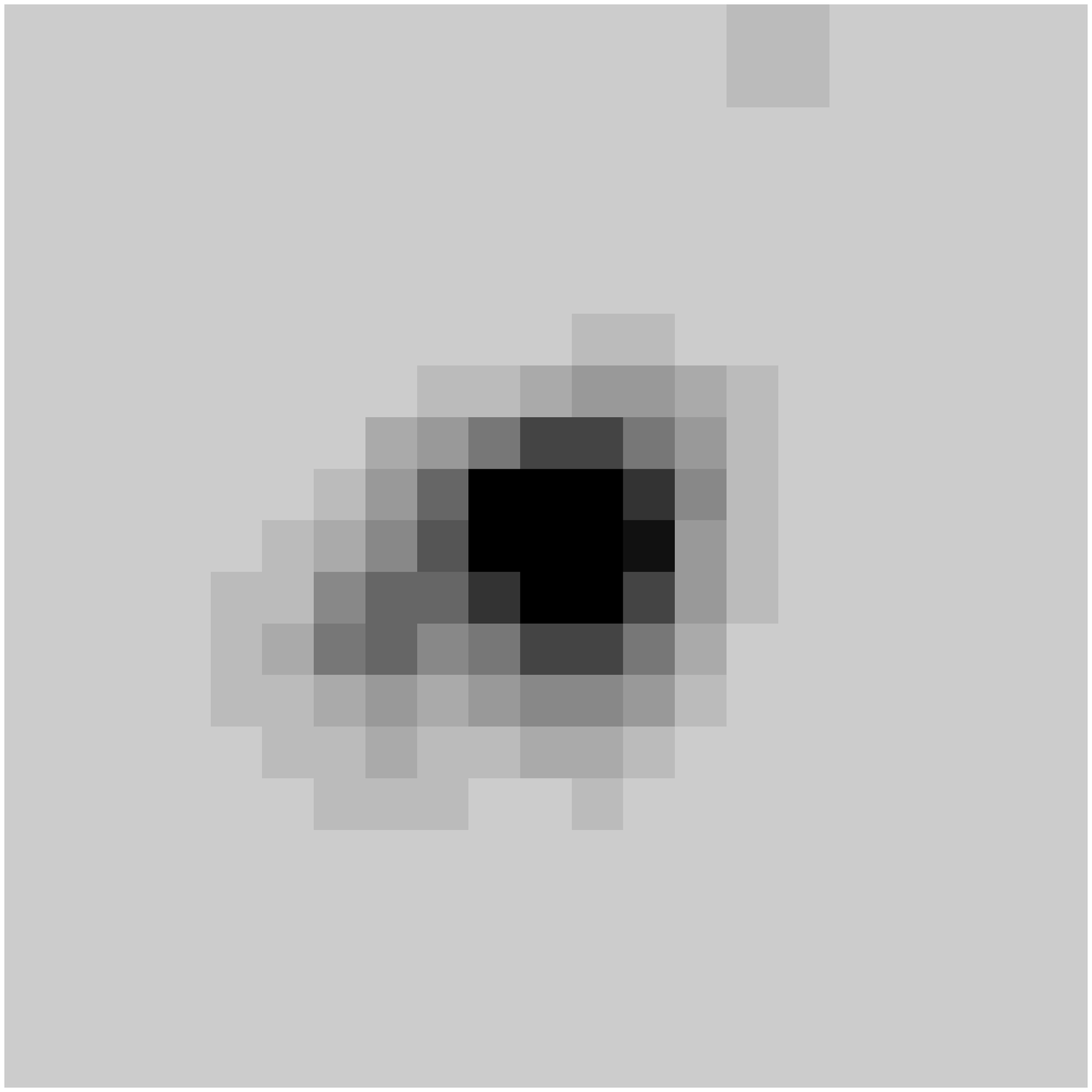}\\
{\bf BLAST 104:} $z_{\rm spec} = 0.547$\\
\\
\\
\includegraphics[width=0.24\textwidth]{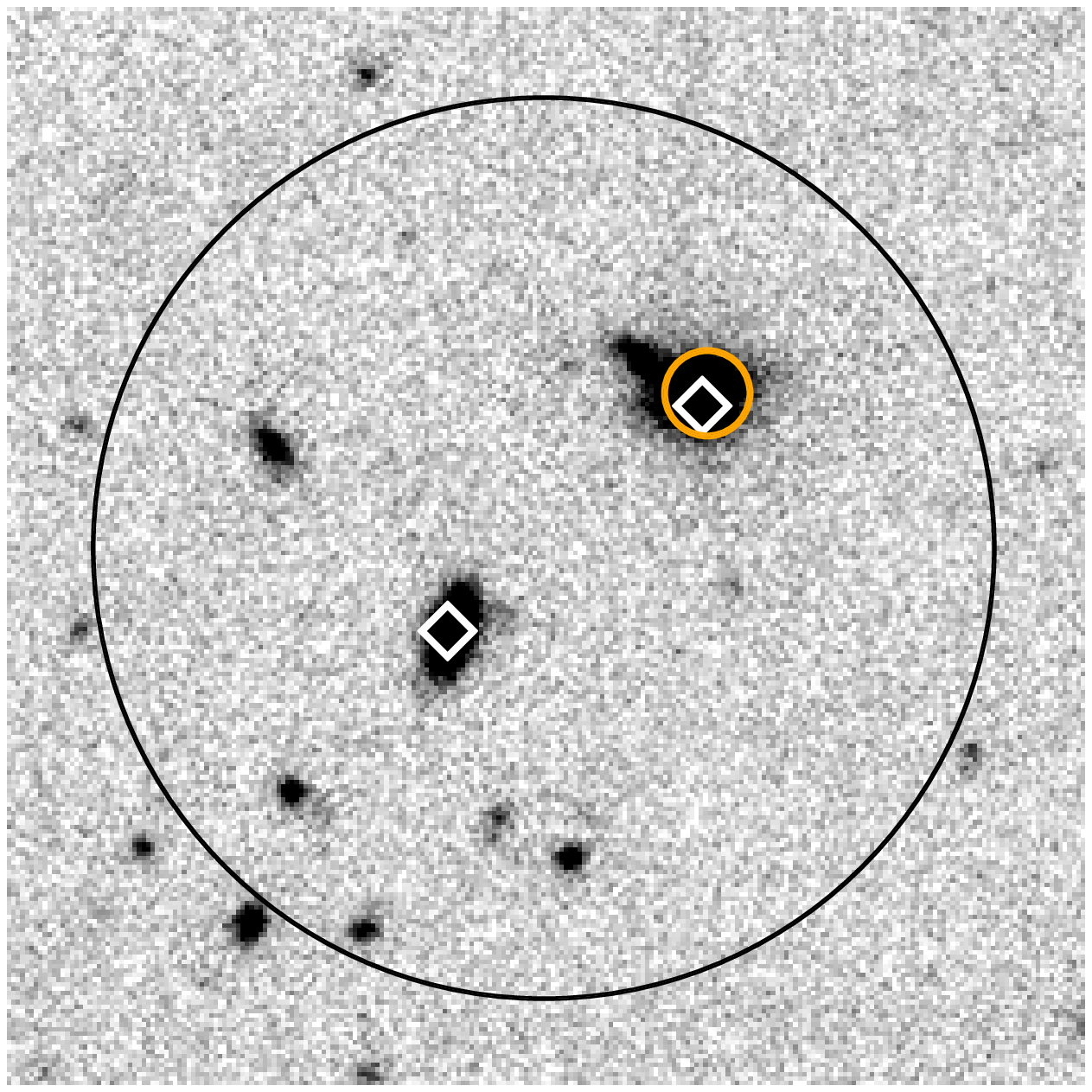}&
\includegraphics[width=0.24\textwidth]{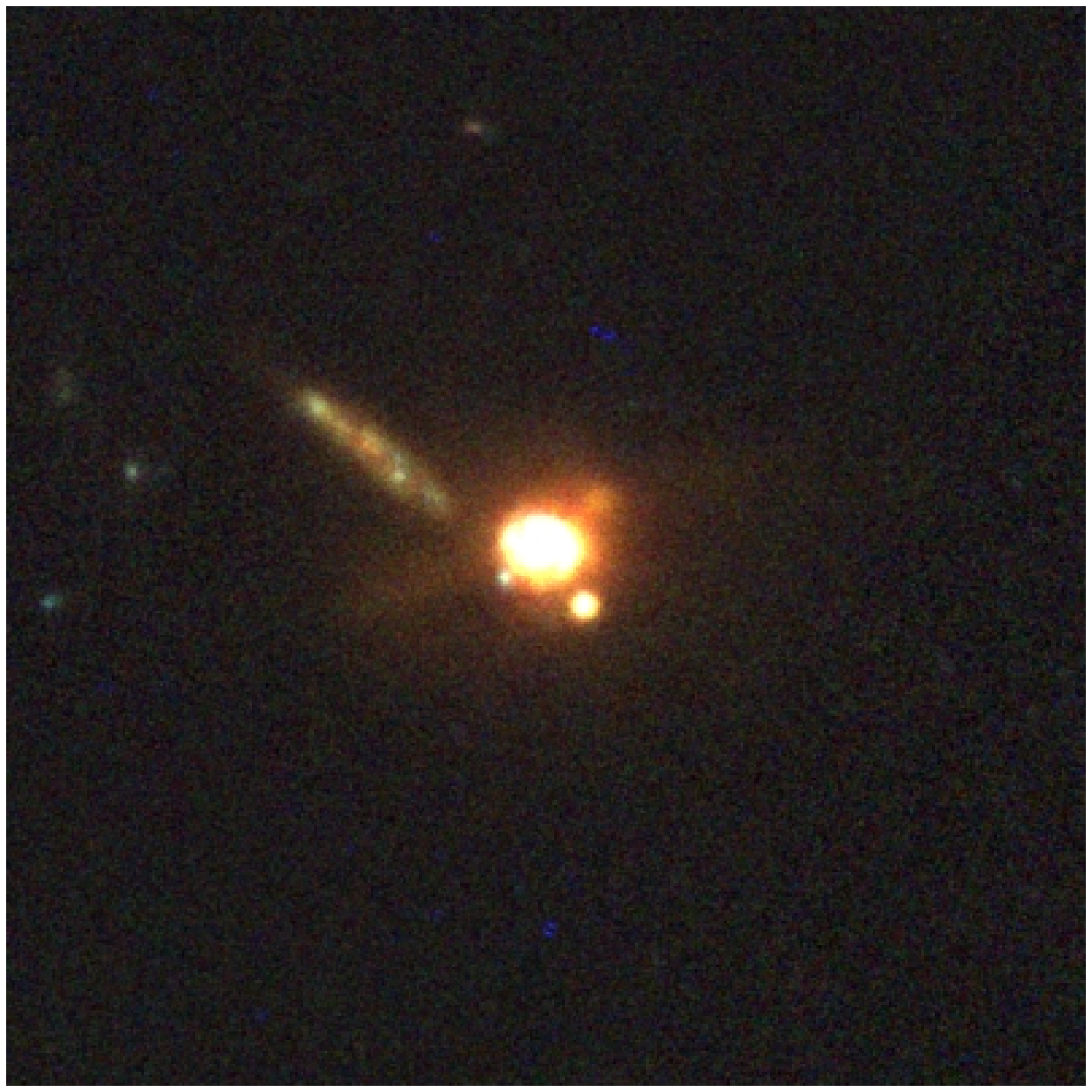}&
\includegraphics[width=0.24\textwidth]{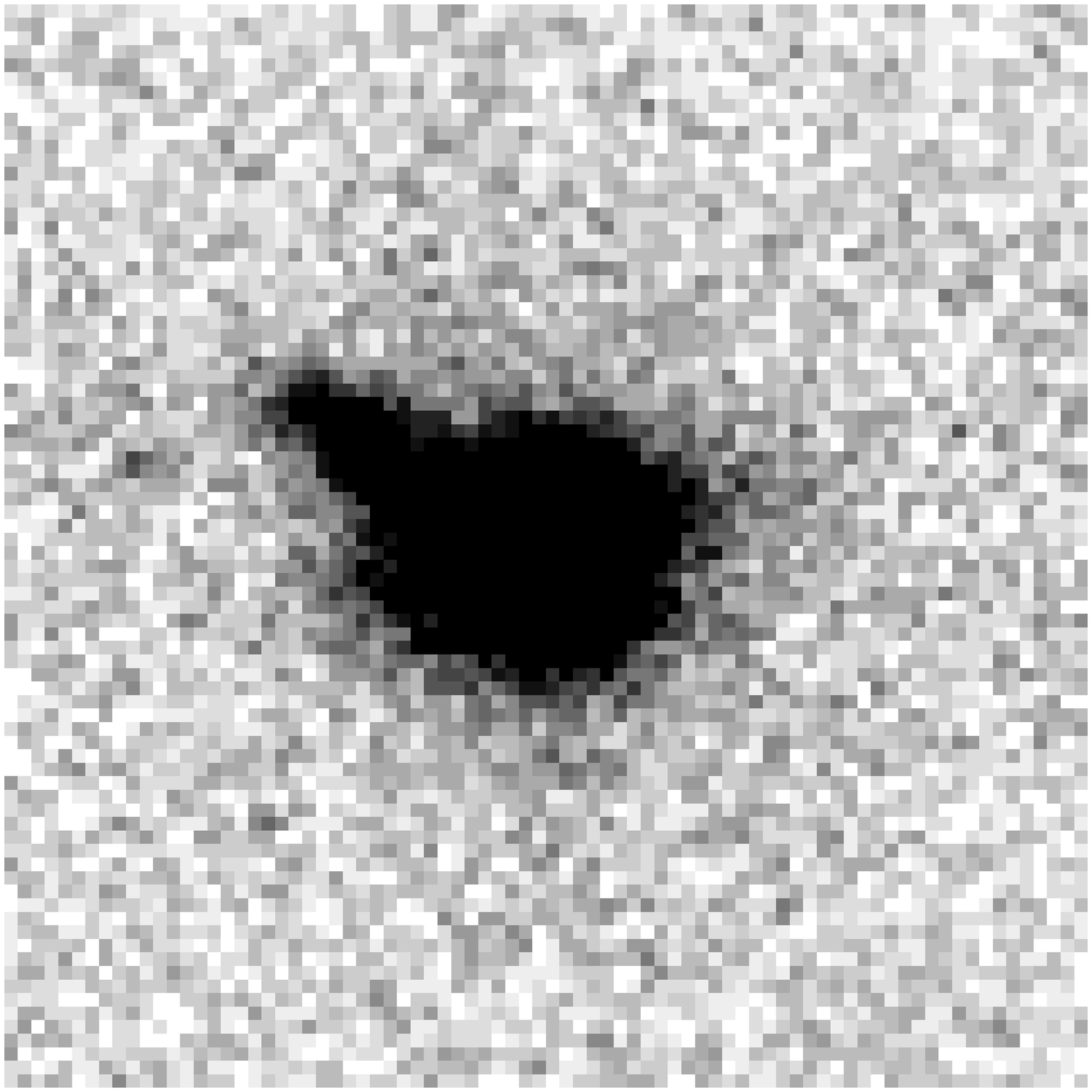}&
\includegraphics[width=0.24\textwidth]{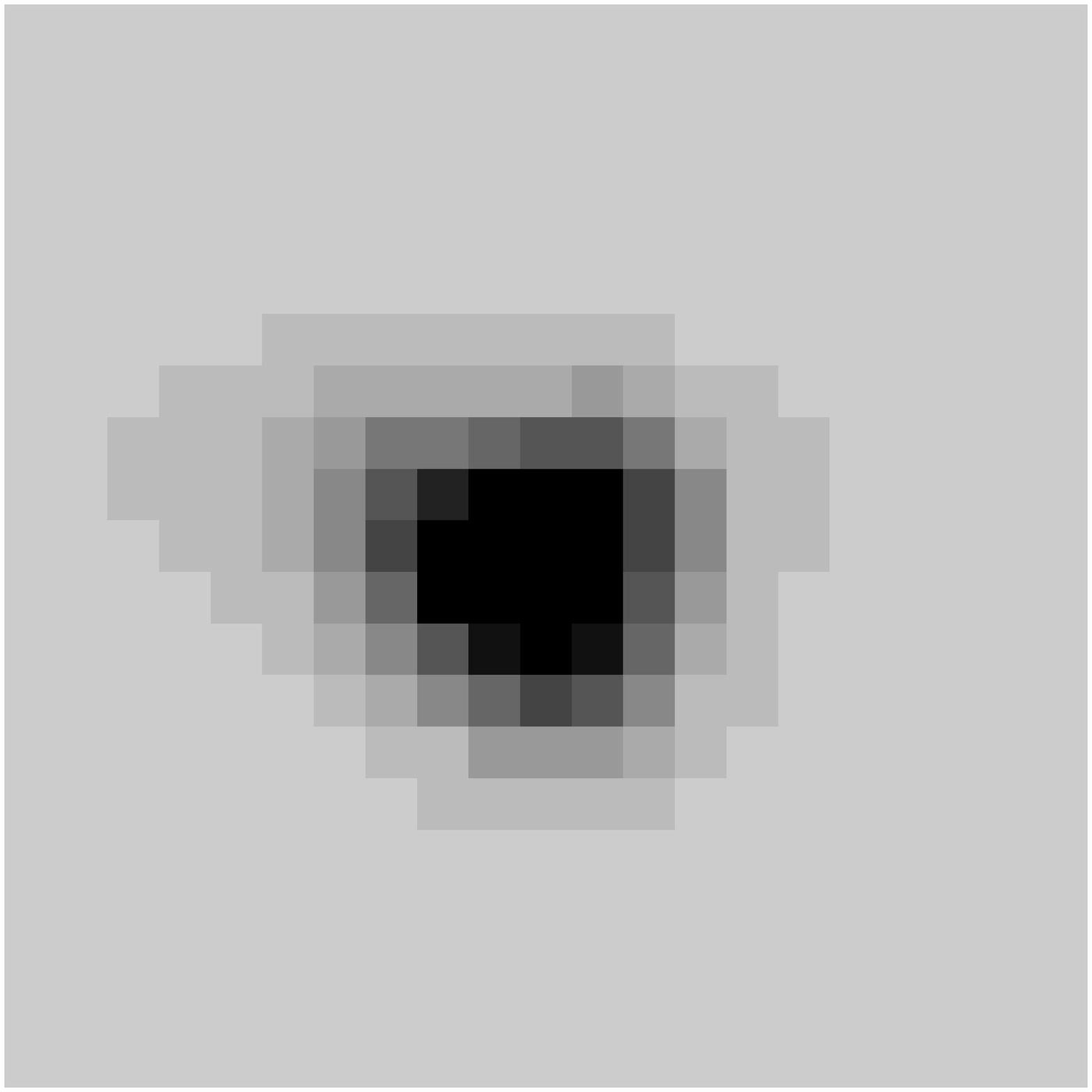}\\
{\bf BLAST 830-1:} $z_{\rm spec} = 0.605$\\
\\
\\
\includegraphics[width=0.24\textwidth]{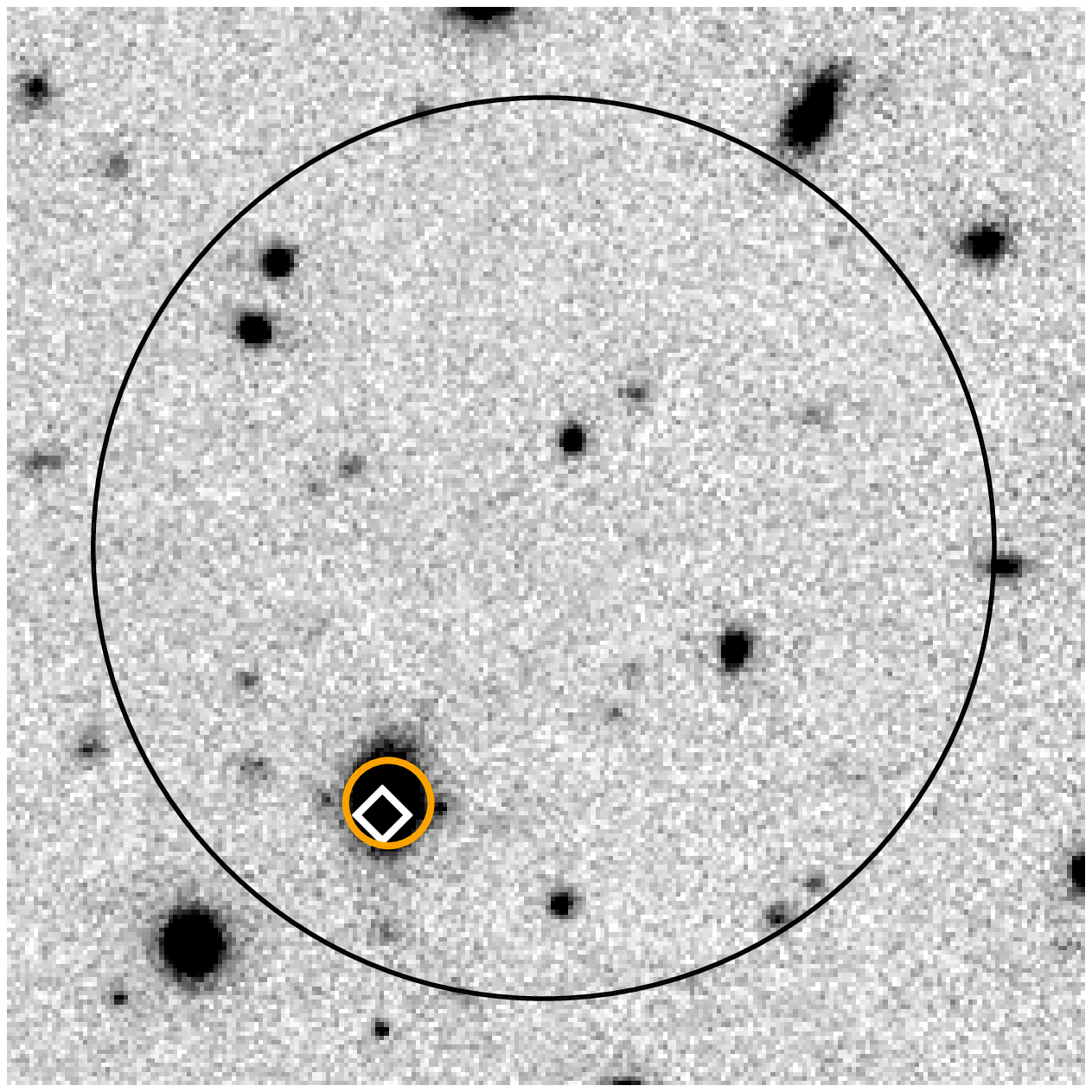}&
\includegraphics[width=0.24\textwidth]{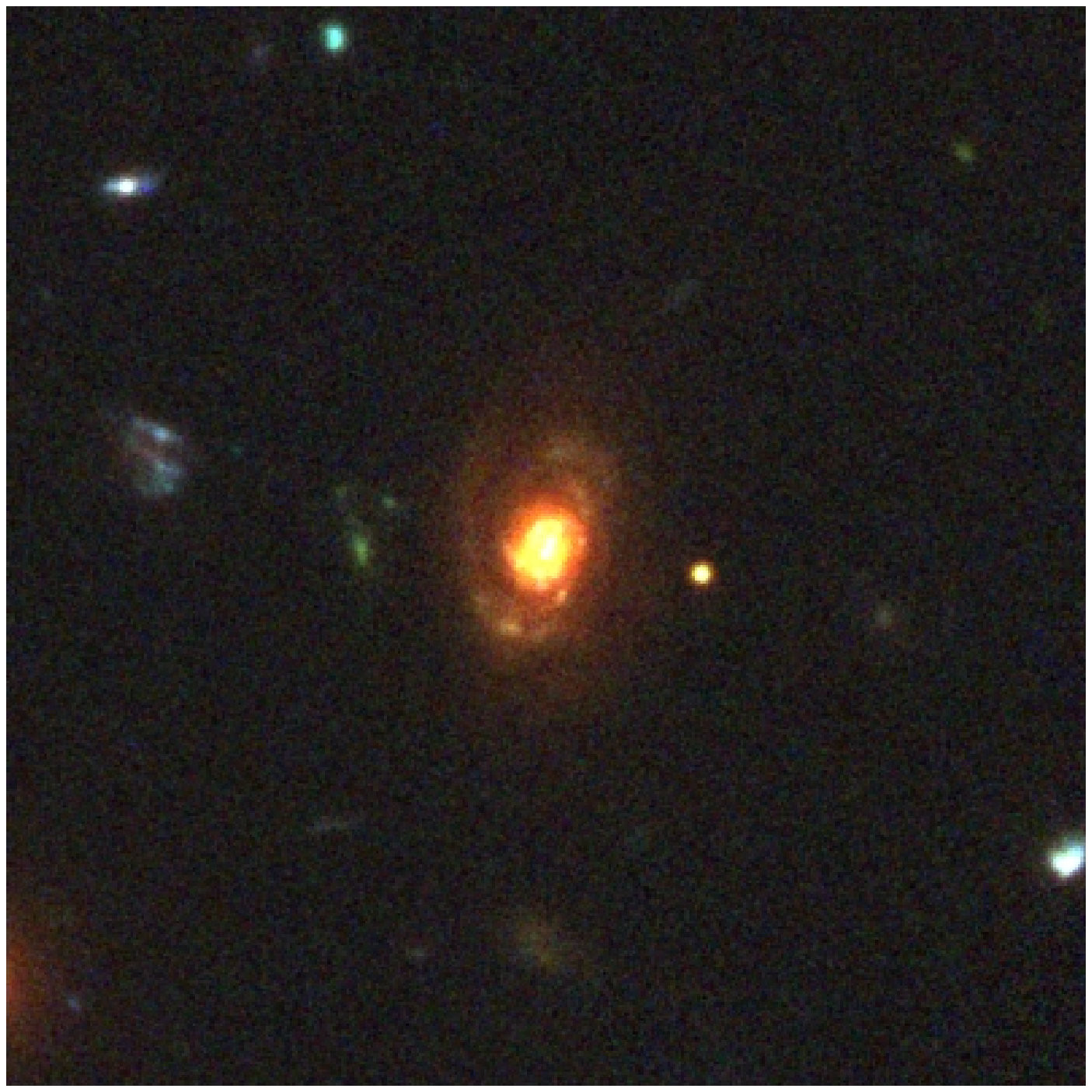}&
\includegraphics[width=0.24\textwidth]{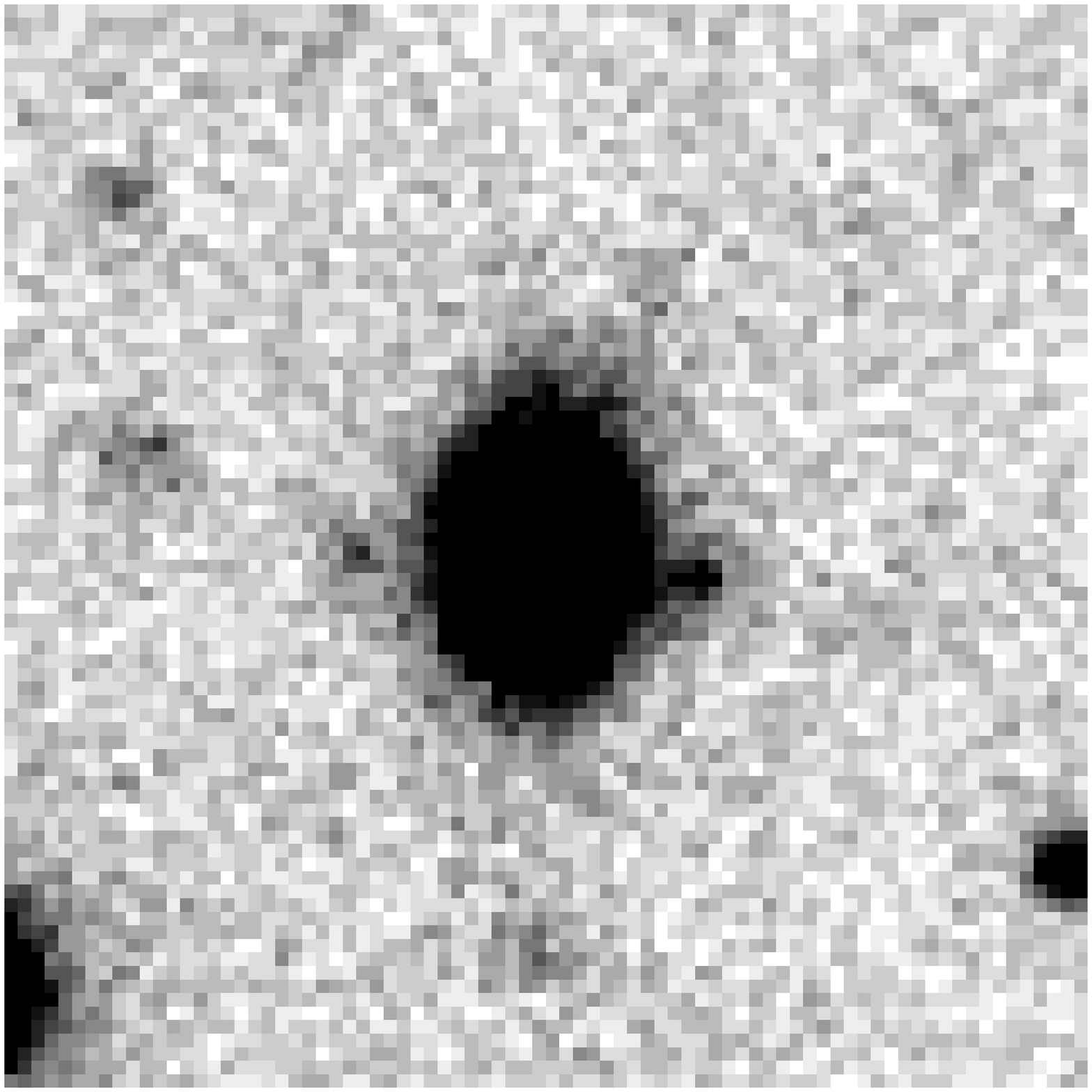}&
\includegraphics[width=0.24\textwidth]{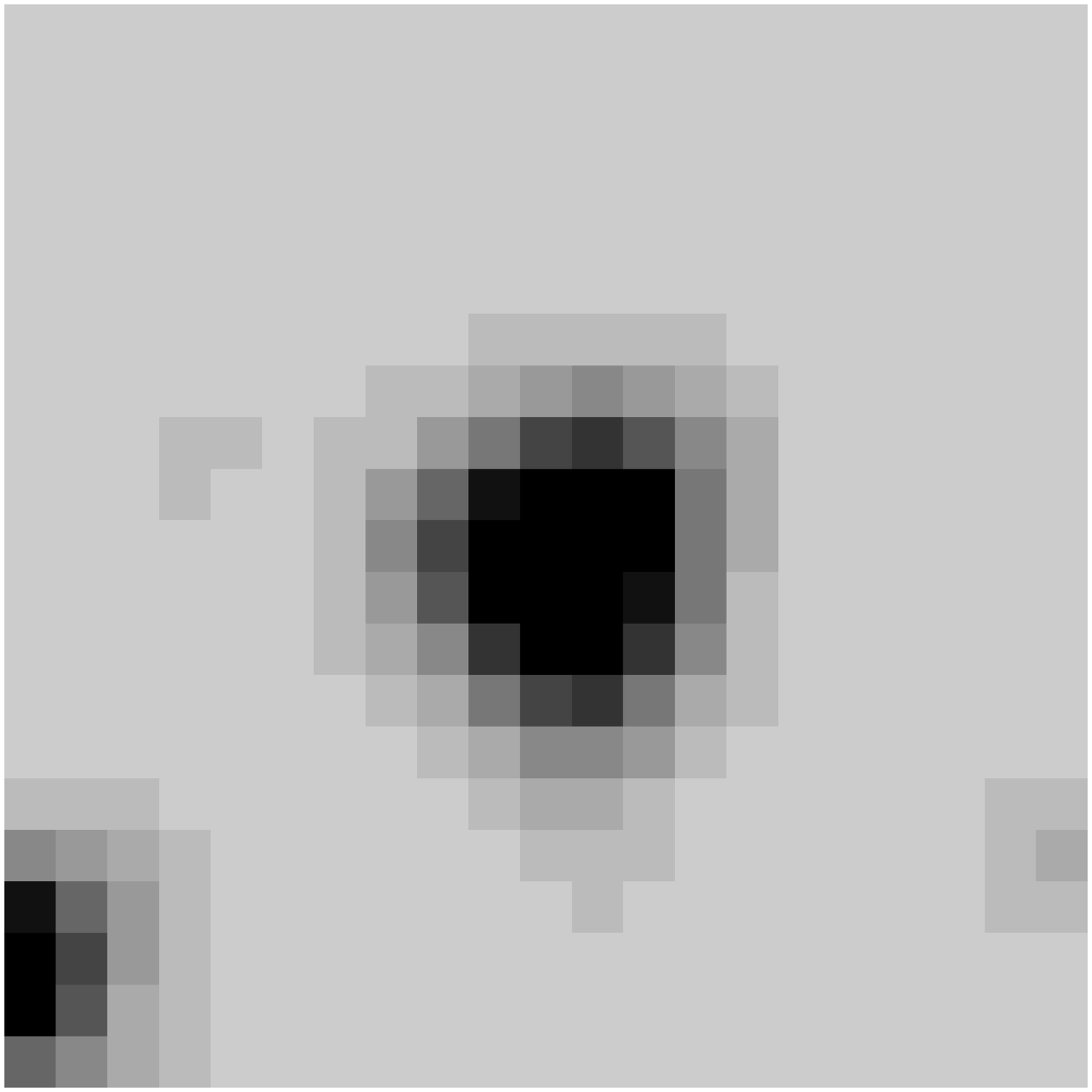}\\
{\bf BLAST 257:} $z_{\rm est} = 0.689$\\
\\
\\
\end{tabular}
\addtocounter{figure}{-1}
\caption{continued}
\end{figure*}

\begin{figure*}
\begin{tabular}{llll}
\\                          
\includegraphics[width=0.24\textwidth]{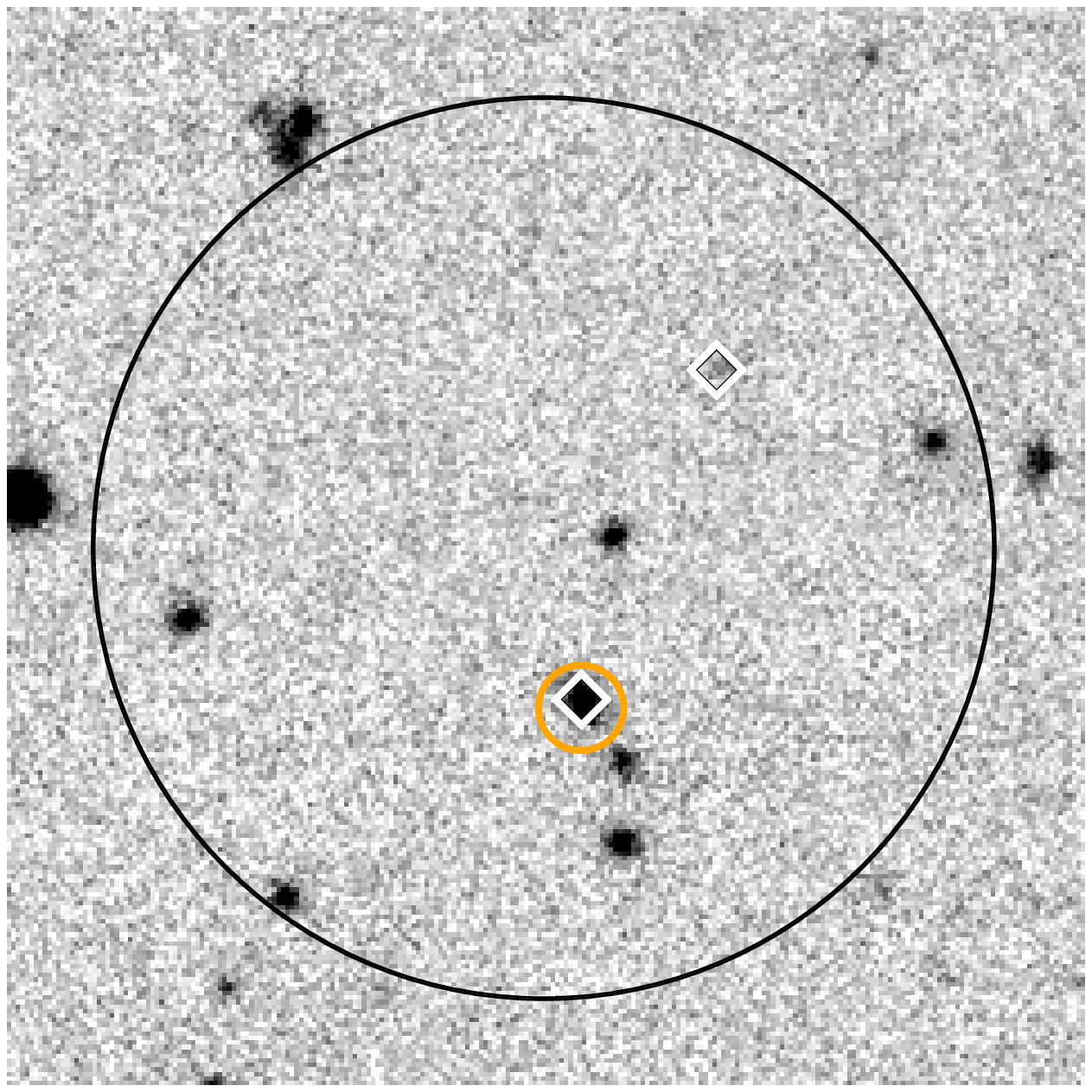}&
\includegraphics[width=0.24\textwidth]{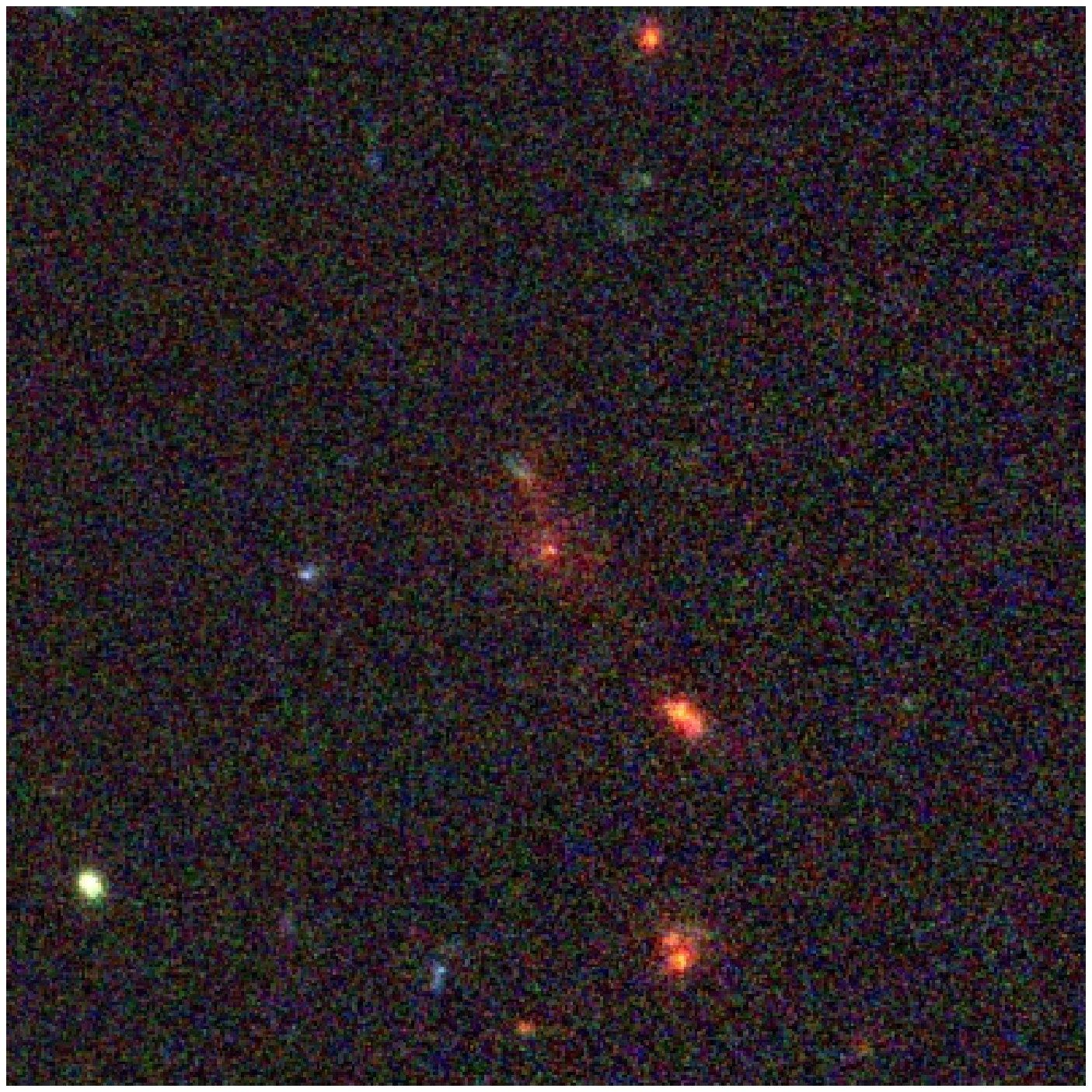}&
\includegraphics[width=0.24\textwidth]{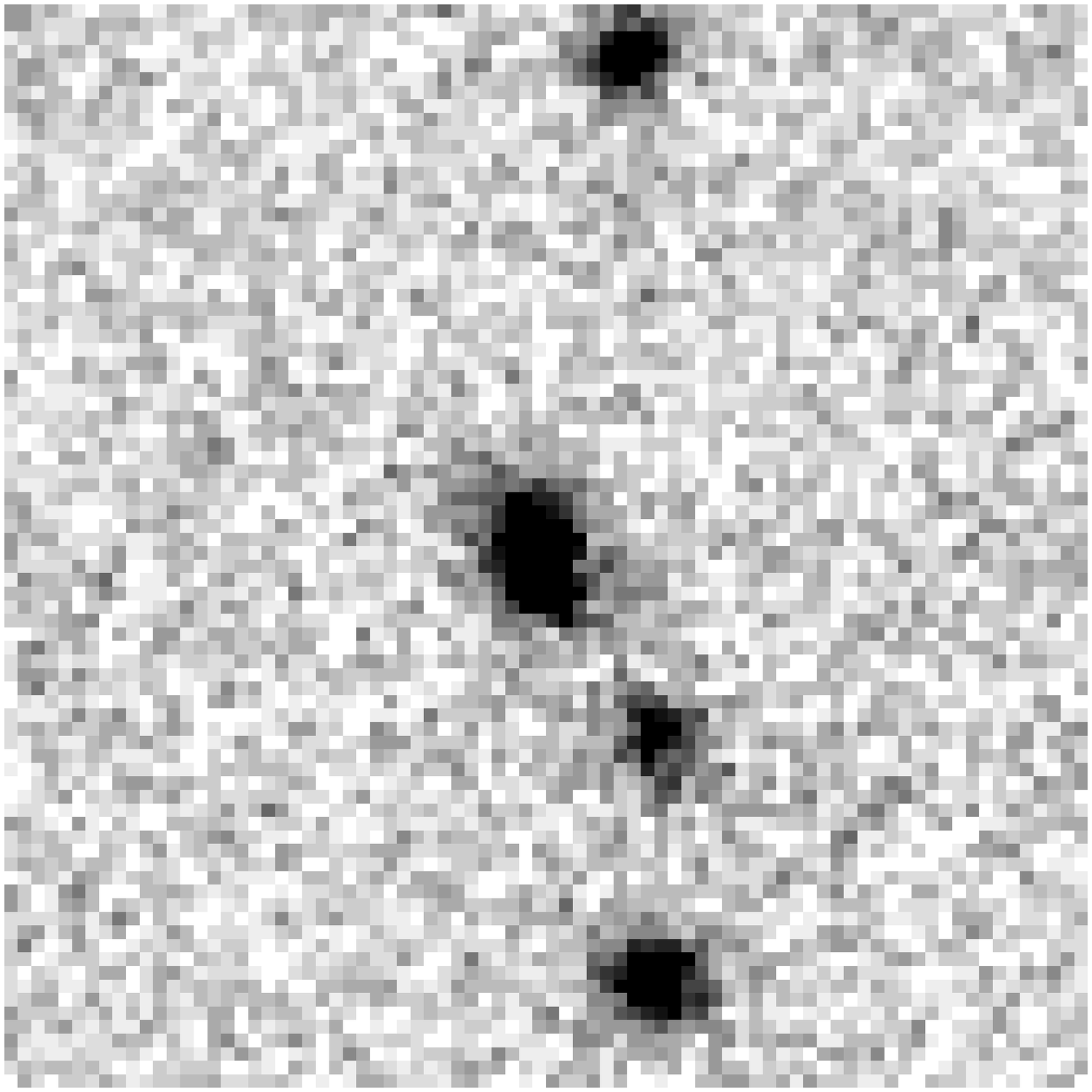}&
\includegraphics[width=0.24\textwidth]{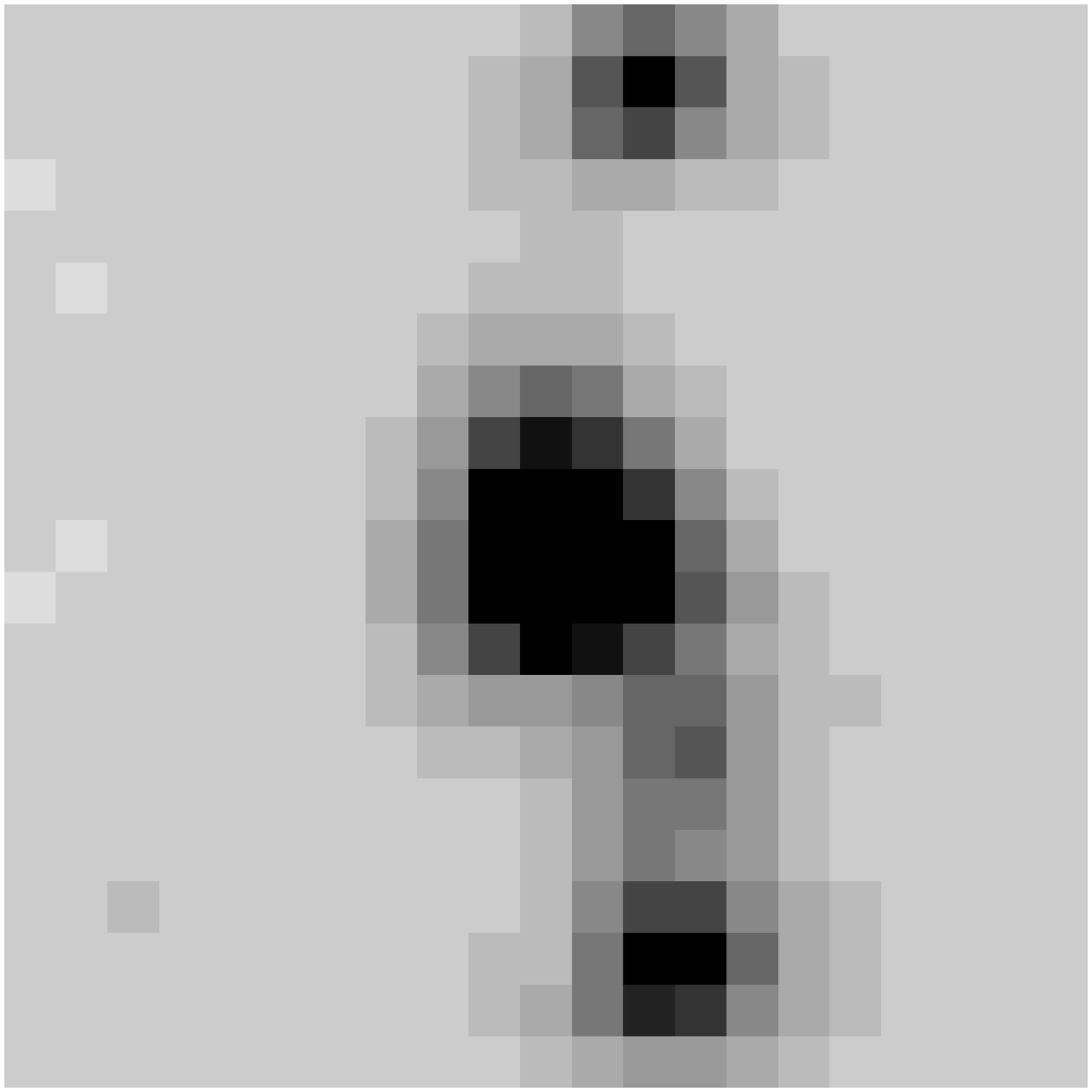}\\
{\bf BLAST 1293:} $z_{\rm spec} = 1.382$\\
\\
\\
\includegraphics[width=0.24\textwidth]{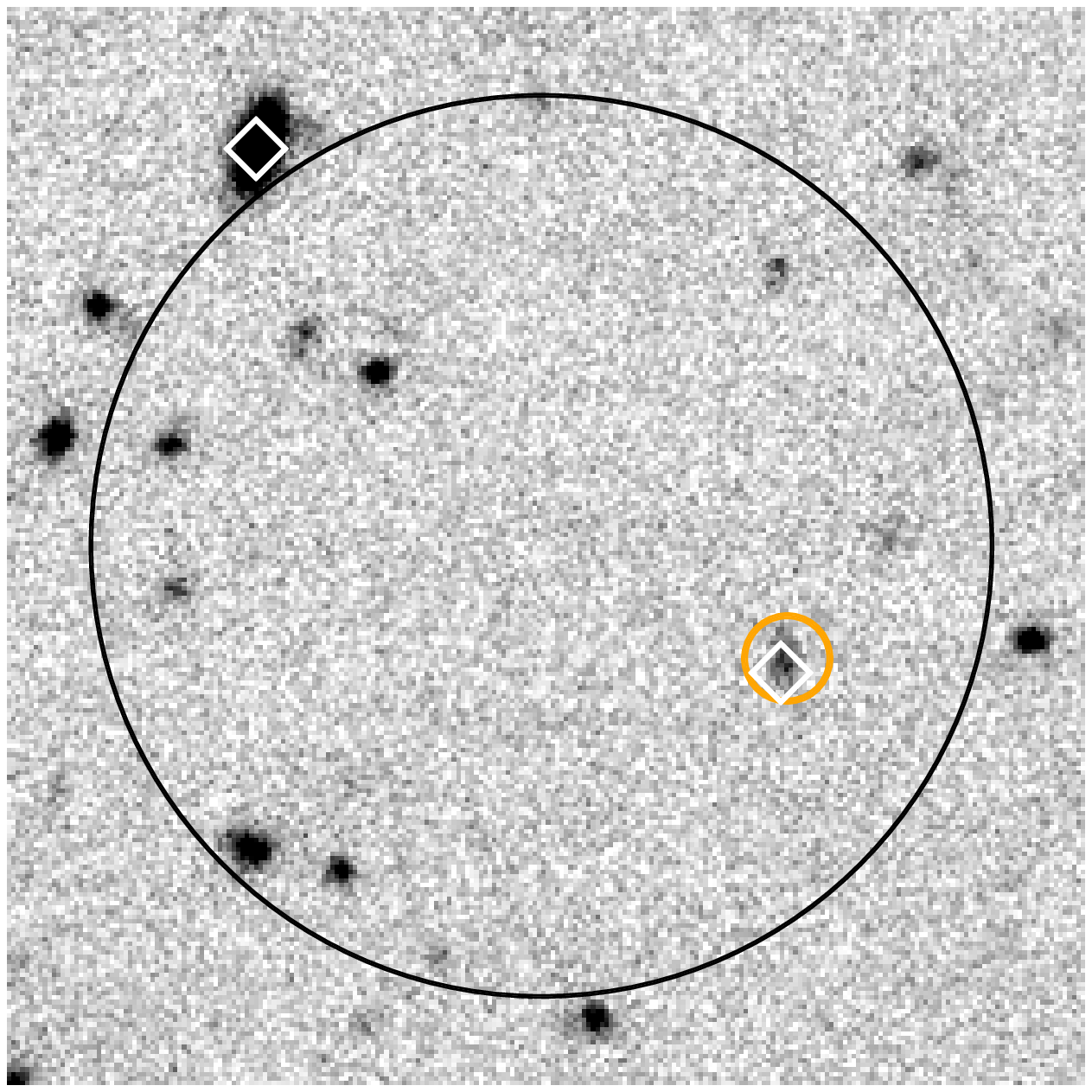}&
\includegraphics[width=0.24\textwidth]{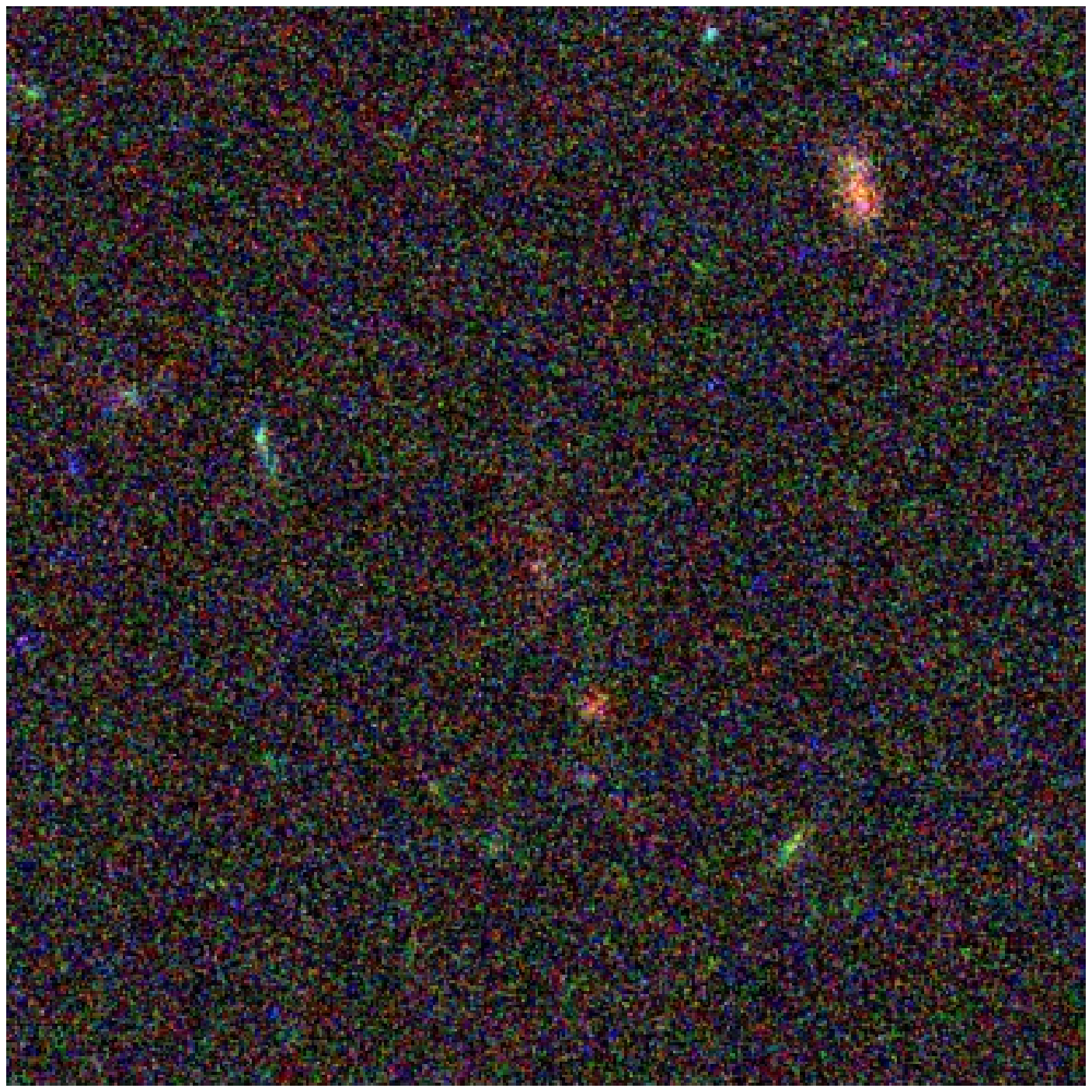}&
\includegraphics[width=0.24\textwidth]{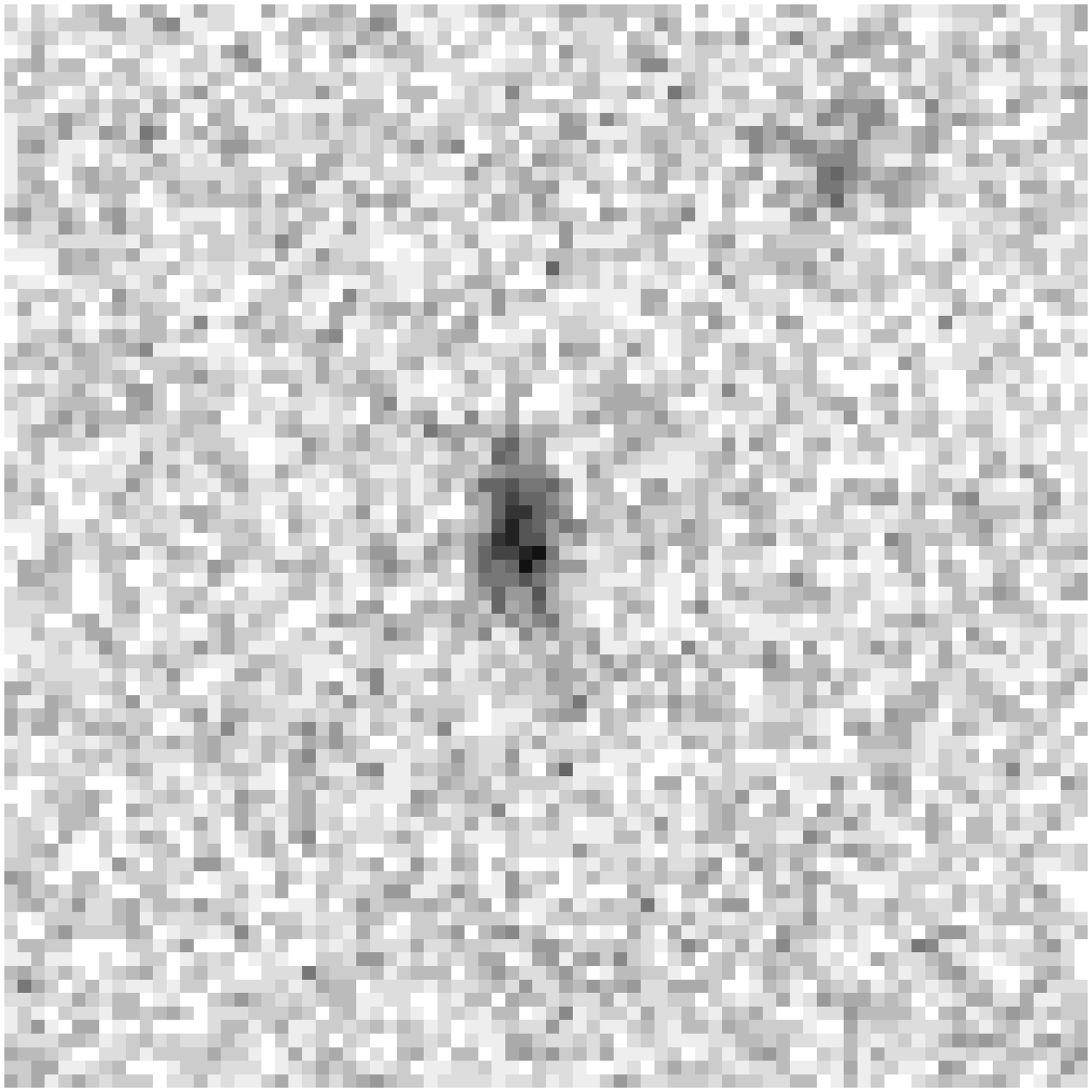}&
\includegraphics[width=0.24\textwidth]{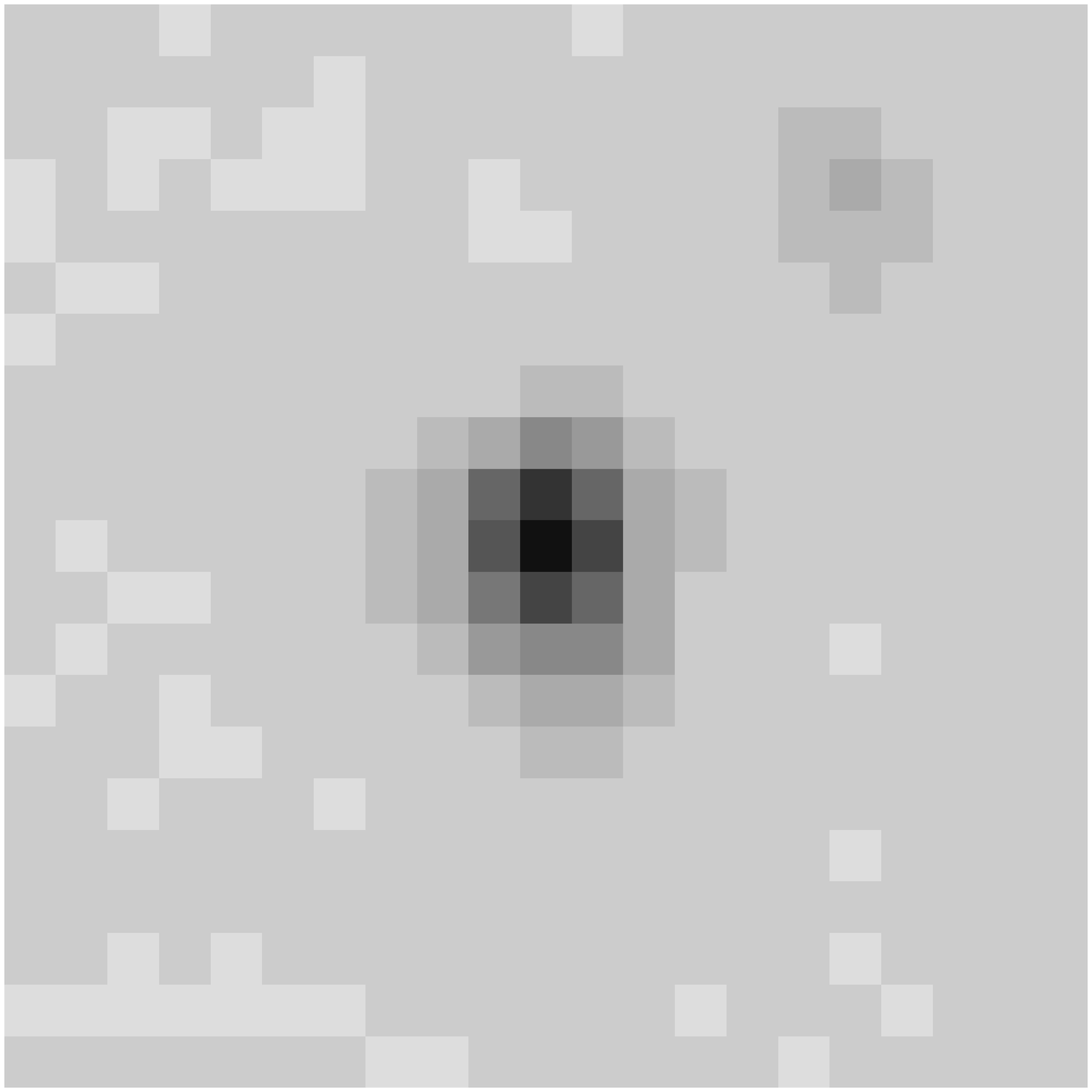}\\
{\bf BLAST 552:} $z_{\rm est} = 1.68$\\ 
\\
\\
\includegraphics[width=0.24\textwidth]{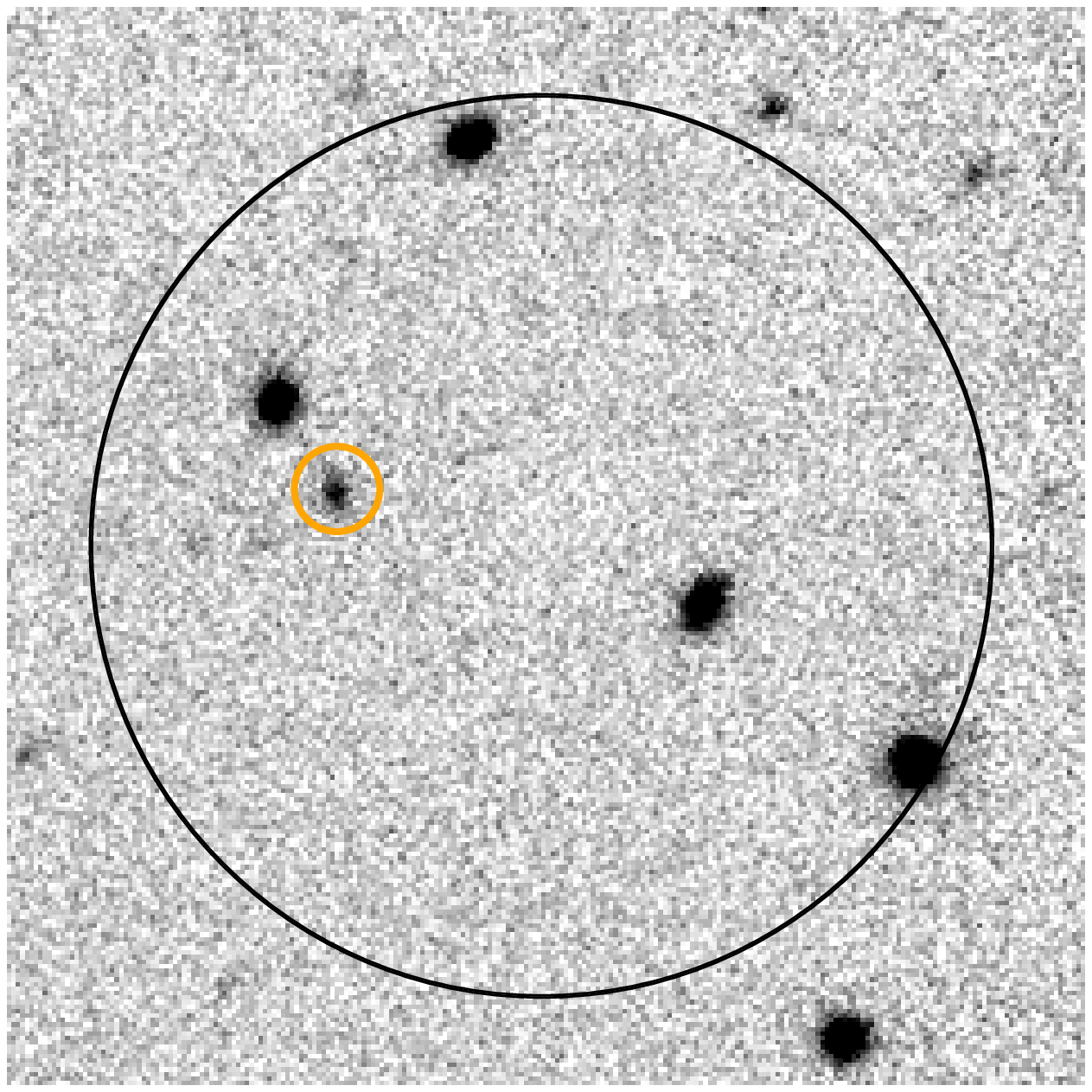}&
\includegraphics[width=0.24\textwidth]{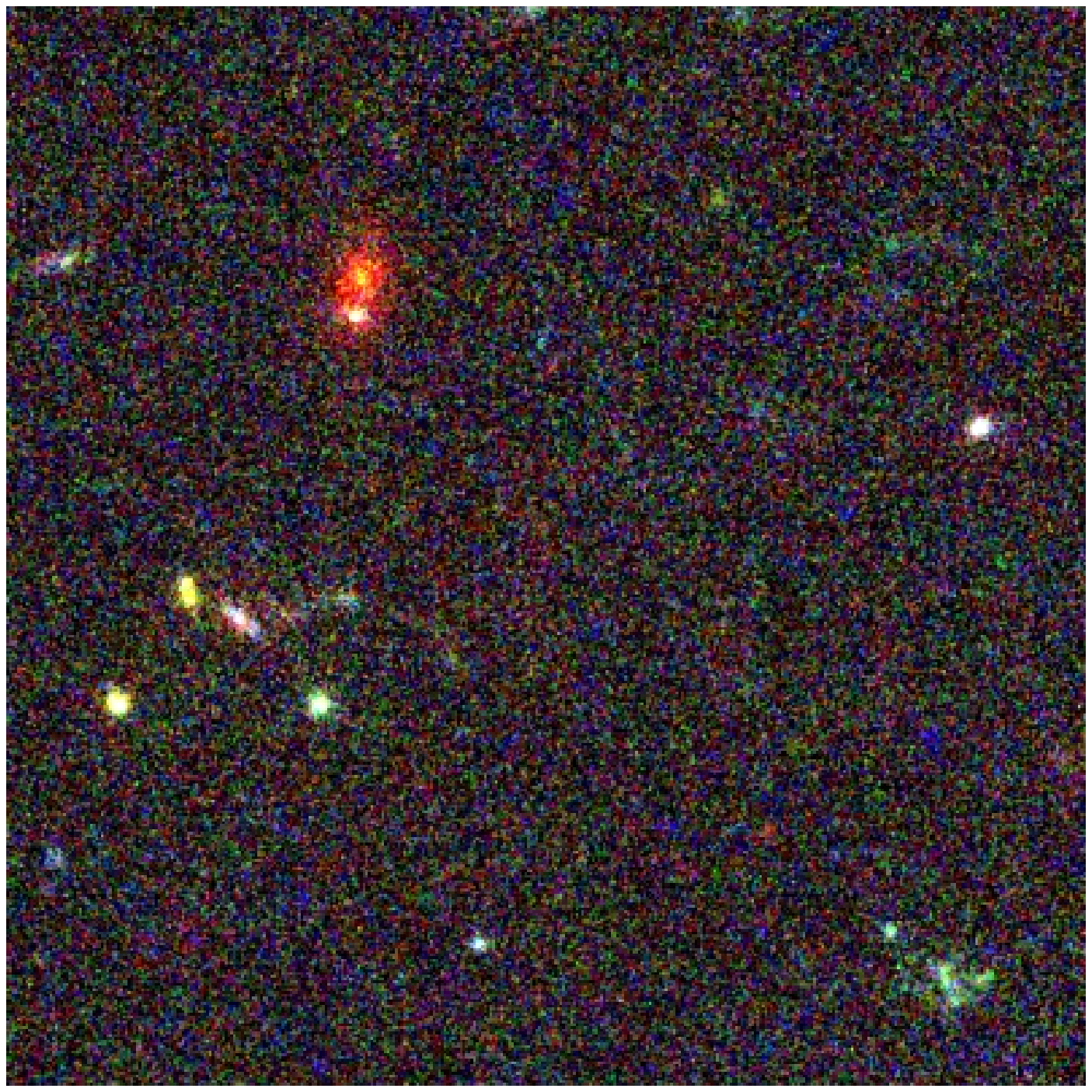}&
\includegraphics[width=0.24\textwidth]{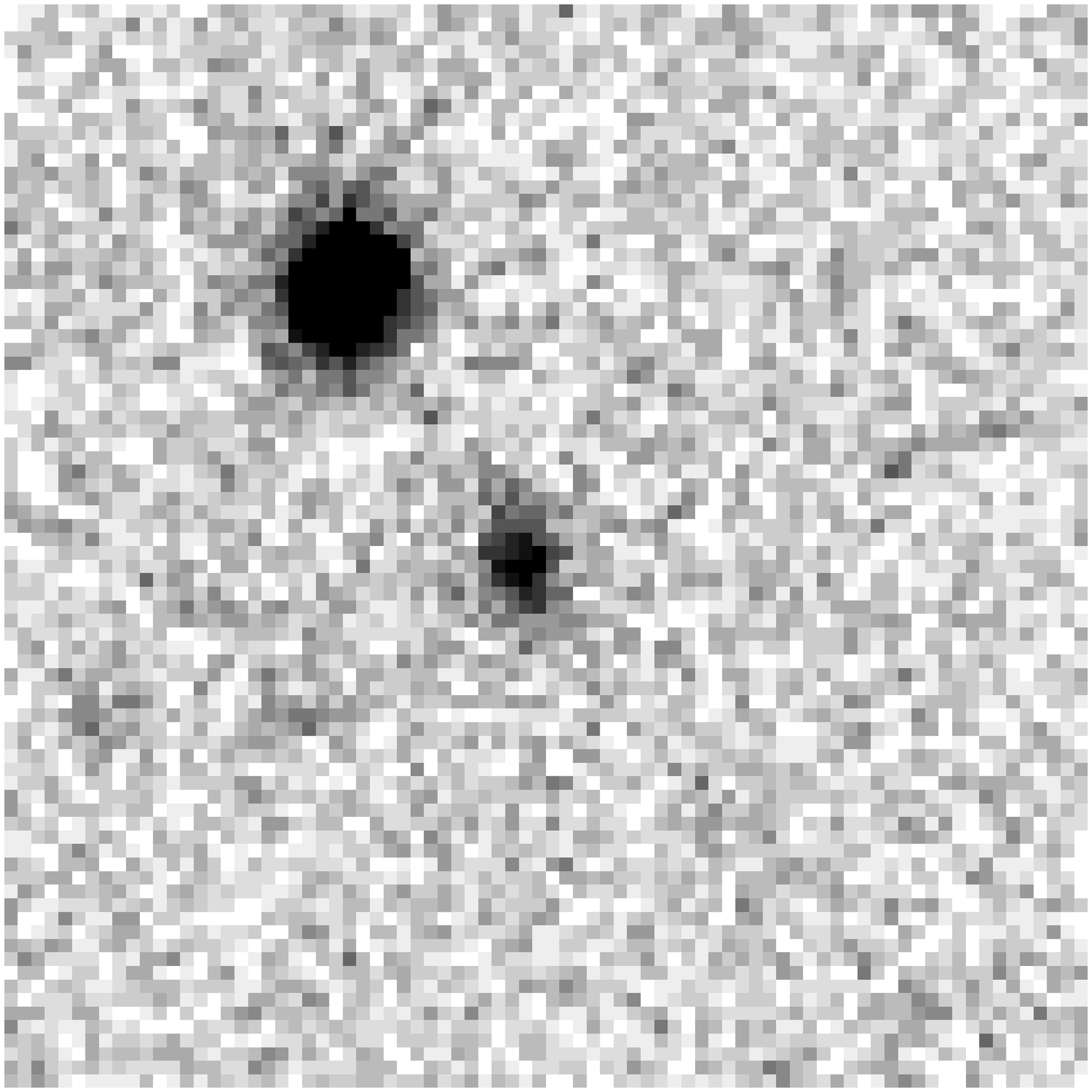}&
\includegraphics[width=0.24\textwidth]{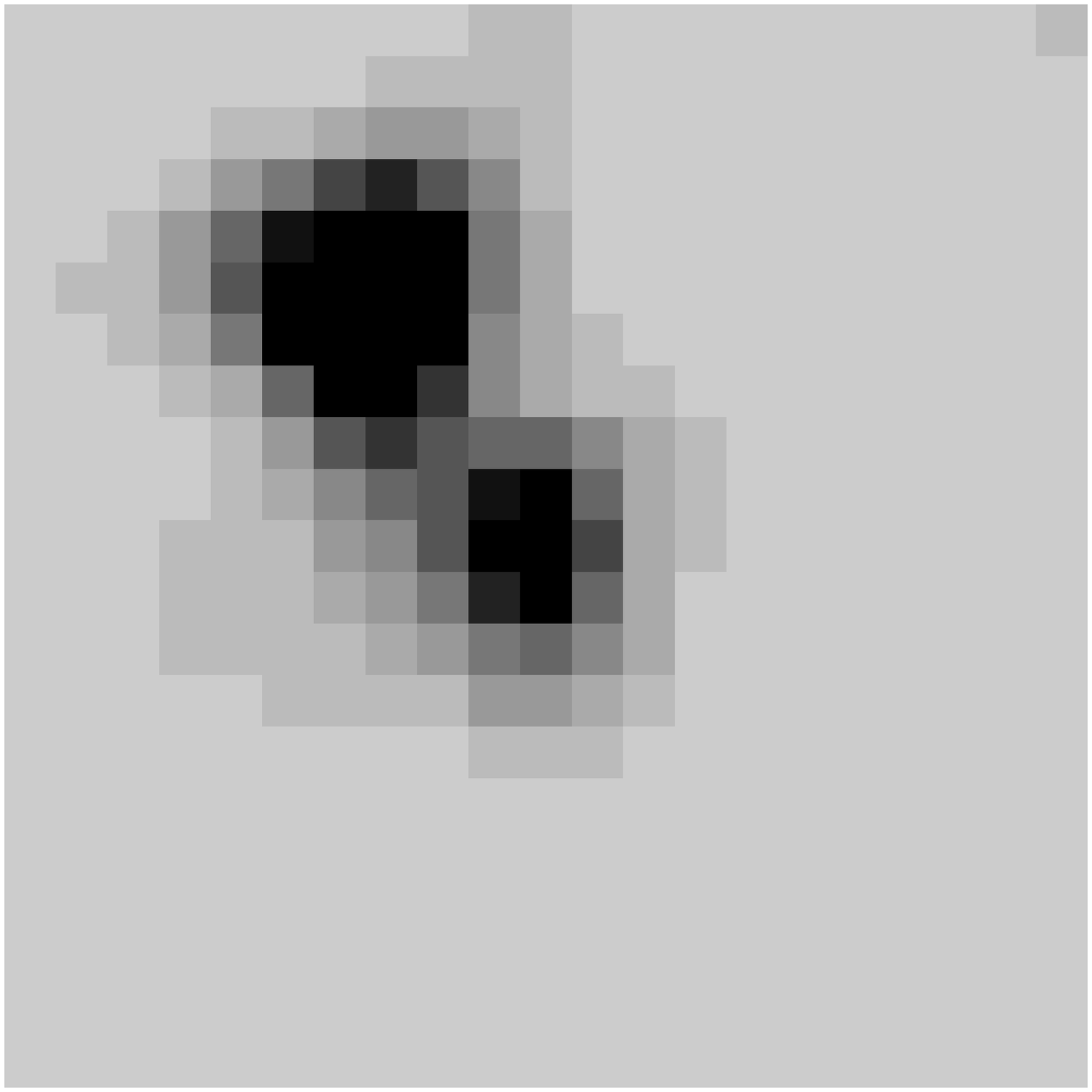}\\
{\bf BLAST 193:} $z_{\rm est} = 1.81$\\
\\
\\
\includegraphics[width=0.24\textwidth]{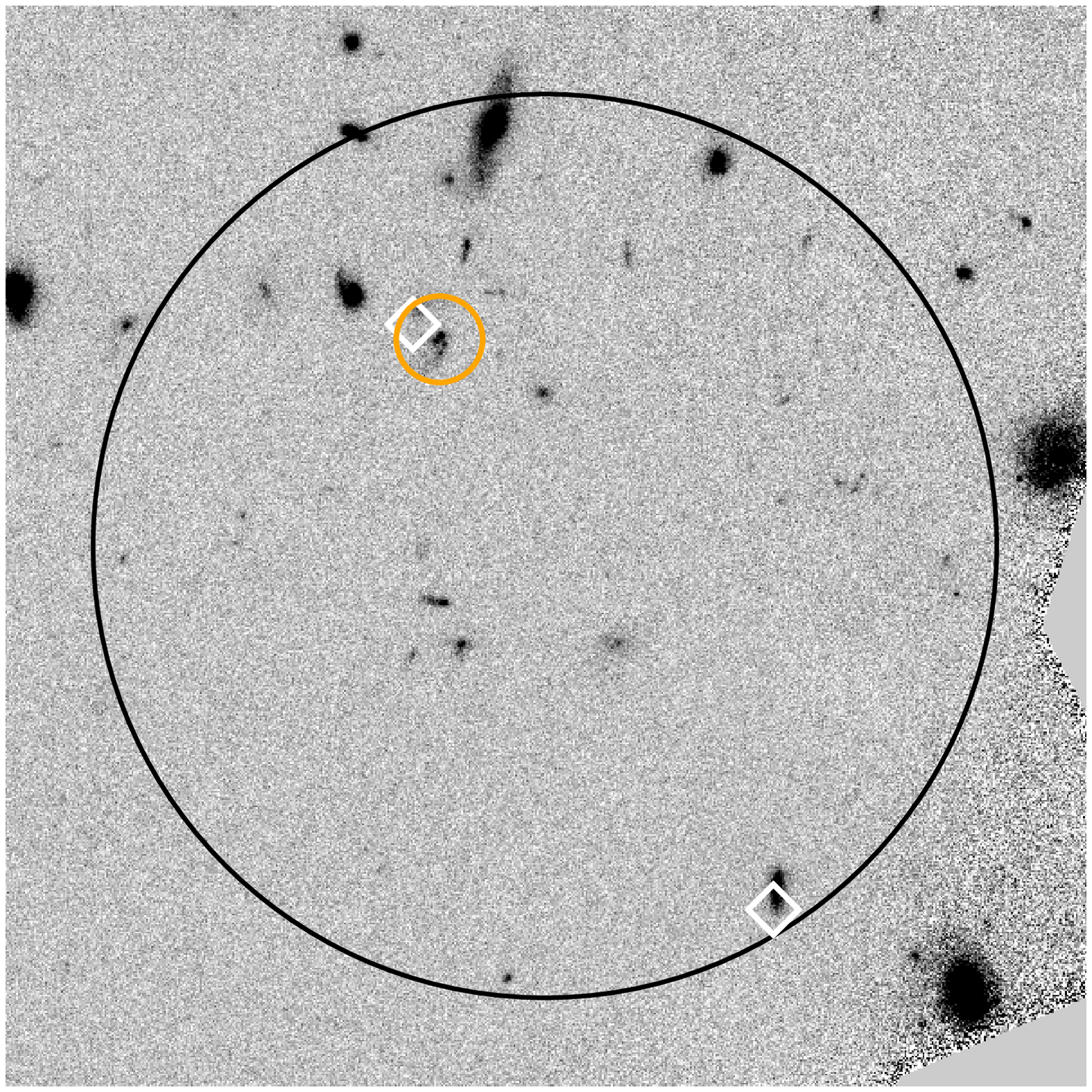}&
\includegraphics[width=0.24\textwidth]{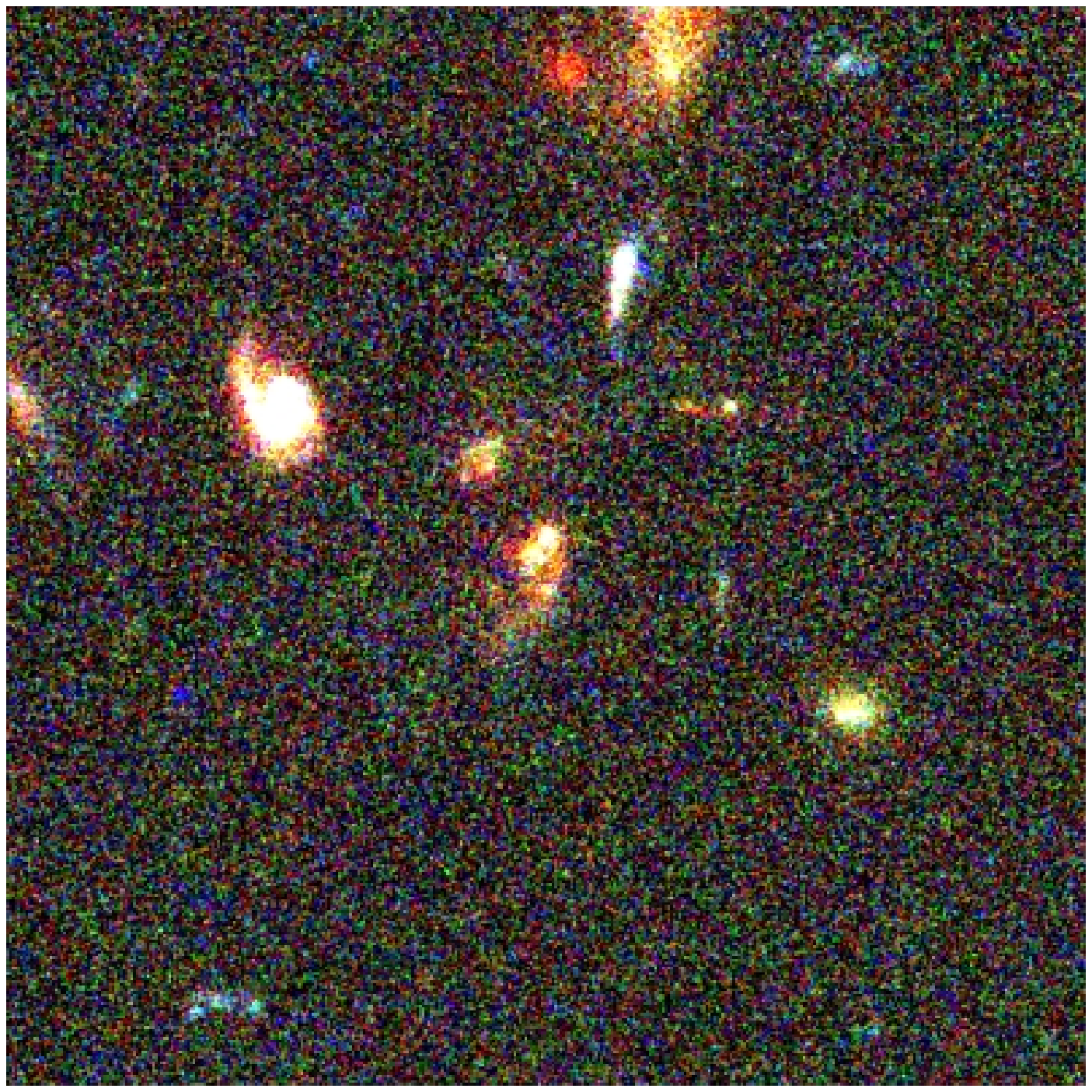}&&
\includegraphics[width=0.24\textwidth]{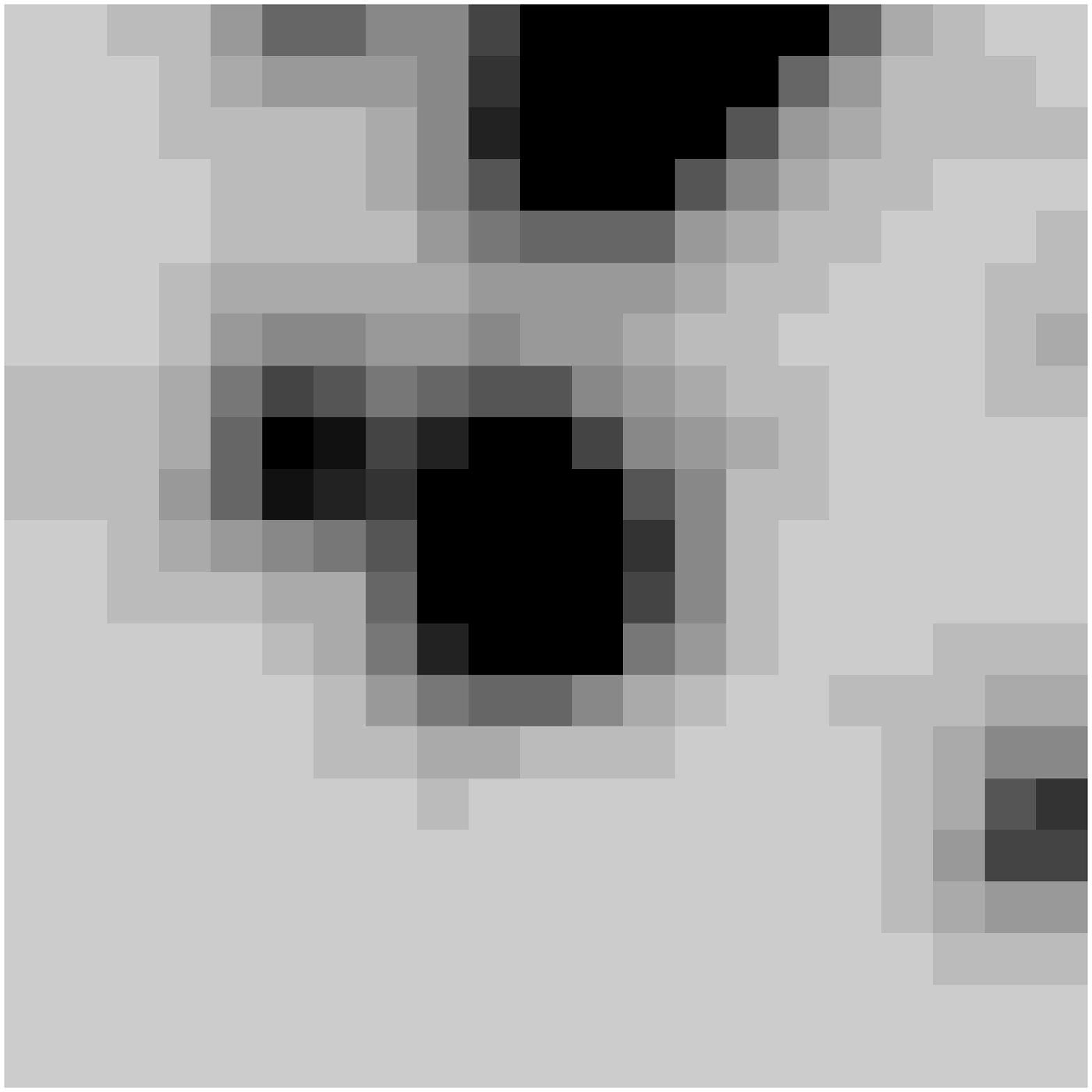}\\
{\bf BLAST 158:} $z_{\rm est} = 1.85$\\
\\
\\
\end{tabular}
\addtocounter{figure}{-1}
\caption{continued}
\end{figure*}

\begin{figure*}
\begin{tabular}{llll}
\\   
\includegraphics[width=0.24\textwidth]{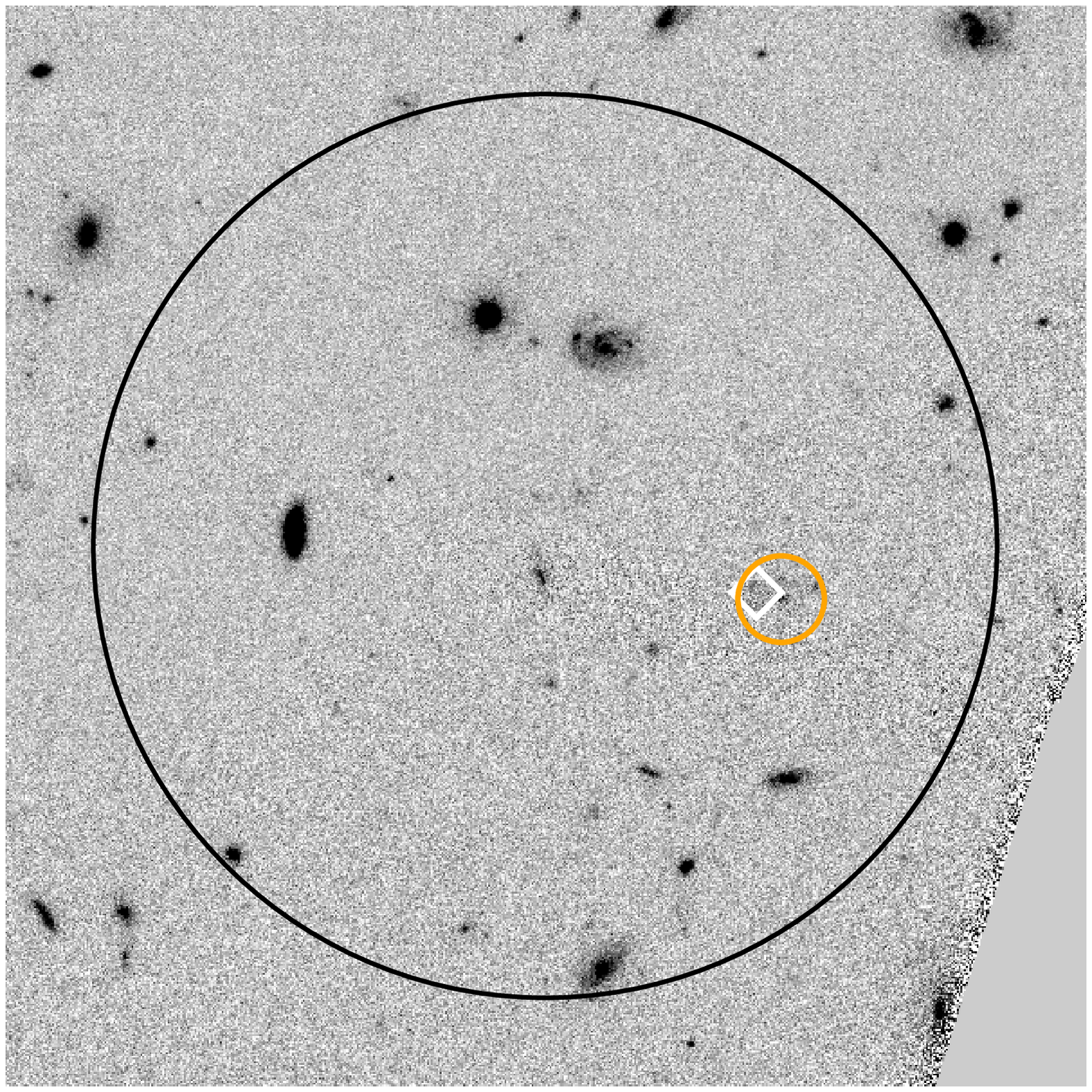}&
\includegraphics[width=0.24\textwidth]{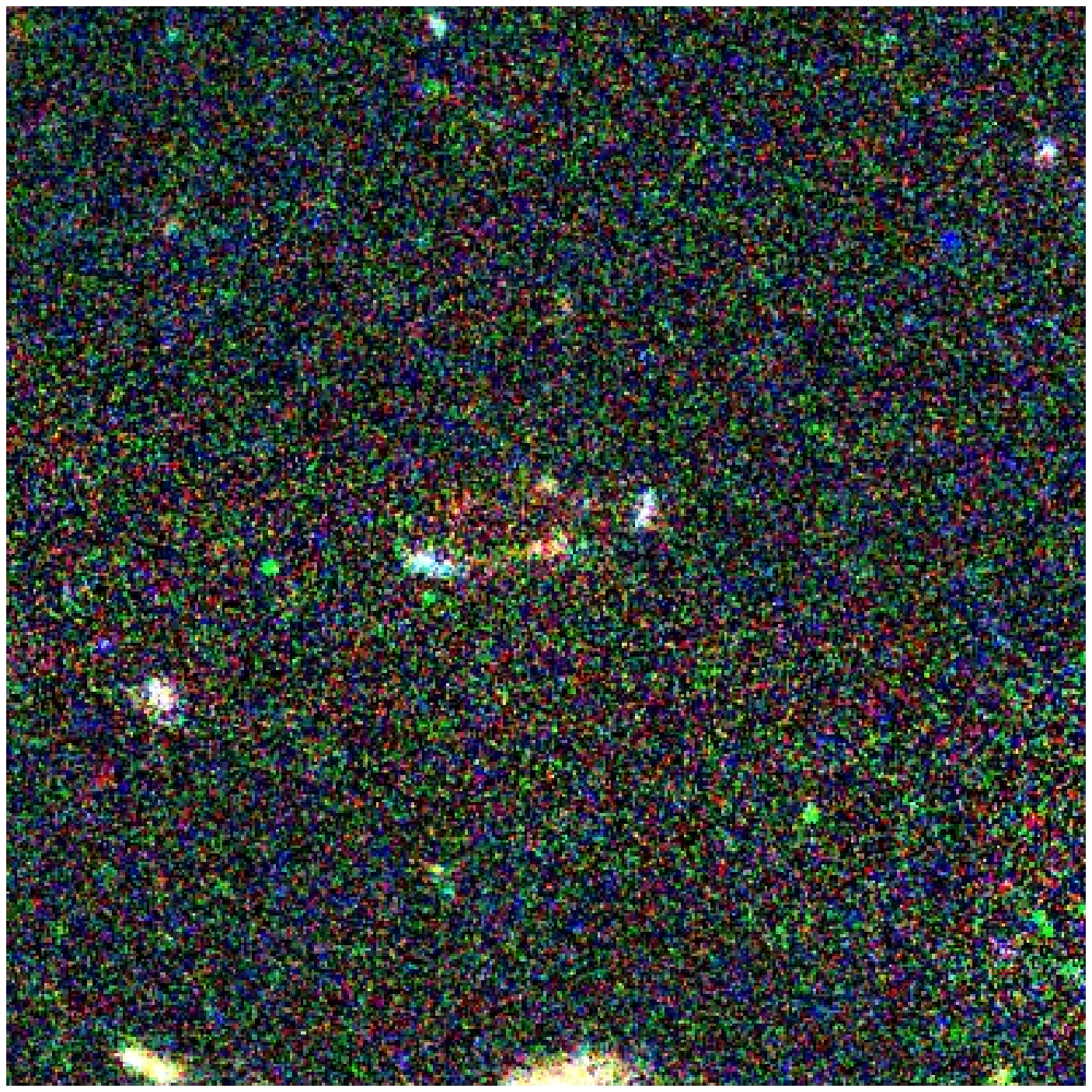}&&
\includegraphics[width=0.24\textwidth]{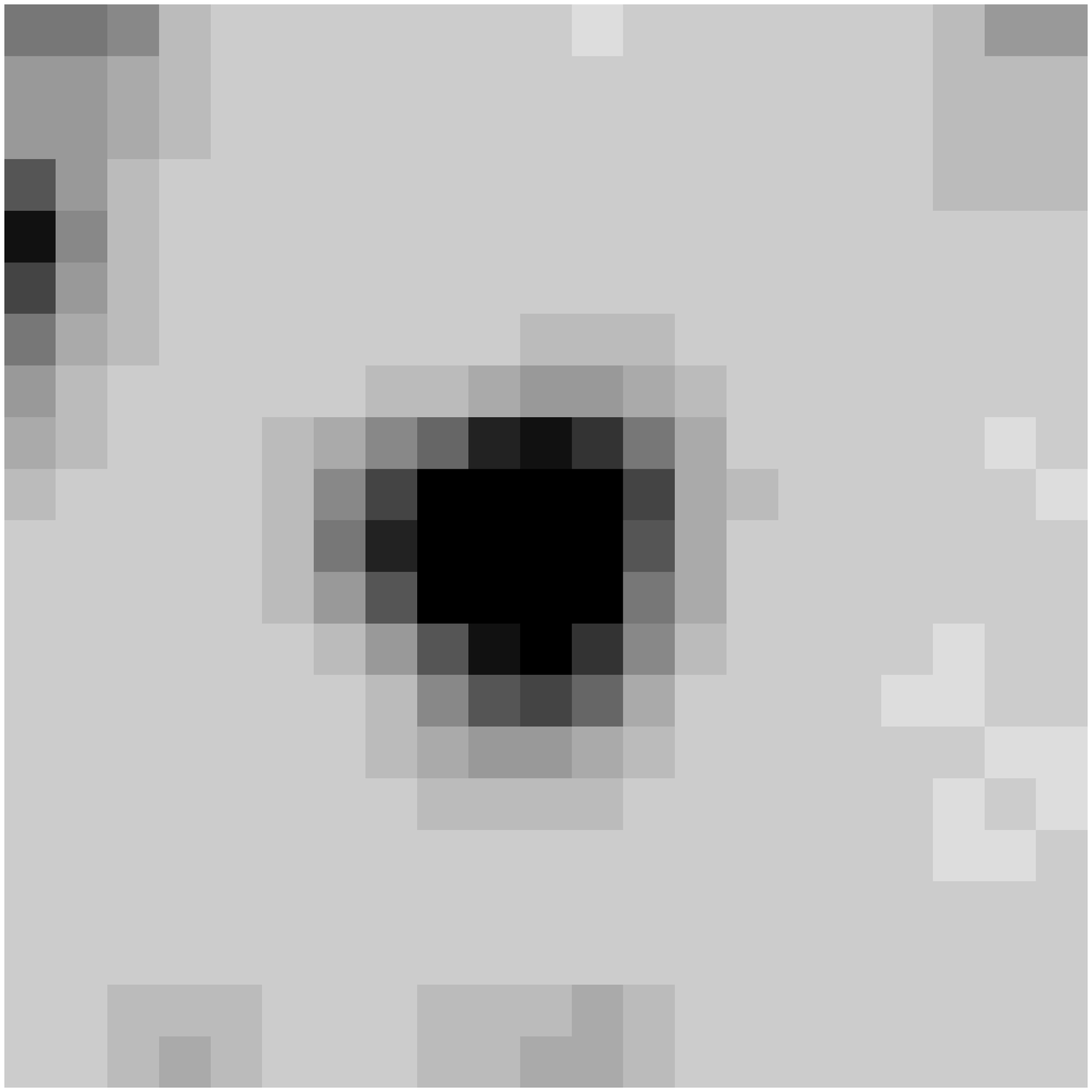}\\
{\bf BLAST 66:} $z_{\rm est} = 1.94$\\
\\
\\  
\includegraphics[width=0.24\textwidth]{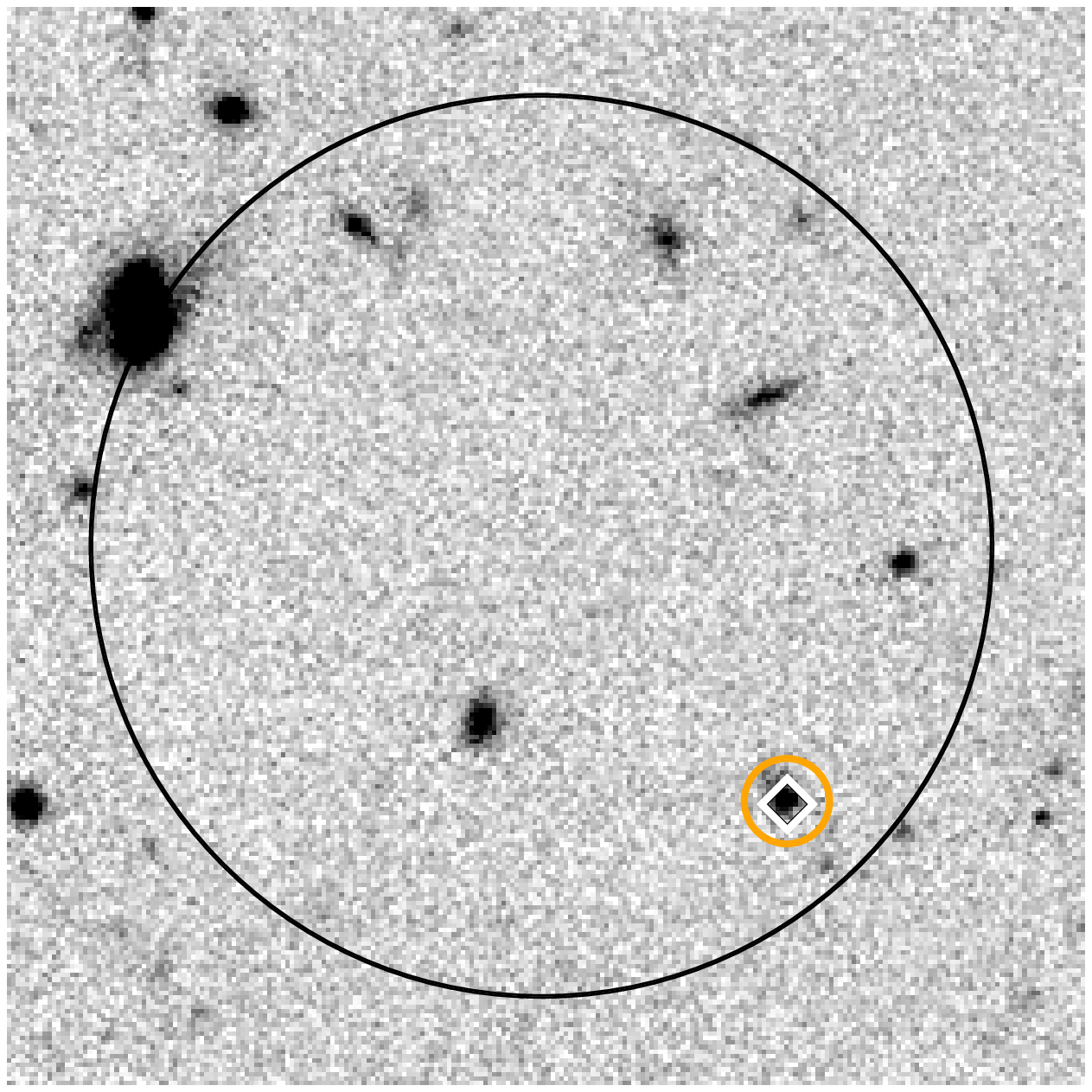}&
\includegraphics[width=0.24\textwidth]{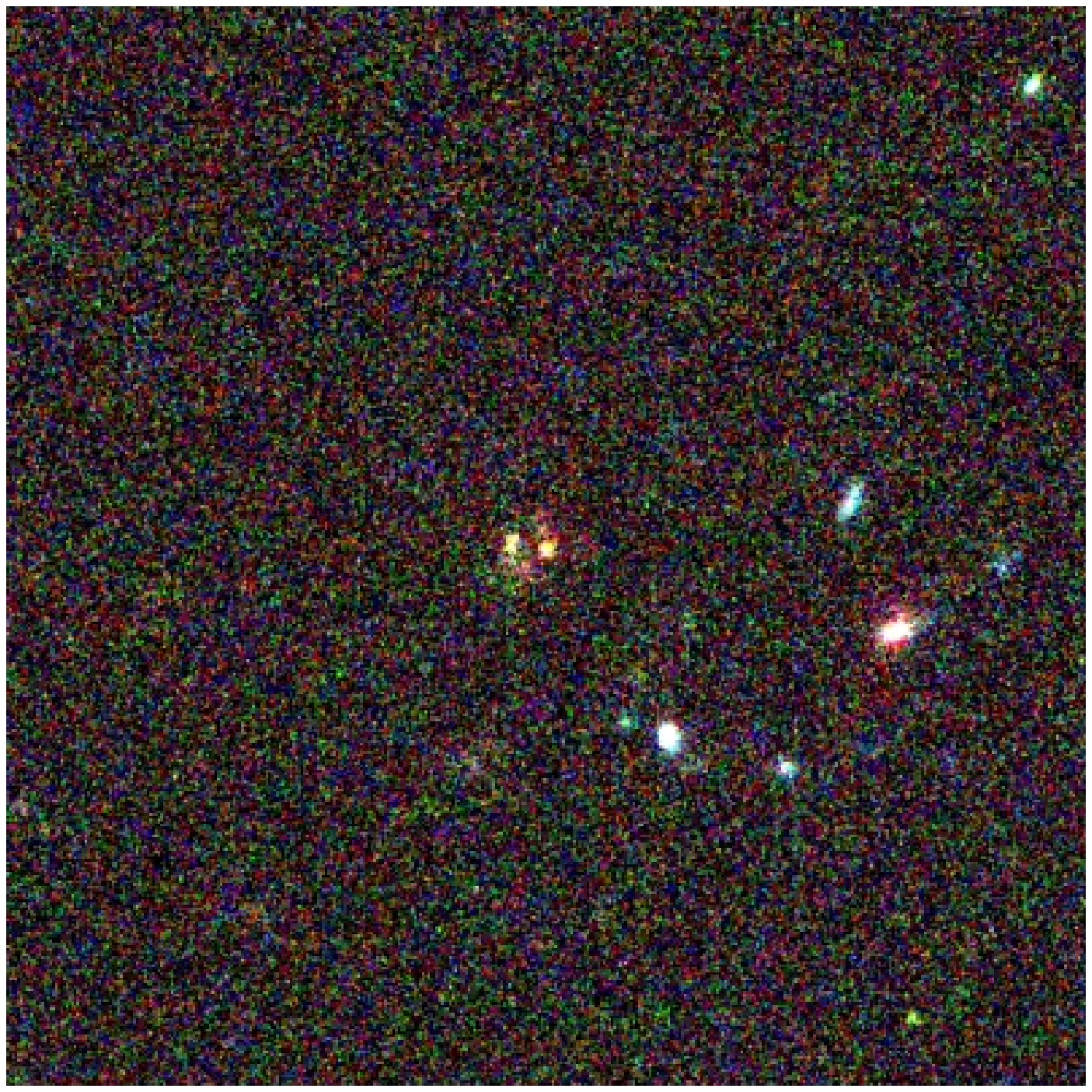}&
\includegraphics[width=0.24\textwidth]{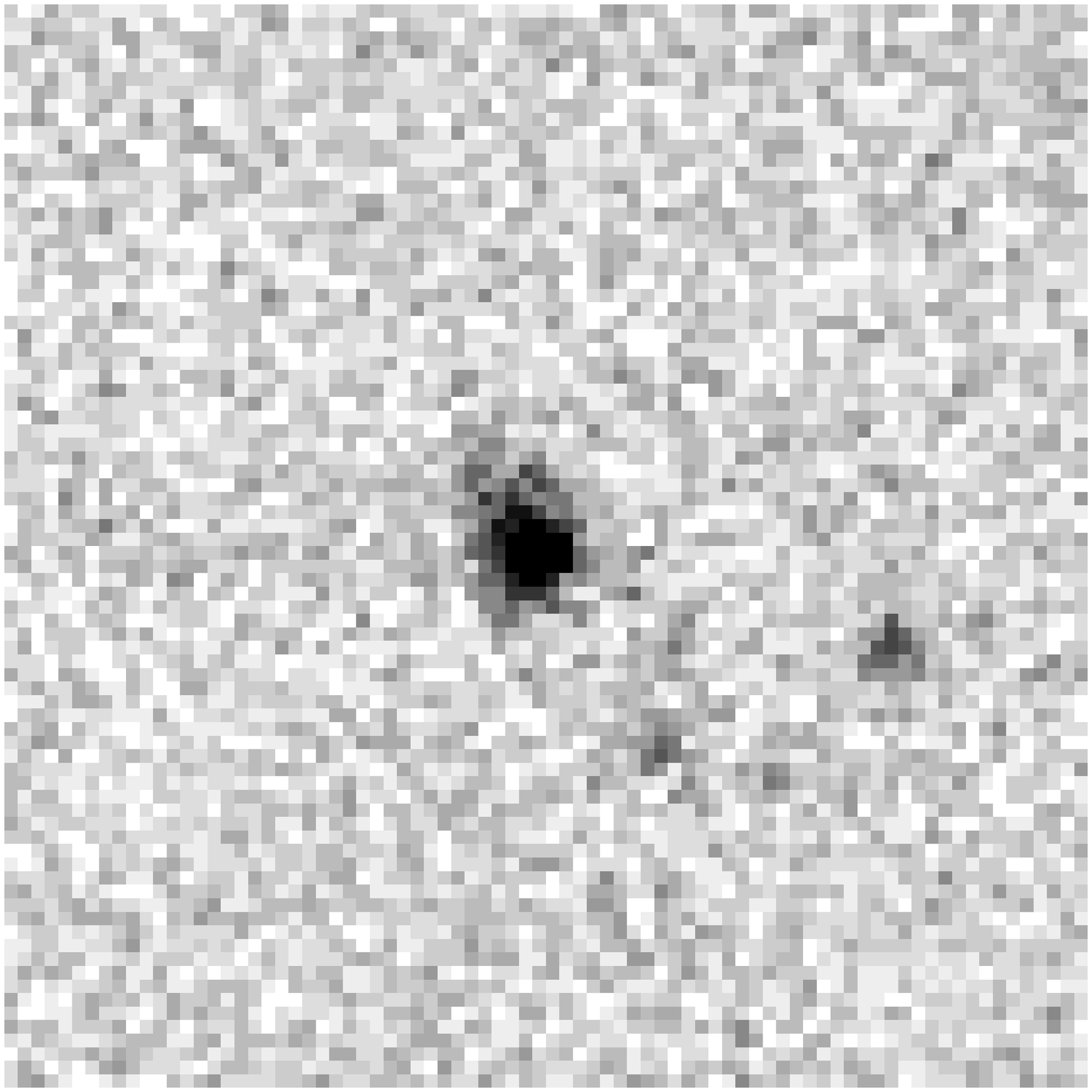}&
\includegraphics[width=0.24\textwidth]{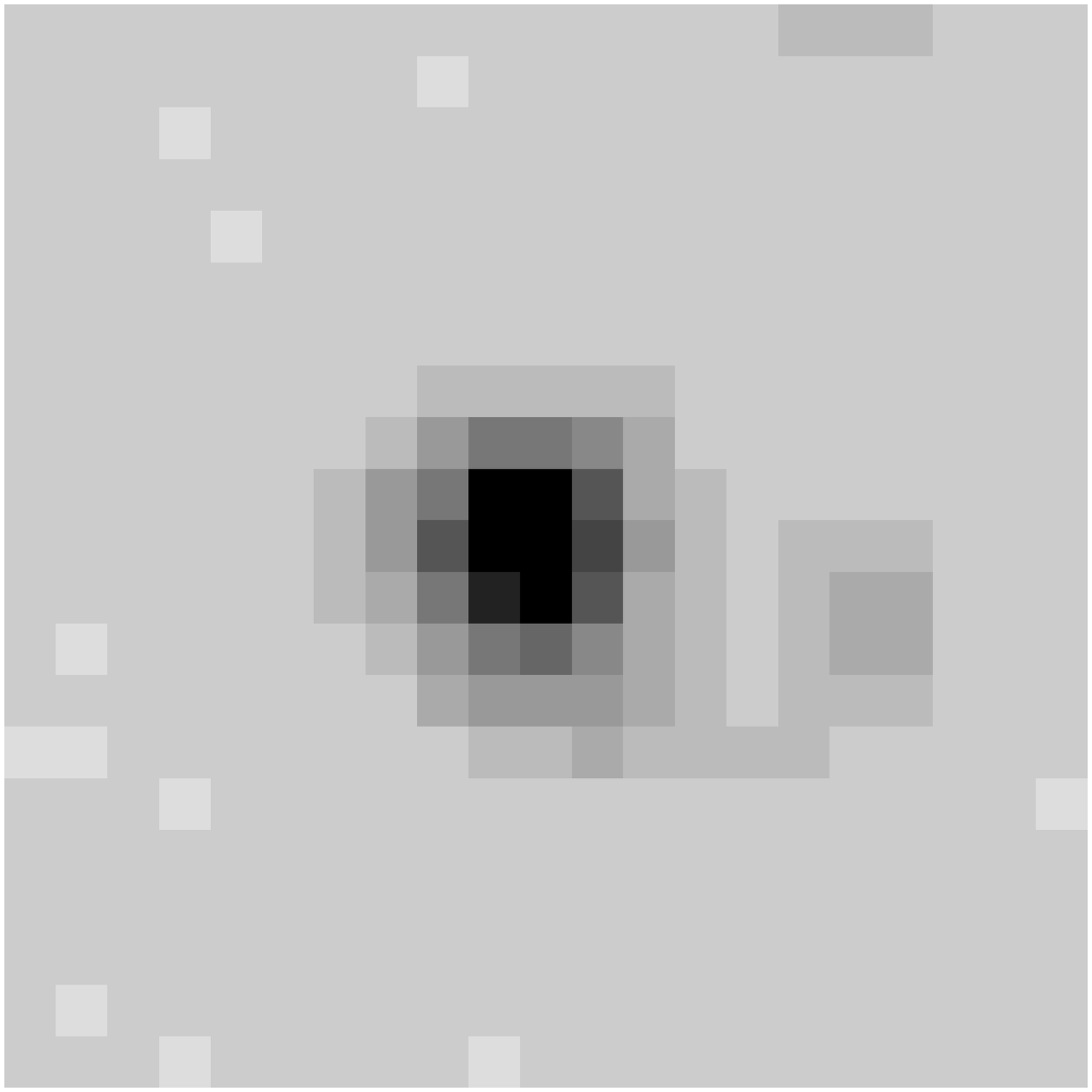}\\
{\bf BLAST 861:} $z_{\rm est} = 1.95$\\
\\
\\ 
\includegraphics[width=0.24\textwidth]{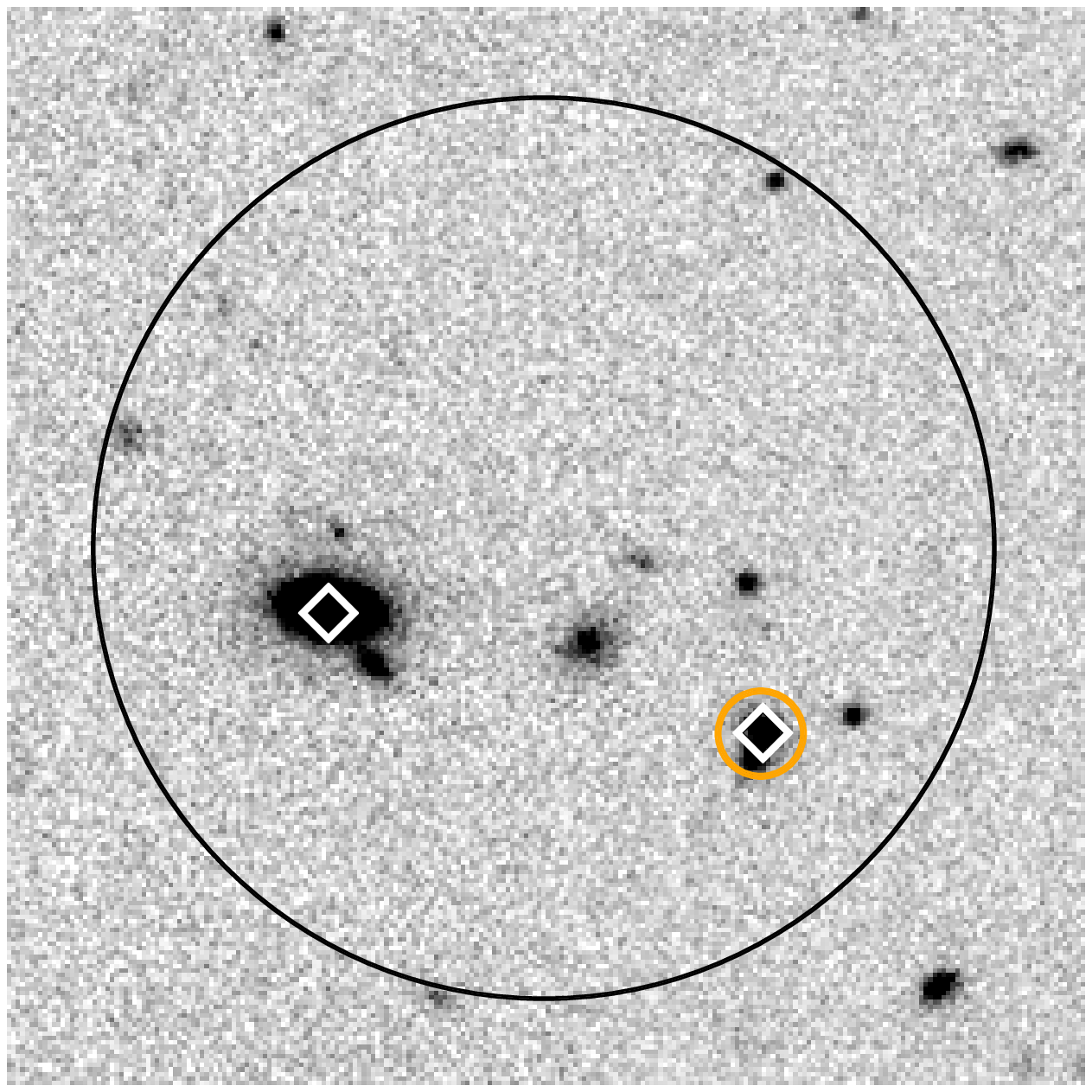}&
\includegraphics[width=0.24\textwidth]{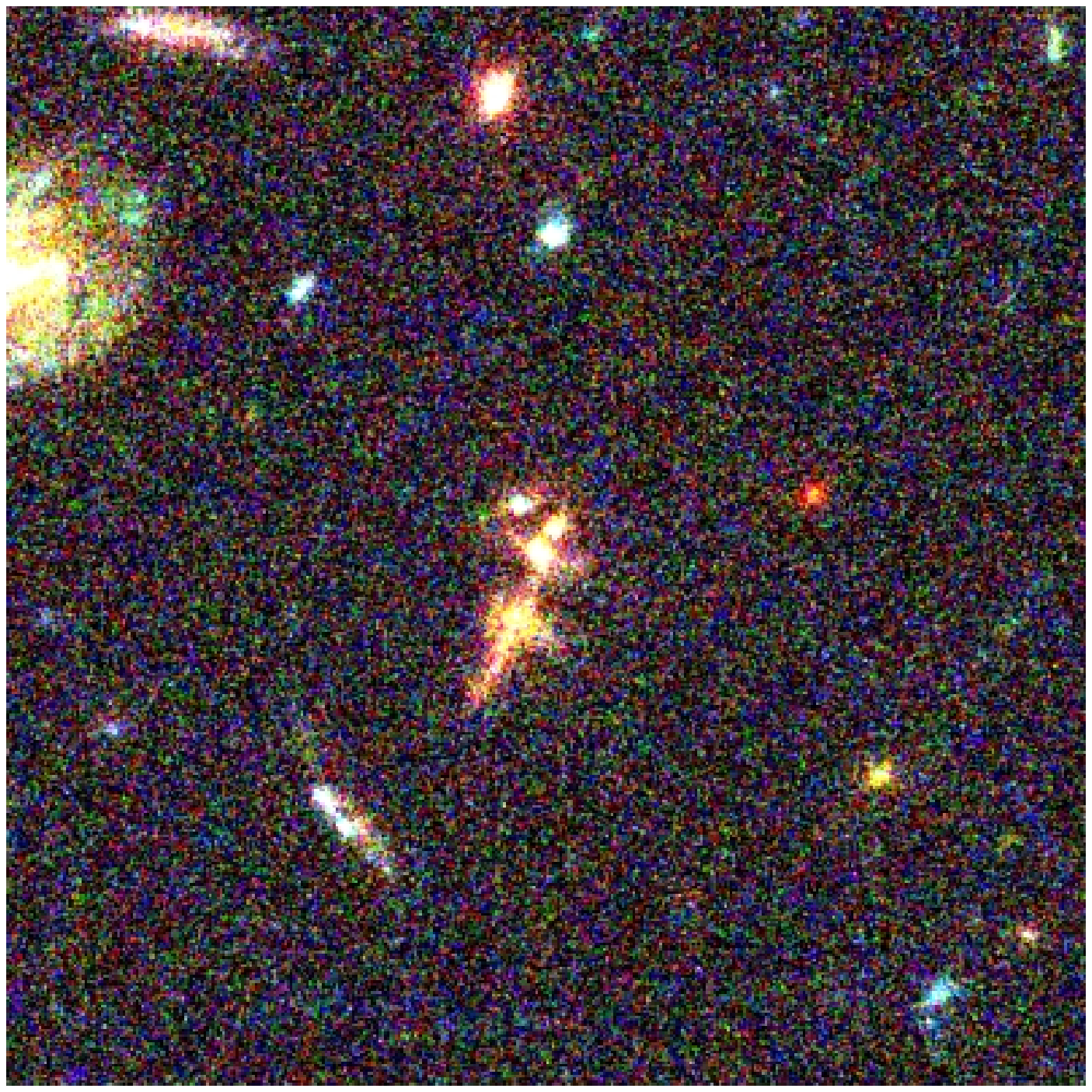}&
\includegraphics[width=0.24\textwidth]{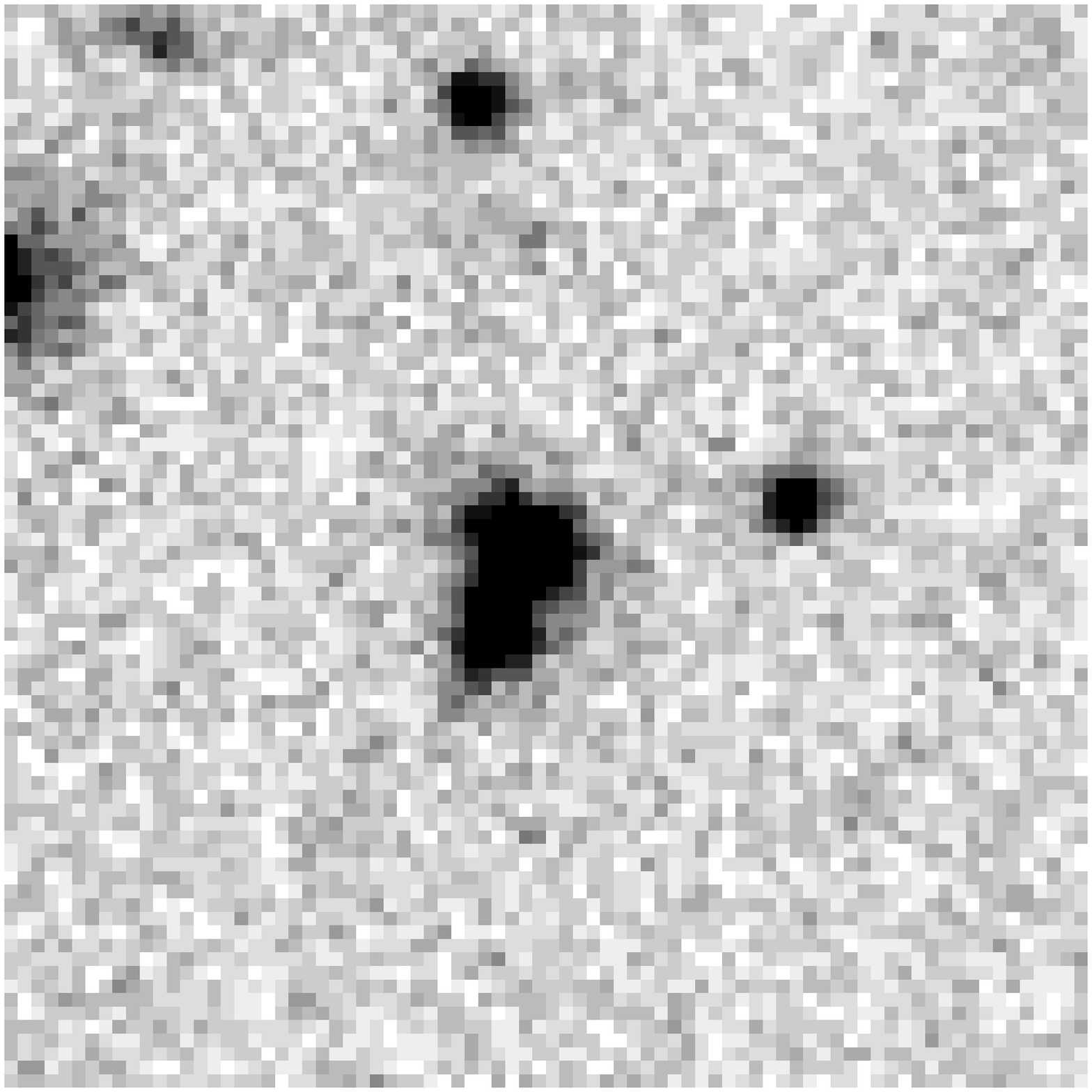}&
\includegraphics[width=0.24\textwidth]{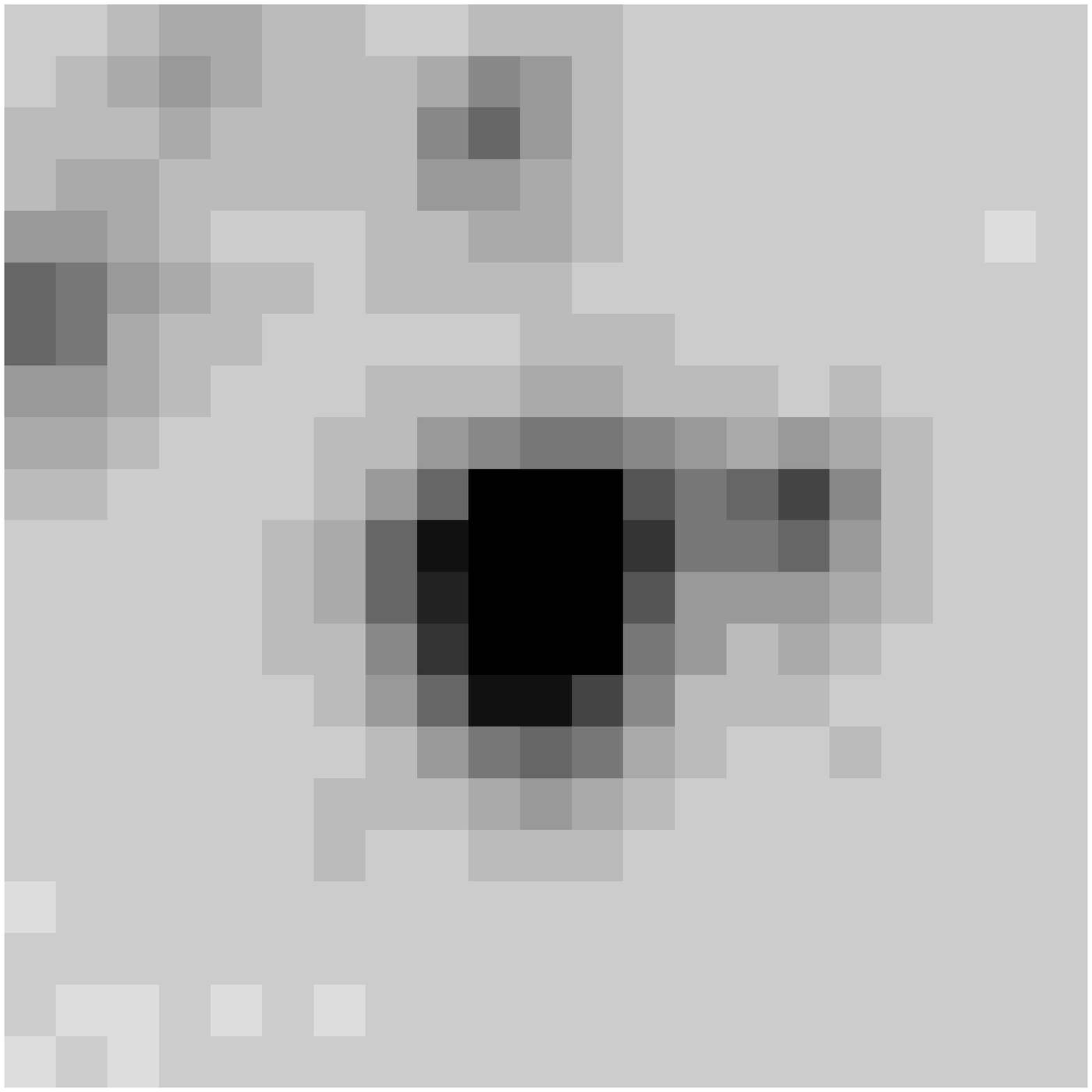}\\
{\bf BLAST 503-1:} $z_{\rm est} = 1.96$\\ 
\\
\\  
\includegraphics[width=0.24\textwidth]{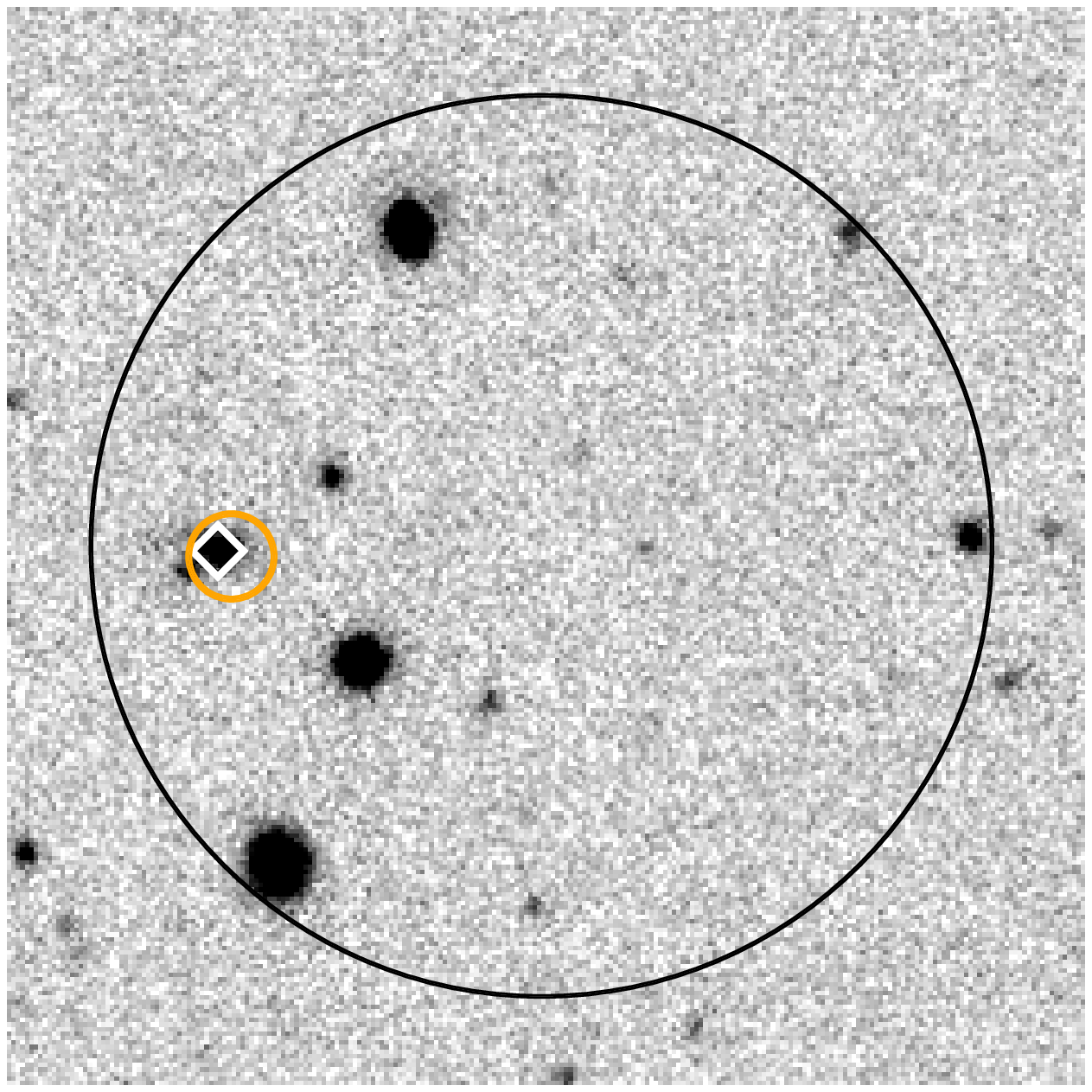}&
\includegraphics[width=0.24\textwidth]{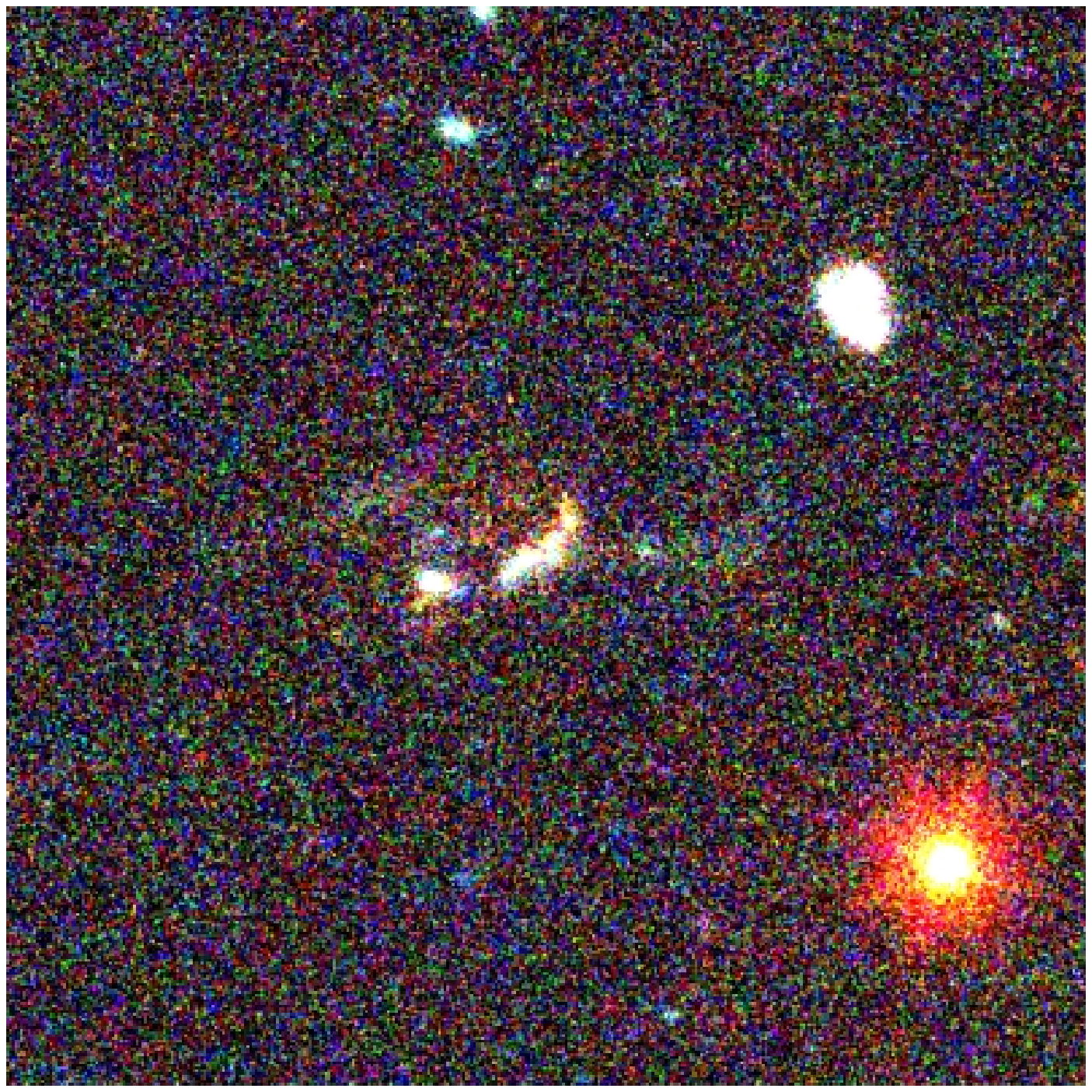}&
\includegraphics[width=0.24\textwidth]{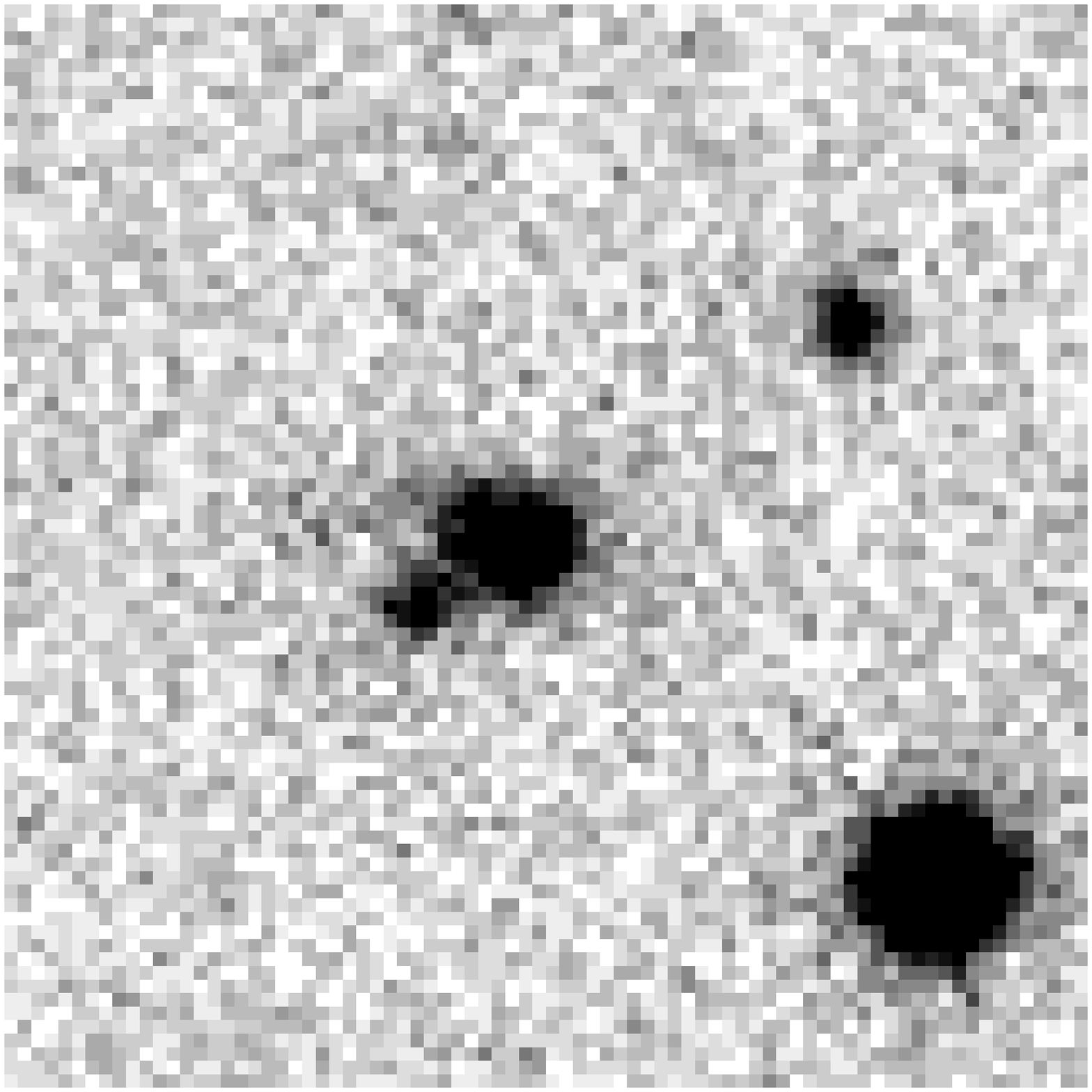}&
\includegraphics[width=0.24\textwidth]{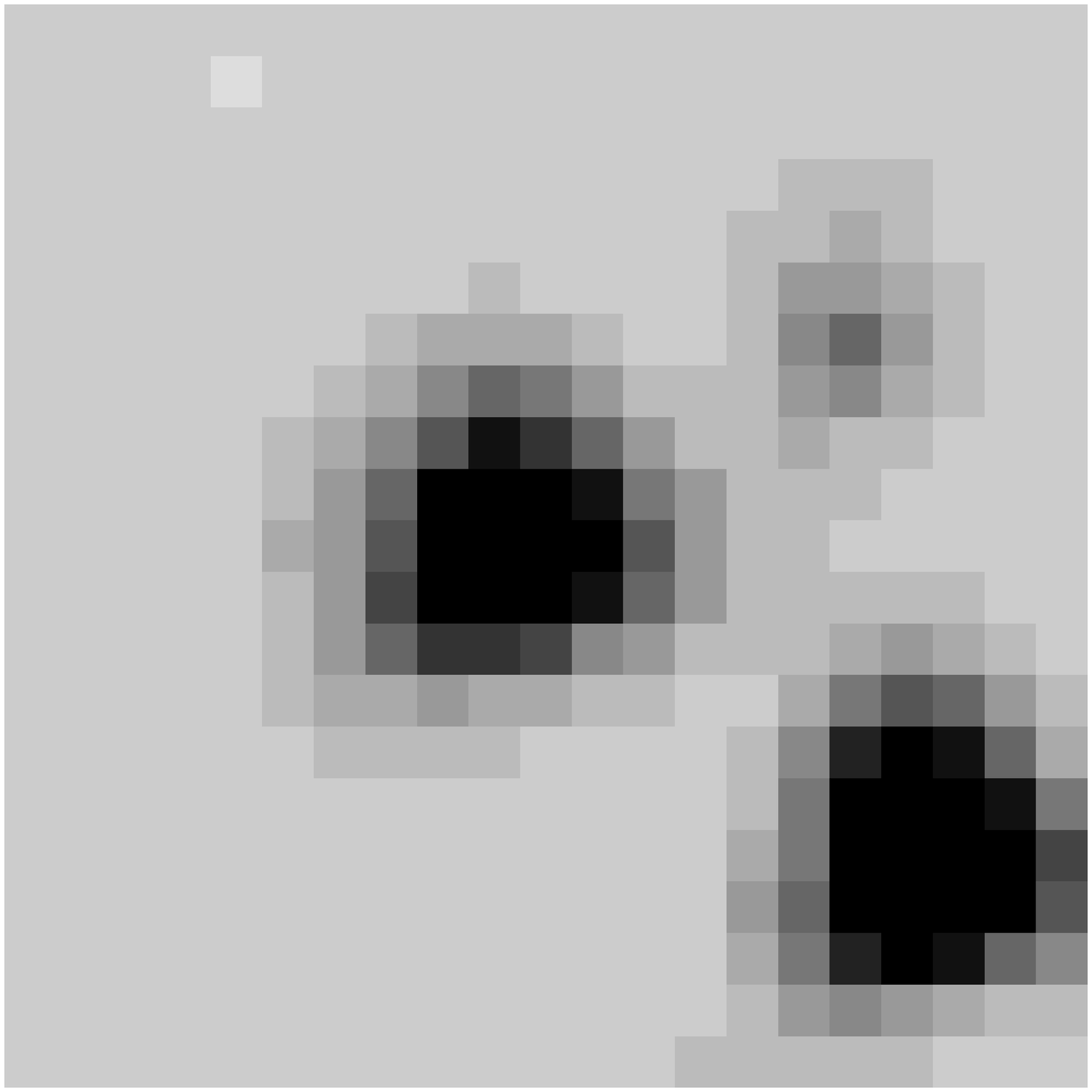}\\
{\bf BLAST 318:} $z_{\rm est} = 2.09$\\
\\
\\
\end{tabular}
\addtocounter{figure}{-1}
\caption{continued}
\end{figure*}

\begin{figure*}
\begin{tabular}{llll}
\\                    
\includegraphics[width=0.24\textwidth]{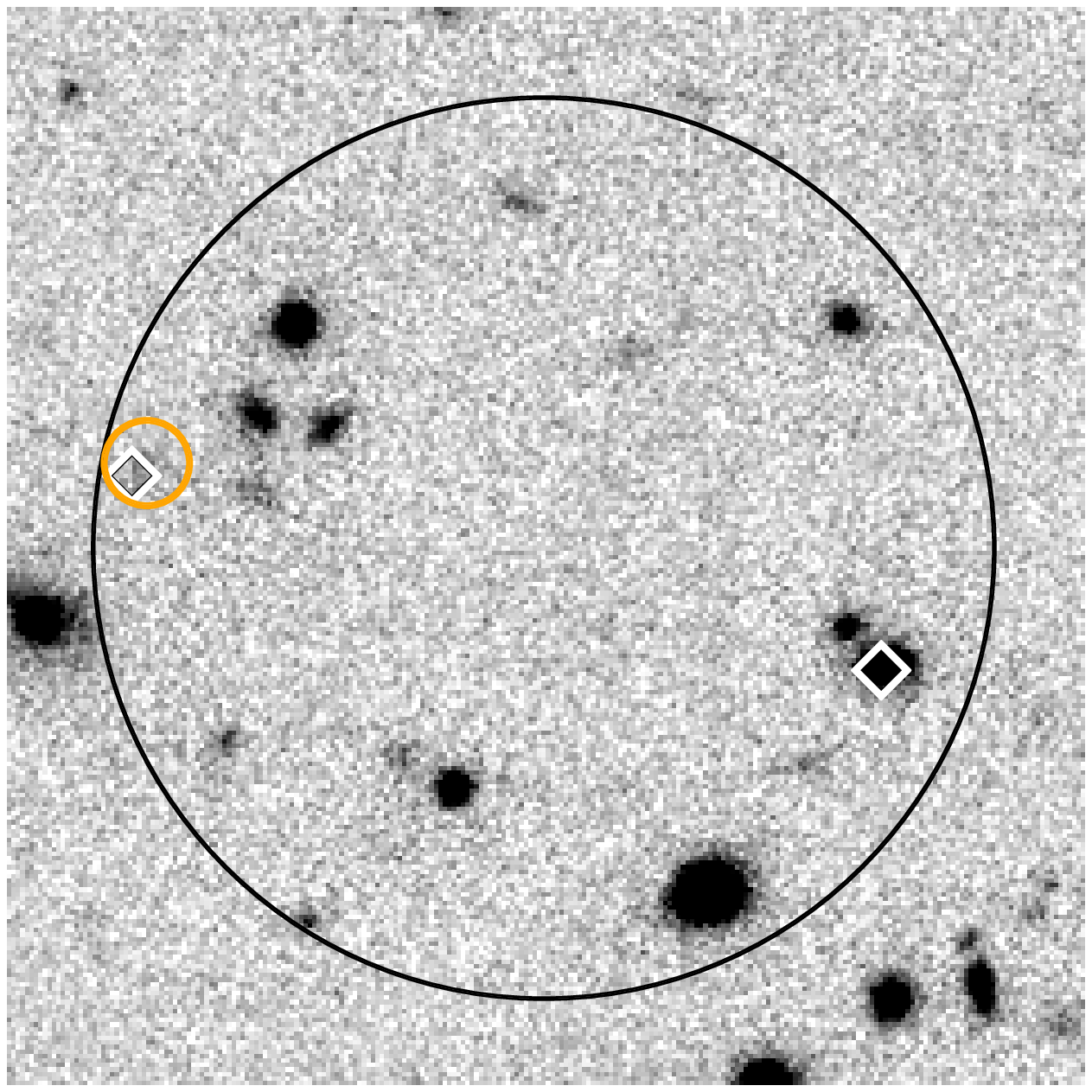}&
\includegraphics[width=0.24\textwidth]{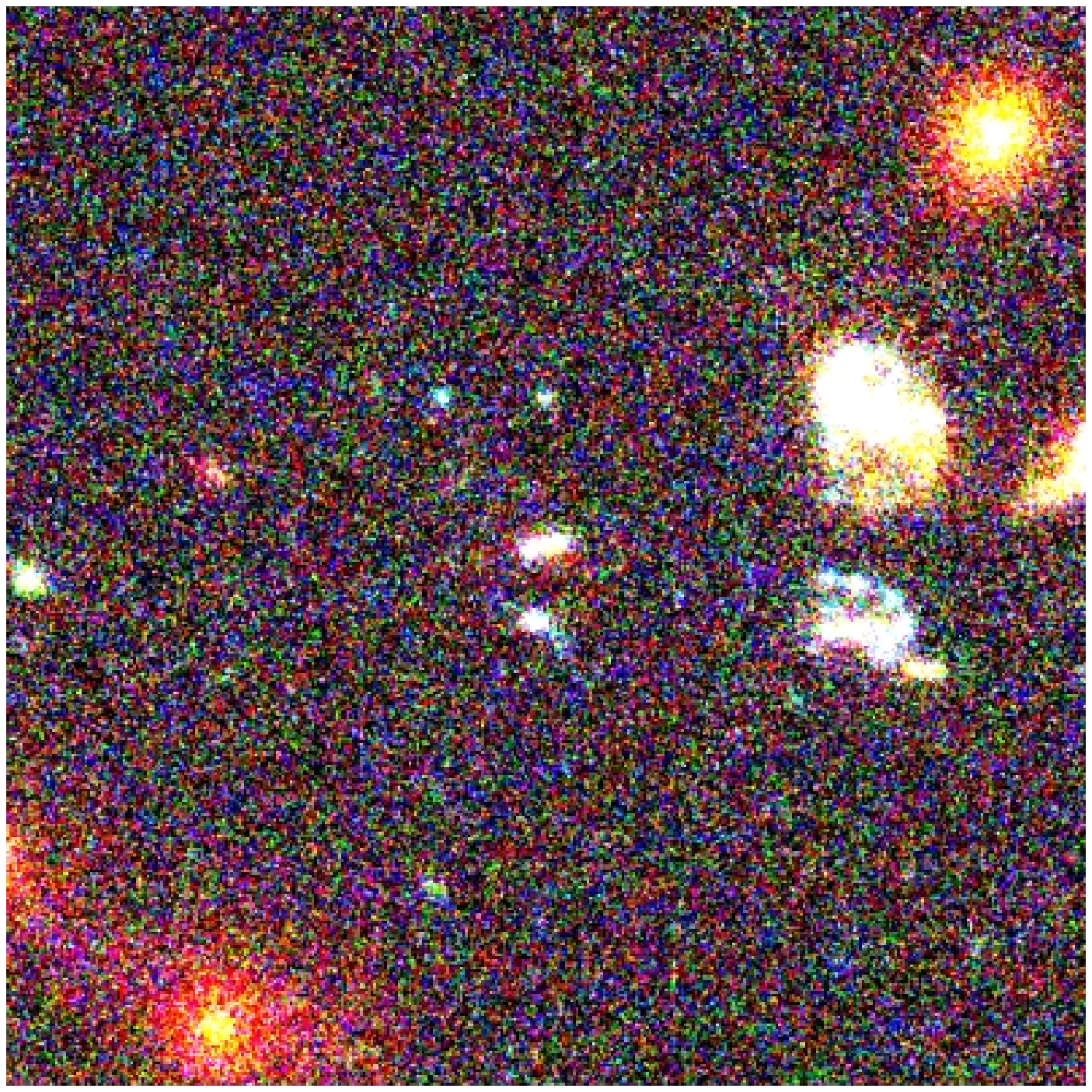}&
\includegraphics[width=0.24\textwidth]{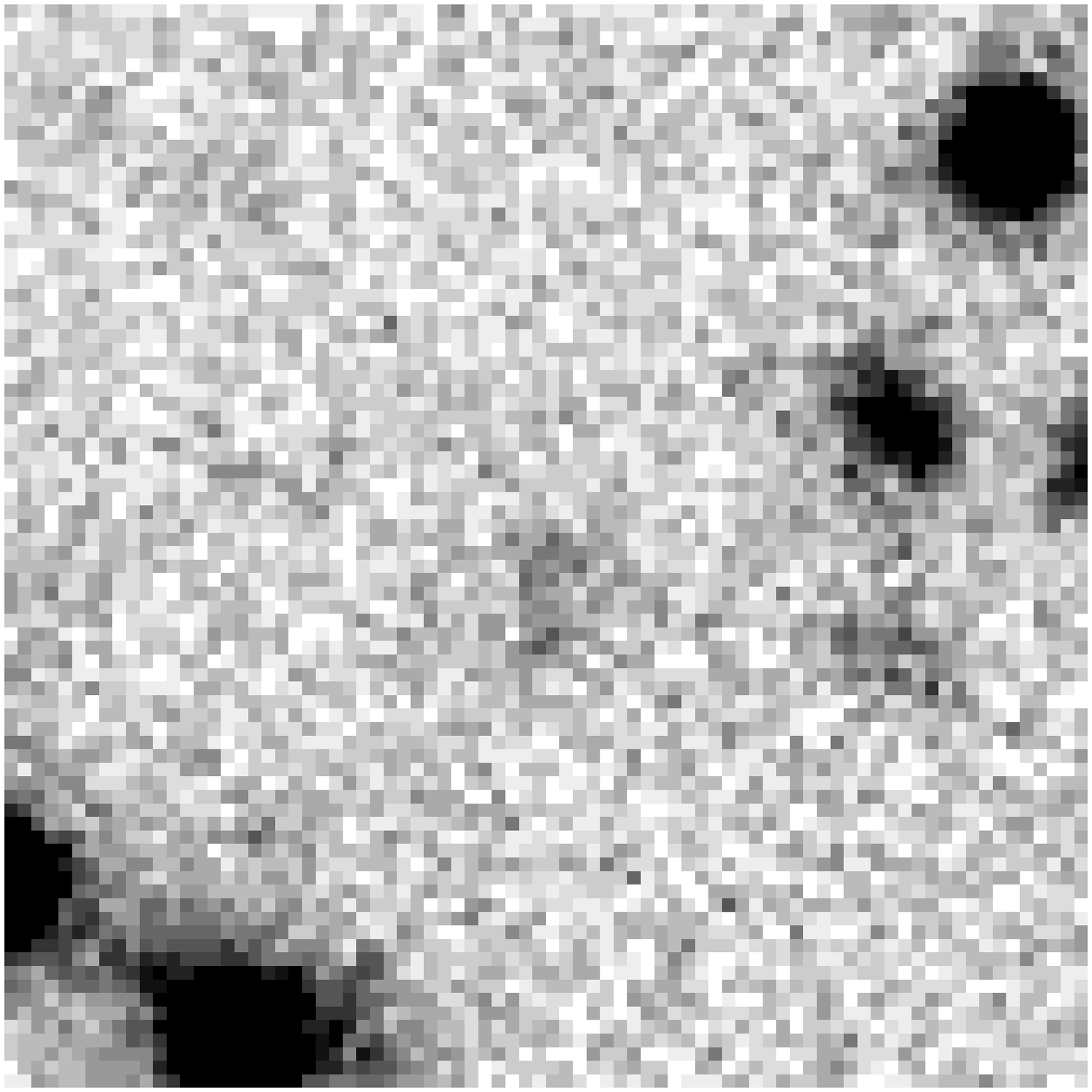}&
\includegraphics[width=0.24\textwidth]{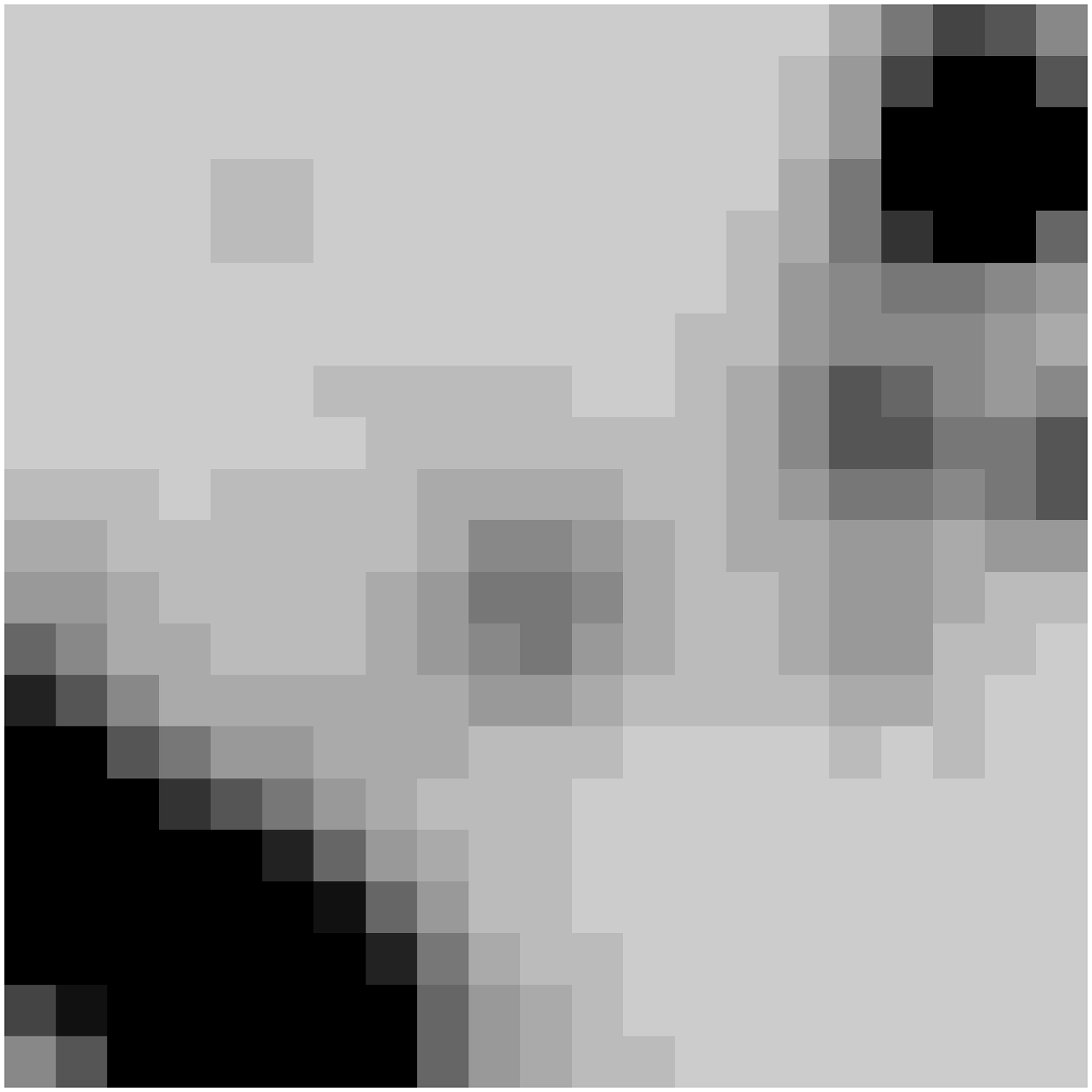}\\
{\bf BLAST 59-1:} $z_{\rm est} = 2.29$\\
\\
\\
\includegraphics[width=0.24\textwidth]{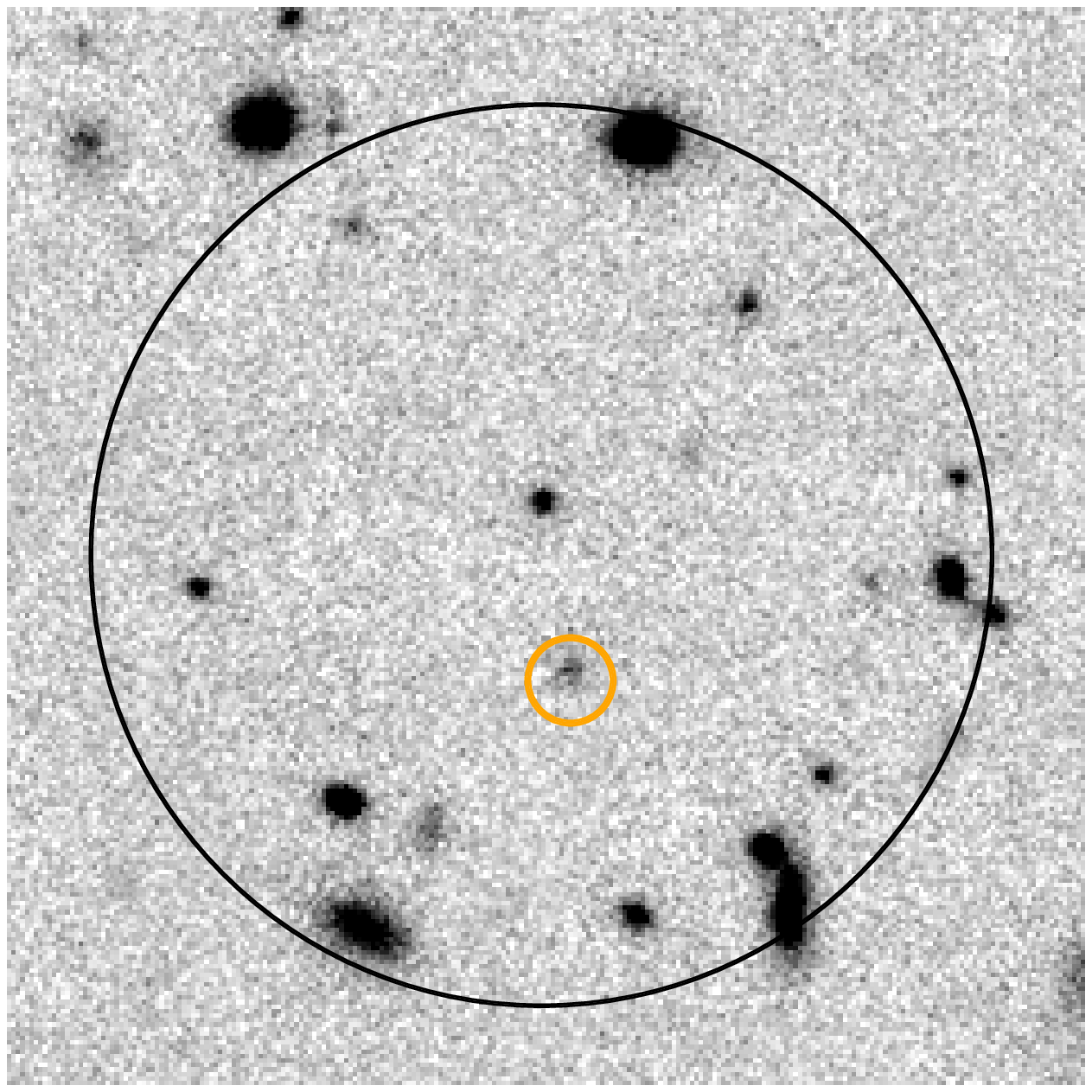}&
\includegraphics[width=0.24\textwidth]{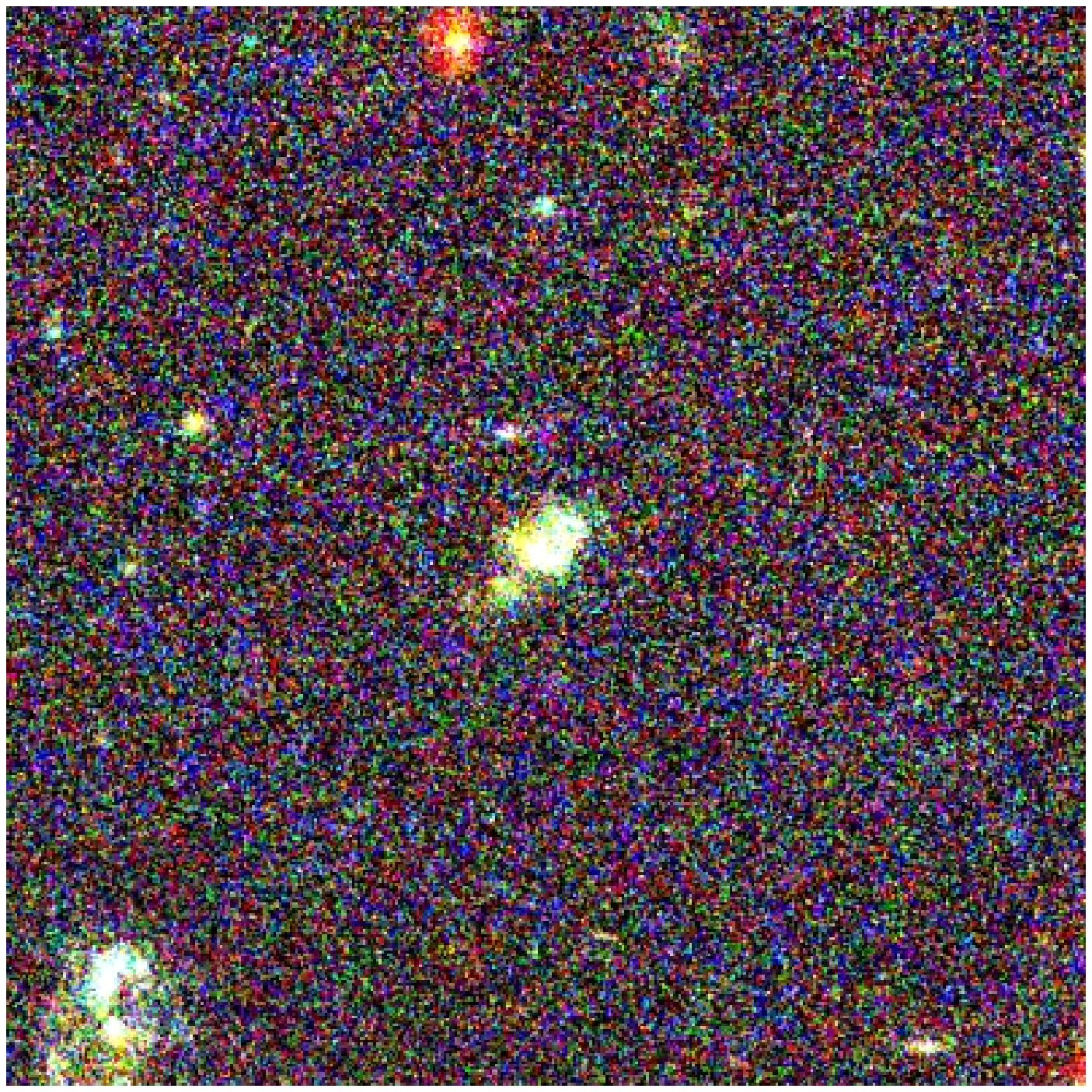}&
\includegraphics[width=0.24\textwidth]{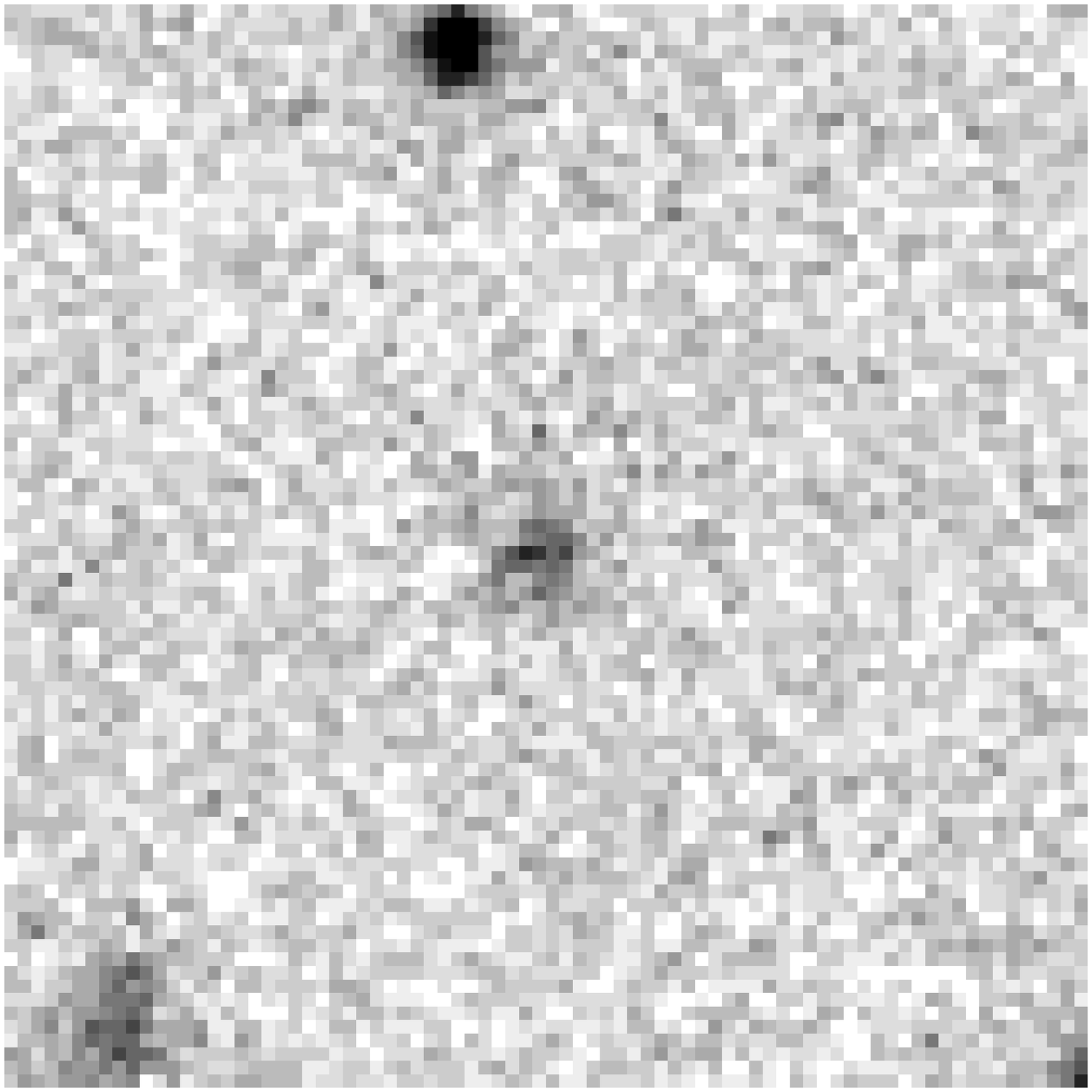}&
\includegraphics[width=0.24\textwidth]{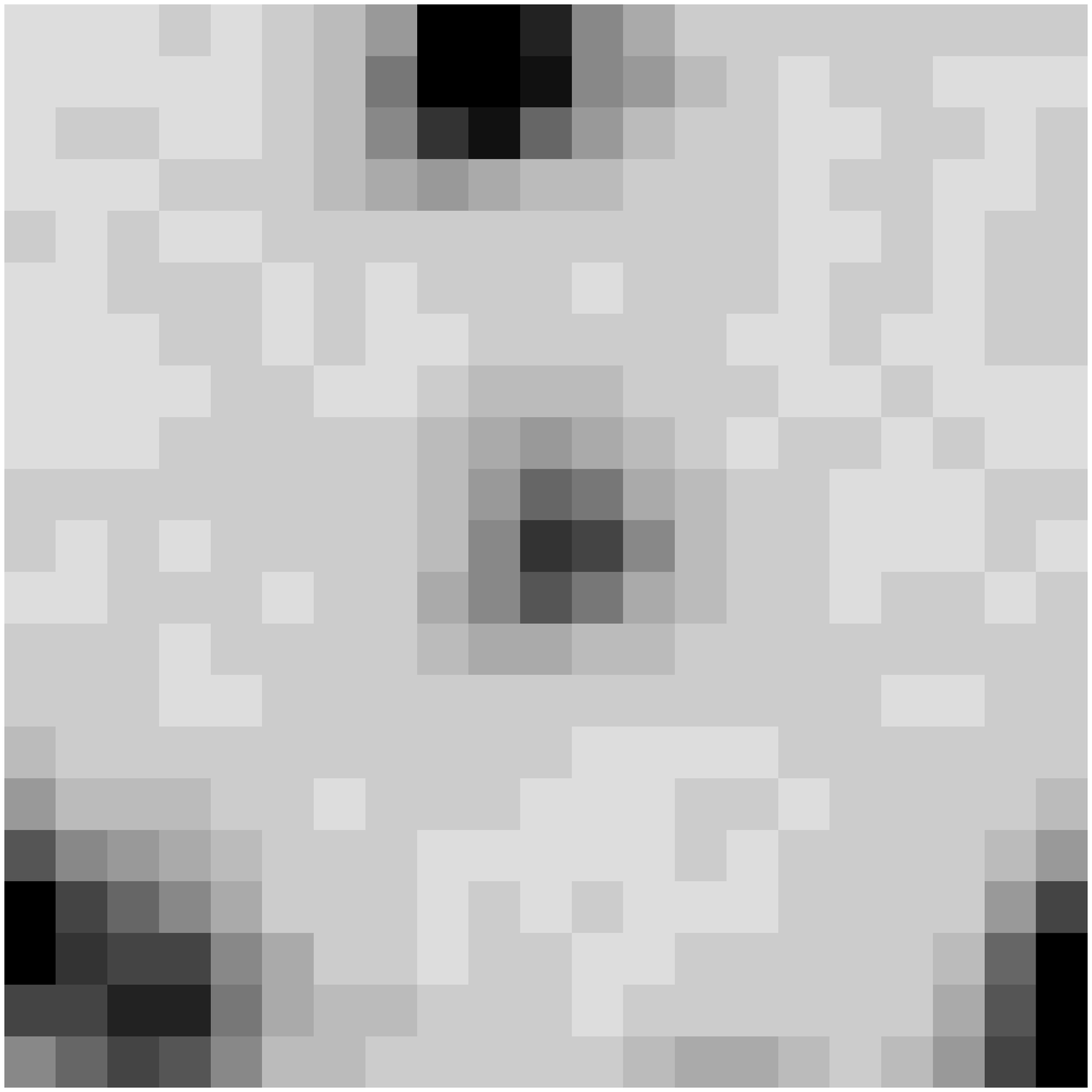}\\
{\bf BLAST 654:} $z_{\rm est} = 2.62$\\ 
\\
\\
\includegraphics[width=0.24\textwidth]{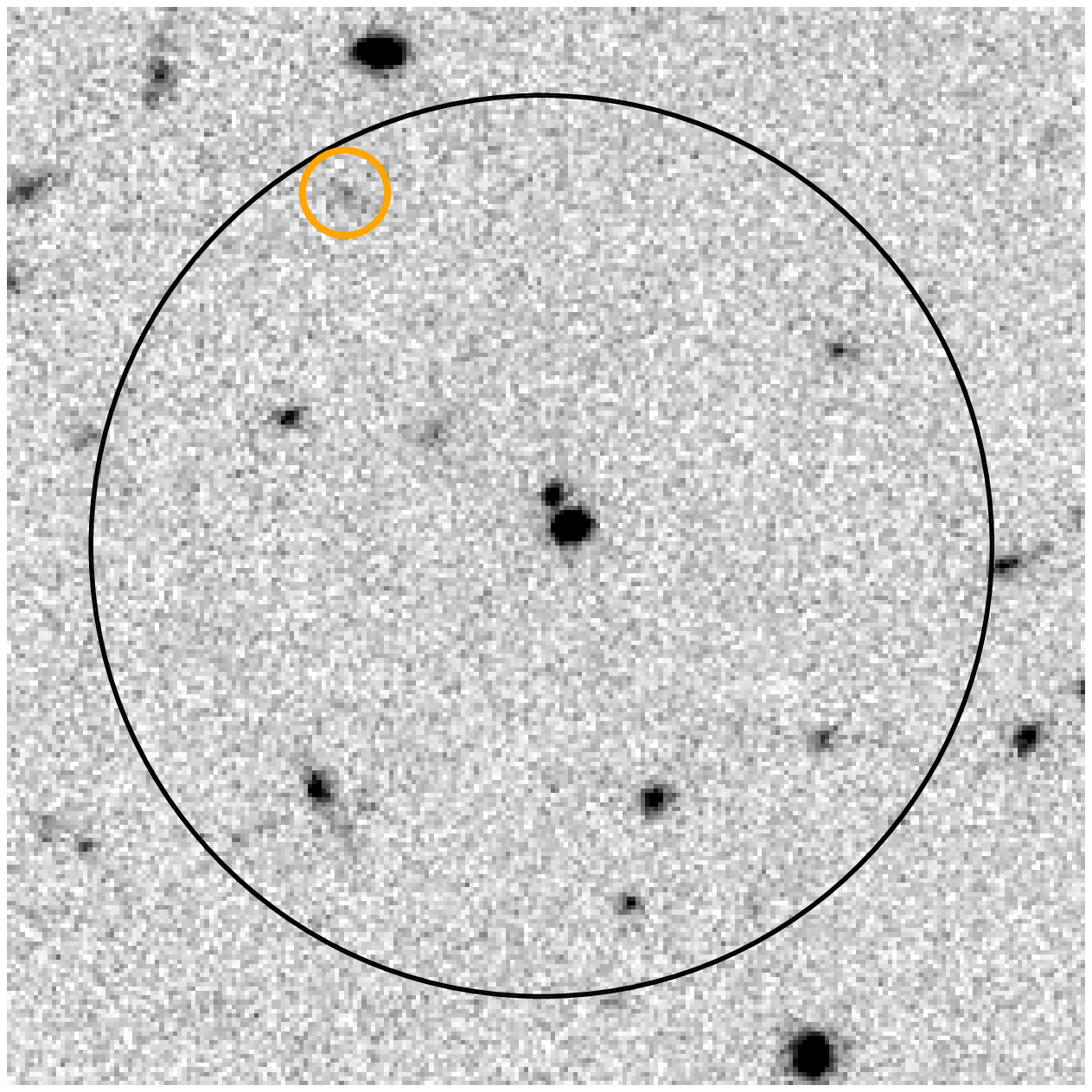}&
\includegraphics[width=0.24\textwidth]{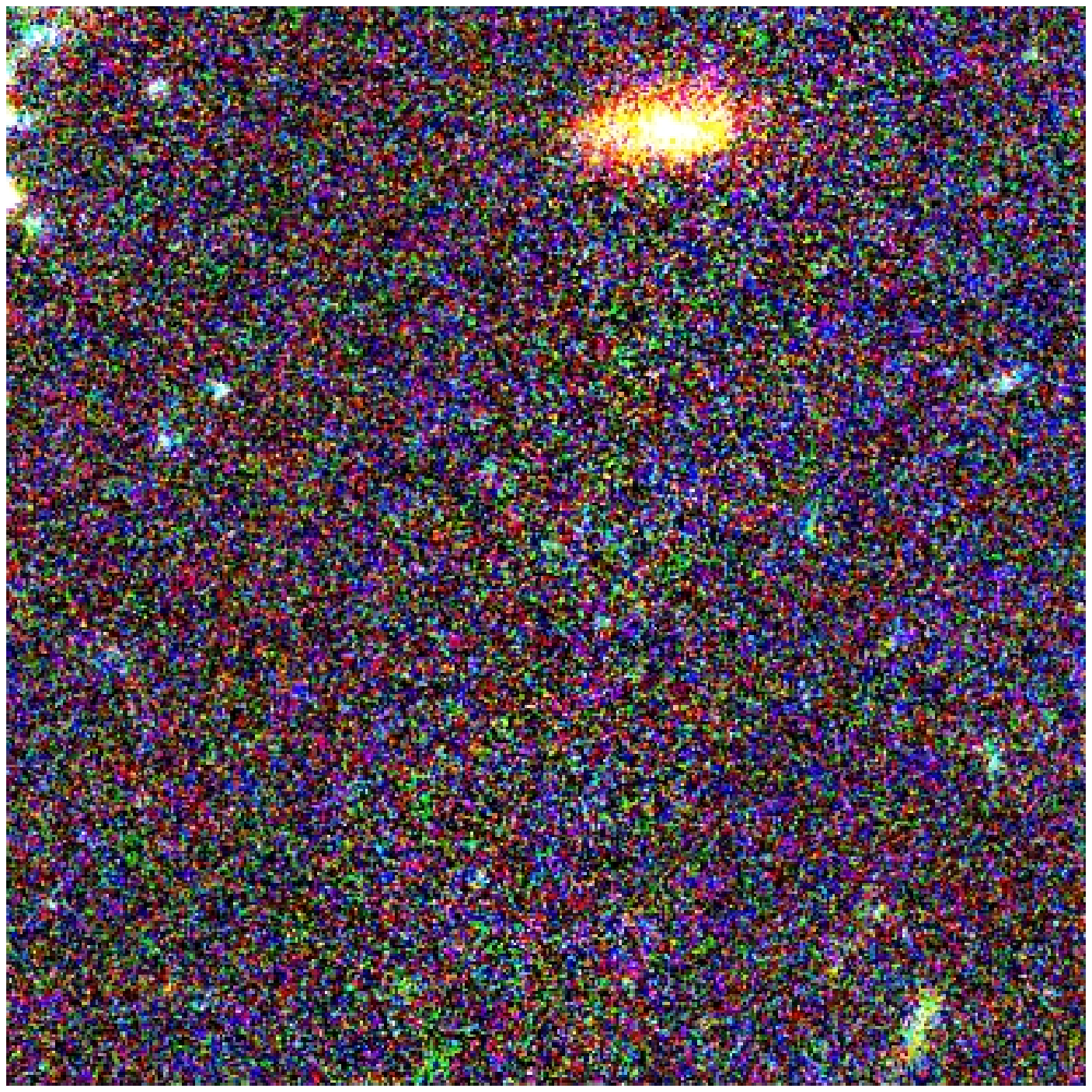}&
\includegraphics[width=0.24\textwidth]{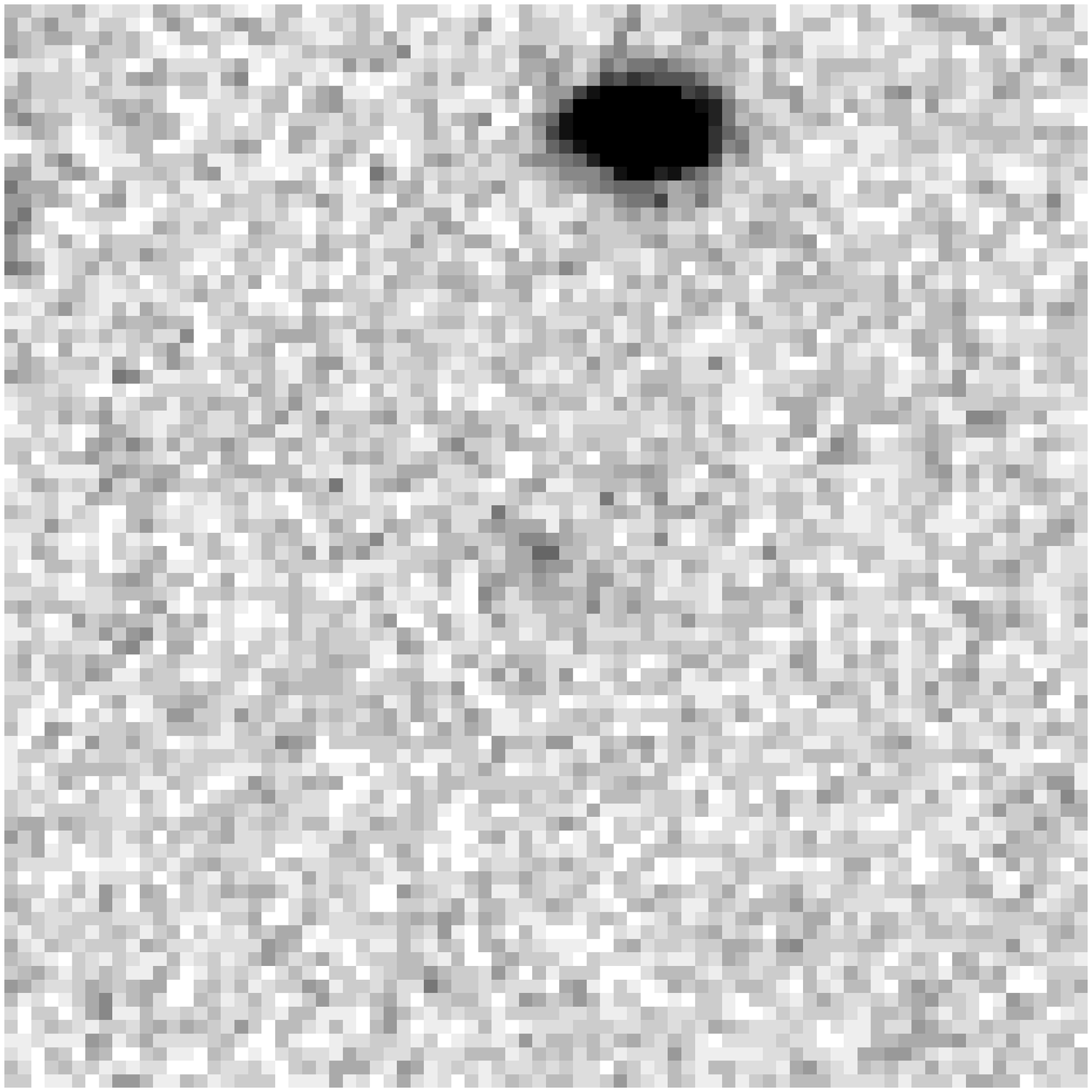}&
\includegraphics[width=0.24\textwidth]{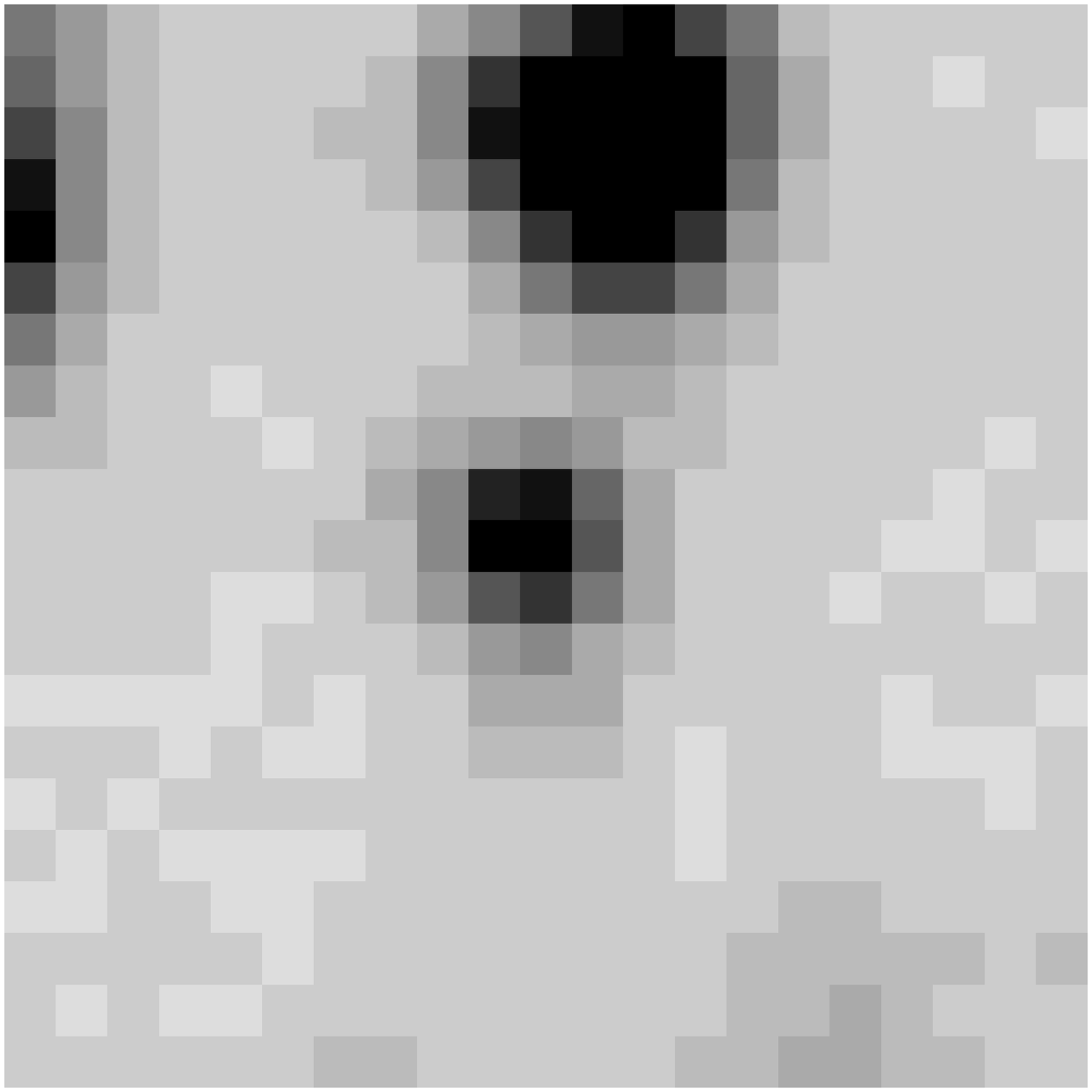}\\
{\bf BLAST 732-1:} $z_{\rm est} = 2.97$\\ 
\\
\\
\includegraphics[width=0.24\textwidth]{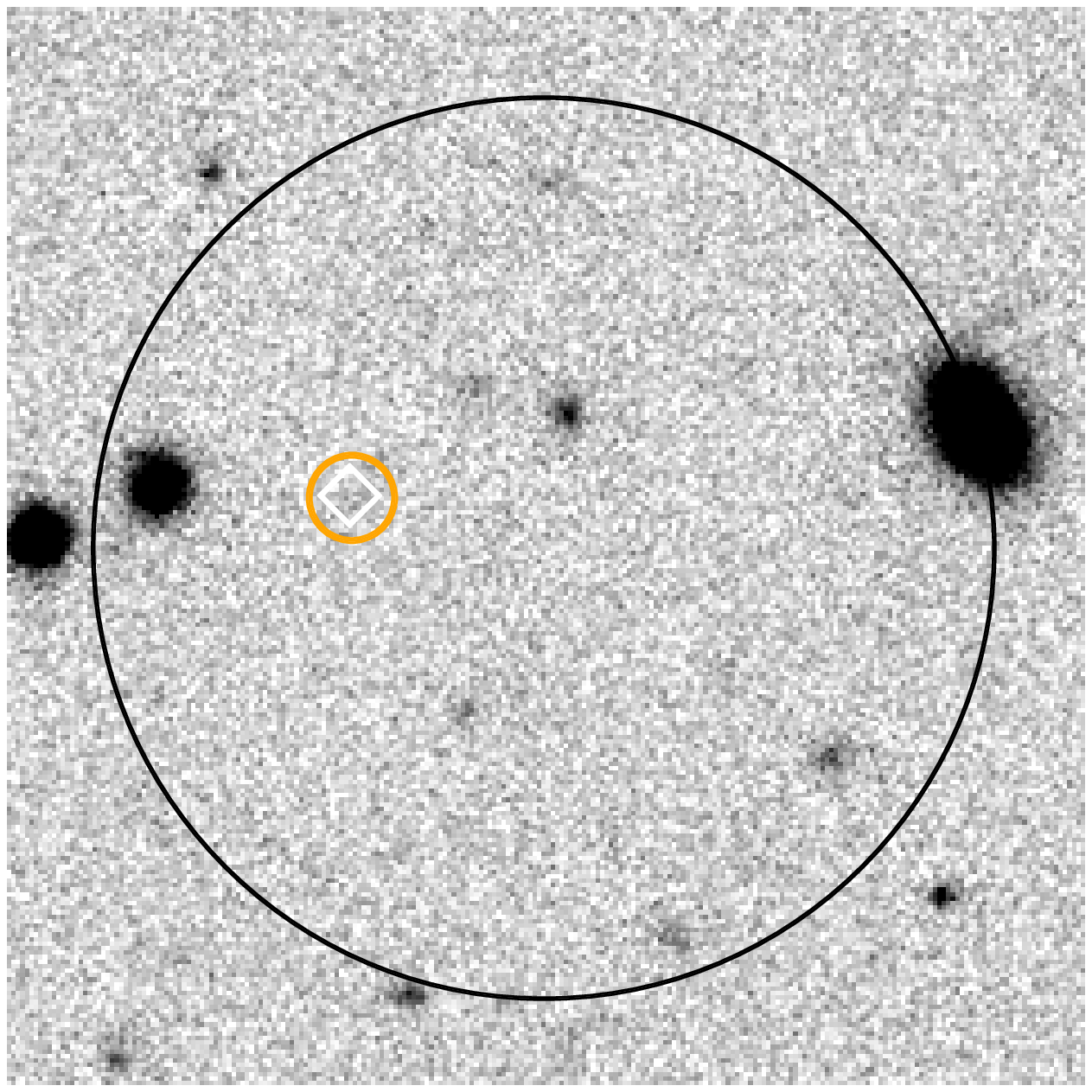}&
\includegraphics[width=0.24\textwidth]{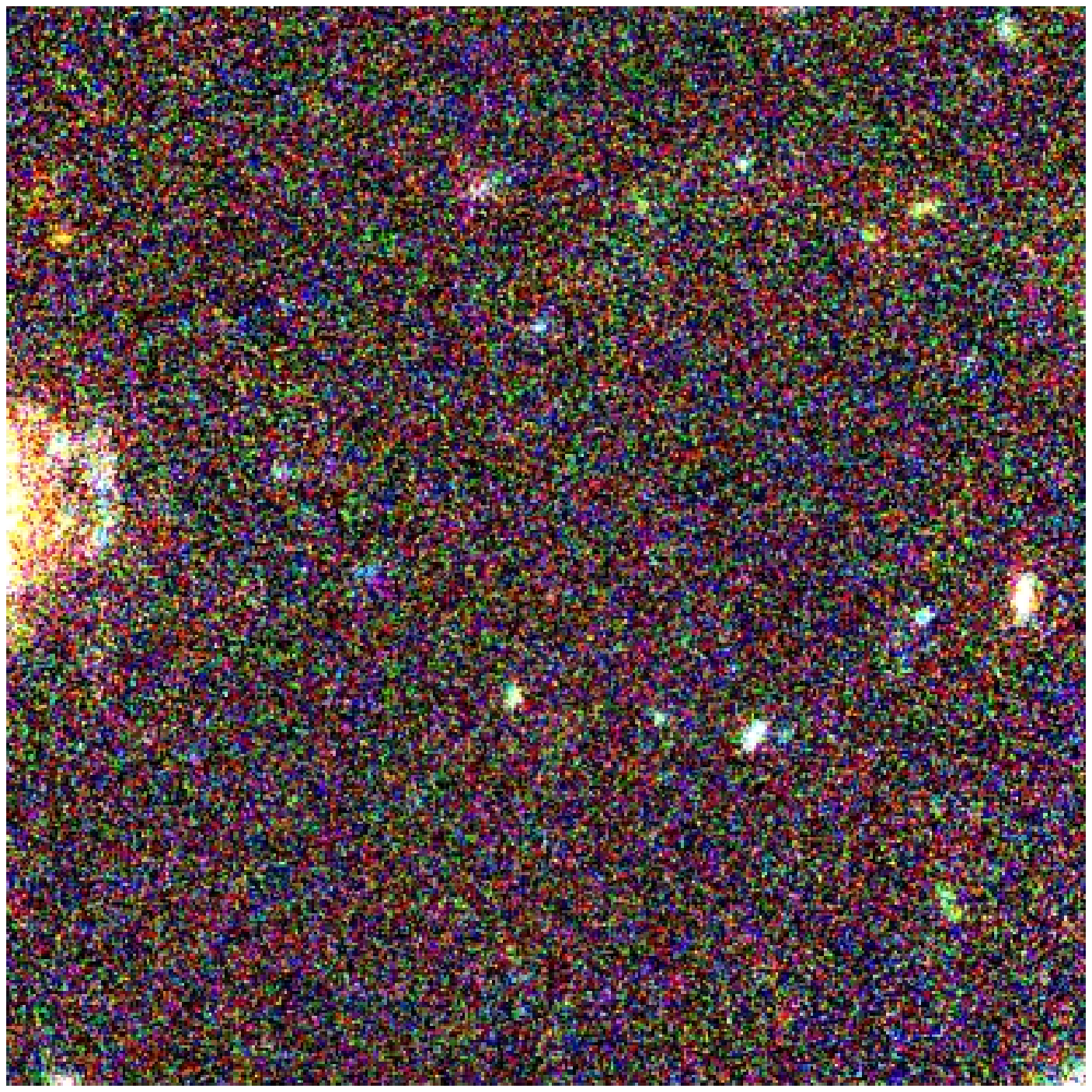}&
\includegraphics[width=0.24\textwidth]{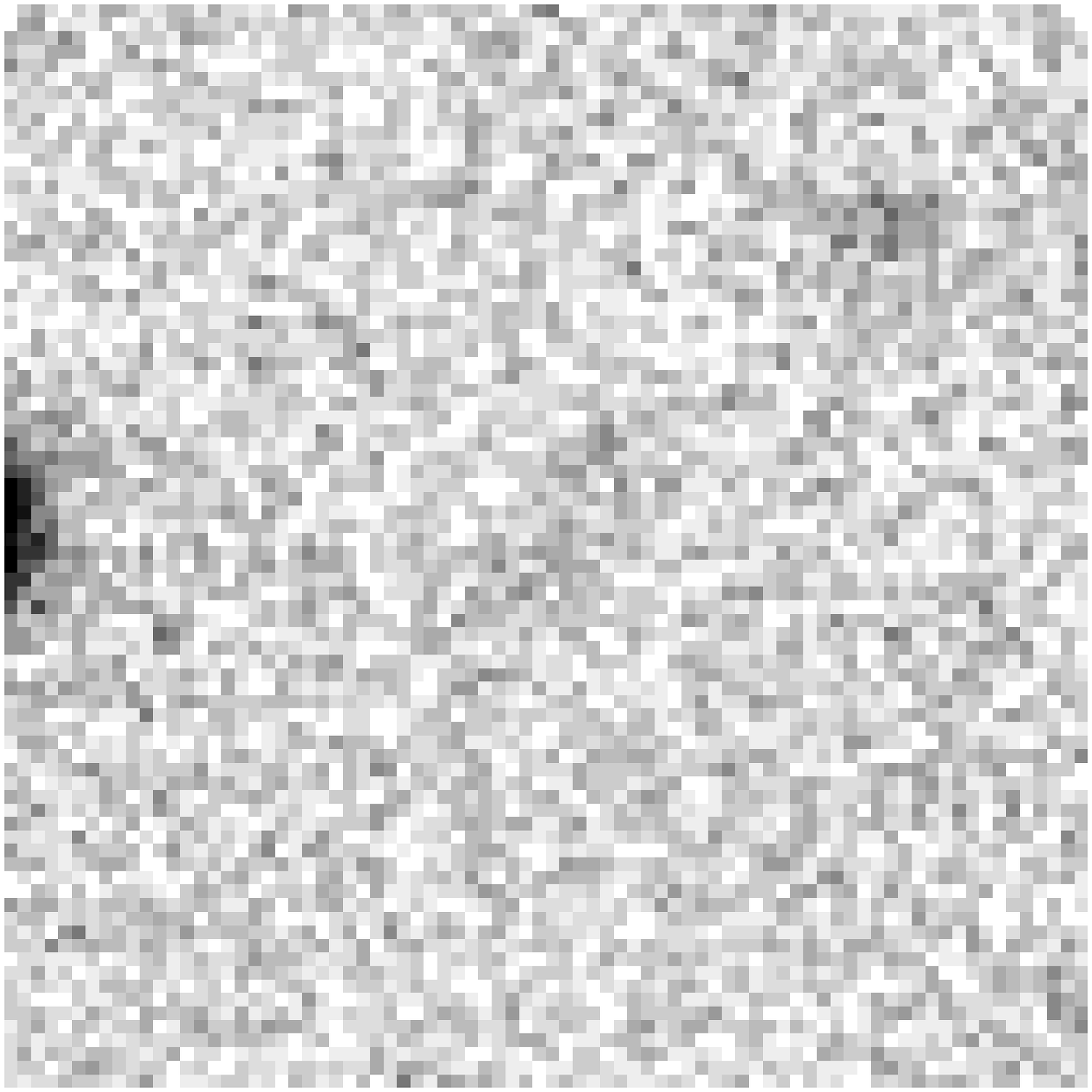}&
\includegraphics[width=0.24\textwidth]{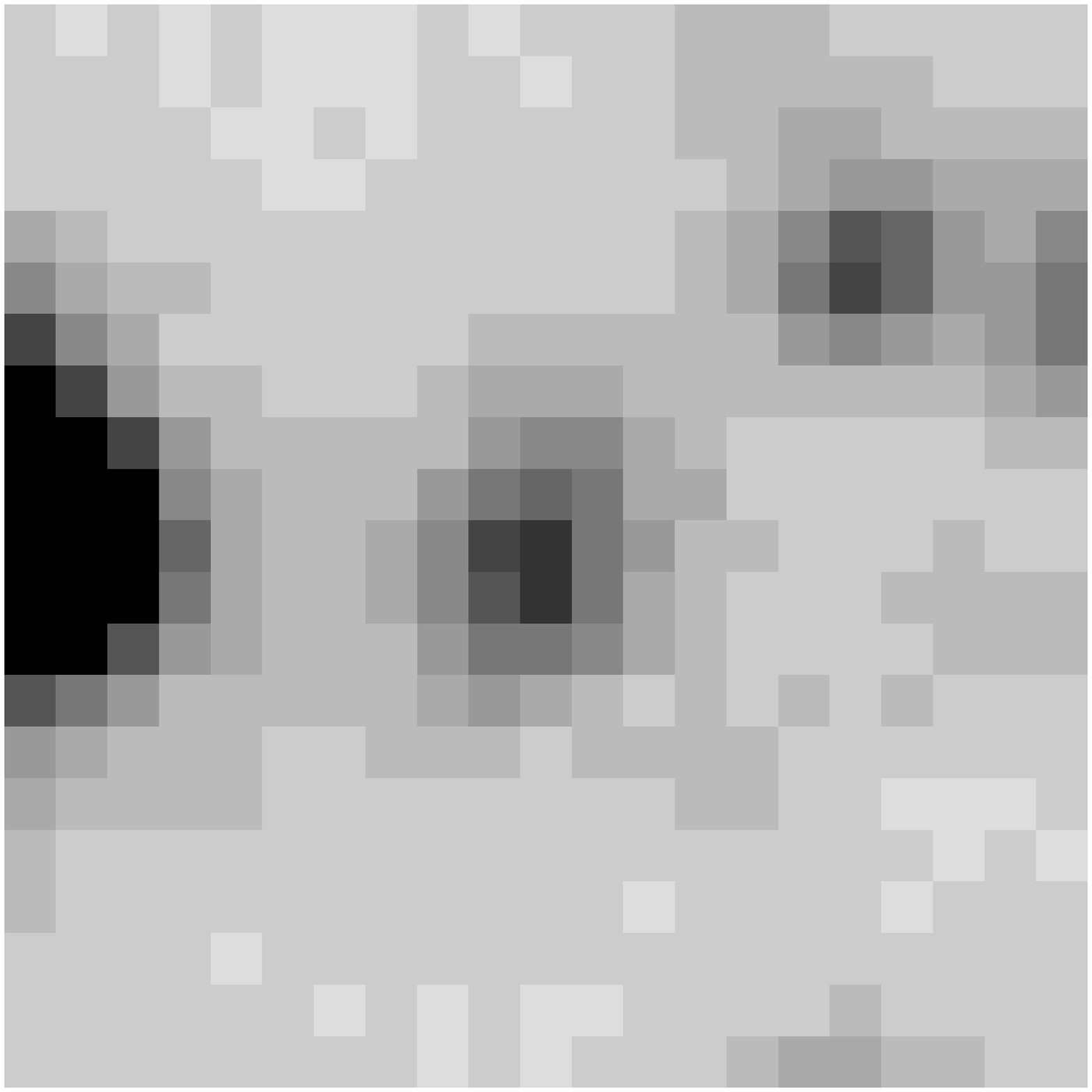}\\
{\bf BLAST 593:} $z_{\rm est} > 2.5$\\ 
\\
\\
\end{tabular}
\addtocounter{figure}{-1}
\caption{continued}
\end{figure*}

\begin{figure*}
\begin{tabular}{llll}
\\
\includegraphics[width=0.24\textwidth]{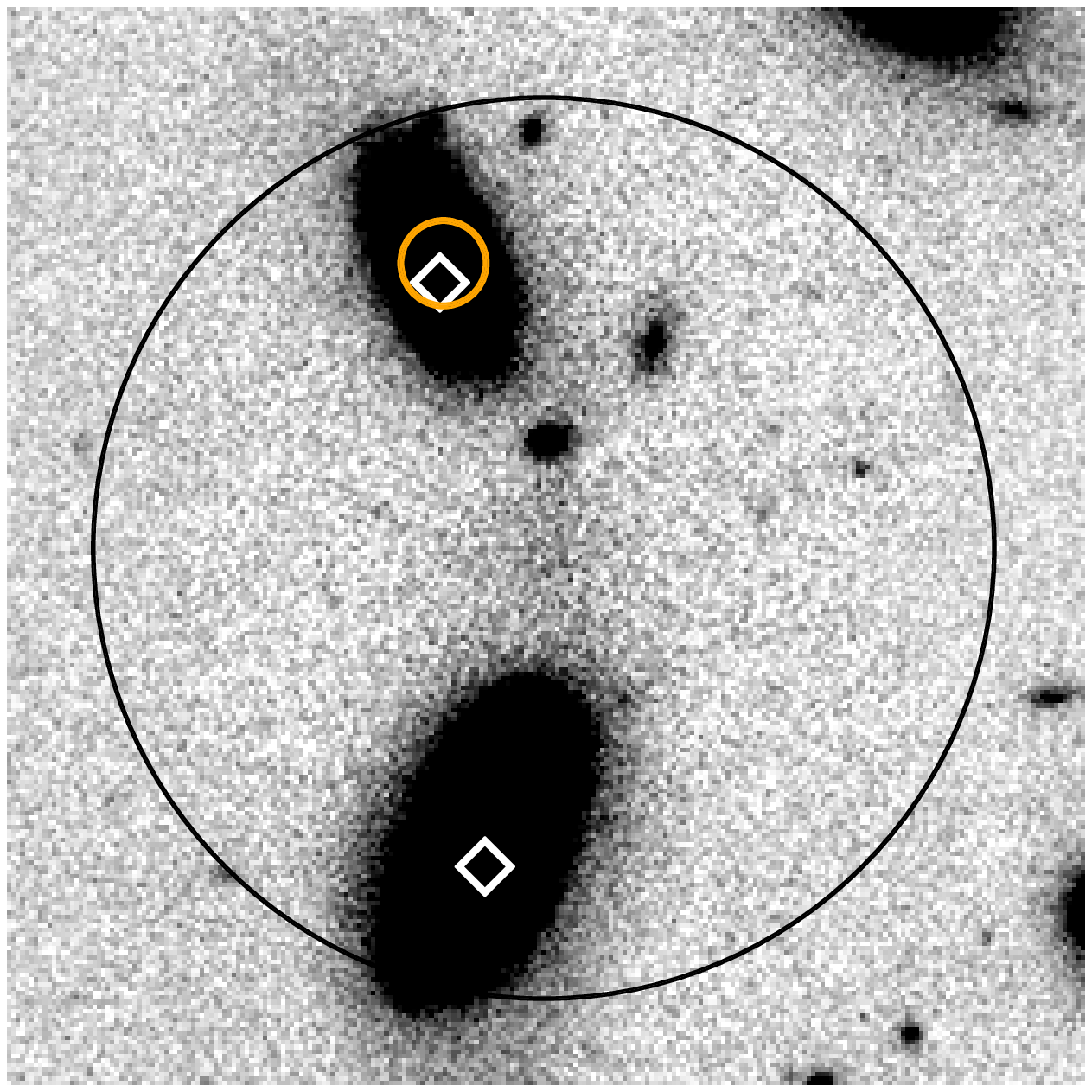}&
\includegraphics[width=0.24\textwidth]{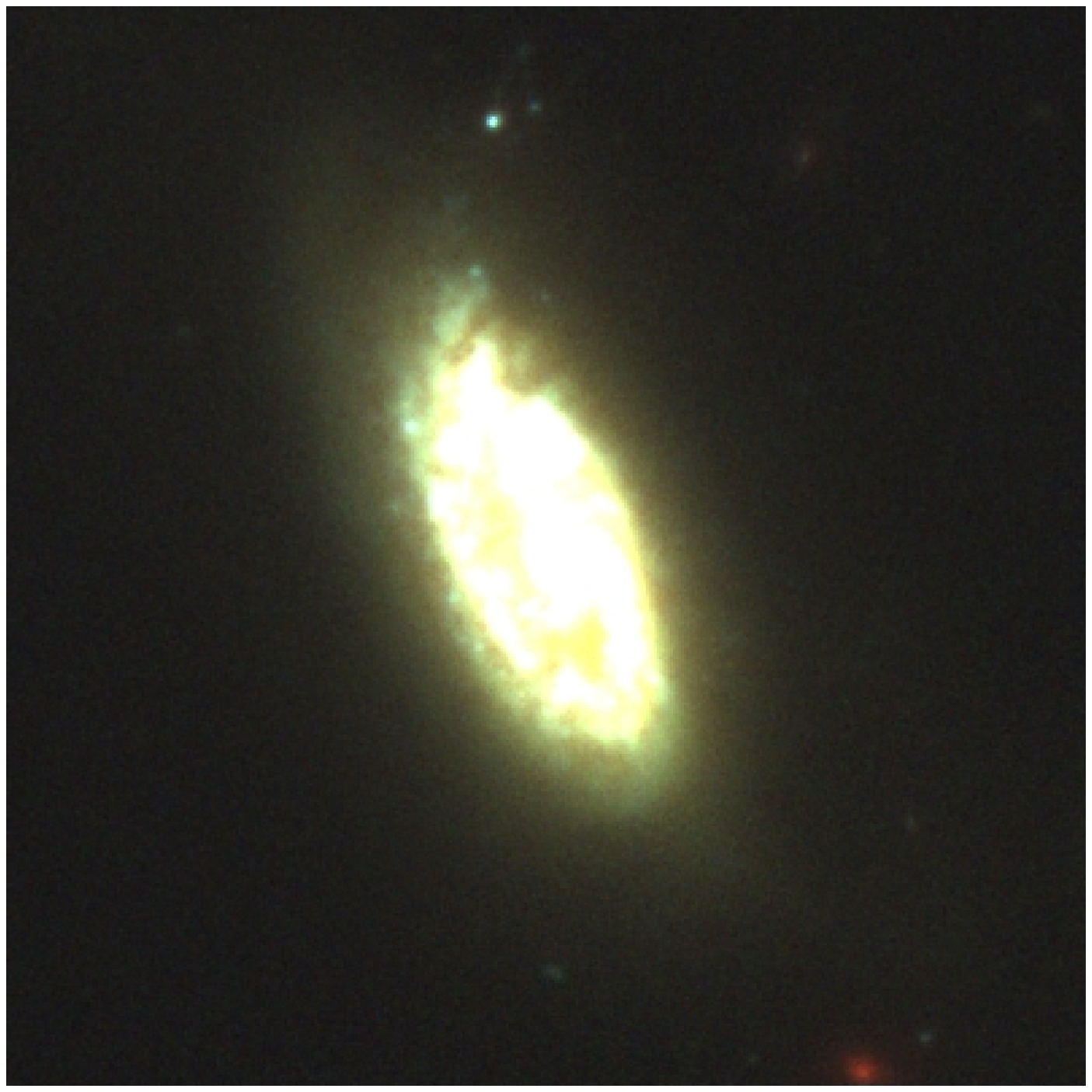}&
\includegraphics[width=0.24\textwidth]{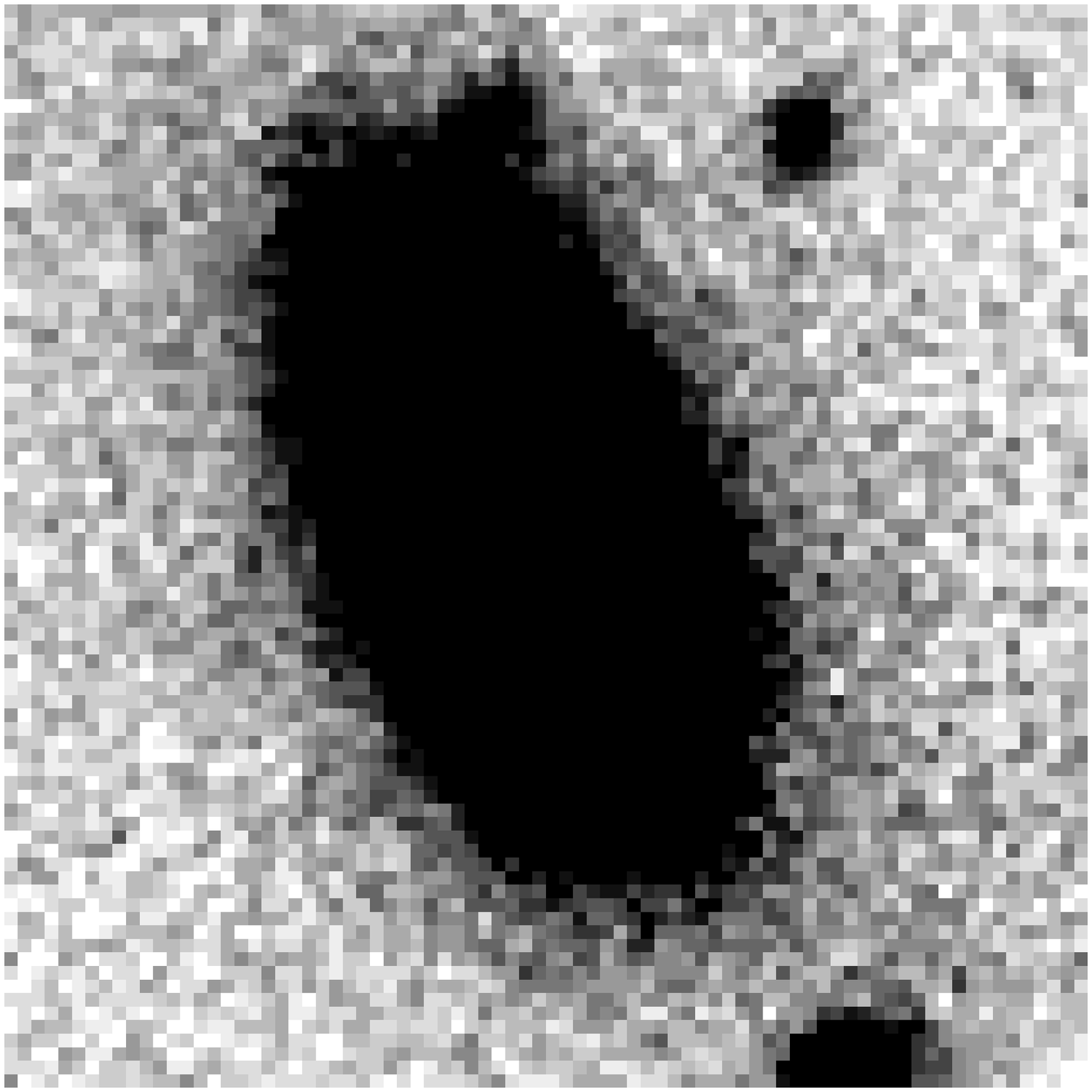}&
\includegraphics[width=0.24\textwidth]{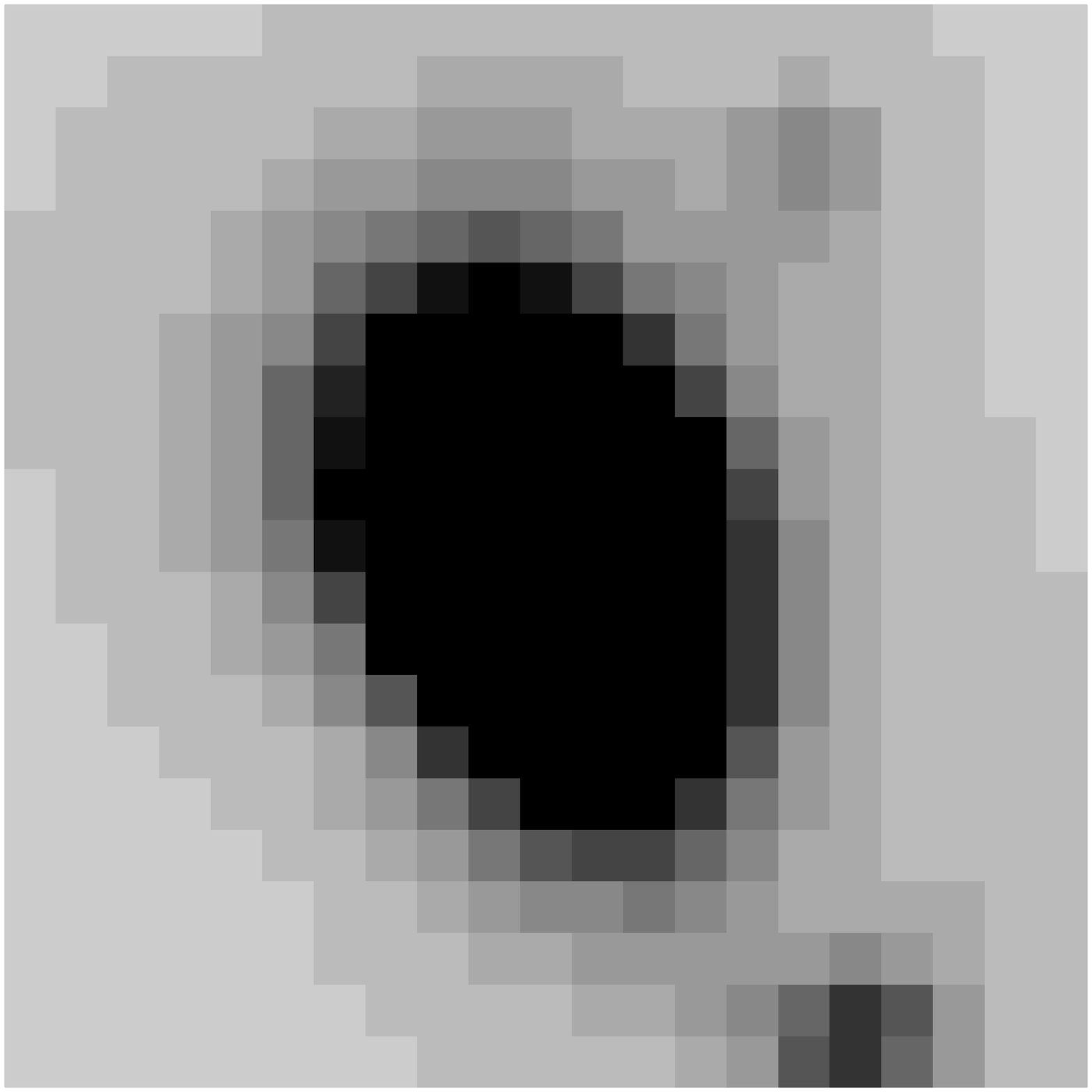}\\
{\bf BLAST 6-2:} $z_{\rm spec} = 0.076$\\ 
\\
\\
\includegraphics[width=0.24\textwidth]{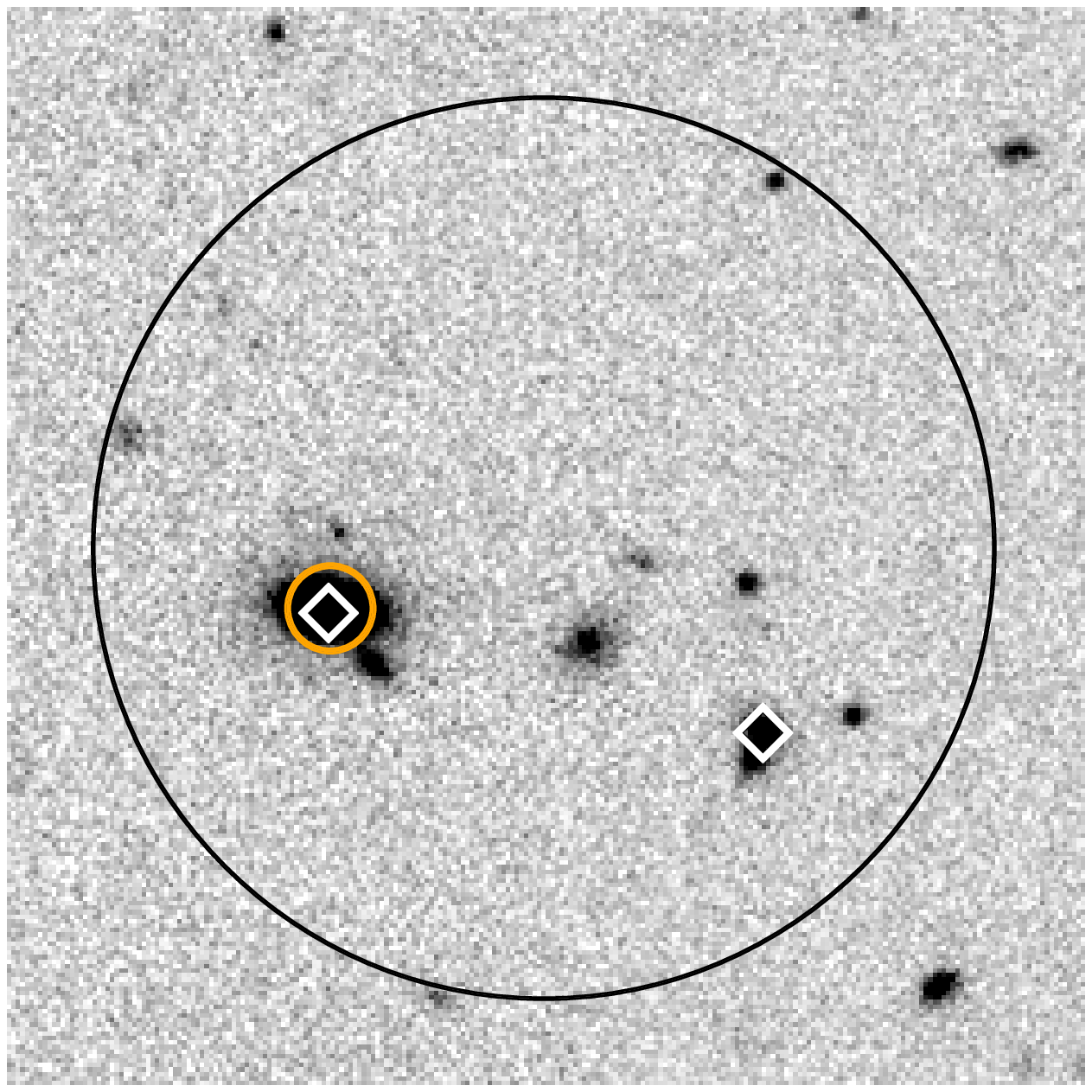}&
\includegraphics[width=0.24\textwidth]{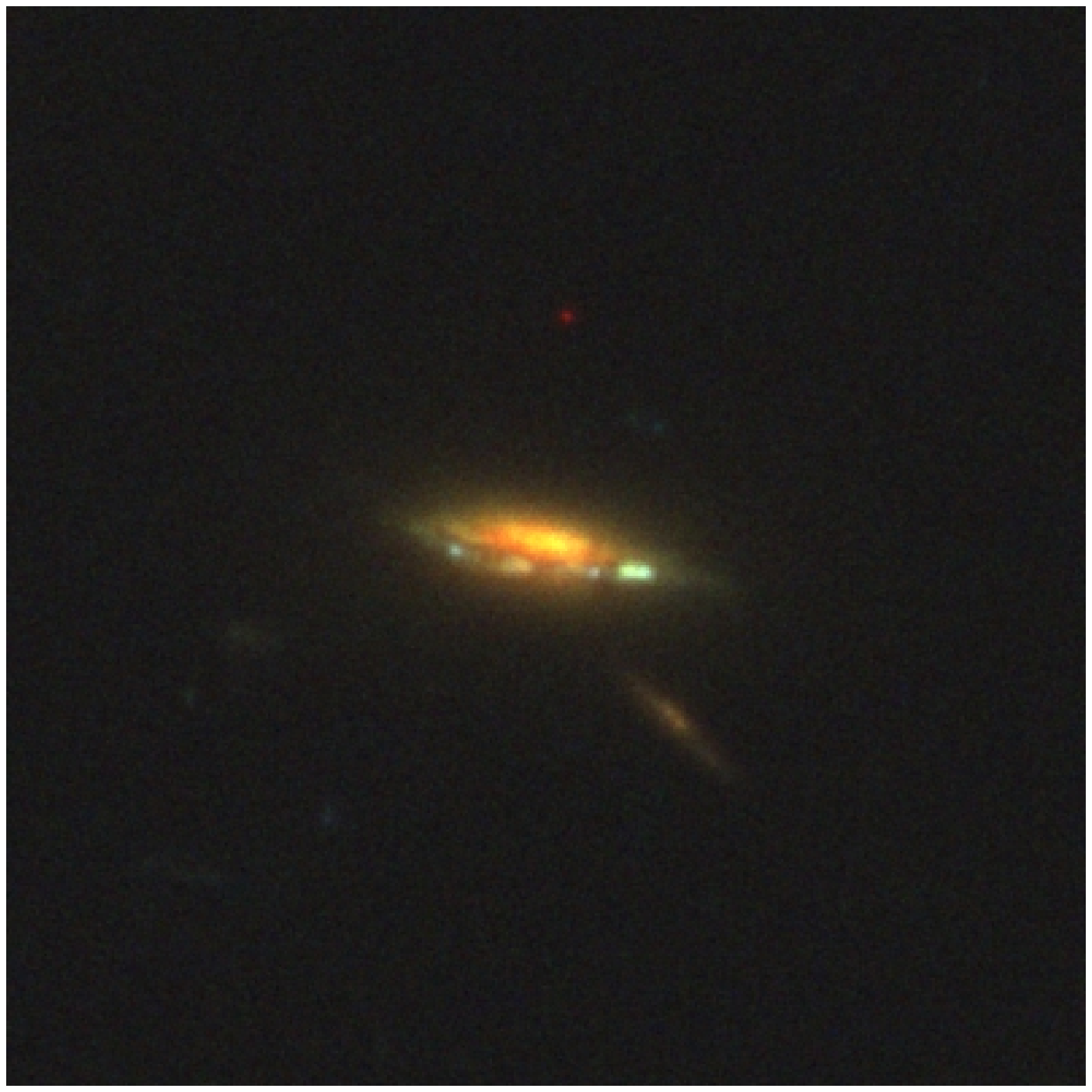}&
\includegraphics[width=0.24\textwidth]{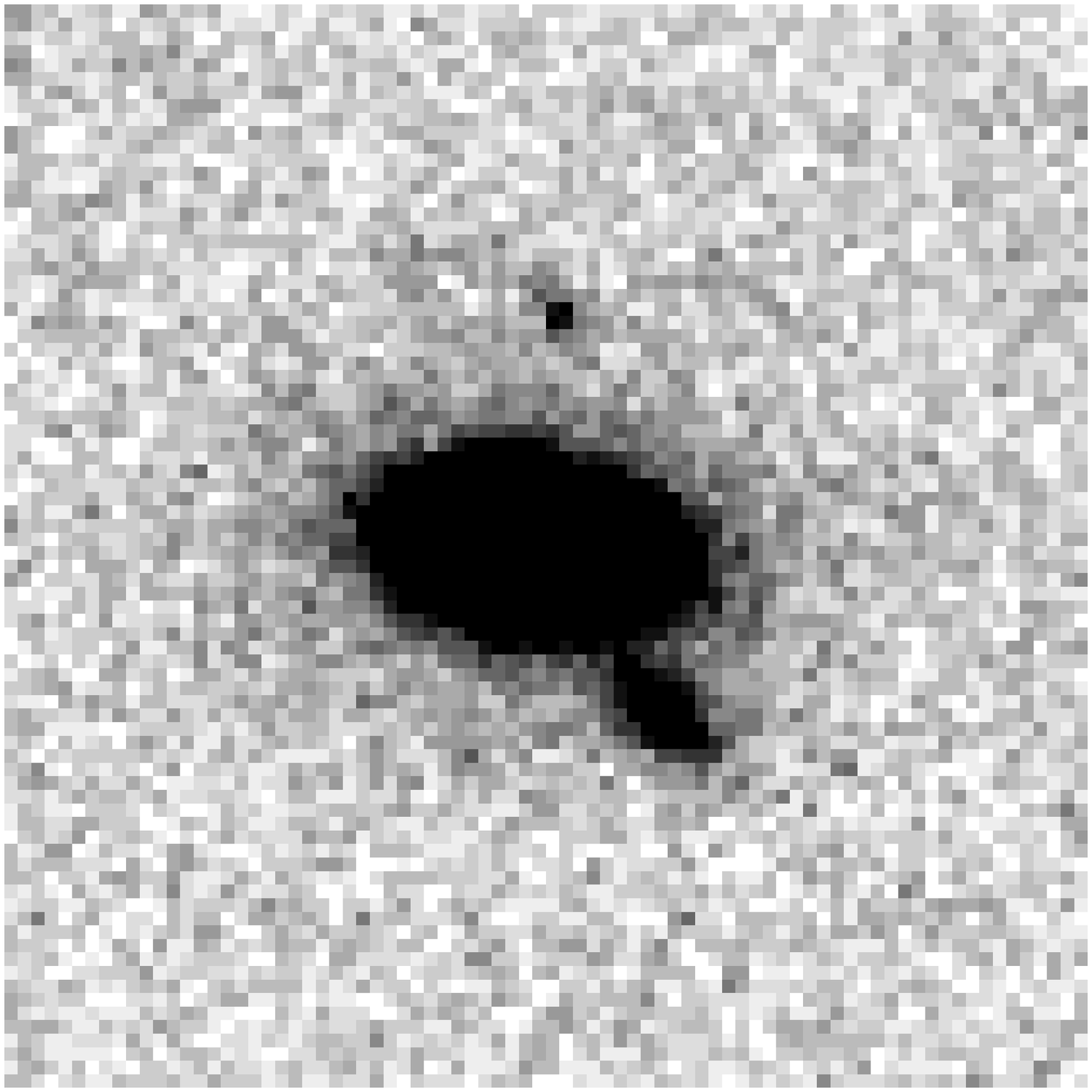}&
\includegraphics[width=0.24\textwidth]{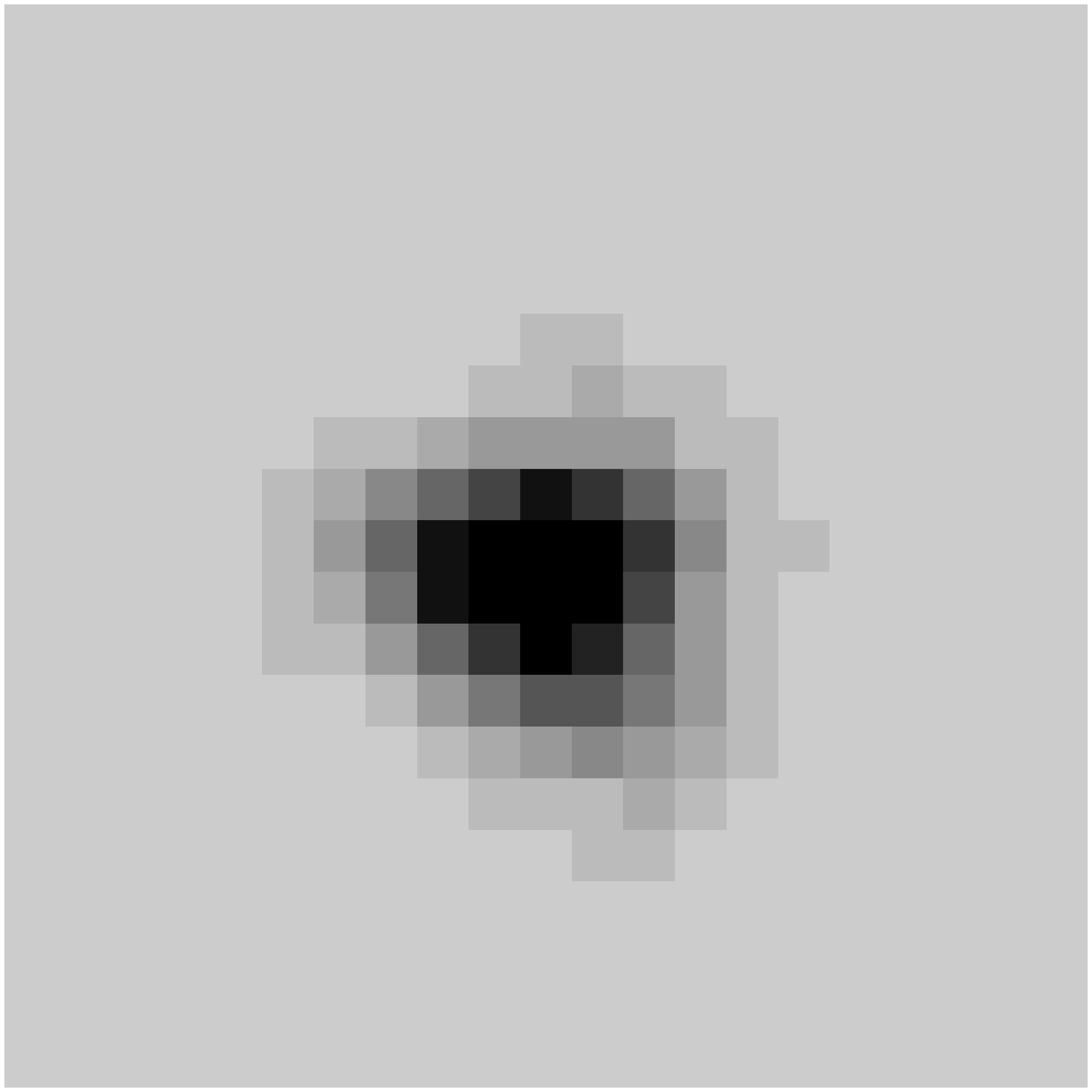}\\
{\bf BLAST 503-2:} $z_{\rm spec} = 0.241$\\ 
\\
\\
\includegraphics[width=0.24\textwidth]{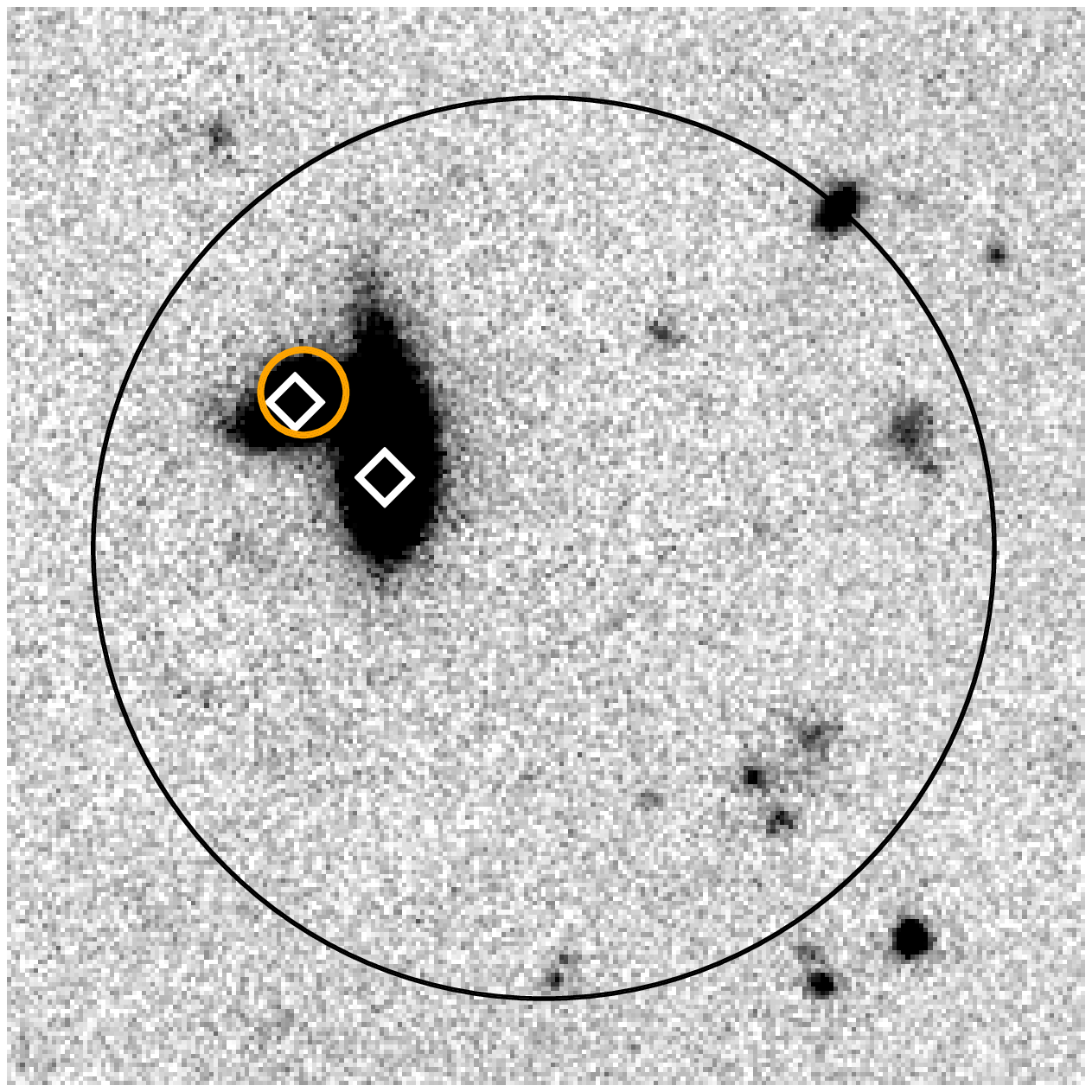}&
\includegraphics[width=0.24\textwidth]{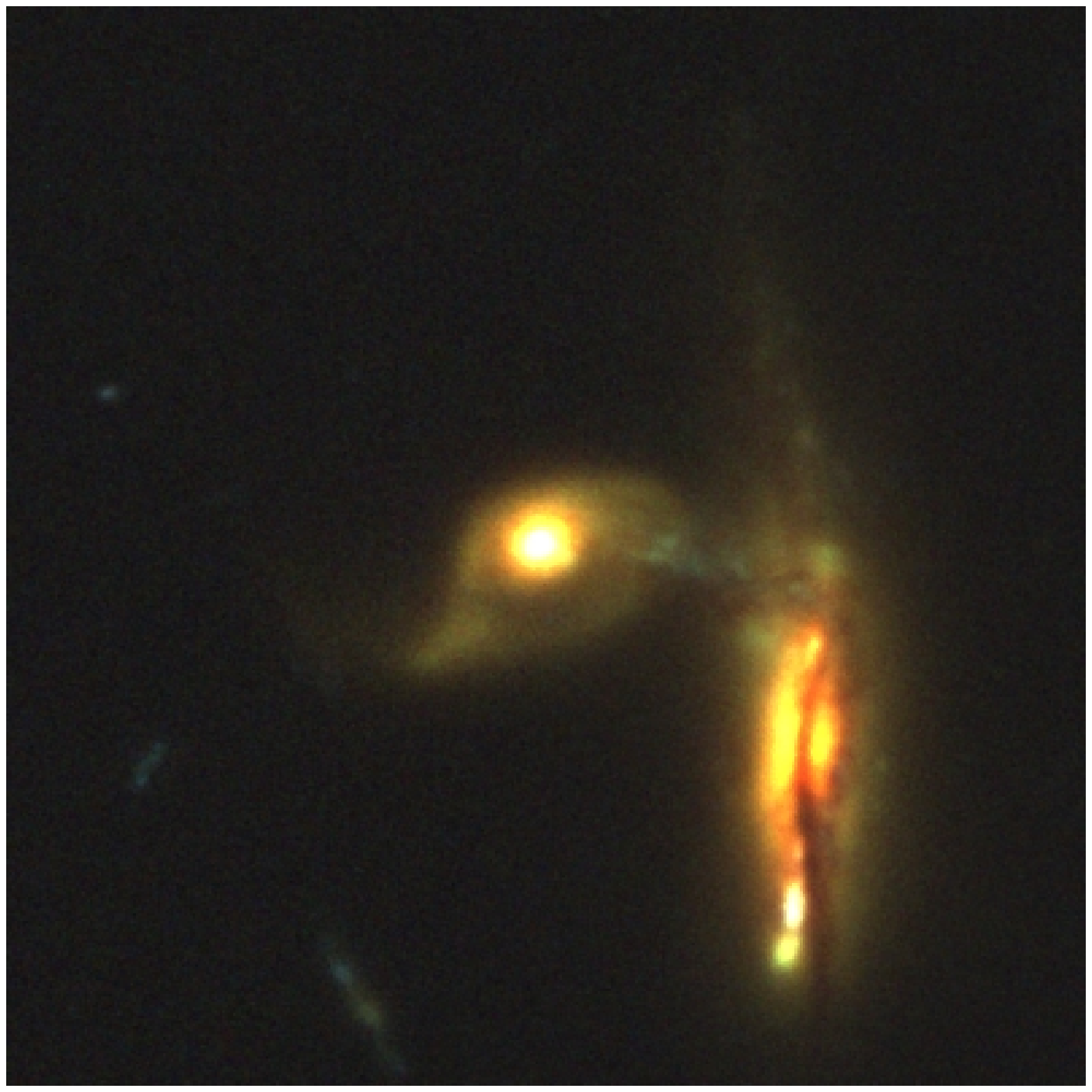}&
\includegraphics[width=0.24\textwidth]{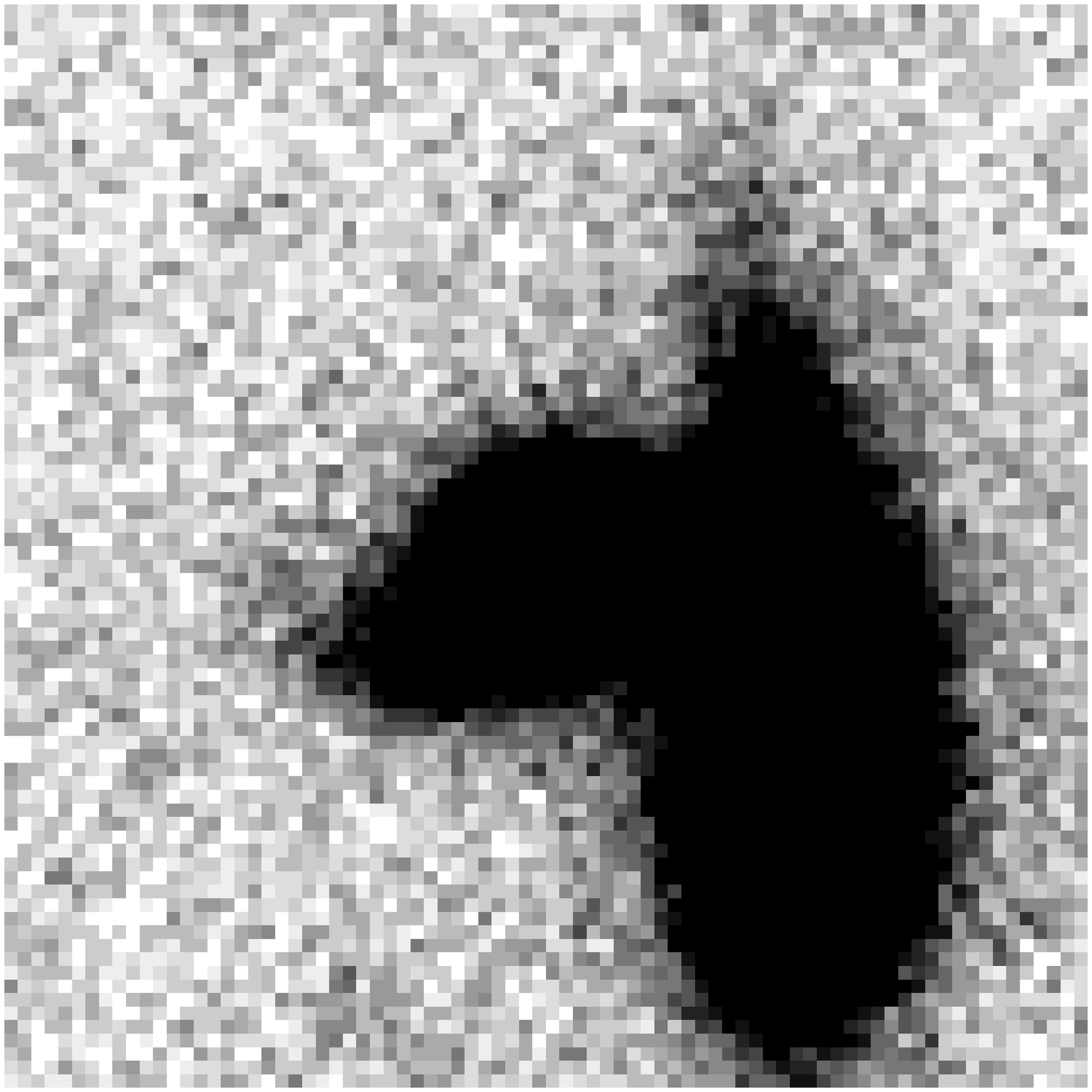}&
\includegraphics[width=0.24\textwidth]{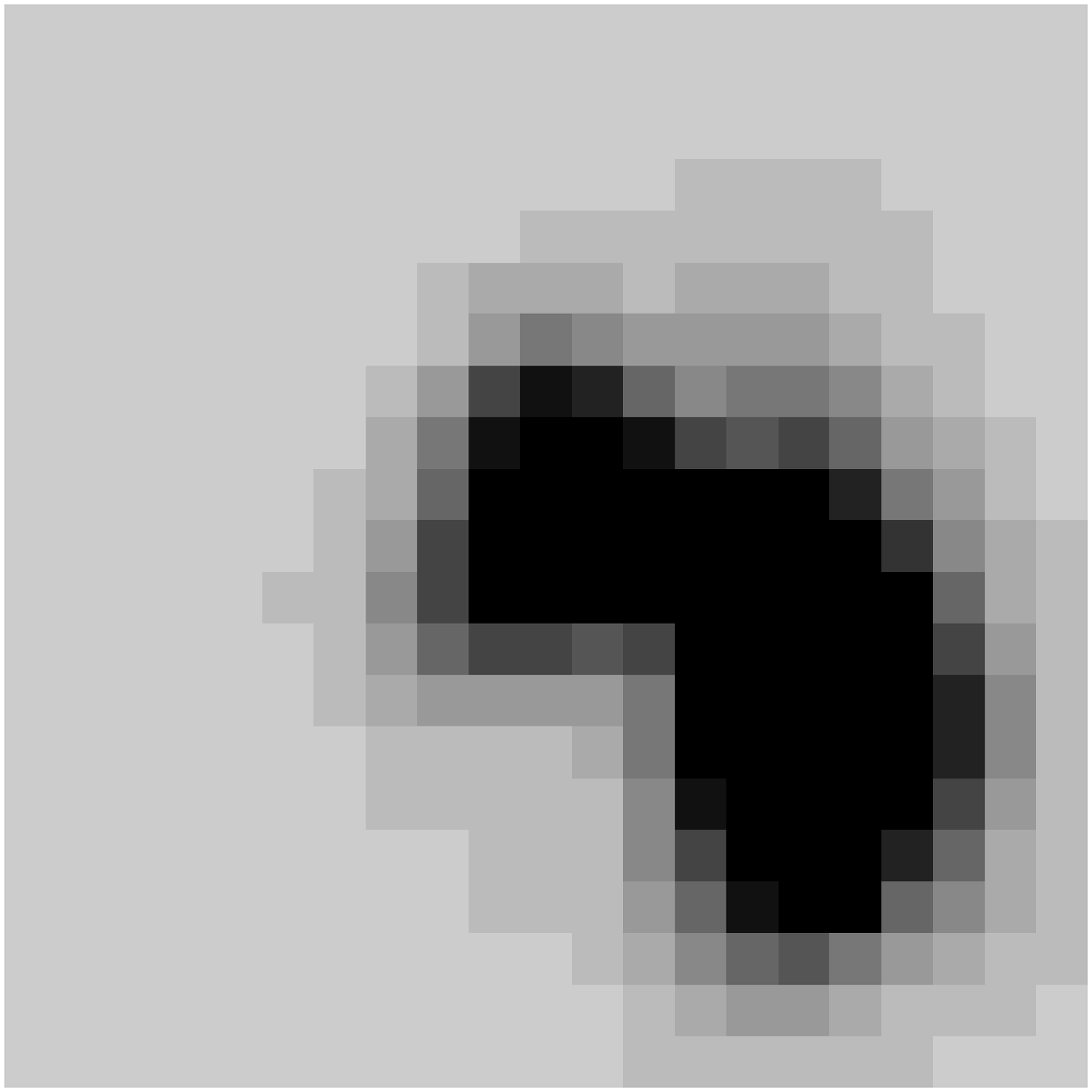}\\
{\bf BLAST 637-2:} $z_{\rm spec} = 0.279$\\
\\
\\
\includegraphics[width=0.24\textwidth]{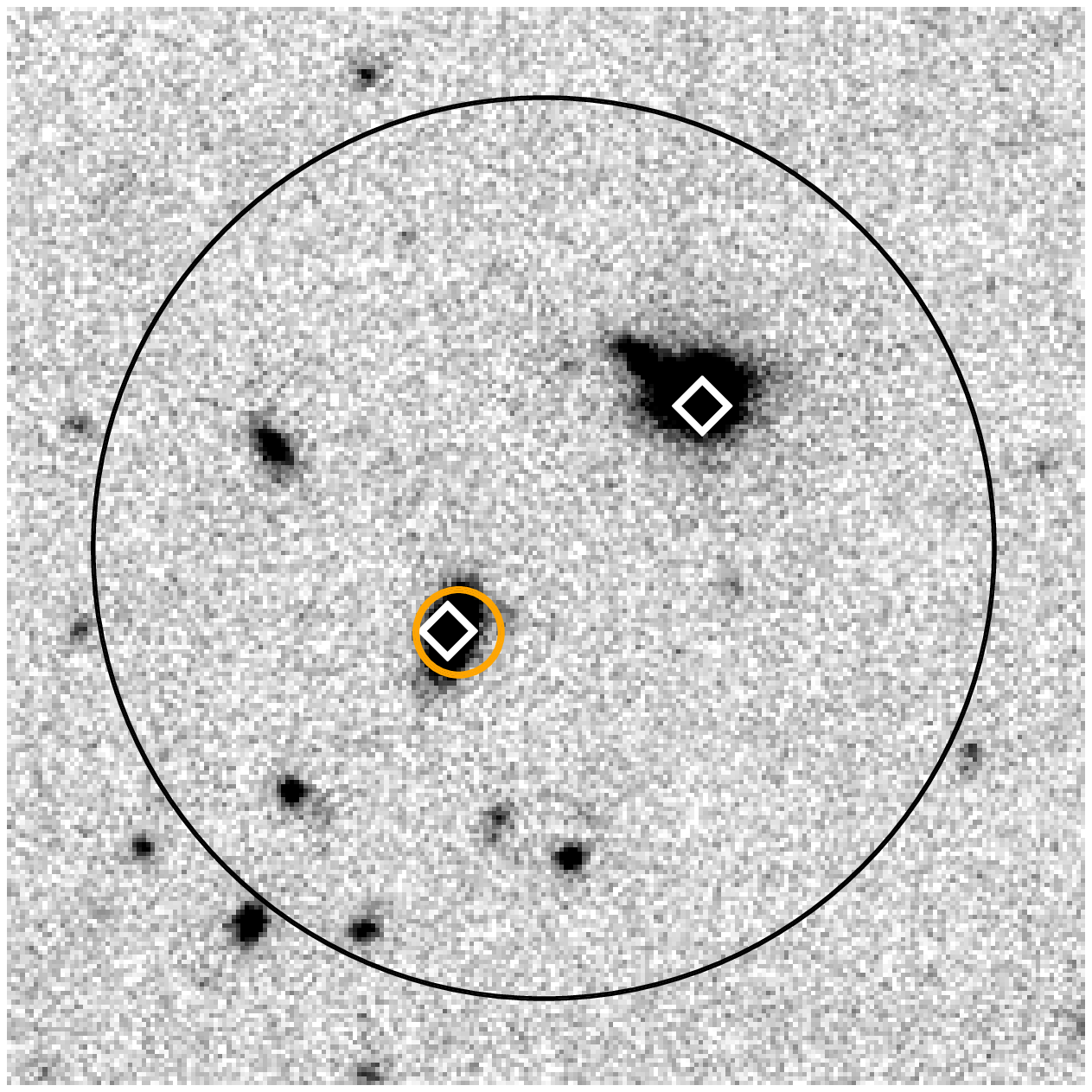}&
\includegraphics[width=0.24\textwidth]{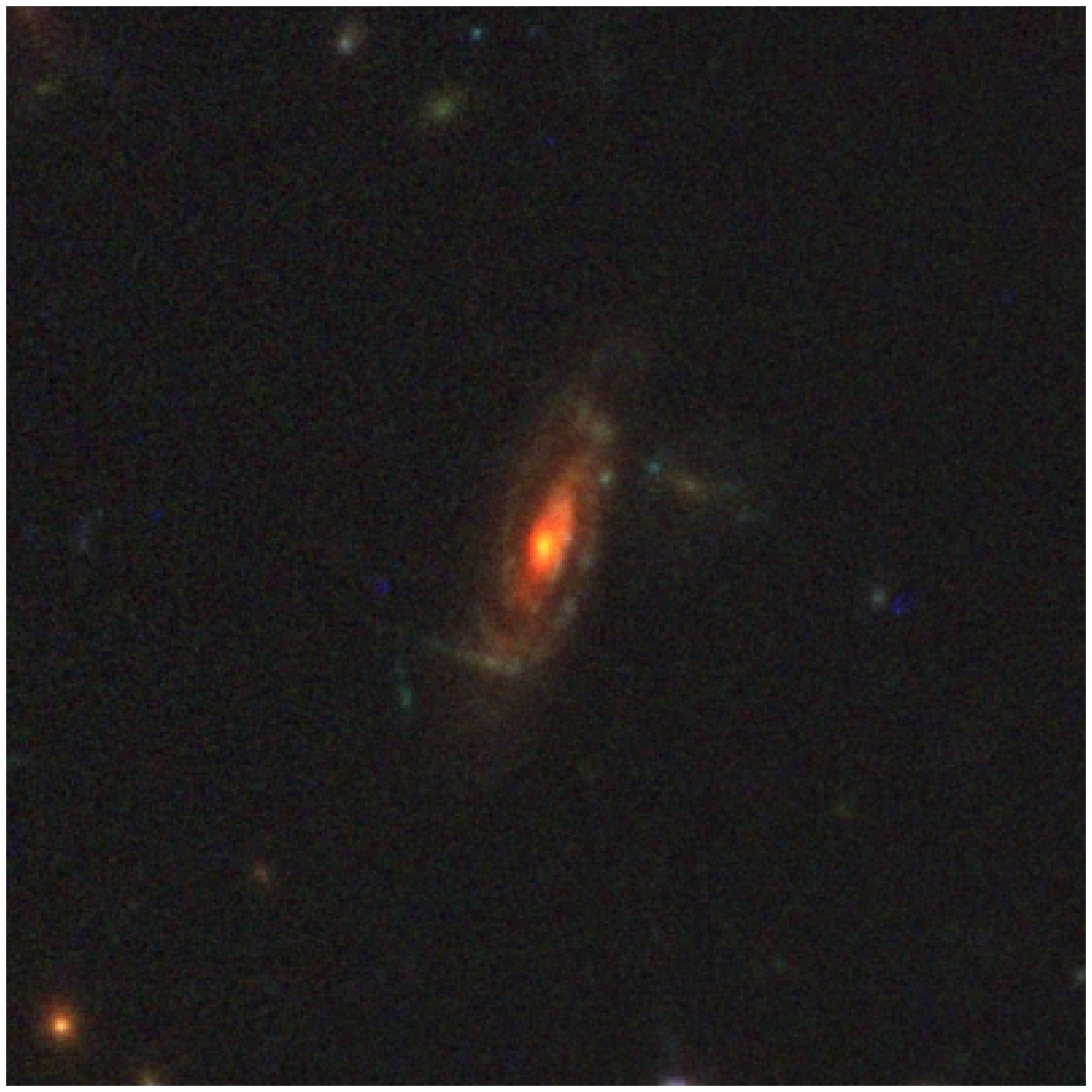}&
\includegraphics[width=0.24\textwidth]{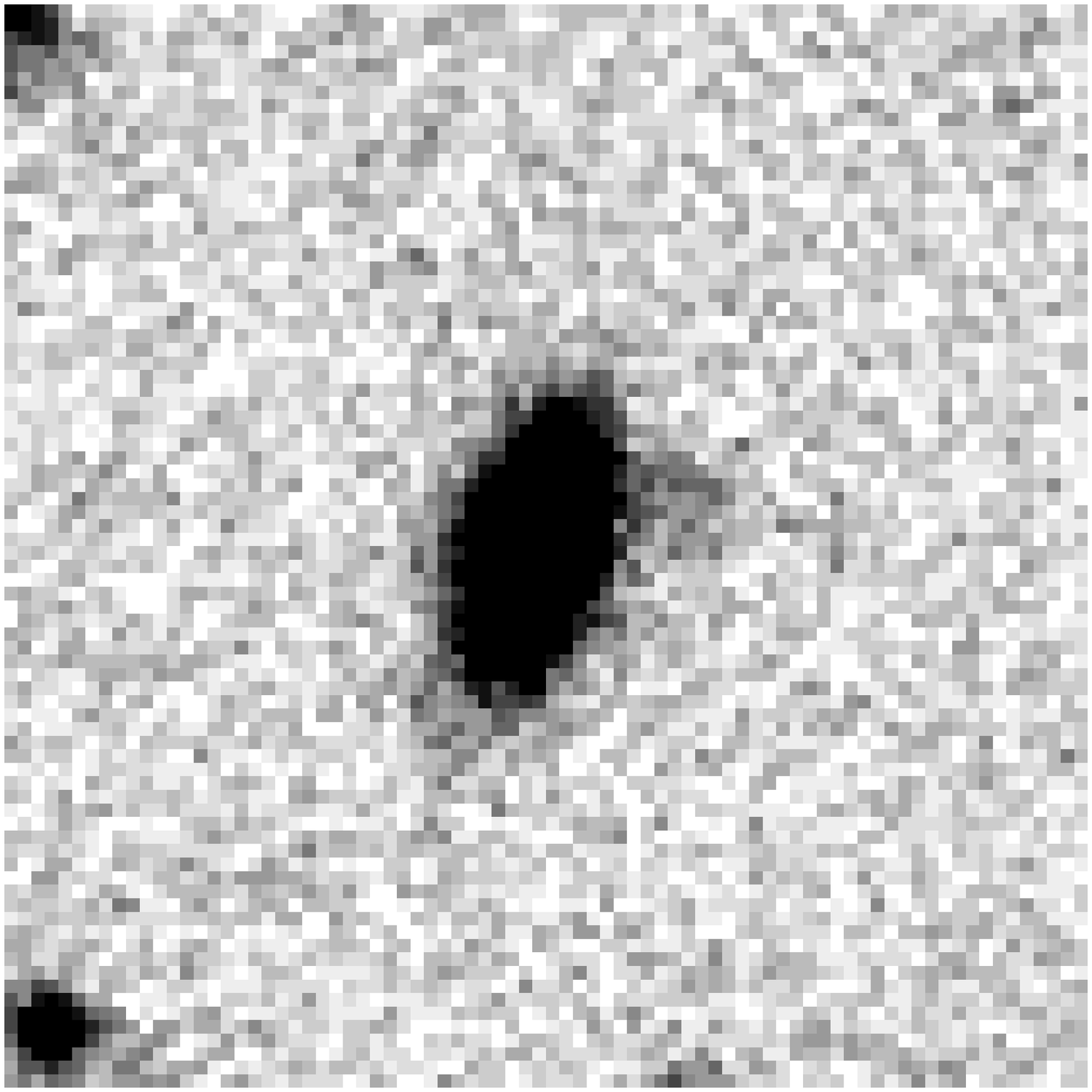}&
\includegraphics[width=0.24\textwidth]{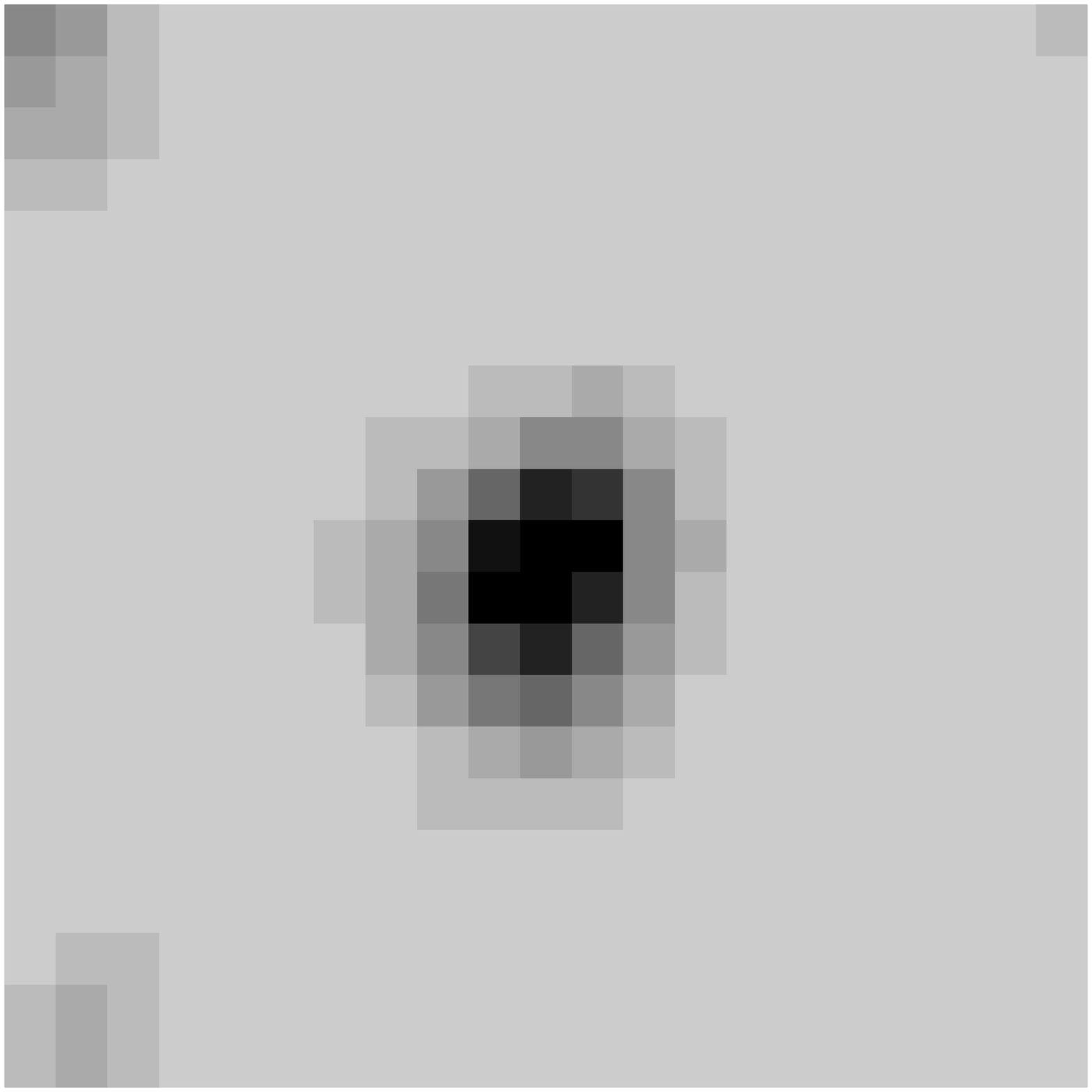}\\
{\bf BLAST 830-2:} $z_{\rm spec} = 0.735$\\ 
\\
\\
\end{tabular}
\caption{Optical/infrared postage stamp images of the 
alternative galaxy identifications for 6 of the BLAST sources, including two alternative candidates 
for BLAST 732. Individual panels are as described in the caption to Fig. 7.}
\end{figure*}

\begin{figure*}
\begin{tabular}{llll}
\\
\includegraphics[width=0.24\textwidth]{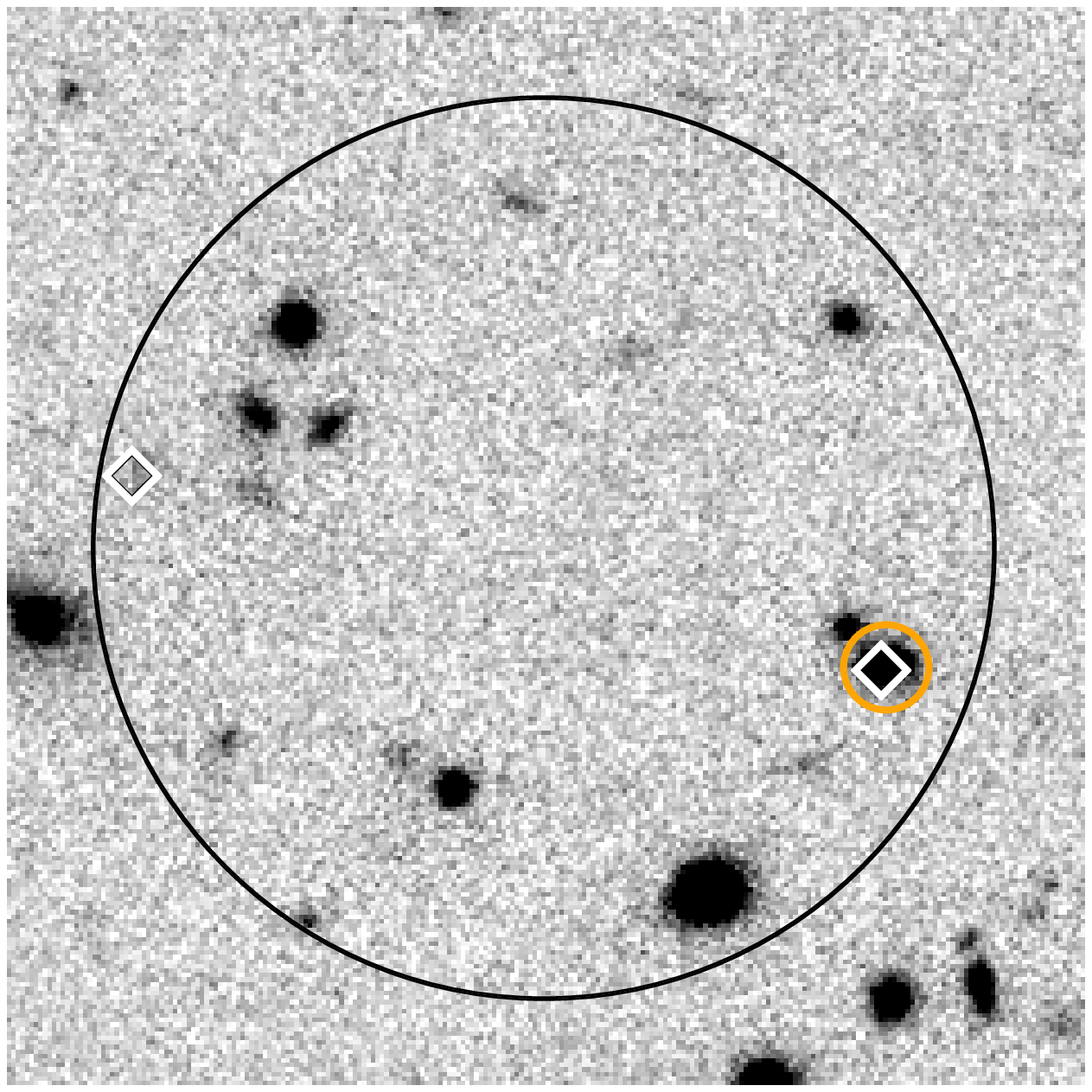}&
\includegraphics[width=0.24\textwidth]{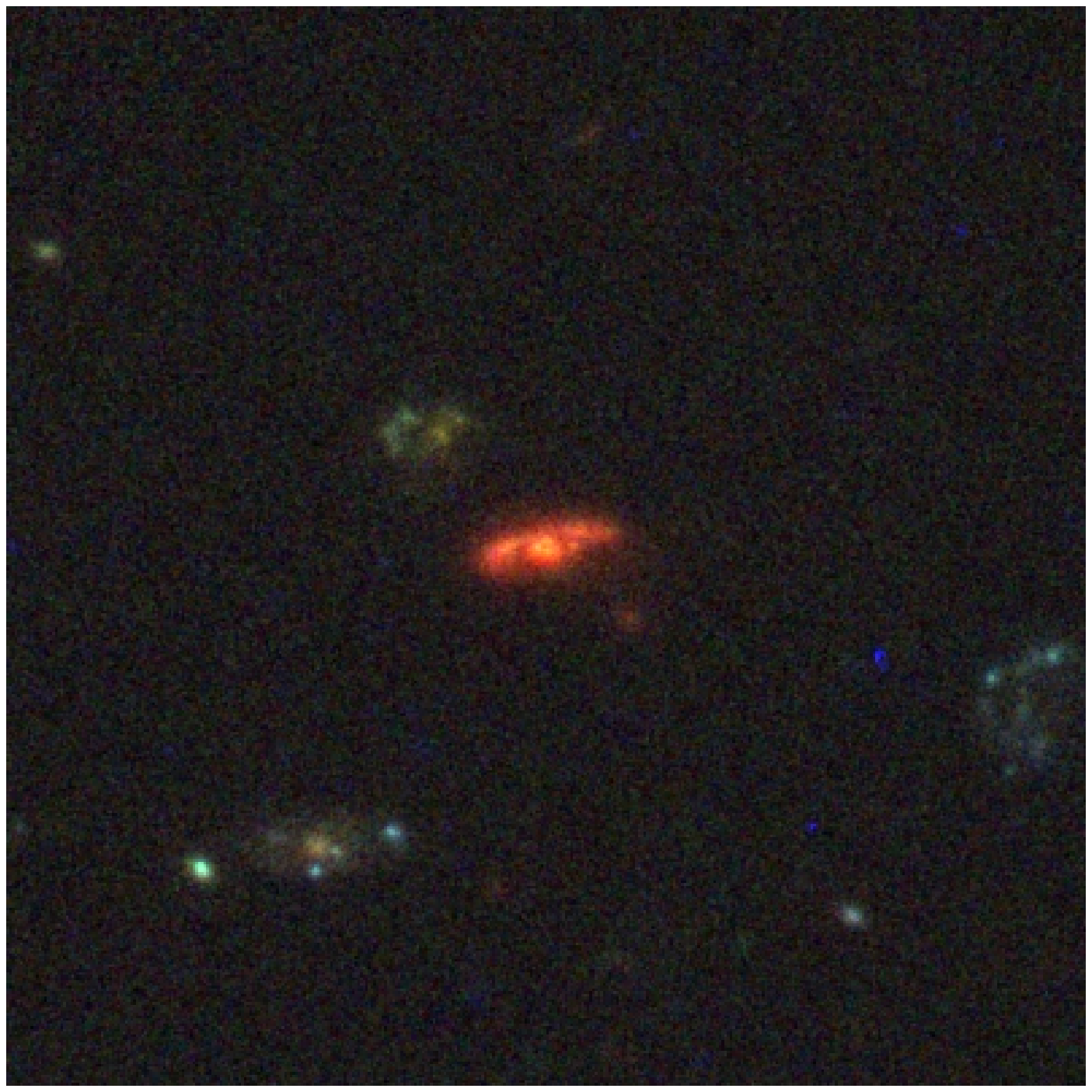}&
\includegraphics[width=0.24\textwidth]{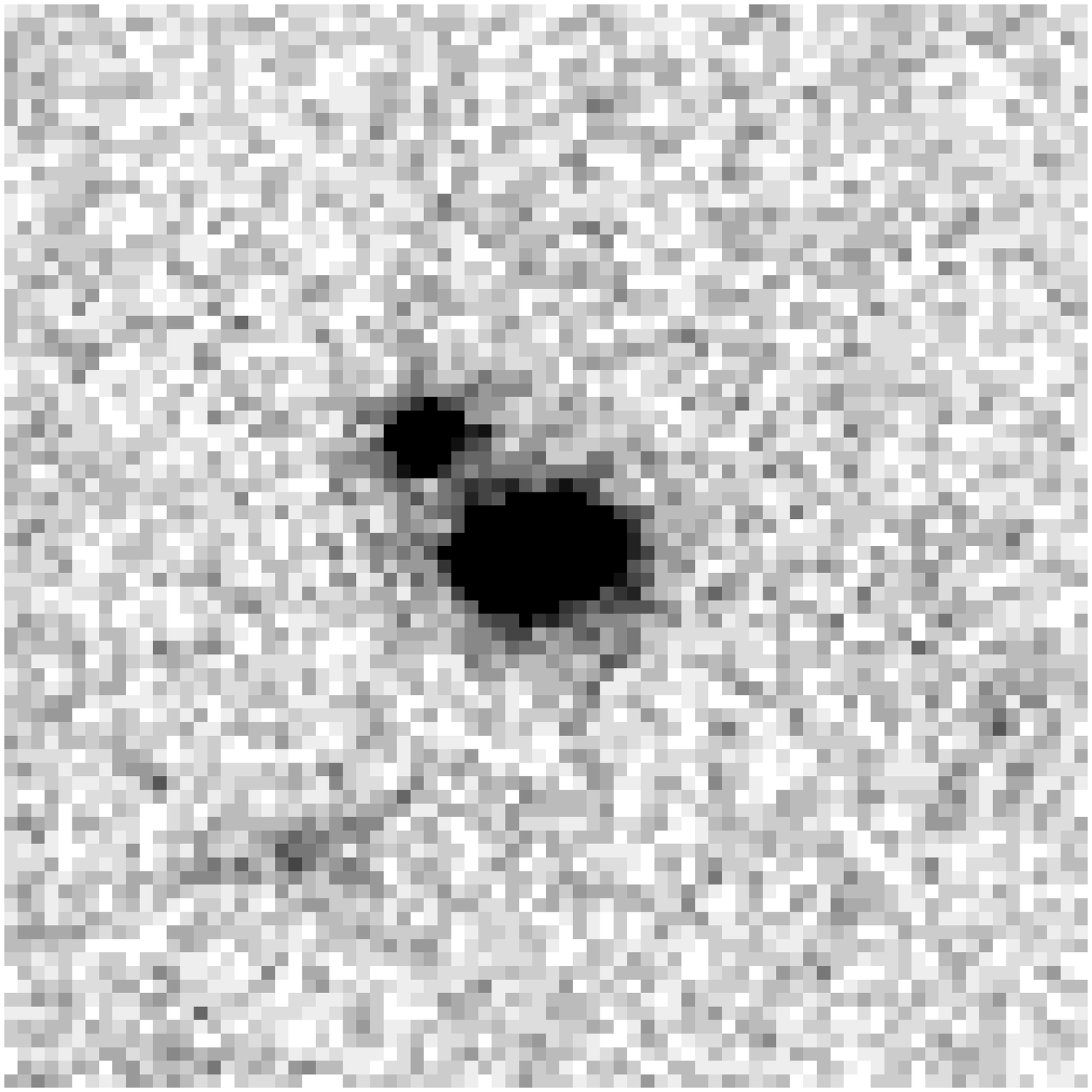}&
\includegraphics[width=0.24\textwidth]{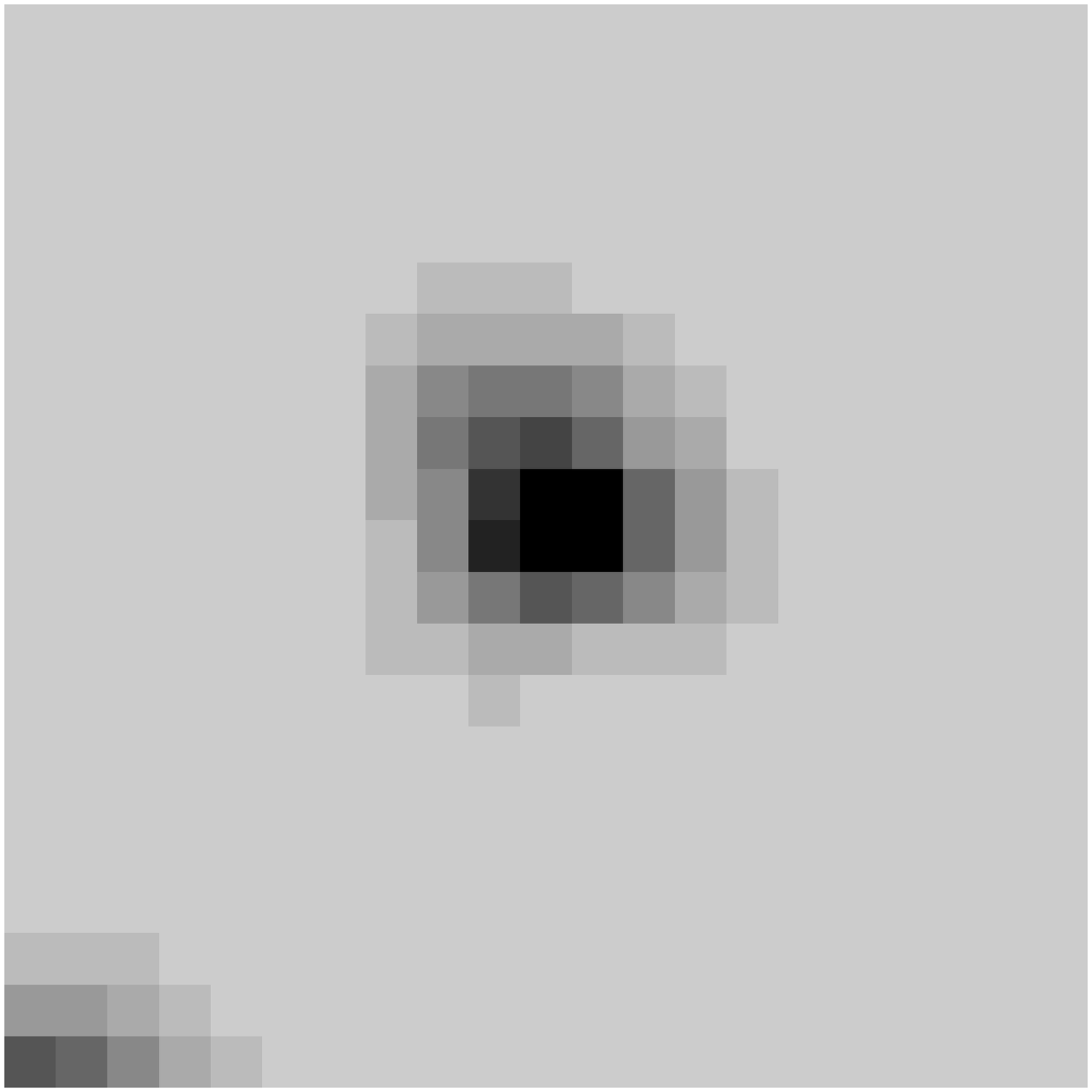}\\
{\bf BLAST 59-2:} $z_{\rm spec} = 1.097$\\
\\
\\
\includegraphics[width=0.24\textwidth]{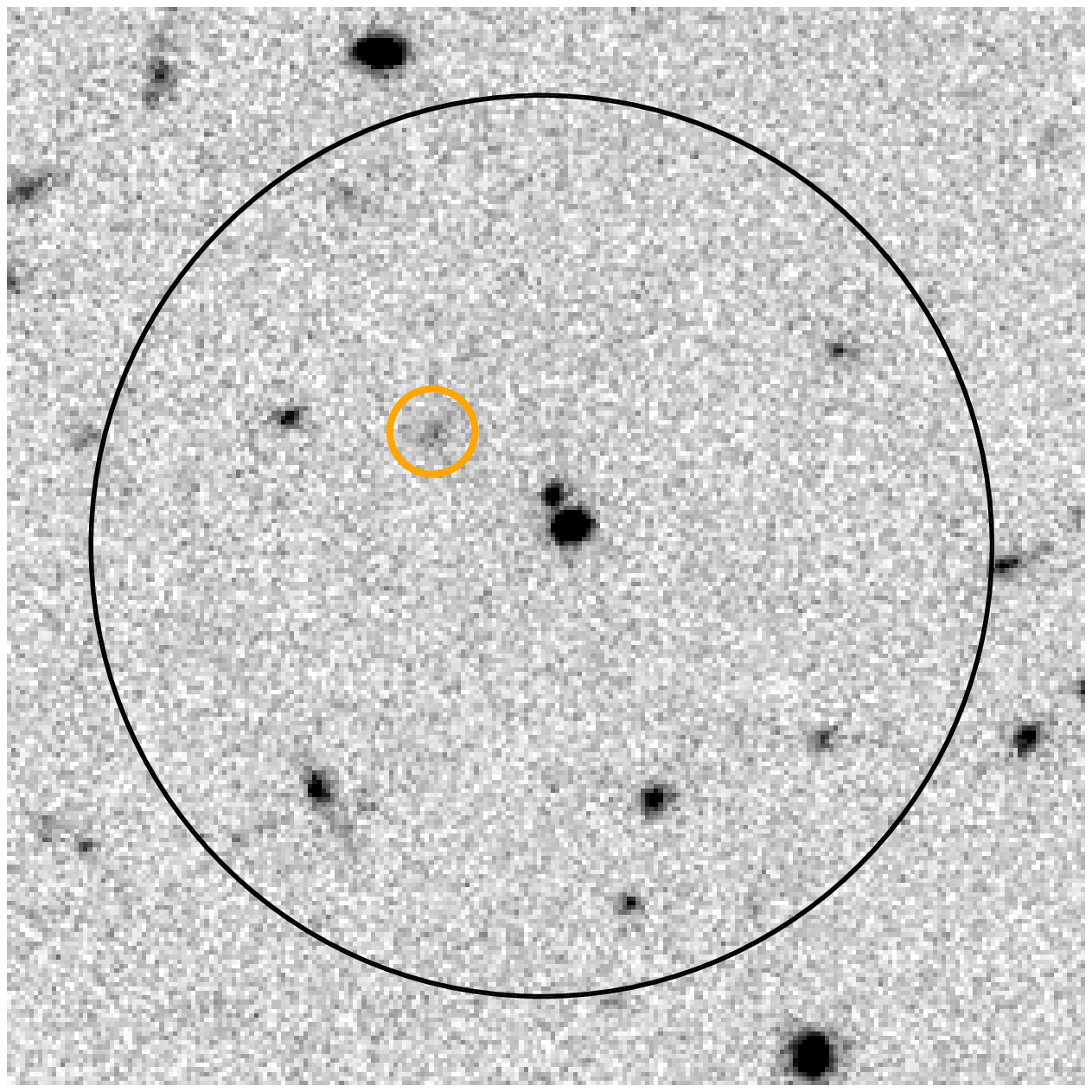}&
\includegraphics[width=0.24\textwidth]{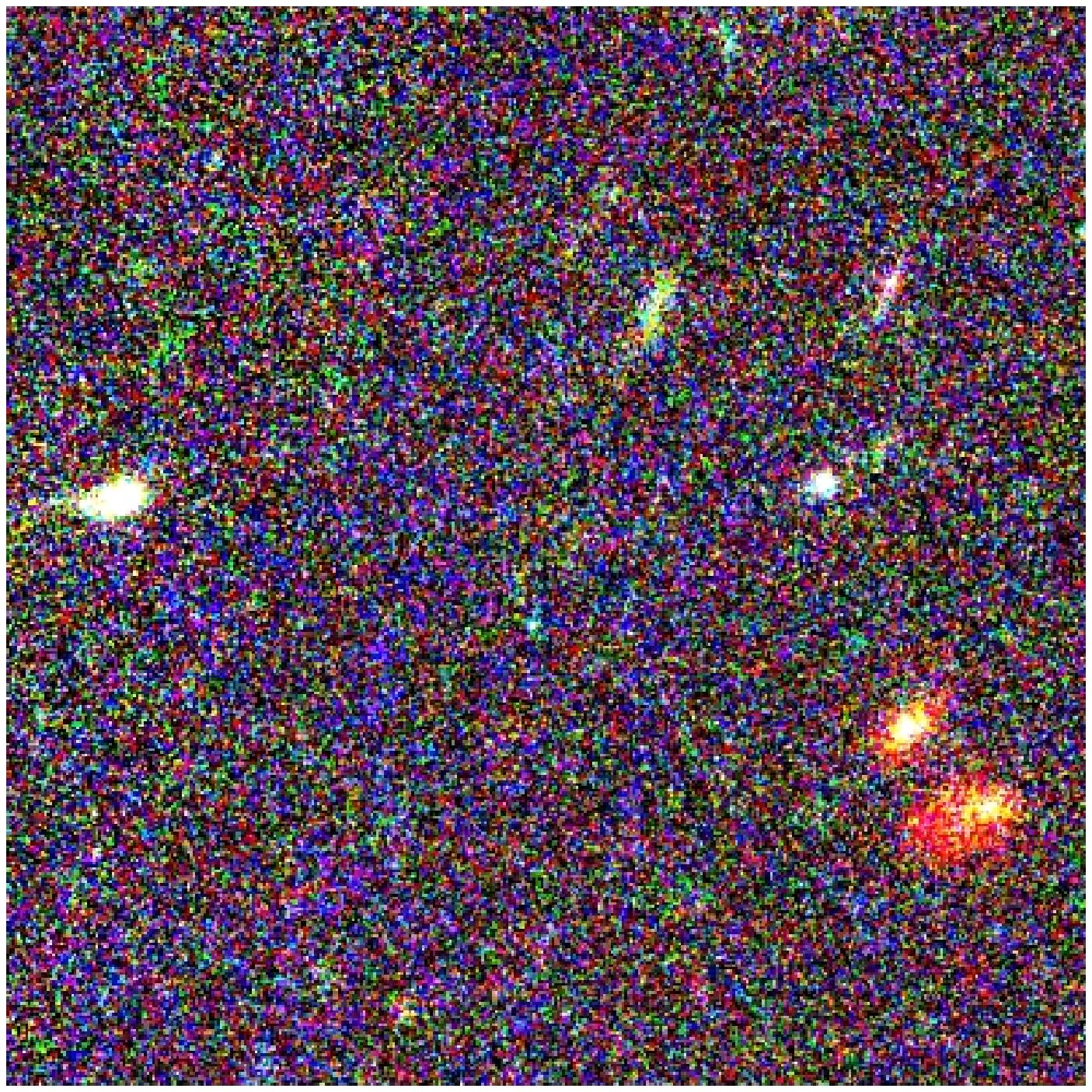}&
\includegraphics[width=0.24\textwidth]{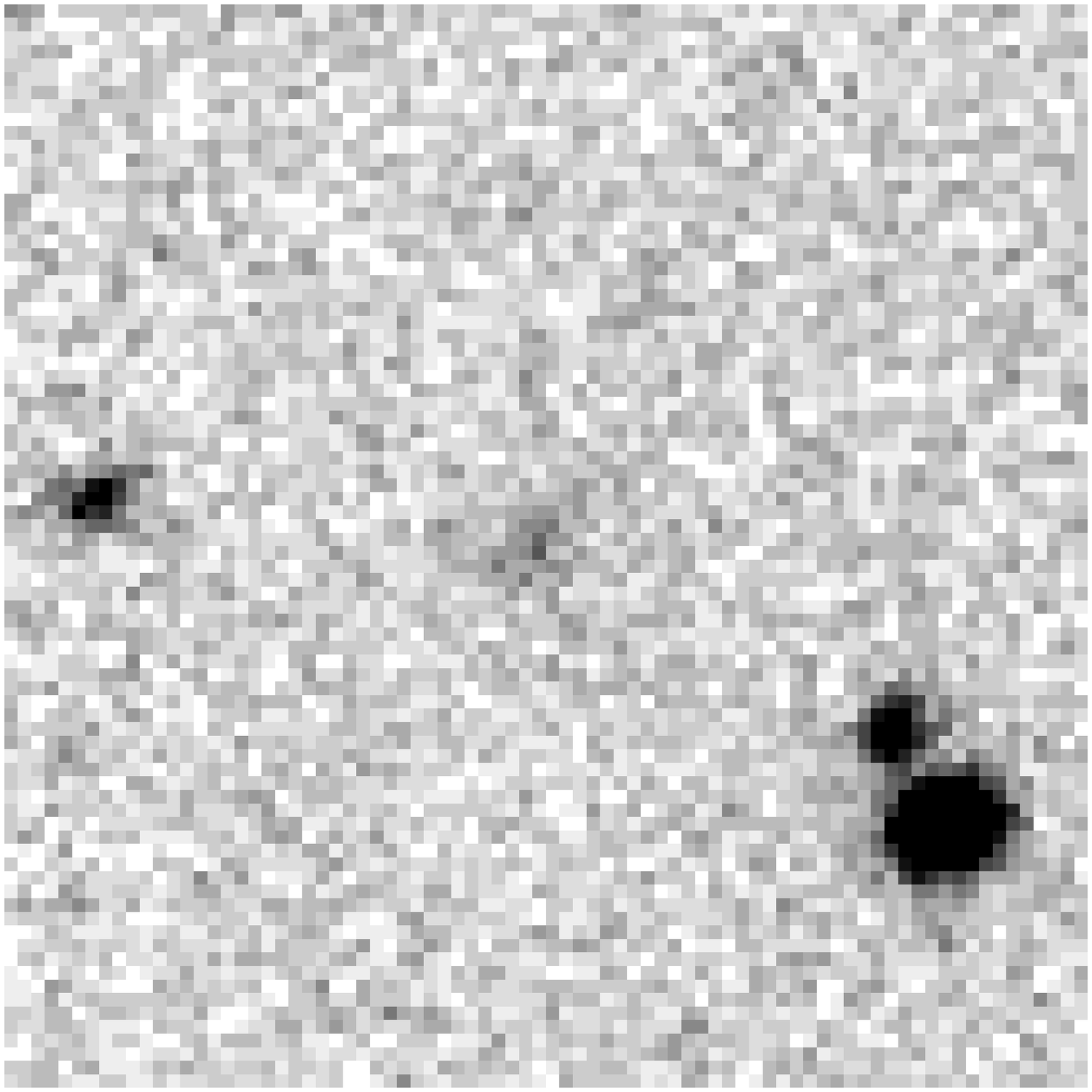}&
\includegraphics[width=0.24\textwidth]{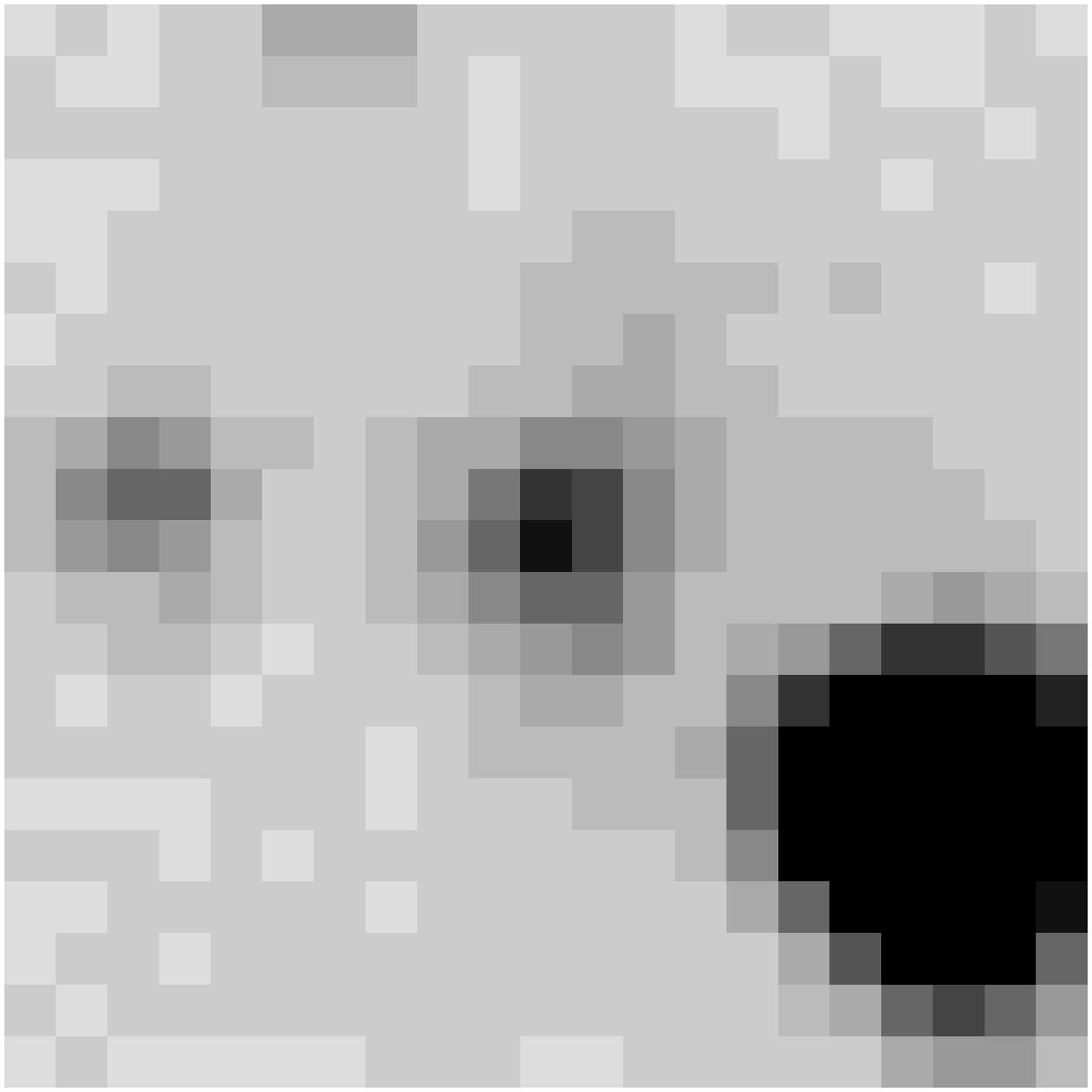}\\
{\bf BLAST 732-2:} $z_{\rm est} = 2.40$\\ 
\\
\\
\includegraphics[width=0.24\textwidth]{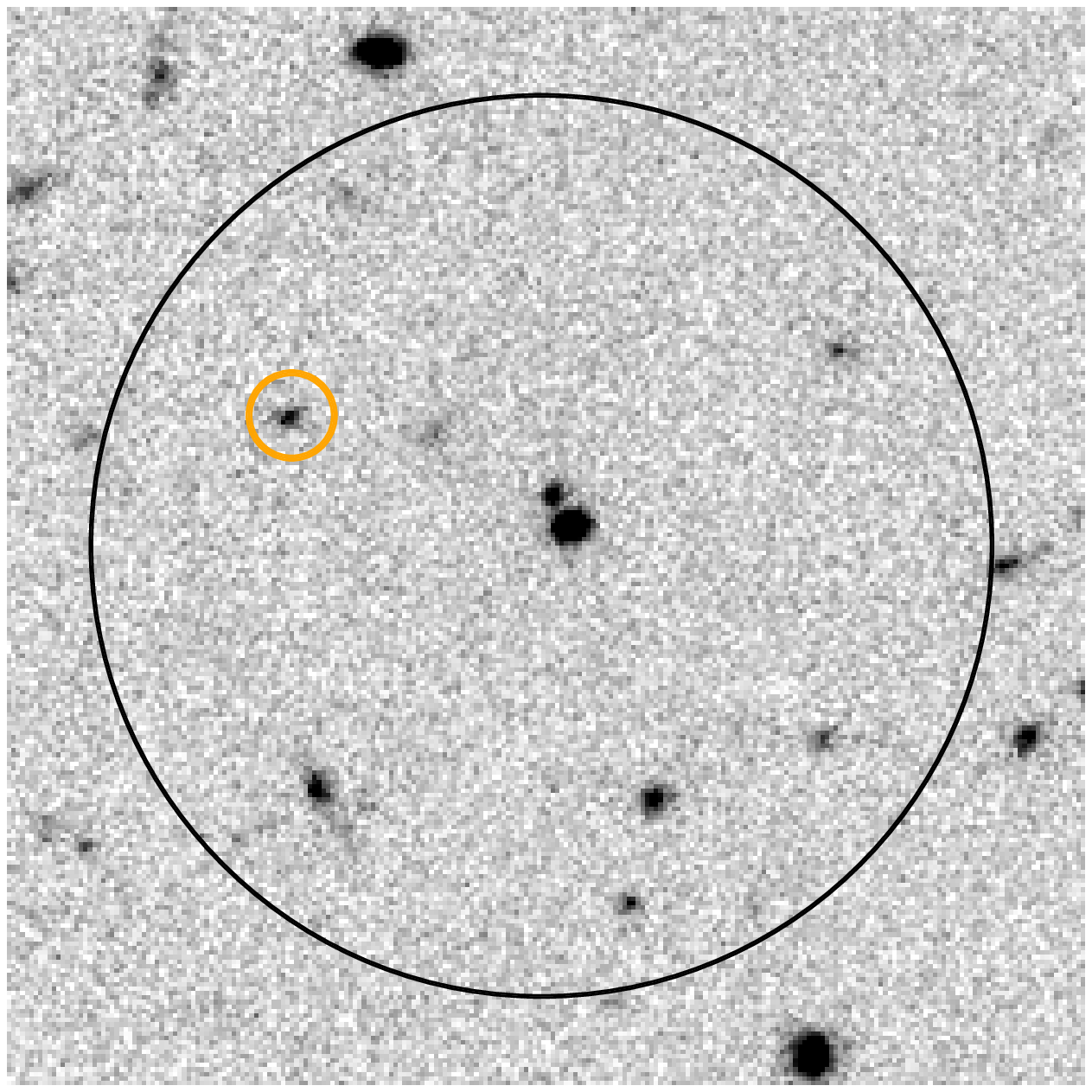}&
\includegraphics[width=0.24\textwidth]{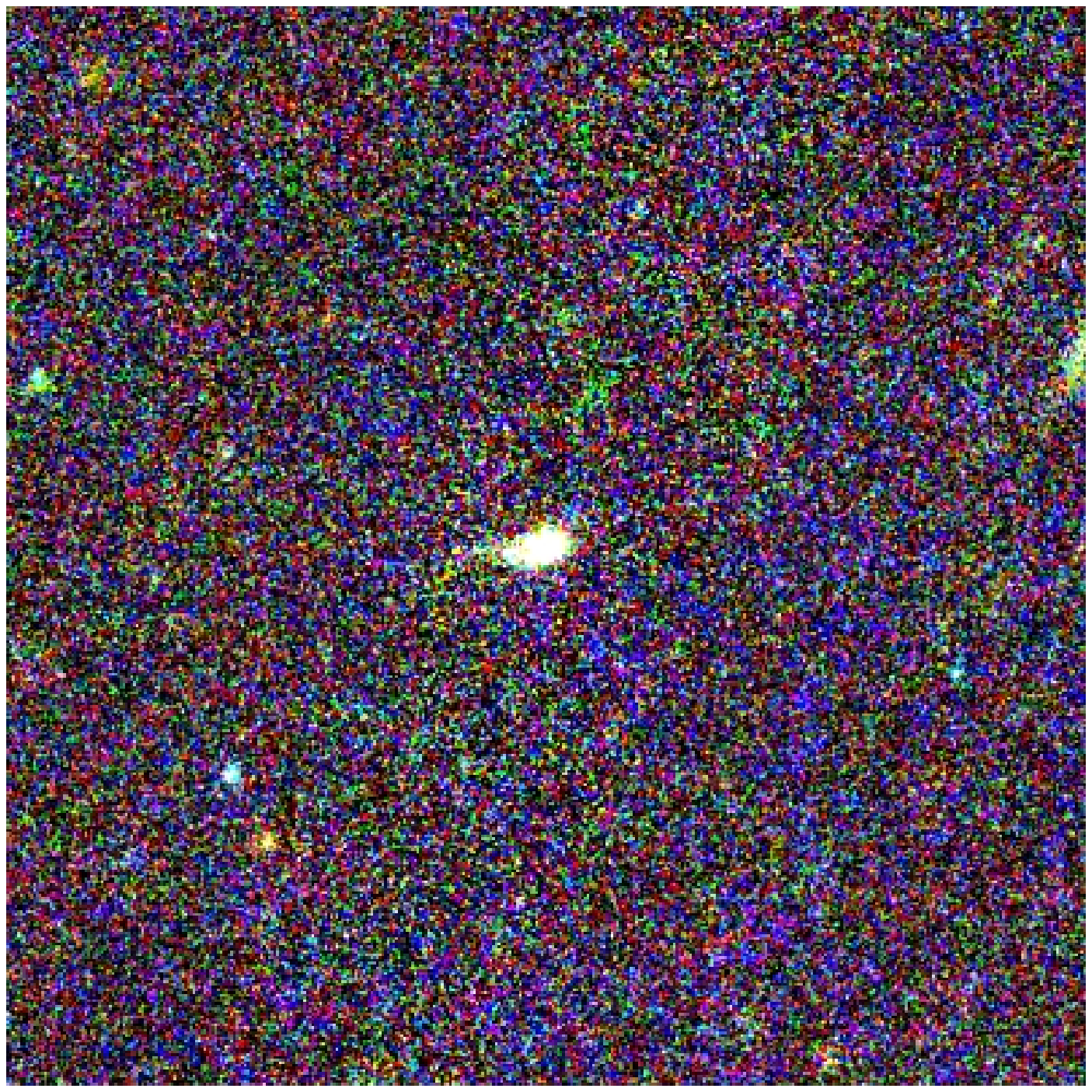}&
\includegraphics[width=0.24\textwidth]{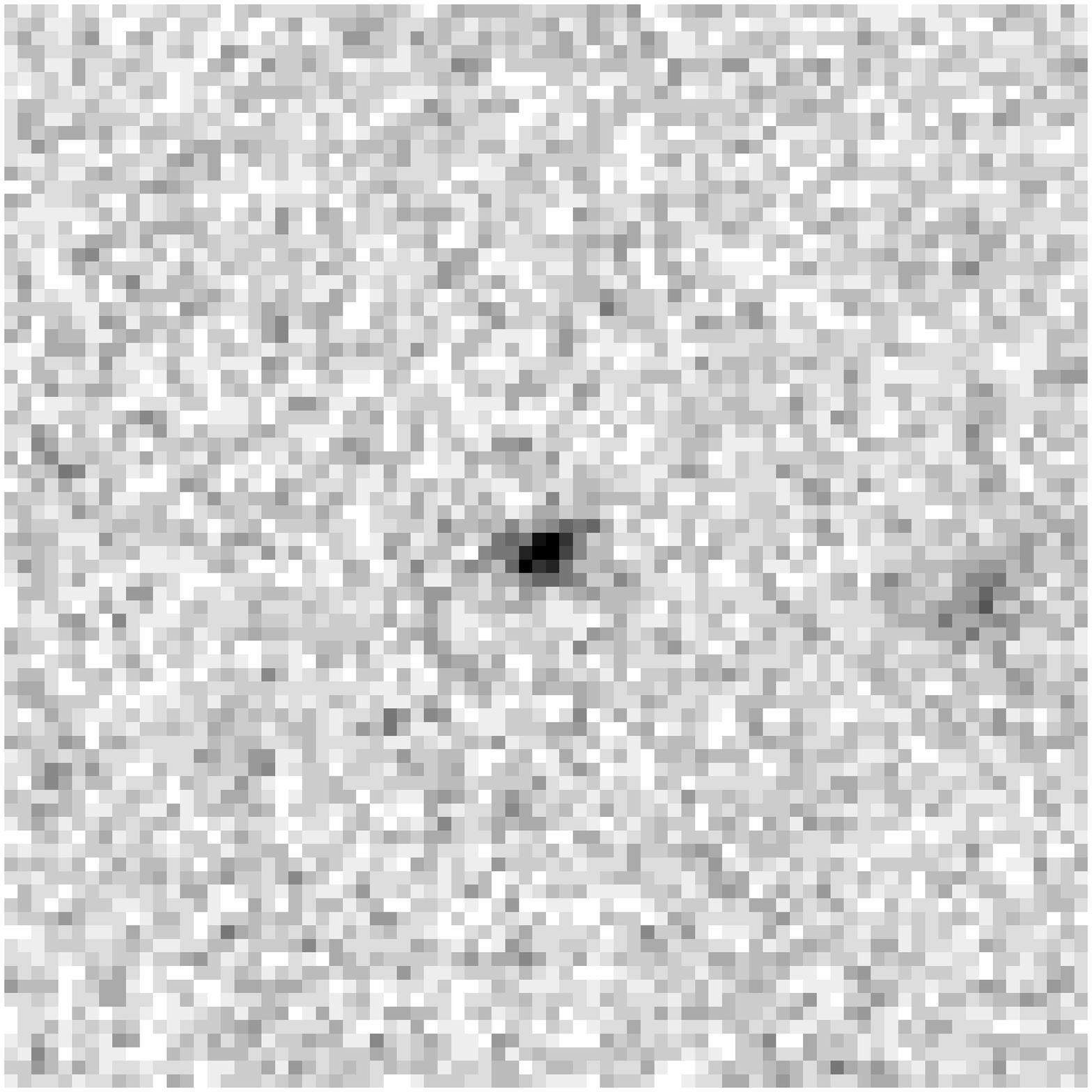}&
\includegraphics[width=0.24\textwidth]{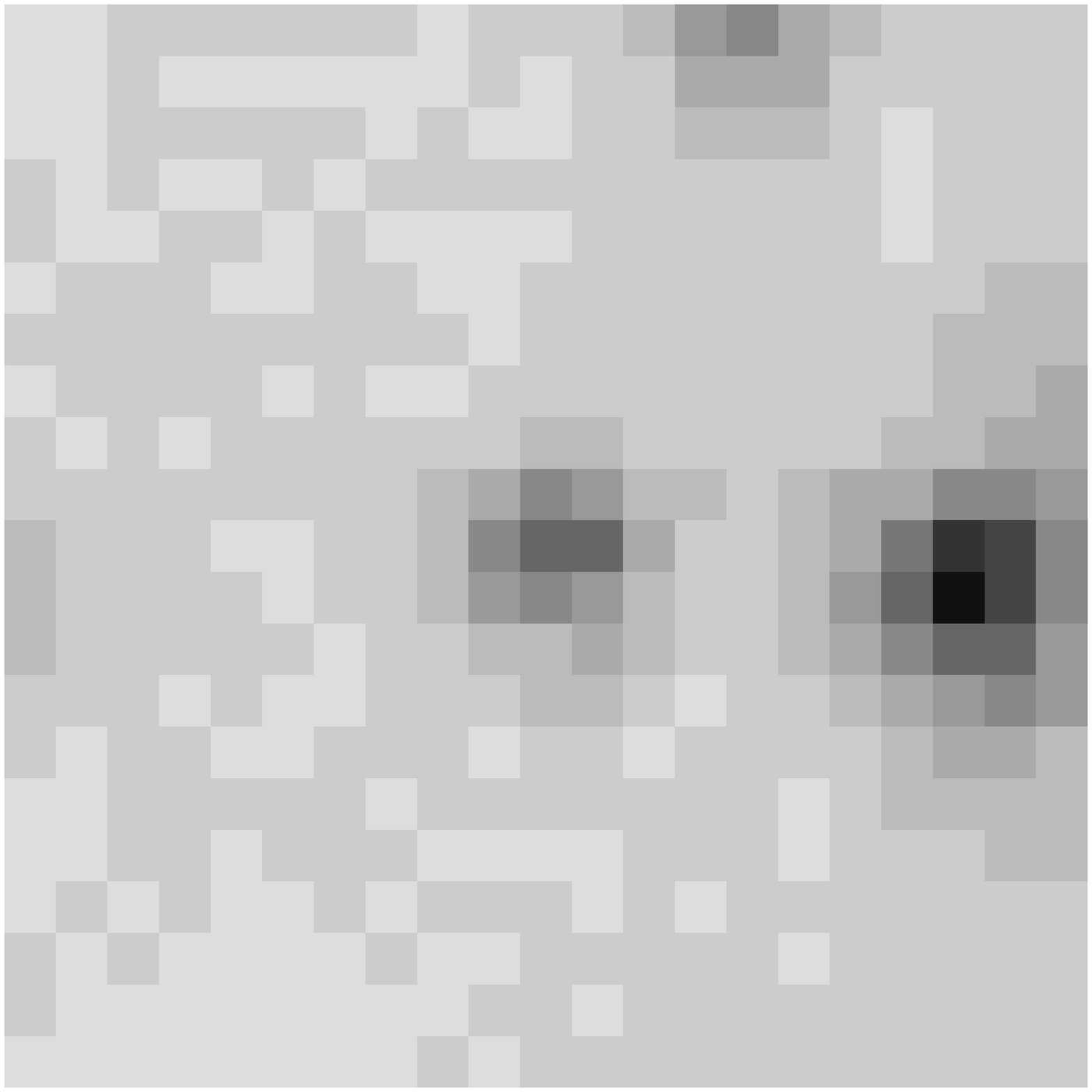}\\
{\bf BLAST 732-3:} $z_{\rm est} = 2.6$\\ 
\\
\\
\end{tabular}
\addtocounter{figure}{-1}
\caption{continued}
\end{figure*}

\clearpage

%%%%%%%%%%%%%%%%%%%%%%%%%%%%%%%%%%%%%%%%%%%%%%%%%%%%%%%%%%%%%%%%%%%%%%%%%%%%
\section{DISCUSSION}
\label{DISCUSSION}
%%%%%%%%%%%%%%%%%%%%%%%%%%%%%%%%%%%%%%%%%%%%%%%%%%%%%%%%%%%%%%%%%%%%%%%%%%%%

\begin{table}
\begin{tabular}{rrccrl}
\hline
BLAST & LABOCA & $d_{B-L}$ & $d_{L-Rad}$ & \phantom{00}$P_{LB}$ & Notes\\
ID\phantom{ST}& ID\phantom{ST}  & /arcsec  & /arcsec & & \\
\hline
59 &  10+34  &  8.9  & 7.7  &   0.004  &Blend\\
66 &  18     &  7.5  & 5.5  &   0.008   \\
158&  79     &  7.4  & 5.0  &   0.011   \\
318&  67     &  12.4 & 3.7  &   0.019   \\
%552&         &  15.4 & 9.4  &   0.035  &3.5-$\sigma$ source\\
593&  12     &  8.7  & 2.4  &   0.008\\
732&  32     & 9.6   & 4.4  &   0.012 & Opt ID\\
\hline
\end{tabular}
\caption{Associations between BLAST 250\,${\rm \mu m}$ and LABOCA LESS 870\,${\rm \mu m}$ sources within the central 150\,arcmin$^2$ 
of GOODS-South. $d_{B-L}$ is the distance, in arcsec, between the BLAST and LABOCA source. For comparison, 
$d_{L-Rad}$ is the distance, in arcsec, between the LABOCA source
and the radio identification listed in Table 2. $P_{LB}$ is the probability that the association between the BLAST and LABOCA source is the result
of chance, calculated in an analogous manner to the $P$ values calculated 
for the BLAST-Radio associations in Table 2, but with a search radius of 18\,arcsec.}
\end{table}

\subsection{LABOCA sub-mm detections and 250/870\,${\rm \mu m}$ flux ratios}

Within the central area of GOODS-South under study here, Weiss et al. (2009) have extracted 10 LABOCA 870\,${\rm \mu m}$ sources with a raw significance 
$> 4\sigma$. Searching around the BLAST positions out to a search radius of 18\,arcsec, we find that seven of these 870\,${\rm \mu m}$ sources 
coincide with 250\,${\rm \mu m}$ sources to within a positional accuracy of $< 13$\,arcsec (including two associated with BLAST 59).
These associations are tabulated in Table 4. We adopted a search radius of 18\,arcsec by adding the positional uncertainty adopted for the 
BLAST sources in Section 3.1.1 ($\sigma_{pos} = 6$\,arcsec) in quadrature with an assumed positional uncertainty in the LABOCA sources of
$\sigma_{pos} = 4$\,arcsec (calculated assuming the 19\,arcsec LABOCA beam, and a typical deboosted S/N = 3), 
and then multiplying by 2.5 to ensure 95\% completeness. 

The LABOCA and BLAST source surface densities are both so much lower than the radio, mid-infrared or optical source surface densities,
that all of these associations are statistically compelling, as demonstrated by the derived values of $P_{LB}$ given in Table 4. The other 3 LABOCA 
sources in the field are not near to any BLAST source, and so the choice of search radius appears to have been sensible.

\begin{figure}
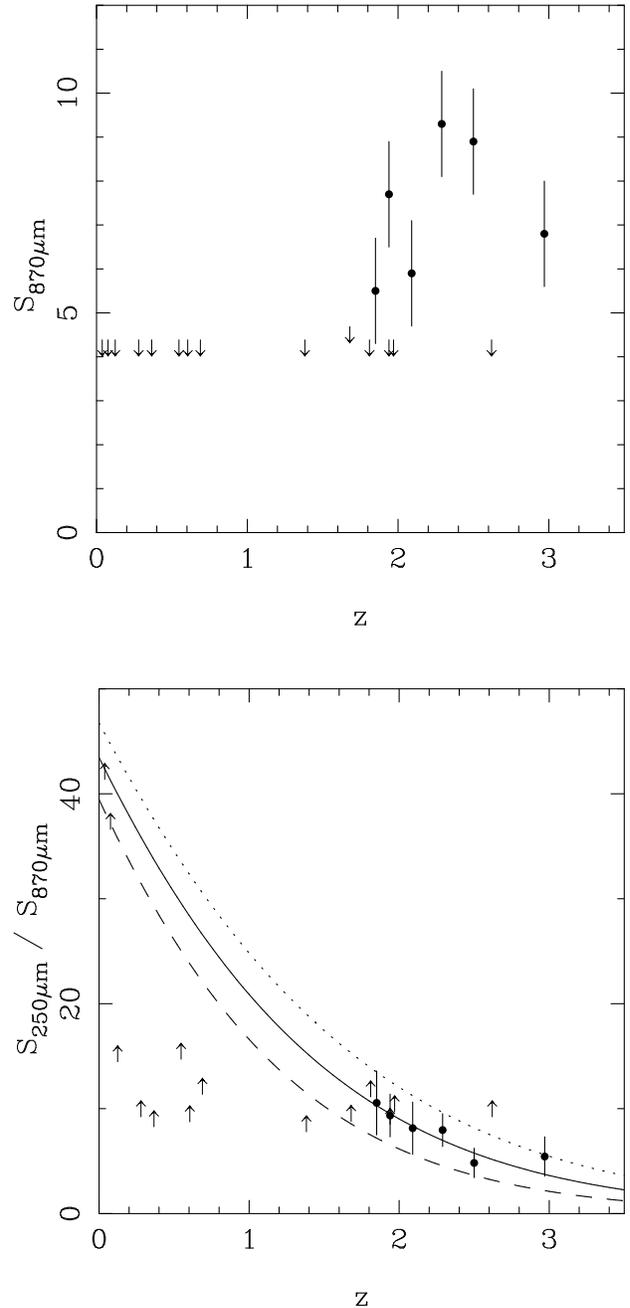

\begin{tabular}{c}
\includegraphics[width=0.46\textwidth]{870.ps}\\
   \\
   \\
\includegraphics[width=0.46\textwidth]{250_870.ps}\\
  \\
  \\
\end{tabular}
\caption{The LABOCA LESS 870\,${\rm \mu m}$ detections and non-detections of the BLAST 250\,${\rm \mu m}$ sources. The upper panel shows 870\,${\rm \mu m}$ 
flux density plotted against redshift (spectroscopic or photometric) as derived from the primary proposed galaxy identification for each BLAST source. All
the 870\,${\rm \mu m}$ detections are confined to the high-redshift BLAST subsample. 
The non-detections are indicated by arrows plotted at the adopted conservative $\simeq 3$-$\sigma$ flux-density limit
of $S_{870} < 4\, {\rm mJy}$.
The lower panel plots the derived 250/870\,${\rm \mu m}$ flux-density ratios (or limits), again versus redshift.
The anticipated trend 
of declining 250/870\,${\rm \mu m}$ flux-density ratio with redshift is clearly
evident. The solid line, derived by redshifting a simple modified black-body spectrum with $T=40$\,K, and $\beta = 1.5$, provides a very good description of the 
data. The dotted curve shows the effect of increasing temperature to $T = 45$\,K, and the dashed line the effect of reducing it to $T = 35$\,K. In calculating 
the observed flux ratios, neither the BLAST nor the LABOCA flux densities 
have been deboosted.}
\end{figure}

The associations listed in Table 4 provide an opportunity to check that our adopted positional uncertainties are indeed reasonable. 
First, the median value of $d_{L-Rad} = 5.0$ implies that (assuming zero uncertainty in the radio positions), the positional
uncertainty in the LABOCA positions is $\sigma_{pos} \simeq 4.5$ arcsec (allowing for the radial weighting in the observed distribution), showing 
that our adoption of 4\,arcsec was at least approximately correct. The median value of the BLAST-LABOCA positional offset is $d_{B-L} = 8.9$, implying $\sigma_{B-L} \simeq
7.5$\,arcsec. Subtracting 4.5 in quadrature from 7.5, leads to the conclusion that $\sigma_{pos} = 6$\,arcsec for the BLAST sources, 
as adopted in Section 3.1.1. Repeating this analysis with 
means instead of medians, yields $\sigma_{pos} = 6.5$\,arcsec.

Working backwards, this empirical check on the uncertainty in the positions of the BLAST sources considered in this paper re-affirms that 
they are indeed genuine $\simeq 4\sigma$ sources prior to flux deboosting (including the contribution of confusion to the noise).

The more scientifically interesting aspect of these LABOCA detections is that all of them are associated with {\it high-redshift} BLAST sources.
The 870\,${\rm \mu m}$ flux densities of the LABOCA detections are listed in Table 3. From examination of this Table it can be 
seen that the LABOCA detections are confined to proposed identifications with $z > 1.5$. This point is illustrated in
the upper panel of Fig. 9, where we have also adopted a conservative $\simeq 3\sigma$ flux-density limit
of $S_{870} < 4\, {\rm mJy}$ for the non-detections. Not surprisingly, it can be seen 
that the redshift distribution of the LABOCA-detected BLAST sources is consistent with 
that which is displayed by galaxy samples selected at 850$\mu m$ with SCUBA.

The 870$\mu m$ detections thus provide strong, independent support for most of our proposed high-redshift
BLAST identifications, especially since the identifications and redshifts were established without 
any reference to the LABOCA data. However, given the existence of the LABOCA data, one can then also ask whether the LABOCA {\it non}-detections
of 4 of our proposed $z > 1.5$ identifications cast doubt on their reality. To check this we have
therefore plotted, in the lower panel of Fig. 9, the 250/870\,${\rm \mu m}$ flux-ratios of the BLAST galaxies versus
redshift. Here, despite the fact that only lower limits are available at low redshift, the anticipated trend 
of declining 250/870\,${\rm \mu m}$ flux-density ratio with redshift is evident. It is also clear that, with one exception (BLAST 654), the non detection 
of the high-redshift candidates is not sufficiently deep to cast serious doubt on the proposed redshifts.

The solid curve shown in the lower panel of Fig. 9 is the predicted decline in colour with increasing redshift  
for a simple modified black-body spectrum with an assumed rest-frame temperature $T = 40$\,K and dust emissivity emissivity index $\beta = 1.5$.
This offers a remarkably good description of the high-redshift data and is at least consistent with the lower redshift limits.
In other words, in retrospect 
we should not be surprised that the 250\,${\rm \mu m}$-selected galaxies at $z < 1$ have not been detected in the LABOCA LESS map, nor that 
the deep 250\,${\rm \mu m}$ imaging analysed here has reached sufficient depth to allow us to detect the sub-mm galaxy population at $z \simeq 2$.
The dashed and dotted curves simply illustrate the effect of increasing or decreasing the assumed dust temperature by 5\,K, keeping 
$\beta = 1.5$. Given the mismatch between the BLAST and LABOCA beams, and the fact that neither set of flux densities 
has been deboosted in constructing Fig. 9, we do not want to over-emphasize the limited extent to which one can determine the temperature
of the dust emission from the 250/870\,${\rm \mu m}$ flux-ratios of the BLAST galaxies. However, at the very least, 
Fig. 9 provides a basic sanity check on the plausibility of our inferred galaxy identifications and redshifts.

\subsection{The redshift distribution for the 250\,${\rm \mu m}$ sample}

\begin{figure}
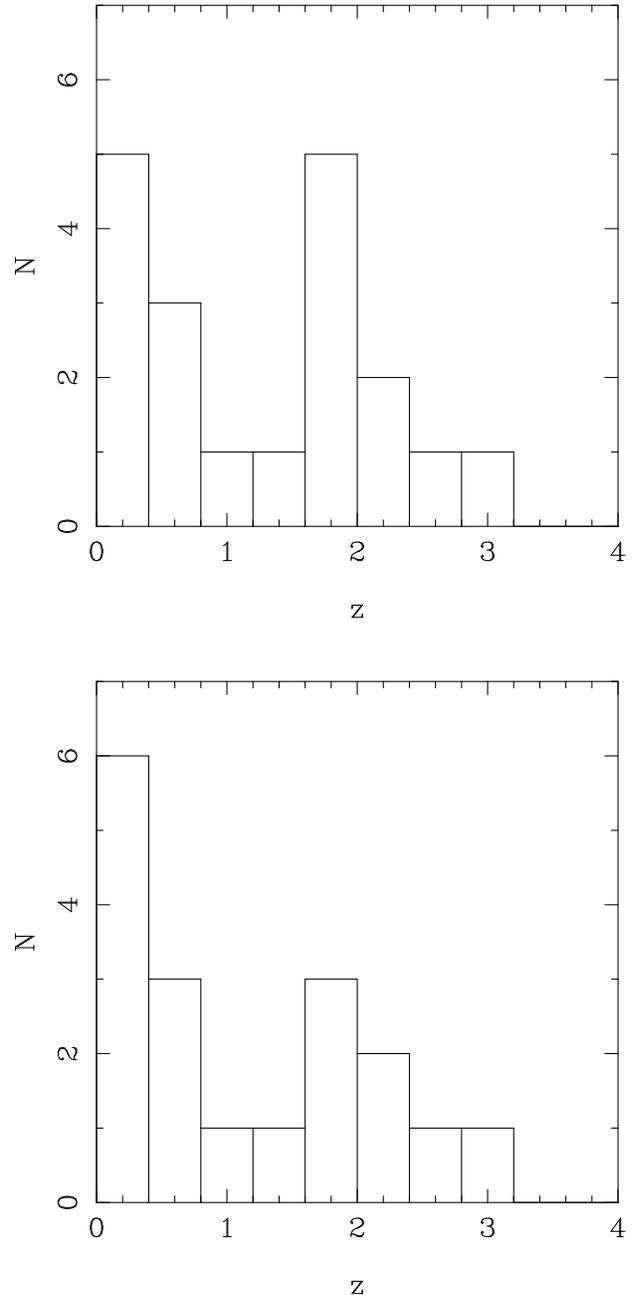

\begin{tabular}{c}
\includegraphics[width=0.46\textwidth]{redshift_hist.ps}\\
  \\
  \\
\includegraphics[width=0.46\textwidth]{redshift_hist2.ps}\\
  \\
  \\
\end{tabular}
\caption{An exploration of the robustness of the basic form of the redshift distribution for the 250\,${\rm \mu m}$ sources 
in GOODS-South. Sources 654 and 552 have been rejected from both distributions for the reasons detailed in the text. 
The upper panel shows the redshift distribution derived from the primary
identifications (plus 59-2) listed in Table 3. In the lower panel we have replaced the primary identifications by the secondary identifications 
in the 5 appropriate cases, and in addition have removed the other source which lacks either a radio or a LABOCA counterpart
(BLAST 193). Despite these 6 changes, the two distributions are statistically indistinguishable. In the upper panel
there are 10/19 galaxies at $z > 1$. In the lower plot there are 8/18 sources at $z > 1$. A median redshift $z \simeq 1$ 
is clearly a fair description of the data presented here, given the uncertainties in identifications and redshifts.}
\end{figure}

The inferred redshift distribution of our 20-source BLAST 250\,${\rm \mu m}$ sample is presented in Fig. 10. We show two alternative versions, 
in order to
illustrate the extent to which its basic form is robust against changes in ambiguous identifications. 
BLAST 654 and 552 are excluded from both plots, the former because its 250/870 \,${\rm \mu m}$ limit appears anomalously high for its 
proposed redshift, the latter because its non detection with LABOCA, combined with the low significance of its radio identification
and a non detection at 24\,${\rm \mu m}$ casts serious doubt on the robustness of this identification.
In the upper panel we plot the histogram derived from the primary galaxy 
identifications/redshifts listed in Table 3 (we include both 59-1 and 59-2). In this case 10/19 sources lie at $z > 1$, and the distribution appears 
almost bi-modal. In the lower panel we have replaced the primary identifications with the secondary/alternative galaxy ID 
for all potentially ambiguous cases (i.e. BLAST sources 6, 257, 503, 732, 830) and have also simply ejected BLAST 193 from the sample
(because it possesses neither a VLA radio identification, nor a LABOCA sub-millimetre detection). This more sceptical approach still leaves 
8/18 sources at $z > 1$, of which 6 have been detected by LABOCA.

\begin{figure*}
\includegraphics[width=1.0\textwidth]{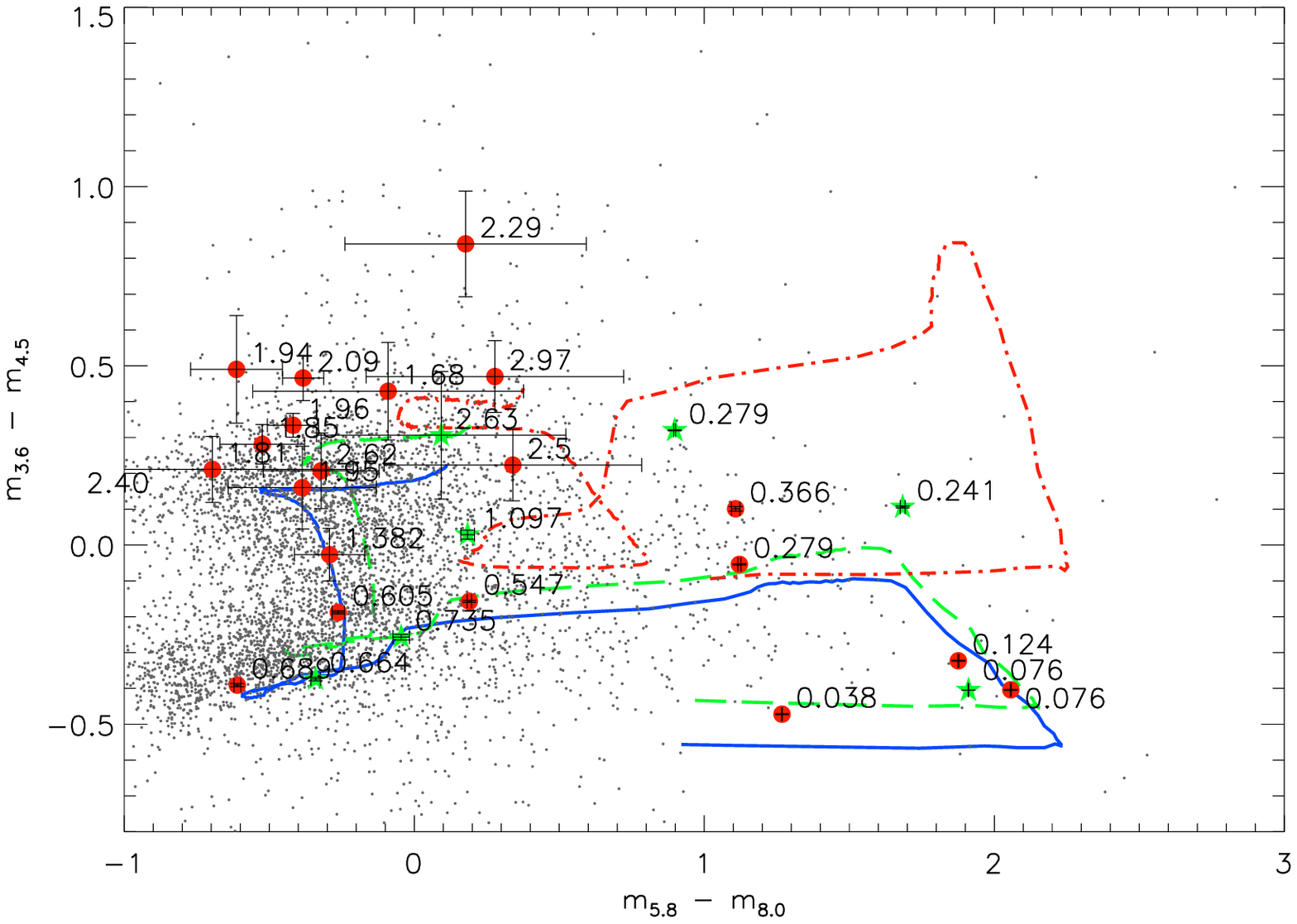}
\caption{The location of the adopted primary (red circles) and secondary (green stars) 
galaxy counterparts for the BLAST 250\,${\rm \mu m}$ sources on the 
{\it Spitzer} IRAC colour-colour plane. The small grey points indicate the positions
of field galaxies within GOODS-South, as derived from all galaxies in the GOODS-MUSIC
catalogue down to limiting IRAC AB magnitudes of $m_{3.6} = 24.0,\ 
m_{4.5} = 23.4,\ m_{5.8} = 22.0,\ m_{8.0} = 22.0$. 
Each BLAST identification is labelled 
with its redshift. The three tracks are derived by simulating the effect 
of redshifting, from  $z = 0$ to $z = 3.5$, the spectral energy distributions 
of two well-known local star-forming galaxies (the Virgo Sc galaxy VCC1972, and M82)
and the more highly obscured ULIRG Apr220. The blue solid track is for VCC1972, the green dashed track 
is for M82, and the red dot-dash track  is for Arp220, with the SEDs 
for all three galaxies taken from the multi-wavelength fits derived by 
Devriendt et al. (1999). The locations of the BLAST primary identifications
are in general very well described by the VCC1972 or M82 tracks, which intially 
move rapidly in observed $5.8-8.0\,{\rm \mu m}$ colour as the 8${\rm \mu m}$ PAH 
feature moves out of the longest IRAC filter, and later move rapidly
up in $3.6-4.5\,{\rm \mu m}$ colour as the 1.6${\rm \mu m}$ peak 
in the starlight moves into the $4.5\,{\rm \mu m}$ filter. Very few of the BLAST 
galaxies lie on the Arp220/ULIRG track, with the vast majority being far better 
described by the less reddened star-forming galaxies. At low redshift this 
tallies well with the appearance of the BLAST identifications shown in Fig. 7, 
which shows that they are generally isolated dusty star-forming discs. 
Moreover, certainly at $z < 1$, 250\,${\rm \mu m}$ selection, especially 
over the relatively small field of GOODS-South will be biased towards cooler SEDs 
which peak longwards of $\lambda = 100\, {\rm \mu m}$. This is true for star-forming 
disc galaxies, but not true for the hotter ULIRGs selected at 60\,${\rm \mu m}$ 
with {\it IRAS}. Perhaps more surprising (given their `ULIRG-like' star-formation rates) 
is the fact that the $z \simeq 2$ BLAST/LABOCA galaxies also appear relatively unobscured 
at IRAC wavelengths.}

\end{figure*}

For our sample, these two alternative redshift distributions are statistically indistinguishable, and we conclude that, at least in GOODS-South, 
$\simeq 50$\% of the 250\,${\rm \mu m}$ sources with $S_{250} > 35\,{\rm mJy}$ lie at $z > 1$. 
Any evidence for bimodality is statistically insignificant,
but the median redshift, and the fact that $\simeq 1/3$ of the sample lies at $z > 1.5$, appear to be robust results.
The precise form of the 
distribution is of course aflicted here both by small-number statistics, and by the potential effects of cosmic variance. It will therefore be interesting  to
see the extent to which the redshift distribution derived here applies to the much larger samples of 250\,${\rm \mu m}$ sources which should
be uncovered with {\it Herschel}.

It is interesting to assess how our results on this small, deep, near-complete BLAST subsample compare with those reported for the 
full 5-$\sigma$ BLAST sample of the wider 9 degree$^2$ of the Extended Chandra Deep Field South by Dye et al. (2009). This comparison must 
be done with care, and the limitations understood, because Dye et al. were studying a larger but brighter sample, with on average much poorer
supporting data. Dye et al. attempted both radio and 24\,${\rm \mu m}$ identifications for 350 BLAST sources, secured identifications for 198
of them, and in the end secured redshifts for 74 of these (i.e. for $\simeq $1/5th of the sample). It is not clear what 
conclusions can be drawn from the resulting redshift distribution, given the low-level of completeness and the mix of flux-density limits. 

More instructive for the present comparison is the redshift distribution derived by Dye et al. for the $>5\sigma$ BLAST sources in the 
central region of the ECDFS, with the best supporting data from the FIDEL survey. Here Dye et al. identified 50\% of the BLAST sources 
($\simeq 75$\% of the 250\,${\rm \mu m}$-selected
sources) at either radio or mid-infrared wavelengths. The median redshift reported in this sub-area by Dye et al. 
is $z_{med} \simeq 0.95$. Given that the redshift information includes some COMBO 17 redshifts which (as discussed here for, e.g. BLAST 66) are undoubtedly under-estimates,
given that the Dye et al. redshift distribution is more incomplete than that presented here, and given that the effective 
250\,${\rm \mu m}$ flux density limit of the Dye et al. deep sample is still $>50$\% higher than in the present study, we conclude 
there is no real evidence  
for any inconsistency between our redshift distribution and that derived by Dye et al. (2009). The only visible difference is that
the high-redshift tail in the present deep study extends to somewhat higher redshifts than that reported by Dye et al., but this is unsurprising 
given the above-mentioned selection effects.

\subsection{The mid-infrared colours of the 250\,${\rm \mu m}$ galaxies.}

Finally, in Fig. 11 we briefly investigate the location of the galaxy counterparts for the BLAST 250\,${\rm \mu m}$ sources on the 
{\it Spitzer} IRAC colour-colour plane. This is of interest for a number of reasons. First, it is worthwhile to check how the mid-infrared colours of
the BLAST-selected galaxies compare with those of the general galaxy population in GOODS-South. Second, we can compare 
their colours with expectations based on the redshifted SEDs of known local far-infrared bright galaxies. Third, it is of potential interest 
to check whether the counterparts of 250\,${\rm \mu m}$-selected galaxies can, in future, be reliably isolated on the basis of IRAC colour (a point of importance 
given the potentially extensive sky coverage which could be provided by the {\it Spitzer} warm mission).

We choose to plot $3.6\,{\rm \mu m}-4.5\,{\rm \mu m}$ colour versus $5.8\,{\rm \mu m}-8.0\,{\rm \mu m}$ colour, as advocated by Stern et al. (2005), rather 
than $3.6\,{\rm \mu m}-5.8\,{\rm \mu m}$ colour versus $4.5\,{\rm \mu m}-8.0\,{\rm \mu m}$ colour as advocated by some other authors (e.g. Lacy et al. 2004; 
Pope et al. 2006). We do this to avoid unhelpful correlations, and because only $3.6\,{\rm \mu m}-4.5\,{\rm \mu m}$ colours will be available from the 
{\it Spitzer} warm mission.

In Fig. 11 we plot the location of the adopted primary and secondary 
galaxy counterparts for the BLAST 250\,${\rm \mu m}$ sources, superimposed on the positions
of field galaxies within GOODS-South, as derived from all galaxies in the GOODS-MUSIC
catalogue down to limiting IRAC AB magnitudes of $m_{3.6} = 24.0,\ 
m_{4.5} = 23.4,\ m_{5.8} = 22.0,\ m_{8.0} = 22.0$. 
In addition, we show tracks in the colour-colour diagram produced by  
redshifting (from $z = 0$ to $z=3.5$) the spectral energy distributions of two well-known local 
star-forming galaxies (the 
Virgo Sc galaxy VCC1972, and the starburst galaxy M82), along with the analogous 
predicted track for the highly-obscured ULIRG Apr\,220.

The main points we can learn from this diagram can be summarised as follows. 
First, the primary 
BLAST galaxy identifications are very well described by the two star-forming galaxy SEDs, with the red data points 
essentially mapping out the redshift evolution of the tracks. The rapid horizontal movement at low redshift ($z = 0 - 0.5$)
is caused by the polycyclic aromatic hydrocarbon (PAH) features moving through the longer two IRAC bands. The subsequent rapid vertical 
movement from $3.6\,{\rm \mu m}-4.5\,{\rm \mu m} < 0$ to  $3.6\,{\rm \mu m}-4.5\,{\rm \mu m} > 0$ is the result of the stellar 
photospheric peak at $\lambda_{rest} \simeq 1.6 {\rm \mu m}$
moving through the shorter two IRAC bands at $z \simeq 1.5$.

Second, it can be seen that very few of our primary identifications appear to be analogues of Arp\,220, although at the very highest redshifts all three 
templates become indistinguishable. Our low-redshift BLAST galaxies stay relatively blue in $3.6\,{\rm \mu m}-4.5\,{\rm \mu m}$ colour, indicating their star-light is not 
swamped by the mid-infrared power-law produced by very hot dust (AGN or starburst heated), as is seen in Arp\,220 and other local ULIRGs. Even at $z \simeq 2$
most of the BLAST/LABOCA galaxies have colours expected from starlight (albeit in some cases reddened to $3.6\,{\rm \mu m}-4.5\,{\rm \mu m} > 0.25$).
Our highest-redshift candidates do lie in the region of the diagram expected for $z \simeq 3$, but at this point essentially all tracks converge on the same location
in the diagram. 

It is interesting to speculate why the Arp\,220 track is such a poor description of the locus occupied by the 
BLAST sources. At low redshift it is perhaps unsurprising, as starburst galaxies like M82 and Sc star-forming
discs both have (relatively) cool SEDs which peak longward of $\lambda_{rest} \simeq 100\,{\rm \mu m}$, whereas the ultra-luminous ULIRGS
first uncovered by {\it IRAS} at $60\,{\rm \mu m}$ peak at shorter wavelengths. A $250\,{\rm \mu m}$ survey is thus biased towards the 
selection of these cooler more `normal' starbursts at low redshift, especially since highly-obscured ULIRGs are  
rather rare objects. The appearance of the low-redshift BLAST galaxies in Fig. 7 (i.e. star-forming discs) is thus consistent with their IRAC colours.
 
Perhaps more surprising (given their `ULIRG-like'
star-formation rates) is the fact that
the $z \simeq 2$ BLAST/LABOCA galaxies also appear relatively unobscured 
at IRAC wavelengths. However, this is arguably consistent with their 250/870\,${\rm \mu m}$ flux-ratios as discussed in Section 6.1, 
which appear to be consistent with a relatively cool dust temperature of 40\,K (see Fig. 9).

Finally we revisit the claim made by Yun et al. (2008) that a simple and robust sub-mm galaxy candidate selection criterion is 
provided by $3.6\,{\rm \mu m} - 4.5\,{\rm \mu m} > -0.2$. It is obvious from Fig. 11 that this criterion is in fact 
satisfied by essentially any galaxy at $z > 1.5$, and thus cannot be used to isolate any special sub-class of source at these redshifts.

These issues, along with the other physical characteristics of the BLAST galaxies (size, morphology, stellar mass, star-formation 
history) will be explored in more detail by Targett et al. (in preparation).

\section{SUMMARY}
We have identified and investigated the nature of the 20 brightest 250\,${\rm \mu m}$ sources detected by
BLAST within the central 150 arcmin$^2$ of the GOODS-South field. Aided by the available deep VLA imaging, 
reaching $S_{\rm 1.4} \simeq 40 {\rm \mu Jy}$ (4-$\sigma$), we have identified 
secure radio counterparts for 17/20 of the 250\,${\rm \mu m}$ sources. The resulting enhanced 
positional accuracy of $\simeq 1$ arcsec has then allowed us to exploit the deep optical ({\it HST}), 
near-infrared (VLT) and mid-infrared ({\it Spitzer}) imaging of GOODS-South to establish secure galaxy 
counterparts for the 17 radio-identified sources, and plausible galaxy candidates for the 
3 radio-unidentified sources. 

Confusion is a serious issue for surveys with such large beams, and in many cases there may be no such thing as a unique, single counterpart.
We can nevertheless expect our chosen counterparts to be significant, and often dominant contributors to the BLAST flux densities, and we give several arguments
to support this claim. It is therefore still reasonable to pursue the properties of this `brightest counterpart' sample.
For all of our 20 chosen identifications we 
have been able to determine spectroscopic (8) or robust photometric (12) redshifts. The result 
is the first near-complete redshift distribution for a deep 250\,${\rm \mu m}$-selected 
galaxy sample. This reveals that 250\,${\rm \mu m}$ surveys reaching detection limits 
$S_{\rm 250 \mu m} \simeq 40$\,mJy contain not only low-redshift spirals/LIRGs, but also the extreme $z \simeq 2$ dust-enshrouded starburst galaxies 
previously discovered at longer sub-millimetre wavelengths. Inspection of the LABOCA 870\,${\rm \mu m}$ imaging 
of the GOODS-South field yields detections of 7 of the proposed $z > 1$ BLAST sources, and reveals 250/870\,${\it \mu m}$ flux ratios consistent  
with a standard 40K modified black-body fit with a dust emissivity index $\beta = 1.5$.
Thus, at least in GOODS-South, we infer that 
$\simeq 50$\% of the 250\,${\rm \mu m}$ sources with $S_{250} > 40\,{\rm mJy}$ lie at $z > 1$.
 
Based on their IRAC colours, we find that virtually all of the BLAST 
galaxy identifications appear better described as analogues of the M82 starburst galaxy, or Sc star-forming discs rather than 
highly obscured ULIRGs. This is perhaps as expected at low-redshift, where the 250\,${\rm \mu m}$ BLAST selection function is biased towards SEDs which peak longward of 
$\lambda_{rest} = 100\,{\it \mu m}$. However, it also appears largely true at $z \simeq 2$.

\section*{ACKNOWLEDGEMENTS}

JSD acknowledges the support of the Royal Society through a Wolfson Research Merit Award, and the support of the 
European Research Concil through the award of an Advanced Grant. 
We acknowledge the support of NASA through grant numbers NAG5-12785, NAG5-13301, and NNGO-6GI11G, the NSF Office of Polar Programs, the 
Canadian Space Agency, the Natural
Sciences and Engineering Research Council (NSERC) of Canada, and the UK Science and Technology Facilities Council (STFC). 
This work is based in part on observations made with the {\it Spitzer} Space Telescope, which is operated by the Jet Propulsion Laboratory, 
California Institute of Technology under a contract with NASA. APEX is operated by the Max-Planck-Institut fur Radioastronomie, the European Southern Observatory, 
and the Onsala Space Observatory. This work is based in part on observations made with the NASA/ESA Hubble Space Telescope, obtained from the Data Archive at the Space Telescope 
Science Institutewhich is operated by the Association of Universities for Research in Astronomy, Inc., under NASA contract NAS 5-26555.
MC acknowledges the award of a STFC Advanced Fellowship. IRS acknowledges support from STFC. 
JLW acknowledges the support of an STFC Studentship.

{}

\appendix

\section{Optical-infrared photometry}

Table A1 provides the optical--infrared photometry which was used to derive the BLAST galaxy photometric redshifts.

\begin{landscape}

\begin{table}
\begin{center}
\scriptsize
\renewcommand{\arraystretch}{0.9}
\begin{tabular}{lcccccccccccccc}
\hline
BLAST & RA & Dec & $B_{435}$ & $V_{606}$ & $i_{775}$ & $z_{850}$ & $J$ & $H$ & $K$ & $S_{3.6}$ & $S_{4.5}$ & $S_{5.6}$ & $S_{8.0}$ & $S_{24}$\\
ID    & (J2000) & (J2000) &       &          &           &          &             &            &             &                      &           &          &\\
\hline
4     & 53.146168 & -27.925831 & 17.59$\pm$0.01 & 16.57$\pm$0.01 & 15.97$\pm$0.01 & 15.62$\pm$0.01 & 15.11$\pm$0.01 &                & 14.83$\pm$0.01 & 15.41$\pm$0.01 & 15.88$\pm$0.01 & 15.38$\pm$0.01 & 14.11$\pm$0.01 & 7090$\pm$709 \\  			
6-1   & 53.124493 & -27.740086 & 17.57$\pm$0.01 & 16.86$\pm$0.01 & 16.46$\pm$0.01 & 16.30$\pm$0.01 & 16.01$\pm$0.01 & 15.87$\pm$0.01 & 15.92$\pm$0.01 & 16.25$\pm$0.01 & 16.65$\pm$0.01 & 16.22$\pm$0.01 & 14.16$\pm$0.01 & 7590$\pm$759 \\  				
109   & 53.074467 & -27.849800 & 20.36$\pm$0.01 & 19.10$\pm$0.01 & 18.41$\pm$0.01 & 18.07$\pm$0.01 & 17.48$\pm$0.01 & 17.11$\pm$0.01 & 17.00$\pm$0.01 & 17.43$\pm$0.01 & 17.75$\pm$0.01 & 17.83$\pm$0.01 & 15.95$\pm$0.01 & 1270$\pm$127 \\				
637-1 & 53.183559 & -27.862030 & 21.11$\pm$0.01 & 19.61$\pm$0.01 & 18.82$\pm$0.01 & 18.46$\pm$0.01 & 17.59$\pm$0.01 &                & 16.78$\pm$0.01 & 17.65$\pm$0.01 & 17.70$\pm$0.01 & 18.21$\pm$0.01 & 17.09$\pm$0.01 & \phantom{0}594$\pm$\phantom{0}36 \\				
983   & 53.136730 & -27.768833 & 22.49$\pm$0.02 & 21.10$\pm$0.01 & 20.39$\pm$0.01 & 20.03$\pm$0.01 & 19.41$\pm$0.01 & 18.86$\pm$0.01 & 18.30$\pm$0.01 & 18.87$\pm$0.01 & 18.77$\pm$0.01 & 19.13$\pm$0.01 & 18.02$\pm$0.01 & \phantom{0}611$\pm$\phantom{0}10 \\				
104   & 53.150726 & -27.825510 & 22.77$\pm$0.01 & 21.49$\pm$0.01 & 20.62$\pm$0.01 & 20.24$\pm$0.01 & 19.67$\pm$0.01 & 19.06$\pm$0.01 & 18.92$\pm$0.01 & 19.09$\pm$0.01 & 19.25$\pm$0.01 & 19.29$\pm$0.01 & 19.10$\pm$0.01 & \phantom{0}744$\pm$\phantom{00}7 \\				
830-1 & 53.055141 & -27.711374 & 21.83$\pm$0.01 & 20.92$\pm$0.01 & 20.06$\pm$0.01 & 19.84$\pm$0.01 & 19.25$\pm$0.01 &                & 18.63$\pm$0.01 & 18.80$\pm$0.01 & 18.99$\pm$0.01 & 18.87$\pm$0.01 & 19.13$\pm$0.02 & \phantom{0}774$\pm$\phantom{0}10 \\				
257   & 53.075142 & -27.831411 & 23.23$\pm$0.02 & 21.68$\pm$0.01 & 20.49$\pm$0.01 & 20.06$\pm$0.01 & 19.43$\pm$0.01 & 18.98$\pm$0.01 & 18.53$\pm$0.01 & 18.71$\pm$0.01 & 19.10$\pm$0.01 & 19.23$\pm$0.01 & 19.84$\pm$0.01 & \phantom{0}394$\pm$\phantom{00}5 \\				
1293  & 53.193047 & -27.890892 & 26.44$\pm$0.20 & 26.11$\pm$0.11 & 25.03$\pm$0.08 & 24.27$\pm$0.05 & 22.98$\pm$0.09 &                & 21.45$\pm$0.06 & 20.07$\pm$0.05 & 20.10$\pm$0.05 & 20.00$\pm$0.07 & 20.29$\pm$0.10 & \phantom{0}187$\pm$\phantom{00}4 \\				
552   & 53.052277 & -27.718325 & 28.34$\pm$0.54 & 27.39$\pm$0.21 & 26.23$\pm$0.14 & 25.74$\pm$0.11 & 24.15$\pm$0.09 &                & 22.84$\pm$0.06 & 22.09$\pm$0.08 & 21.66$\pm$0.11 & 22.07$\pm$0.92 & 22.16$\pm$1.09 & \phantom{0}290$\pm$\phantom{00}4 \\				
193   & 53.053574 & -27.778013 & 28.13$\pm$0.42 & 28.70$\pm$0.65 & 27.18$\pm$0.31 & 26.27$\pm$0.17 & 23.98$\pm$0.13 & 23.29$\pm$0.07 & 22.58$\pm$0.06 & 21.53$\pm$0.04 & 21.32$\pm$0.08 & 21.25$\pm$0.42 & 21.95$\pm$0.98 & 	         \\        		
158   & 53.089821 & -27.940001 & 25.73$\pm$0.09 & 24.85$\pm$0.05 & 24.05$\pm$0.04 & 23.59$\pm$0.04 &                &                &                & 19.83$\pm$0.04 & 19.55$\pm$0.04 & 19.64$\pm$0.07 & 20.17$\pm$0.13 & \phantom{0}586$\pm$\phantom{00}6 \\
66    & 53.020180 & -27.779888 & 27.73$\pm$0.18 & 27.13$\pm$0.11 & 26.56$\pm$0.15 & 25.43$\pm$0.11 &                &                &                & 20.84$\pm$0.11 & 20.35$\pm$0.11 & 20.21$\pm$0.11 & 20.83$\pm$0.12 & \phantom{0}620$\pm$\phantom{00}7 \\
861   & 53.198280 & -27.747866 & 27.17$\pm$0.15 & 26.22$\pm$0.11 & 25.52$\pm$0.12 & 24.92$\pm$0.07 & 23.44$\pm$0.09 & 22.72$\pm$0.11 & 22.30$\pm$0.08 & 20.99$\pm$0.08 & 20.83$\pm$0.08 & 20.55$\pm$0.15 & 20.93$\pm$0.21 & \phantom{0}299$\pm$\phantom{00}4 \\
503-1 & 53.157238 & -27.833452 & 25.90$\pm$0.05 & 25.26$\pm$0.02 & 24.55$\pm$0.02 & 24.18$\pm$0.02 & 23.15$\pm$0.04 & 22.46$\pm$0.03 & 22.13$\pm$0.03 & 21.07$\pm$0.02 & 20.73$\pm$0.02 & 20.80$\pm$0.05 & 21.21$\pm$0.08 & \phantom{0}251$\pm$\phantom{00}5 \\
318   & 53.179874 & -27.920731 & 25.06$\pm$0.06 & 24.54$\pm$0.04 & 24.05$\pm$0.05 & 23.80$\pm$0.04 & 22.90$\pm$0.08 &                & 21.63$\pm$0.05 & 20.56$\pm$0.04 & 20.09$\pm$0.05 & 19.90$\pm$0.05 & 20.28$\pm$0.05 & \phantom{0}554$\pm$\phantom{00}5 \\
59-1  & 53.079300 & -27.870552 & 25.90$\pm$0.08 & 25.55$\pm$0.06 & 24.94$\pm$0.06 & 24.75$\pm$0.07 & 24.48$\pm$0.15 &                & 23.78$\pm$0.14 & 23.29$\pm$0.11 & 22.45$\pm$0.10 & 22.48$\pm$0.30 & 22.30$\pm$0.29 & \phantom{0}148$\pm$\phantom{00}5 \\
654   & 53.136547 & -27.885691 & 25.41$\pm$0.07 & 24.51$\pm$0.03 & 24.29$\pm$0.05 & 24.21$\pm$0.05 & 23.81$\pm$0.08 &                & 23.10$\pm$0.09 & 22.73$\pm$0.06 & 22.53$\pm$0.08 & 22.50$\pm$0.23 & 22.82$\pm$0.32 & 	         \\
732-1 & 53.181458 & -27.777472 & 29.94$\pm$0.50 & 28.23$\pm$0.12 & 27.98$\pm$0.10 & 27.60$\pm$0.15 & 26.54$\pm$0.07 &  25.36$\pm$0.06& 24.04$\pm$0.18 & 22.25$\pm$0.12 & 21.78$\pm$0.14 & 21.44$\pm$0.21 & 21.16$\pm$0.20 & \phantom{00}85$\pm$\phantom{00}4 \\
593   & 53.199965 & -27.904560 &   $>27.3$              &  $>27.3$               &   $>26.5$              &  $>26.2$              &     $>25.5$           &    $>24.7$            & 24.10$\pm$0.20 & 21.93$\pm$0.06 & 21.71$\pm$0.13 & 21.53$\pm$0.22 & 21.19$\pm$0.22 & \phantom{00}49$\pm$\phantom{00}3 \\
6-2   & 53.124950 & -27.734684 & 18.16$\pm$0.01 & 17.39$\pm$0.01 & 16.95$\pm$0.01 & 16.76$\pm$0.01 & 16.45$\pm$0.01 & 16.28$\pm$0.01 & 16.32$\pm$0.01 & 16.73$\pm$0.01 & 17.13$\pm$0.01 & 16.86$\pm$0.01 & 14.95$\pm$0.01 & 4660$\pm$466 \\
503-2 & 53.161751 & -27.832291 & 21.99$\pm$0.01 & 20.69$\pm$0.01 & 19.96$\pm$0.01 & 19.64$\pm$0.01 & 18.90$\pm$0.01 & 18.35$\pm$0.01 & 17.93$\pm$0.01 & 18.87$\pm$0.01 & 18.77$\pm$0.01 & 19.23$\pm$0.01 & 17.55$\pm$0.01 & \phantom{0}543$\pm$\phantom{00}5 \\
637-2 & 53.184448 & -27.861420 & 21.64$\pm$0.01 & 20.19$\pm$0.01 & 19.47$\pm$0.01 & 19.15$\pm$0.01 & 18.48$\pm$0.01 &                & 17.92$\pm$0.01 & 18.44$\pm$0.01 & 18.17$\pm$0.01 & 17.96$\pm$0.01 & 17.06$\pm$0.01 &              \\
830-2 & 53.057747 & -27.713591 & 23.41$\pm$0.02 & 22.36$\pm$0.01 & 21.27$\pm$0.01 & 20.83$\pm$0.01 & 20.04$\pm$0.01 &                & 19.22$\pm$0.01 & 19.56$\pm$0.01 & 19.82$\pm$0.01 & 19.88$\pm$0.02 & 19.92$\pm$0.02 & \phantom{0}333$\pm$\phantom{00}7 \\
59-2  & 53.071545 & -27.872448 & 24.97$\pm$0.06 & 23.87$\pm$0.02 & 22.65$\pm$0.01 & 21.80$\pm$0.01 & 21.03$\pm$0.01 &                & 19.93$\pm$0.01 & 19.68$\pm$0.01 & 19.66$\pm$0.01 & 19.97$\pm$0.02 & 19.79$\pm$0.02 & \phantom{0}288$\pm$\phantom{00}6 \\
732-2 & 53.180542 & -27.779686 & 28.22$\pm$1.09 & 27.07$\pm$0.17 & 27.81$\pm$0.72 & 27.32$\pm$0.52 & 26.08$\pm$0.54 & 24.88$\pm$0.30 & 23.69$\pm$0.09 & 22.70$\pm$0.08 & 22.39$\pm$0.16 & 21.90$\pm$0.51 & 21.80$\pm$0.65 &              \\
732-3 & 53.182018 & -27.779537 & 25.34$\pm$0.07 & 24.72$\pm$0.03 & 24.51$\pm$0.06 & 24.39$\pm$0.05 & 23.95$\pm$0.08 & 23.36$\pm$0.08 & 23.30$\pm$0.07 & 23.29$\pm$0.07 & 23.16$\pm$0.09 & 23.01$\pm$0.26 & 24.19$\pm$0.73 &              \\
\hline
\end{tabular}
\caption{Optical ({\it HST} ACS, near-infrared (VLT ISAAC) and mid-infrared ({\it Spitzer} IRAC + MIPS) photometry for the BLAST galaxy counterparts. The table lists 
the primary identifications, ranked by redshift (as shown in Figure 7), and then the secondary identifications (as shown in Figure 8). All measurements 
are given in AB magnitudes, except for the 24${\rm \mu m}$ photometry, which is given in micro-Jy (the 24${\rm \mu m}$ information was not used in the derivation 
of photometric redshifts, and is simply given here for completeness). Non-detections are given as 2$\sigma$ limiting magnitudes in 2-arcsec diameter apertures. 
Where no entry is given, this means that the available deep imaging coverage at that particular wavelength did not extend to include the object in question. Note that the 
optical and $J$ and $H$ photometry for 732-1 comes from the ultra-deep imaging of the HUDF provided by {\it HST} ACS and WFC3 imaging (McLure et al. 2010).}

\end{center}
\end{table}

\end{landscape}

\end{document}